\documentclass[a4paper,10pt]{article}
\usepackage{jheppub} % for details on the use of the package, please
\usepackage{graphicx,floatflt,amssymb,epsfig}
\usepackage[utf8]{inputenc}
\usepackage[inline]{enumitem}
\usepackage{mathrsfs}
\usepackage{amssymb}
\usepackage{amsmath}
\usepackage{color}
\usepackage{graphicx}
\usepackage{subfig}
\usepackage{multirow}
\usepackage{lipsum}
\usepackage{mathtools}
\usepackage{lscape}
\usepackage{xcolor}
\usepackage{wasysym}
\usepackage{float}
\usepackage{pstricks}
\usepackage[toc,page]{appendix}
\usepackage{pifont}
\usepackage{braket}
\usepackage{ragged2e}
\usepackage{slashed}
\usepackage{rotating}%To rotate tables and figures
\usepackage{caption} 
\renewcommand{\arraystretch}{1.5}

\allowdisplaybreaks[2]
\def\Lb{{\Lambda_b}}

\newcommand{\nn}{\nonumber}

\newcommand\doublecheck{\textbf{\textcolor{blue}{\checked\kern-0.9 em\checked}}}
\definecolor{schrift}{RGB}{120,0,0}

\newcommand{\para}{\parallel}

\newcommand{\gfermi}{G_\text{F}}

\newcommand{\bea}{\begin{eqnarray}}
\newcommand{\eea}{\end{eqnarray}}

\usepackage[normalem]{ulem}

\definecolor{darkgreen}{cmyk}{1,0,1,0.4}
\definecolor{pink}{cmyk}{0.4,1,0.3,0}

\def \lb{\Lambda_b}
\def \lc{\Lambda_c}

\allowdisplaybreaks

\title{ \boldmath An Imperative study of the angular observables in $\Lambda_b \to \Lambda_c^{+}(\to \Lambda \pi^{+})\ell^-\bar{\nu}_{\ell}$ decay and probing the footprint of new physics}

\author[a]{Soumitra Nandi}
\emailAdd{soumitra.nandi@iitg.ac.in}

\author[a]{Shantanu Sahoo}
\emailAdd{shantanu\_sahoo@iitg.ac.in}

\author[a]{Ria Sain}
\emailAdd{riasain@rnd.iitg.ac.in}

\affiliation[a]{Indian Institute of Technology Guwahati, North Guwahati, Guwahati 781039, Assam, India }		

\abstract{
	We study the four-fold angular distribution of the semileptonic $\Lambda_b \to \Lambda_c^{+}(\to \Lambda \pi^{+})\ell^- \bar{\nu}_{\ell}$ decay and find out analytical expressions for various asymmetric and angular observables in the standard model (SM) and the relevant new physics (NP) scenarios. Using the available inputs on the form factors from the lattice, we predict the values with uncertainties of all these observables in the SM. We have considered NP effects only in $b\to c\tau^-\bar{\nu}_{\tau}$ transitions, and constrain the Wilson coefficients of the model-independent beyond the SM operators from the available data on $B\to D^{(*)}\ell\nu_{\ell}$ and $\Lambda_b \to \Lambda_c^{+}\ell^- \bar{\nu}_{\ell}$ decays. In this study, we focus on analysing the contributions to these decays in one- and two-operator scenarios. Furthermore, we test the new physics sensitivities (one or two-operator scenarios) of the different angular and asymmetric observables in $\Lambda_b \to \Lambda_c^+ (\to \Lambda \pi^{+}) \tau^- \bar{\nu}_{\tau}$ decays and discuss the phenomenology. We observe that it will be possible to distinguish the different NP effects from one another by measuring these observables.}

\arxivnumber{}

\begin{document}
	\maketitle

	\section{Introduction}
	Over the past few years, the flavour changing charged current (FCCC) decay $b\to c\ell^-\bar{\nu}$ ($\ell =\tau, \mu, e$) has been extensively studied in both experimental and theory fronts and particular attention has been given on the exclusive semileptonic decays of B-meson \cite{HeavyFlavorAveragingGroup:2022wzx}. The semileptonic $B\to D^{(*)}\mu^-\bar{\nu}$ or $B\to D^{(*)} e^-\bar{\nu}$ decays are useful for the extraction of Cabibbo-Kobayashi-Maskawa (CKM) matrix element $|V_{cb}|$ ~\cite{Bigi:2017njr,Jaiswal:2017rve,Gambino:2019sif,Jaiswal:2020wer,Iguro:2020cpg,Biswas:2022yvh,Martinelli:2021onb,Martinelli:2021myh,Ray:2023xjn}. In addition, we define the ratios of the decay rates $R(D^{(*)}) = \Gamma(B\to D^{(*)}\tau^-\bar{\nu})/\Gamma(B\to D^{(*)}\mu^- (e^-) \bar{\nu})$ which are potentially sensitive to NP interactions \cite{Fajfer:2012jt,Datta:2012qk,Bhattacharya:2016zcw,Bardhan:2016uhr,Celis:2016azn,Jaiswal:2017rve,Huang:2018nnq,Azatov:2018knx,Bhattacharya:2018kig,PhysRevD.98.095018,Angelescu:2018tyl,Iguro:2018fni,Murgui:2019czp,Shi:2019gxi,Becirevic:2019tpx,Biswas:2021pic,Iguro:2022yzr,Ray:2023xjn,Bhatta:2023odk}. In the SM, these ratios conserve the lepton flavour universality (LFU). Therefore, the ratios $R(D)$ and $R(D^*)$ could be useful for testing LFU violating NP scenarios. The measurements of these ratios show discrepancies with the respective SM estimates~\cite{BaBar:2012obs,BaBar:2013mob,Belle:2015qfa,LHCb:2015gmp, Belle:2016ure,Belle:2016dyj,Belle:2017ilt,LHCb:2017rln,LHCb:2017smo,Belle:2019rba,LHCb:2023zxo,HeavyFlavorAveragingGroup:2022wzx}, for a recent study see the refs.\cite{Fedele:2022iib,Ray:2023xjn}. Apart from $R(D^{(*)})$, measurements are available on a few angular observables, like the $D^*$- and $\tau$-polarisation asymmetries $F_L^{D^*}$ \cite{LHCb:2023ssl} and $P^{\tau}(D^*)$ \cite{Belle:2017ilt}, respectively.
 
The baryonic decay $\Lambda_b \to \Lambda_c \ell^- \bar{\nu}$ will provide complementary information. We can extract $|V_{cb}|$ from the measurements of the rates in $\Lambda_b \to \Lambda_c \mu^-(e^-) \bar{\nu}$ decays. At the same time, we could define the ratio $R(\Lambda_c)$, similar to those defined in $B\to D^{(*)}\ell^-\bar{\nu}$ decays, which are sensitive to the interactions beyond the SM \cite{Detmold:2015aaa,Shivashankara:2015cta,Li:2016pdv,Datta:2017aue,DiSalvo:2018ngq,Ray:2018hrx,Penalva:2019rgt,Ferrillo:2019owd,Mu:2019bin,Boer:2019zmp,Hu:2020axt,Becirevic:2022bev,Fedele:2022iib,Karmakar:2023rdt}. The Large Hadron Collider beauty (LHCb) collaboration has produced copious amounts of $\Lambda_b$, which provide information on its semileptonic decay. For instance, LHCb has already measured the semitauonic branching fraction and $R(\Lambda_c)$ which are given in ref. \cite{LHCb:2022piu}.
%\begin{equation}
%\mathcal{B}( \Lambda_b \to \Lambda_c \tau^- \nu_\tau) = (1.50 \pm 0.16 \pm 0.25 \pm 0.23)\%.
%\end{equation}
We can obtain more information as more data will come in the subsequent runs. Hence, it is an appropriate time to study a full 4-body angular distribution for $\Lambda_b \to \Lambda_c \ell^- \bar{\nu}_{\ell}$ followed by $\Lambda_c \to \Lambda \pi$, in both standard model (SM) as well as in the presence of new physics (NP). From a complete angular analysis, we can define many observables other than $R(\Lambda_c)$, which could be potentially sensitive to new interactions beyond the SM. Hence, the precise predictions of those observables in the SM are very relevant. In addition, we need to test their NP sensitivities. Furthermore, the correlations among those observables in different NP scenarios could have distinct features, a comparative study of which might help distinguish the effect of one NP operator from the others. More broadly, such a study could help distinguish the effects of different NP models. 
  
In this article, we have done an angular analysis for $\Lambda_{b}\to\Lambda_{c}^+ (\to \Lambda \pi^+)\ell^- \bar{\nu}_{\ell}$ decay and obtained different angular and asymmetric observables. Using the lattice inputs on the form factors \cite{Detmold:2015aaa}, we have predicted all these observables in the SM integrated over the full-$q^2$ region or integrated a few small $q^2$-bins. Furthermore, we have tested the sensitivities of these observables to all the beyond the SM (BSM) independent set of effective operators relevant for $b\to c\tau^-\bar{\nu}$ transitions. For this test, we need benchmark values of the Wilson coefficient (WC) for each BSM operator. We have obtained the values of these WCs from a simultaneous analysis of the available data on $R(D^{(*)})$, $R(\Lambda_c)$, $F_L^{D^*}$ and the branching fraction $\mathcal{B}( \Lambda_b \to \Lambda_c \tau^- \nu_\tau) $. We have done a frequentist analysis considering one or two operator contributions at a time. The fit results give us an estimate of the order of magnitudes of the allowed values of the WCs. The two-operator scenarios could explain all the data quite comfortably as compared to the one-operator scenarios. In addition, we have studied the sensitivities of the observables to all these one- or two-operator BSM scenarios, and we have found some exciting outcomes. Finally, we provide the predictions of all the observables in all the scenarios integrated in a few small $ q^2$ bins and have predicted the full $q^2$ integrated observables.   
	
The paper is structured as follows. In section~\ref{sec:frmlsm}, we have discussed the formalism used to calculate the decay amplitudes. Furthermore, in this section, we have introduced the observables that we will obtain from the four-fold angular distribution. Section~\ref{sec:BGLparamfitres} has outlined the analysis for obtaining the $q^2$-shapes of the form factors. In section \ref{sec:SMprd}, we have predicted all the angular observables alongside $R(\Lambda_c)$ in the SM. In section~\ref{sec:NPtst}, we have tested the NP sensitivities of all the relevant observables. Finally, We have concluded in section~\ref{sec:summary}.

	%%%%%%%%%%%%%%%%%%%%%%%%%%%%%%%%%%%%%%%%%%%%
\section{Formalism}\label{sec:frmlsm}
In this section, we will derive the expression for the full angular distribution of the differential decay rate of $\Lambda_{b} \to \Lambda_c^+(\to \Lambda \pi^+) \ell^- \bar{\nu}_{\ell}$. The angular coefficients in this rate will be written in terms of helicity amplitudes, which are functions of the kinematical variables and hadronic form factors. With explicit expressions for the angular coefficients in hands, we will define the relevant observables for further discussion.  

\begin{figure}[ht!]
	\centering
	\includegraphics[scale=.55]{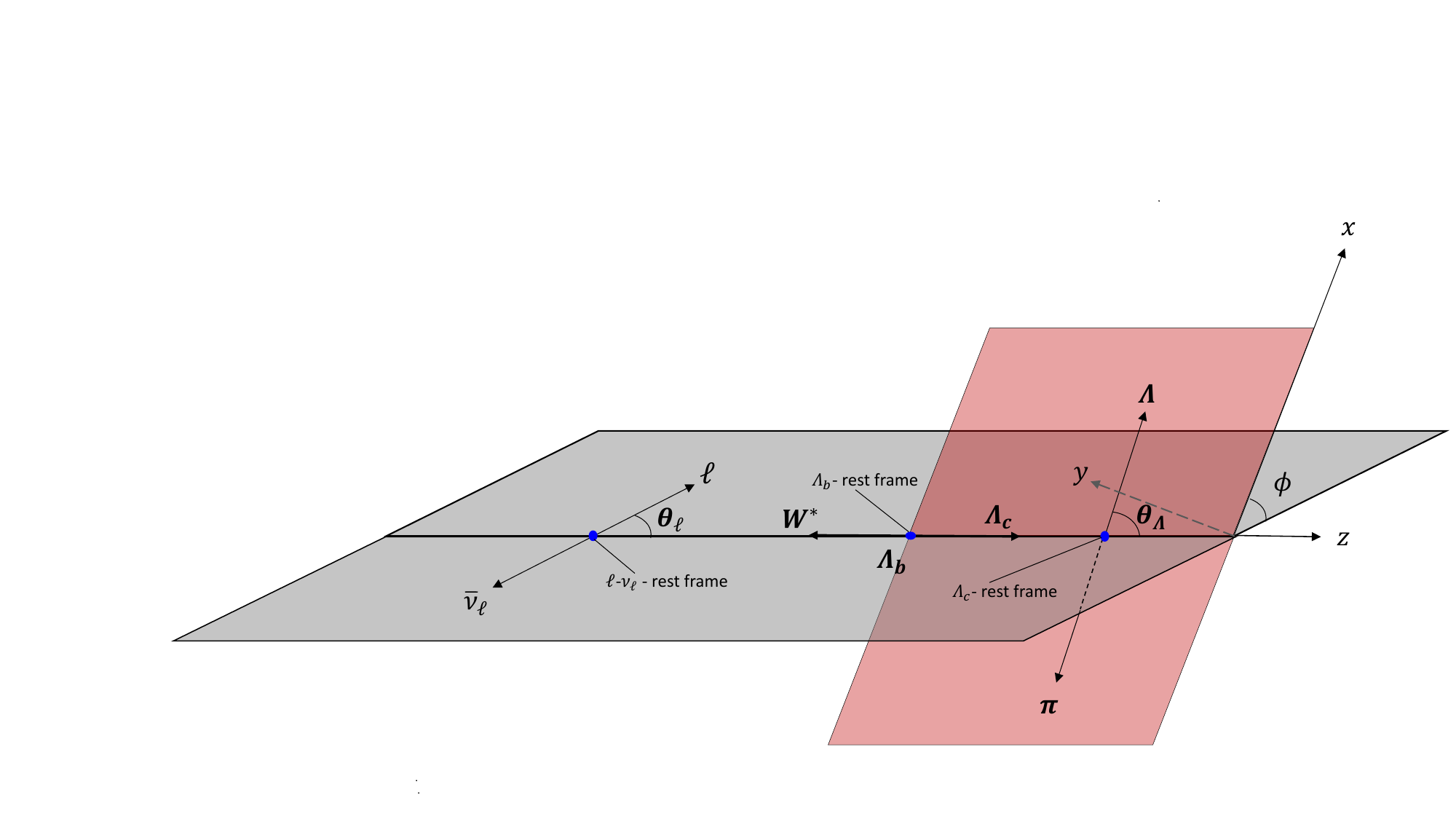}
	\caption{ The schematic diagram of the 4-body $\Lambda_{b} \to \Lambda_{c}^+(\to \Lambda \pi^+) \ell~ \bar{\nu}_{\ell}$ decay.}
	\label{fig:diagram_full}
\end{figure}

The four-fold differential decay rate of $\Lambda_{b} \to \Lambda_{c}^+(\to \Lambda \pi^+) \ell \,\bar{\nu}_{\ell}$ decay, with a unpolarized $\Lambda_{b}$ baryon, is fully parameterized in terms of the di-lepton invariant mass square, $q^2$, and the three angles as introduced in fig.~\ref{fig:diagram_full}. 

\begin{itemize}
	\item $\theta_{\ell}$: The angle between the leptons and the direction to the $\Lambda_{c}$ baryon in the virtual W-boson rest frame.
	\item $ \theta_{\Lambda}$: The angle between the $\Lambda$ baryon and direction of the $\Lambda_{c}$ baryon in the $\Lambda_{c}$ rest frame.
	\item $\phi$: The azimuthal angle between the two decay planes spanned by \text{$W^*$-$\ell$} and $\Lambda_{c}$-$\Lambda$ system in the $\Lambda_{b}$ rest frame.
\end{itemize}
Towards the end of this section, we will present the four-fold decay distribution and the related angular observables. In the following subsections, we will discuss the theory framework to calculate the amplitude of the decay rate in the SM and in the presence of the BSM operators in a model-independent framework.  
	%%%%%%%%%%%%%%%%%%%%%%%%%%%%%%%%%%%%%%%%%%%%%%%%
\subsection{Decay amplitude and transition matrix elements}
We can write the invariant amplitude for the $\Lambda_{b} \to \Lambda_c^+(\to \Lambda \pi^+) \ell^- \bar{\nu}_{\ell}$ decay as
\begin{equation}
\mathcal{M} = \sum_{\lambda_{\lc}} \langle \Lambda_c \ell\nu| {\mathcal{H}}_{eff}^{b\to c\ell \nu} | \Lambda_b \rangle   \frac{i}{k^2- m_{\Lambda_c}^2 + i m_{\Lambda_c} \Gamma_{\Lambda_c} } \langle \Lambda \pi | {\mathcal{H}}_{eff}^{\Delta S=1} | \Lambda_c \rangle.
\end{equation}
In this equation, we have used the Breit-Wigner formula to define the $\Lambda_c$ propagator, and $\lambda_{\lc}$ is the helicity of the $\Lambda_c$ baryon. In the appendix, in eq. \ref{eq:heffds1}, we have defined the effective Hamiltonian ${\mathcal{H}}_{eff}^{\Delta S=1}$. Furthermore, in the subsection \ref{subsec:lambdactolambda} in the appendix, we have given the detail of the calculation of the matrix element $\langle \Lambda \pi | {\mathcal{H}}_{eff}^{\Delta S=1} | \Lambda_c \rangle$. 

The effective Hamiltonian ${\mathcal{H}}_{eff}^{b\to c\ell \nu}$ for the semileptonic flavor changing charged current transition $b\to c\ell^-\bar{\nu}_\ell$ can be written as \cite{Dassinger:2008as,Sakaki:2013bfa,Bhattacharya:2018kig}
%\begin{align}\label{eq:heff}
%\mathcal{H}_{eff}^{b\to c\ell \nu} &=  \frac{G_F V_{cb}}{\sqrt{2}}\Bigg\{
%\Big[(1+C_{V_1}) \bar{c}\gamma_\mu (1-\gamma_5)  b + C_{V_2} \bar{c}\gamma_\mu (1+\gamma_5) b\Big] \bar{\ell} \gamma^\mu(1-\gamma_5) \nu_{\ell}  \nn \\ 
%& + \Big[C_{S_1} \bar{c} (1+\gamma_5) b   + C_{S_2} \bar{c}(1-\gamma_5)b \Big] \bar{\ell} %(1-\gamma_5)\nu_{\ell} + \Big[C_T \bar{c}\sigma^{\mu %\nu}(1-\gamma_5)b\Big]\bar{\ell}\sigma_{\mu \nu}(1-\gamma_5)\nu_{\ell} \Bigg\}
%\end{align} 
\begin{align}\label{eq:heff}
{\mathcal{H}}_{eff}^{b\to c\ell \nu}&=\frac{4\, G_F V_{cb}}{\sqrt{2}}[(1 + C_{V_1})\mathcal{O}_{V_1}+C_{V_2} \mathcal{O}_{V_2}+C_{S_1}\mathcal{O}_{S_1}+C_{S_2}\mathcal{O}_{S_2}+C_T \mathcal{O}_T]. 
\end{align} 
In the above equation, the four-Fermi operators are given by,
\begin{eqnarray}\label{eq:effopr}
\mathcal{O}_{V_1} &=& (\bar{c}_L\gamma^\mu b_L)(\bar{\ell}_L\gamma_\mu\nu_{\ell L}),~~~~~~
\mathcal{O}_{V_2} = (\bar{c}_R\gamma^\mu b_R)(\bar{\ell}_L\gamma_\mu\nu_{\ell L}), \nonumber
\\
\mathcal{O}_{S_1} &=& (\bar{c}_L b_R)(\bar{\ell}_R\nu_{\ell L}),~~~~~~~~~~
\mathcal{O}_{S_2} = (\bar{c}_R b_L)(\bar{\ell}_R\nu_{\ell L}), \nonumber
\\
\mathcal{O}_{T}&=& (\bar{c}_R\sigma^{\mu\nu}b_L)(\bar{\ell}_R\sigma_{\mu\nu}\nu_{\ell L}).
\end{eqnarray}
and $G_F$ is the Fermi constant and we use $\sigma_{\mu \nu} = i[\gamma_\mu, \gamma_\nu]/2$. In the above equation, we have parametrised the WC of each BSM four-fermi operator as $C_i$. We will obtain the effective Hamiltonian in the SM by letting $C_i = 0$. In practice, all the information about short-distance physics is contained in the WCs, which are calculable within a model-dependent framework. However, in a model-independent framework like ours, these WCs are the free parameters, and we either have to guess benchmark values for them or extract them from the data. 
	%\subsection{Decay process}
%	Using the quark level effective Hamiltonian for the process to be studied $\Lambda_{b}(p_{\lb})\rightarrow \Lambda_{c}(p_{\lc})+\tau^{-}(p_{\tau})+\bar{\nu}_{\tau}(p_{\bar{\nu}_\tau})$ where the daughter baryon $\Lambda_{c}$ decays to $ \Lambda \pi^+$.
	%The process under consideration is
	%$$\Lambda_{b}(p_{\lb})\rightarrow\tau^{-}(p_{\tau})+\bar{\nu}_{\tau}(p_{\bar{\nu}_\tau})+\Lambda_{c}(p_{\lc}).$$ 
	%%%%%%%%%%%%%%%%%%%%%%%%%%%%%%%%%%%%%%
	
We have obtained the expression for the decay amplitude of the semileptonic $\Lambda_{b}\to\Lambda_{c}^+ \ell^- \bar{\nu}_{\ell}$ decays in the SM and beyond using the effective Hamiltonian given in eq. \ref{eq:heff}. In the absence of QED corrections, we can factorize the matrix element of the semileptonic operators into the hadronic and leptonic parts, and the amplitude can be written as a convolution of the hadronic and leptonic tensors. Therefore, following the helicity amplitude formalism, we can express the decay amplitude $\mathcal{M}_{\lambda_{\lc}}^{\lambda_\tau}$ in terms of leptonic and hadronic helicity amplitudes which is as given below 
	\begin{align}\label{eq:helicityamp}
		\mathcal{M}^{\lambda_\ell}_{\lambda_{\Lambda_c}}=& \frac{G_F V_{cb}}{\sqrt{2}} \Biggl[H^{SP}_{\lambda_{\Lambda_c},\lambda=0}L^{\lambda_\ell}+\sum_{\lambda}\eta_{\lambda}H^{VA}_{\lambda_{\Lambda_c},\lambda}L^{\lambda_\ell}_{\lambda}+\sum_{\lambda,\lambda^{\prime}} \eta_\lambda \eta_{\lambda^{\prime}} H^{T,{\lambda_{\lb}}}_{\lambda_{\Lambda_c},\lambda ,\lambda^{\prime}}L^{\lambda_\ell}_{\lambda,\lambda^\prime}\Biggr].
	\end{align}	
	Here, ($\lambda$, $\lambda^\prime$) indicate the helicity of the virtual vector boson, $\lambda_{\Lambda_c}$ and $\lambda_\tau$ are the helicities of the  $\Lambda_c$ baryon and $\ell$ lepton, respectively, and $\eta_\lambda=1$ for $\lambda=t$ and $\eta_\lambda=-1$ for $\lambda=0,\pm 1$. The semileptonic decay $\Lambda_b \to \Lambda_{c} \ell\bar{\nu}_{\ell}$ can be considered as two subsequent 2-body decays, such as the decay $\Lambda_b \to \Lambda_{c} W^*$, followed by a subsequent decay of the off-shell $W^* \to \ell \nu_{\ell}$. The off-shell $W^*$ has four helicities, with two angular momentum $J=0,1$ in the rest frame of $W^*$, namely $\lambda = \pm 1,0$ ($J=1$) and $\lambda =0$ ($J=0$). It is only the off-shell $W^*$ which has time-like polarization. To distinguish the two $\lambda=0$ states, we adopt the notation $\lambda=0$ for $J=1$ and $\lambda=t$ for $J=0$\footnote{As the spin-0 component $\lambda$=t has the property $\bar{\epsilon}_{\mu}(t)\propto q_{\mu}$, does not has effects on semileptonic decay in the lepton massless limit.}. 

\paragraph{\bf \underline{Hadronic helicity amplitudes}:}  
In eq. \ref{eq:helicityamp}, we have defined the hadronic helicity amplitudes as 
\begin{align}\label{eq:defhelamp}
	H^{SP}_{\lambda_{\Lambda_c},\lambda=0}= & H^S_{\lambda_{\Lambda_c},\lambda=0}+H^P_{\lambda_{\Lambda_c},\lambda=0}, \nn\\
	H^{VA}_{\lambda_{\Lambda_c},\lambda}= & H^V_{\lambda_{\Lambda_c},\lambda}-H^A_{\lambda_{\Lambda_c},\lambda}, \nn \\
	H^{T,{\lambda_{\lb}}}_{\lambda_{\lc},\lambda ,\lambda^{\prime}}= & H^{T1,{\lambda_{\lb}}}_{\lambda_{\lc},\lambda ,\lambda^{\prime}}-H^{T2,{\lambda_{\lb}}}_{\lambda_{\lc},\lambda ,\lambda^{\prime}}, 
\end{align}
Here, $H^S_{\lambda_{\Lambda_c},\lambda=0}$ and $H^P_{\lambda_{\Lambda_c},\lambda=0}$ are the helicity amplitudes of the scalar and pseudoscalar quark currents defined in eq. \ref{eq:heff}. The helicity amplitudes $H^V_{\lambda_{\Lambda_c},\lambda}$ and $H^A_{\lambda_{\Lambda_c},\lambda}$ are associated with the vector and axial-vector quark currents, respectively. And that for the tensor quark currents are defined in terms of $H^{T1,{\lambda_{\lb}}}_{\lambda_{\lc},\lambda ,\lambda^{\prime}}$ and $H^{T2,{\lambda_{\lb}}}_{\lambda_{\lc},\lambda ,\lambda^{\prime}}$. 

The general expressions for these scalar, pseudoscalar, vector, axial-vector and tensor helicity amplitudes are given as 	
\begin{align}\label{eq:defhelampexpanded}
		H^S_{\lambda_{\Lambda_c},\lambda=0}=&(C_{S_1}+C_{S_2}) \bra{\lc}\bar{c} b\ket{\lb},\nn\\
		H^P_{\lambda_{\Lambda_c},\lambda=0}=&(C_{S_1}-C_{S_2})\bra{\lc}\bar{c}\gamma_5 b\ket{\lb}, \nn \\
		H^V_{\lambda_{\Lambda_c},\lambda}=&(1+C_{V_1}+C_{V_2})\,\epsilon^*_{\mu}(\lambda)\bra{\lc}\bar{c}\gamma^{\mu} b\ket{\lb}, \nn\\
		H^A_{\lambda_{\Lambda_c},\lambda}=&(1+C_{V_1}-C_{V_2})\,\epsilon^*_{\mu}(\lambda)\bra{\lc}\bar{c}\gamma^{\mu}\gamma_5 b\ket{\lb},  \nn \\
		H^{T1,{\lambda_{\lb}}}_{\lambda_{\lc},\lambda ,\lambda^{\prime}}=&C_T\: \epsilon^*_{\mu}(\lambda)\epsilon^*_{\nu}(\lambda^{\prime})\bra{\lc}\bar{c}i\sigma^{\mu \nu} b\ket{\lb},\nn\\
		H^{T2,{\lambda_{\lb}}}_{\lambda_{\lc},\lambda ,\lambda^{\prime}}=&C_T\:\epsilon^*_{\mu}(\lambda)\epsilon^*_{\nu}(\lambda^{\prime})\bra{\lc}\bar{c}i\sigma^{\mu \nu}\gamma_5 b\ket{\lb}.
\end{align}
Here, $\epsilon^{\mu}(\lambda)$ are the polarization vectors of the virtual vector boson.

We can express the hadronic matrix elements in the above equation in terms of the respective $\Lambda_b \to \Lambda_c$ form factors. The definitions of these hadronic matrix elements in terms of the form factors are given in the appendix in subsection \ref{app:hadronicmatrix}. Using these definitions and substituting eq. \ref{eq:defhelampexpanded} in eq. \ref{eq:defhelamp}, we obtain the helicity amplitudes (non-vanishing) as 
\begin{align}
	H^{SP}_{1/2,0}&=f_0(q^2)(C_{S_1}+C_{S_2}) \frac{\sqrt{Q_+}}{m_b-m_c}(m_{\lb}-m_{\lc})-g_0(q^2) (C_{S_1}-C_{S_2})\frac{\sqrt{Q_-}}{m_b+m_c}(m_{\lb}+m_{\lc}), \\
	H^{SP}_{-1/2,0}&= f_0(q^2)(C_{S_1}+C_{S_2})\frac{\sqrt{Q_+}}{m_b-m_c}(m_{\lb}-m_{\lc})+g_0(q^2) (C_{S_1}-C_{S_2})\frac{\sqrt{Q_-}}{m_b+m_c}(m_{\lb}+m_{\lc}),			
\end{align}
\begin{align}
	H^{VA}_{1/2,0}=&f_+(q^2) (1+C_{V_1}+C_{V_2})\frac{\sqrt{Q_-}}{\sqrt{q^2}}(m_{\lb}+m_{\lc})-g_+(q^2) (1+C_{V_1}-C_{V_2})\frac{\sqrt{Q_+}}{\sqrt{q^2}}(m_{\lb}-m_{\lc}), \\
	H^{VA}_{1/2,+1}= &-f_\perp(q^2) (1+C_{V_1}+C_{V_2})\sqrt{2Q_-} +g_\perp(q^2) (1+C_{V_1}-C_{V_2})\sqrt{2Q_+}, \\		
	H^{VA}_{1/2,t}= &f_0(q^2)(1+C_{V_1}+C_{V_2})\frac{\sqrt{Q_+}}{\sqrt{q^2}}(m_{\lb}-m_{\lc})-g_0(q^2)(1+C_{V_1}-C_{V_2})\frac{\sqrt{Q_-}}{\sqrt{q^2}}(m_{\lb}+m_{\lc}), \\
	H^{VA}_{-1/2,0}= &f_+(q^2) (1+C_{V_1}+C_{V_2}) \frac{\sqrt{Q_-}}{\sqrt{q^2}}(m_{\lb}+m_{\lc})+g_+(q^2) (1+C_{V_1}-C_{V_2})\frac{\sqrt{Q_+}}{\sqrt{q^2}}(m_{\lb}-m_{\lc}), \\			
	H^{VA}_{-1/2,-1}= &-f_\perp(q^2) (1+C_{V_1}+C_{V_2})\sqrt{2Q_-} -g_\perp(q^2) (1+C_{V_1}-C_{V_2})\sqrt{2Q_+}, \\		
	H^{VA}_{-1/2,t}= &f_0(q^2) (1+C_{V_1}+C_{V_2})\frac{\sqrt{Q_+}}{\sqrt{q^2}}(m_{\lb}-m_{\lc}) +g_0(q^2)(1+C_{V_1}-C_{V_2})\frac{\sqrt{Q_-}}{\sqrt{q^2}}(m_{\lb}+m_{\lc}),			
\end{align}
and 
\begin{align}
	H^{T,-1/2}_{-1/2,t,0}=&-C_T\Big[-h_+(q^2)\sqrt{Q_-}+\widetilde{h}_+(q^2)\sqrt{Q_+}\Big], \\[10pt]
	H^{T,+1/2}_{+1/2,t,0}=&C_T\Big[h_+(q^2)\sqrt{Q_-}+\widetilde{h}_+(q^2)\sqrt{Q_+}\Big],\\[10pt]
	H^{T,-1/2}_{+1/2,t,+1}=& -C_T\frac{\sqrt{2}}{\sqrt{q^2}}\Big[h_\perp(q^2)(m_{\lb}+m_{\lc})\sqrt{Q_-}+\widetilde{h}_\perp(q^2)(m_{\lb}-m_{\lc})\sqrt{Q_+}\Big],\\[10pt]
	H^{T,+1/2}_{-1/2,t,-1}=&-C_T\frac{\sqrt{2}}{\sqrt{q^2}}\Big[h_\perp(q^2)(m_{\lb}+m_{\lc})\sqrt{Q_-}-\widetilde{h}_\perp(q^2)(m_{\lb}-m_{\lc})\sqrt{Q_+}\Big],\\[10pt]
	H^{T,-1/2}_{+1/2,0,+1}=&-C_T\frac{\sqrt{2}}{\sqrt{q^2}}\Big[h_\perp(q^2)(m_{\lb}+m_{\lc})\sqrt{Q_-}+\widetilde{h}_\perp(q^2)(m_{\lb}-m_{\lc})\sqrt{Q_+}\Big],\\[10pt]
	H^{T,+1/2}_{-1/2,0,-1}=&C_T\frac{\sqrt{2}}{\sqrt{q^2}}\Big[h_\perp(q^2)(m_{\lb}+m_{\lc})\sqrt{Q_-}-\widetilde{h}_\perp(q^2)(m_{\lb}-m_{\lc})\sqrt{Q_+}\Big],\\[10pt]
	H^{T,+1/2}_{+1/2,+1,-1}=&-C_T\Big[h_+(q^2)\sqrt{Q_-}+\widetilde{h}_+(q^2)\sqrt{Q_+}\Big],\\[10pt]
	H^{T,-1/2}_{-1/2,+1,-1}=&-C_T\Big[h_+(q^2)\sqrt{Q_-}-\widetilde{h}_+(q^2)\sqrt{Q_+}\Big].
\end{align}
In the above equations, $f_i(q^2)$, $g_i(q^2)$, $h_j(q^2)$, $\tilde{h}_j(q^2)$ are the hadronic form factors with $i=0, +, \perp$ and $j=+, \perp$, respectively, and we have defined $Q_{\pm}$=$(m_{\lb} \pm m_{\lc})^2-q^2$.  

	%%%%%%%%%%%%%%%%%%%%%%%%%%%%%%%%%%%%%%%%%%%%%%%%%%%
	\paragraph{\bf \underline{Leptonic helicity amplitudes}:}  
	The leptonic helicity amplitudes in eq. \ref{eq:helicityamp} are obtained as  
	\begin{align}\label{eq:leptonhel}
		L^{\lambda_\ell}=&\bra{\ell\bar{\nu}_\ell}\bar{\ell} (1-\gamma_5)\nu_\ell\ket{0} = \bar{u}_\ell (1-\gamma_5) v_{\bar{\nu}_\ell}\nn\\
		L^{\lambda_\ell}_{\lambda}=&\epsilon_\mu (\lambda)\bra{\ell\bar{\nu}_\ell}\bar{\ell}\gamma^\mu (1-\gamma_5)\nu_\ell\ket{0}=\epsilon_\mu (\lambda) \bar{u}_\ell \gamma^\mu(1-\gamma_5) v_{\bar{\nu}_\ell} \nn \\
		L^{\lambda_\ell}_{\lambda ,\lambda^{\prime}}=&-i\epsilon_\mu (\lambda)\epsilon_\nu (\lambda^\prime)\bra{\ell\bar{\nu}_\ell}\bar{\ell}\sigma^{\mu \nu} (1-\gamma_5)\nu_\ell\ket{0}=-i\epsilon_\mu (\lambda)\epsilon_\nu (\lambda^\prime)\bar{u}_\ell \sigma^{\mu \nu} (1-\gamma_5) v_{\bar{\nu}_\ell} .
	\end{align}
	In calculating the lepton helicity amplitudes, we work in the rest frame of the virtual vector boson, the $\ell-\nu_\ell$ dilepton rest frame. We have presented the polarisation vectors of this virtual vector boson in eq.~\ref{eq:polWst} in the appendix~\ref{app:leptonichel}.
	Furthermore, in the same appendix, we have presented the representation of the spinor fields in eq.~\ref{eq:spinors} defined earlier. Following these definitions, from eq.~\ref{eq:leptonhel} we have obtained the expressions of the leptonic helicity amplitudes. For the scalar and pseudoscalar leptonic currents, we have obtained  
	\begin{eqnarray}
		L^{+1/2}=& 2\sqrt{q^2} v, \\
		L^{-1/2}=& 0.
	\end{eqnarray}	
	For the vector and axial-vector leptonic currents, the helicity amplitudes are 
	\begin{align}
		L^{+1/2}_{\pm1}=&\pm \sqrt{2}m_{\ell} v\; \sin(\theta_\ell), \\
		L^{+1/2}_{0}=&-2m_\ell v\; \cos{(\theta_\ell)}, \\
		L^{+1/2}_{t}=&2m_\ell v, \\
		L^{-1/2}_{\pm1}=&\sqrt{2 q^2}v\; (1\pm \cos(\theta_\ell)), \\
		L^{-1/2}_{0}=&2\sqrt{q^2}v \; \sin{(\theta_\ell)}, \\
		L^{-1/2}_{t}=&0,
	\end{align}
	and the same for the tensor leptonic currents are obtained as 
	\begin{align}
		L^{+1/2}_{0,\pm1}=&-\sqrt{2 q^2}v\; \sin(\theta_\ell), \\
		L^{+1/2}_{\pm1,t}=&\mp\sqrt{2 q^2}v\; \sin(\theta_\ell), \\
		L^{+1/2}_{t,0}=&L^{+1/2}_{+1,-1}=-2\sqrt{q^2}v\; \cos(\theta_\ell), \\
		L^{-1/2}_{0,\pm1}=&\mp\sqrt{2}m_\ell v\; (1\pm \cos(\theta_\ell)), \\
		L^{-1/2}_{\pm1,t}=&-\sqrt{2}m_\ell v\; (1\pm \cos(\theta_\ell)), \\
		L^{-1/2}_{t,0}=&L^{-1/2}_{+1,-1}=2m_\ell v\; \sin(\theta_\ell) .
	\end{align}
In the above equations, we have defined $v=\sqrt{1-\frac{m_\ell^2}{q^2}}.$

	%%%%%%%%%%%%%%%%%%%%%%%%%%%%%%%%%%%%%%%%%%%%%%%%%%%%%%%%%%%%%%%%%%%%%
\subsection{Angular distribution and Observables}
Following the geometry given in fig.\ref{fig:diagram_full} for the decay $\Lambda_{b} \to \Lambda_c^+(\to \Lambda \pi^+) \ell^- \bar{\nu}_{\ell}$ we will obtain the detail of the measurable 4-fold angular decay distribution which we can express as 
\begin{equation}\label{eq:diffrate1}
\frac{d^4 \Gamma}{dq^2 d \cos\theta_\ell d\cos\theta_\Lambda d\phi} = N^{\prime} \sum_{\lambda_{\lb}, \lambda_{\ell}, \lambda_{\Lambda}} \sum_{\lambda_{\lc}}
\Bigl|\mathcal{M}_{\lambda_{\Lambda_c},\lambda_\ell}^{\lambda_{\lb}} \mathcal{A}^{\lambda_{ \Lambda_c}}_{ \lambda_{\Lambda}}\Bigr|^2.
\end{equation}
We have obtained the above expression using the narrow width approximation for the Breit-Wigner propagator. In eq.~\ref{eq:helicityamp}, we have defined the helicity amplitude $\mathcal{M}_{\lambda_{\Lambda_c},\lambda_\ell}^{\lambda_{\lb}}$. The helicity amplitudes  $\mathcal{A}^{\lambda_{ \Lambda_c}}_{ \lambda_{\Lambda}}$ for $\Lambda_c^+ \to \Lambda \pi^+$ decay are defined in the subsection~\ref{subsec:lambdactolambda}. We have obtained the following expression for the $N^{\prime}$  
\begin{equation} 	
	N^{\prime} =  \frac{G_F^2 |V_{cb}|^2}{2^{17} \pi^5 \Gamma_{\lc}}\frac{\sqrt{Q_+ Q_-}}{m^3_{\lb}} \frac{\sqrt{r_+ r_-}}{m^3_{\lc}} \Bigl(1-\frac{m_\ell ^2}{q^2}\Bigr),
\end{equation}
with $r_{\pm}$=$(m_{\lc} \pm m_{\Lambda})^2-m_{\pi}^2$. 

After summing over the $\lambda_{\lc}$ and the respective values of $\lambda_{\Lambda}$ and following the definitions of the helicity amplitudes $\mathcal{A}^{\lambda_{ \Lambda_c}}_{ \lambda_{\Lambda}}$ given in the appendix, we expand the r.h.s. of eq.~\ref{eq:diffrate1} and obtain  
 \begin{align}\label{eq:totl_dist}
 	\frac{d^4 \Gamma}{dq^2 d \cos\theta_\ell d\cos\theta_\Lambda d\phi}&= 
 %	&= N^{\prime} \sum_{\lambda_{\lb}, \lambda_{\ell}}  \sum_{\lambda_{\Lambda}}
 %	\Biggl[\mathcal{M}_{\frac{1}{2},\lambda_\ell}^{\lambda_{\lb}} \bigl(\mathcal{M}_{\frac{1}{2},\lambda_\ell}^{\lambda_{\lb}}\bigr)^*
 %	\mathcal{A}^{\frac{1}{2}}_{ \lambda_{\Lambda}} \bigl( \mathcal{A}^{\frac{1}{2}}_{ \lambda_{\Lambda}}\bigr)^*
 %	+ \mathcal{M}_{\frac{1}{2},\lambda_\ell}^{\lambda_{\lb}} \bigl(\mathcal{M}_{-\frac{1}{2},\lambda_\ell}^{\lambda_{\lb}}\bigr)^*
 %	\mathcal{A}^{\frac{1}{2}}_{ \lambda_{\Lambda}} \bigl( \mathcal{A}^{-\frac{1}{2}}_{ \lambda_{\Lambda}}\bigr)^*\nn\\
 %	&+ \mathcal{M}_{-\frac{1}{2},\lambda_\ell}^{\lambda_{\lb}} \bigl(\mathcal{M}_{-\frac{1}{2},\lambda_\ell}^{\lambda_{\lb}}\bigr)^*
 %	\mathcal{A}^{-\frac{1}{2}}_{ \lambda_{\Lambda}} \bigl( \mathcal{A}^{-\frac{1}{2}}_{ \lambda_{\Lambda}}\bigr)^*
 %	+ \mathcal{M}_{-\frac{1}{2},\lambda_\ell}^{\lambda_{\lb}} \bigl(\mathcal{M}_{\frac{1}{2},\lambda_\ell}^{\lambda_{\lb}}\bigr)^*
 %	\mathcal{A}^{-\frac{1}{2}}_{ \lambda_{\Lambda}} \bigl( \mathcal{A}^{\frac{1}{2}}_{ \lambda_{\Lambda}}\bigr)^* \Biggr] \nn\\ 
 	\frac{2^4 \pi m^3_{\lc} N^{\prime} }{\sqrt{r_+ r_-}}  \mathcal{B}(\lc \to \Lambda \pi) \times \nn \\
 	&\quad  \sum_{\lambda_{\lb}, \lambda_{\ell}}
 	\Biggl[\mathcal{M}_{\frac{1}{2},\lambda_\ell}^{\lambda_{\lb}} \bigl(\mathcal{M}_{\frac{1}{2},\lambda_\ell}^{\lambda_{\lb}}\bigr)^*
 	(1+\alpha \cos \theta_{\Lambda}) 
 	+ \mathcal{M}_{\frac{1}{2},\lambda_\ell}^{\lambda_{\lb}} \bigl(\mathcal{M}_{-\frac{1}{2},\lambda_\ell}^{\lambda_{\lb}}\bigr)^*
 	(-\alpha \sin \theta_{\Lambda} e^{i \phi}) \nn \\
  &	+ \mathcal{M}_{-\frac{1}{2},\lambda_\ell}^{\lambda_{\lb}} \bigl(\mathcal{M}_{-\frac{1}{2},\lambda_\ell}^{\lambda_{\lb}}\bigr)^*
 	(1-\alpha \cos \theta_{\Lambda})
 	+ \mathcal{M}_{-\frac{1}{2},\lambda_\ell}^{\lambda_{\lb}} \bigl(\mathcal{M}_{\frac{1}{2},\lambda_\ell}^{\lambda_{\lb}}\bigr)^*
 	(-\alpha \sin \theta_{\Lambda} e^{-i \phi}) \Biggr]	\nn \\
 	& = \frac{3}{8\pi}K(q^2, \cos\theta_\ell, \cos\theta_\Lambda, \phi)
 	\end{align}
In eq.~\ref{eq:totl_dist}, we obtained the first step by summing over $\lambda_{\lc}$ and $\lambda_{\Lambda}$ and the final step is obtained after summing over $\lambda_{\lb}$ and $\lambda_{\ell}$. The function $K(q^2, \cos\theta_\ell, \cos\theta_\Lambda, \phi)$ will depend on the helicity amplitudes defined in eq. \ref{eq:defhelamp}, and the general expression of it is given by 
	\begin{align}
		\label{eq:ang_dist}
		K(q^2, \cos\theta_\ell, \cos\theta_\Lambda, \phi)
		& =
		\big( K_{1ss} \sin^2\theta_\ell +\, K_{1cc} \cos^2\theta_\ell + K_{1c} \cos\theta_\ell\big) \,\cr
		&  + \big( K_{2ss} \sin^2\theta_\ell +\, K_{2cc} \cos^2\theta_\ell + K_{2c} \cos\theta_\ell\big) \cos\theta_\Lambda
		\cr
		&  + \big( K_{3sc}\sin\theta_\ell \cos\theta_\ell + K_{3s} \sin\theta_\ell\big) \sin\theta_\Lambda \cos\phi\cr
		&  + \big( K_{4sc}\sin\theta_\ell \cos\theta_\ell + K_{4s} \sin\theta_\ell\big) \sin\theta_\Lambda \sin\phi \,.
	\end{align}
	%%%%%%%%%%%%%%%%%%%%%%%%%%%	
In the  appendix~\ref{sec:anay_expr}, we have presented the expressions for the angular coefficients $K_i$'s \footnote{With the given spinor and $W^*$ polarization vector, define in ref~\cite{Datta:2017aue}, we could not find the corresponding leptonic and hadronic helicity amplitudes in ref~\cite{Datta:2017aue}.}.

After the integration over the angles $\theta_{\ell}$ and $\phi$, we obtain from eq.~\ref{eq:totl_dist} the following two-fold differential decay rate distribution 
	\begin{equation}\label{eq:d2dq2dcosthga}
		\frac{\rm d^2\Gamma}{\rm d \text{$q^2$}  d\cos\theta_{\Lambda}}=\frac{1}{2}\frac{\rm d\Gamma}{\rm d \text{$q^2$}}(1+\alpha_{\Lambda}~P_{\Lambda_c}(q^2)\cos\theta_{\Lambda})
	\end{equation}
	with the $\Lambda_{c}$ spin polarization $P_{\Lambda_c}(q^2)$, which we have defined as
	\begin{equation}
		P_{\Lambda_c}(q^2)=\frac{{\rm d}\Gamma^{\lambda_{\Lambda_c}=1/2}/{\rm d}q^2-
			{\rm d}\Gamma^{\lambda_{\Lambda_c}=-1/2}/{\rm d}q^2}{{\rm d}\Gamma/{\rm d}q^2}.
	\end{equation}
	The detailed expression of the differential rates ${\rm d}\Gamma^{\lambda_{\Lambda_c}=\pm 1/2}/{\rm d}q^2$ are given in the appendix~\ref{sec:anay_expr}. Integrating eq.~\ref{eq:d2dq2dcosthga} over the angle $\theta_{\Lambda}$, we obtain the following $q^2$ distribution of the decay rate 
	\begin{align}
		\frac{d \Gamma}{dq^2}=2K_{1ss}(q^2)+K_{1cc}(q^2).
	\end{align}
The measurable angular asymmetries are obtained by normalising the $K_i$'s in eq.~\ref{eq:ang_dist} by the total rate  
\begin{align}
	\hat{K_i}(q^2) =\frac{K_i(q^2)}{2K_{1ss}(q^2)+K_{1cc}(q^2)}. 
\end{align}
	
We have defined a couple of other important observables which we have presented in the following enumerated items: 
\begin{enumerate}
	\item The $q^2$ distribution of lepton-polarization asymmetry   
	\begin{equation}
		P_{\ell}^{(\Lambda_c)}(q^2)=\frac{{\rm d}\Gamma^{\lambda_{\ell}=1/2}/{\rm d}q^2-
			{\rm d}\Gamma^{\lambda_{\ell}=-1/2}/{\rm d}q^2}{{\rm d}\Gamma/{\rm d}q^2}\,, 
	\end{equation}
	which is the difference between the $q^2$ distributions of the rates of the left- and right-polarised lepton beams normalised by ${{\rm d}\Gamma/{\rm d}q^2}$. The expressions for $\frac{\rm d \Gamma}{\rm d q^2}^{\lambda_{\ell}=\pm 1/2}$ are given in the appendix~\ref{sec:anay_expr}.

\item We have defined the convexity parameter $C_F^{\ell}(q^2)$ defined by 
\begin{equation}
	C_F^{\ell}(q^2) =\frac{1}{d\Gamma/dq^2}\bigg(\frac{d }{d(\cos \theta_{\ell})}\bigg)^2\bigg(\frac{d^2\Gamma}{d q^2 d \cos \theta_{\ell}} \bigg).
\end{equation}
The detail expression for $C_{F}^{\ell}(q^2)$ in terms of the helicity amplitude are given in the appendix~\ref{sec:anay_expr}. 

\item Among the other important observables, the forward-backward asymmetry concerning the leptonic scattering angle, normalized to the differential rate, is defined as
\begin{equation}
	A^\ell_\text{FB}(q^2) = \frac{3}{2} \, \frac{K_{1c}(q^2)}{2 K_{1ss}(q^2) + K_{1cc}(q^2)},   
\end{equation}
%%%%%%%
which we can obtain from eq.~\ref{eq:ang_dist}, following the integration as given below
\begin{equation}
	A^\ell_\text{FB}(q^2) =\frac{\int_{0}^{2\pi} d\phi\int_{-1}^{1} d(\cos \theta_{\Lambda}) \big[\int_{0}^{1}-\int_{-1}^{0}\big]d(\cos\theta_{\ell}) K(q^2, \cos\theta_\ell, \cos\theta_\Lambda, \phi)}{d \Gamma/dq^2}
\end{equation}

\item The analogous asymmetry for the baryonic scattering angle reads
\begin{align}
	A^{\Lambda_c}_\text{FB}(q^2)
	& = \frac{1}{2} \, \frac{2 K_{2ss}(q^2) + K_{2cc}(q^2)}{2 K_{1ss}(q^2) + K_{1cc}(q^2)}. 
\end{align}
Using eq.~\ref{eq:ang_dist} we can define the above asymmetry as  
\begin{equation}
	A^{\Lambda_c}_\text{FB}(q^2)  =\frac{\int_{0}^{2\pi} d\phi \int_{-1}^{1} d(\cos \theta_{\ell})\big[\int_{0}^{1}-\int_{-1}^{0}\big]d(\cos \theta_{\Lambda}) K(q^2, \cos\theta_\ell, \cos\theta_\Lambda, \phi)}{{d \Gamma/dq^2}}.
\end{equation}

\item 	For $\Lambda_b\to \Lambda^+_c (\to \Lambda \pi^+) \ell^-\bar{\nu}$ decays, one could also define a combined forward-backward asymmetry 
\begin{align}
	A^{\Lambda_c \ell}_\text{FB}(q^2) & = \frac{3}{4} \, \frac{K_{2c}(q^2)}{2 K_{1ss}(q^2) + K_{1cc}(q^2)},
\end{align}
which we have obtained following the integrations as given below

\begin{align}
	A^{\Lambda_c \ell }_\text{FB}(q^2) & =\frac{\int_{0}^{2\pi} d\phi \big[\int_{0}^{1}-\int_{-1}^{0}\big]d(\cos \theta_{\Lambda})  I_\ell(q^2, \cos\theta_\Lambda, \phi) }{d \Gamma/dq^2},
\end{align}
with 
\begin{equation}
I_\ell(q^2, \cos\theta_\Lambda, \phi) = \Big[\int_{0}^{1}   -\int_{-1}^{0}  \Big]d(\cos\theta_{\ell})  K(q^2, \cos\theta_\ell, \cos\theta_\Lambda, \phi).
\end{equation}
Here, $I_{\ell}(q^2, \cos\theta_\Lambda, \phi)$ is defined as the difference of the three-fold decay rates of a lepton moving in the forward and backward regions respectively.  
%\begin{align}
%	A^{\Lambda_c \ell }_\text{FB}(q^2) & =\frac{\int_{0}^{2\pi} d\phi [\int_{0}^{1}-\int_{-1}^{0}]d\cos \theta_{\Lambda}[\int_{0}^{1}-\int_{-1}^{0}]d \cos\theta_{\ell} K(q^2, \cos\theta_\ell, \cos\theta_\Lambda, \phi)}{\int_{0}^{2\pi} d\phi \int_{0}^{1} d\cos \theta_{\Lambda}\int_{-1}^{1}d \cos\theta_{\ell} K(q^2, \cos\theta_\ell, \cos\theta_\Lambda, \phi)}.
%\end{align}

\end{enumerate}
 
After reviewing the other literature, we have summarized our findings as follows:
	\begin{itemize}
		\item Note that the 4-fold decay rate distributions defined in eq.~\ref{eq:totl_dist} in the NP scenarios with vector (V), axial-vector (A), scalar (S) and pseudo-scalar (P) interactions were studied earlier in refs.~\cite{Shivashankara:2015cta} and \cite{Boer:2019zmp}, respectively. However, there are some discrepancies between the analytical expressions of these two references, particularly in the terms proportional to $\sin\theta_{\ell}$. Ref.~\cite{Boer:2019zmp} observed interference terms of the type $\text{V}\times\text{S}$ and $\text{A}\times\text{P}$ in the angular observables $K_{3s}$ and $K_{4s}$, which are missing in the corresponding terms (defined as $C^{int_4}$) of ref.~\cite{Shivashankara:2015cta}. Note that these contributions will be essential in the analysis with two or more operators at a time, which we have done in this analysis. We have independently derived the 4-fold decay distribution in the different NP scenarios mentioned above. We provide the detailed analytical formulas in the appendix. Our findings agree with those given in ref.~\cite{Boer:2019zmp}.
			%\footnote{To cross-check our analytical expressions of the 4-fold decay distributions to those given in \cite{Boer:2019zmp}, we have to redefine the angles $\theta_{\tau}\to \pi-\theta$, $\theta_{\Lambda_c}\to\theta_s$, and $\phi\to-\chi$.}. 
		
		\item The 4-fold decay rate distributions in the presence of a tensor operator were given only in ref.~\cite{Boer:2019zmp}. We have redone the analysis independently and our analytical formulas agrees with those presented in \cite{Boer:2019zmp}.

		\item Our analytical expressions for the observables $\frac{d \Gamma}{dq^2}$  and $A^\ell_\text{FB}(q^2)$ are in agreement with the result of \cite{Datta:2017aue} in the SM and the NP scenarios defined in eq.~\ref{eq:heff}. Furthermore, we have found that our results for $\frac{d^2 \Gamma}{dq^2 d\cos\theta_{\ell}}$ are in complete agreement with that result of~\cite{Li:2016pdv}.
		
		\item In ref.~\cite{Boer:2019zmp}, the authors have defined the angular distribution in terms of the transversality amplitudes. They have not provided any relationship between their transversality amplitudes and the helicity amplitudes that we are using. We have worked out those relations and provided them in the appendix~\ref{Appndx:litreview}. Following those relations, we have reproduced $\frac{d \Gamma}{dq^2}$ given in ref.~\cite{Boer:2019zmp}, and it matches with ours. 
		%We have a factor of 4 extra for each angular observable, which gives the same value after normalizing.
		
\end{itemize}	

We have mentioned the discrepancy between the results of the refs above \cite{Shivashankara:2015cta} and \cite{Boer:2019zmp}. Hence, it is essential to do an independent check. Also, the results for the 4-fold distribution in the presence of a tensor current was not verified, we have also verified the results of ref.~\cite{Boer:2019zmp}.     
	In the following items, we will point out the other important results of our analysis that were unavailable in the literature. 
		\begin{enumerate}
		\item For the first time we have provided predictions of $\langle\frac{\Gamma}{|V_{cb}^2|} \rangle$, $\langle {A^{\ell}_{FB}}(q^2) \rangle$, $\langle {A^{\Lambda_c \ell}_{FB}}(q^2) \rangle$, $\langle {A^{\Lambda_c}_{FB}}(q^2) \rangle$, $\langle {P_{\Lambda_c}}(q^2) \rangle$, $\langle {P_{\ell}^{(\Lambda_c)}}(q^2) \rangle$ and $\langle C_{F}^{\ell}(q^2)\rangle$ in 5 small $q^2$-bins both for $\ell=\mu$ and $\ell=\tau$ in the SM and in the NP scenarios given in eq.~\ref{eq:heff}. These are the very basic requirements for the SM test and looking for NP effects in the measurements. 
		
	\item In addition, we have predicted the observables integrated over the entire kinematically allowed $q^2$ regions both in the SM and in the NP scenarios, which were not available in the literature. In the SM, those predictions for $\langle {P_{\Lambda_c}}(q^2) \rangle$, $\langle {P_{\ell}^{(\Lambda_c)}}(q^2) \rangle$ were not available in the literature. We have predicted the full $q^2$ integrated values for all the observables with the respective 1$\sigma$ errors in all the NP scenarios. A comparison of these estimates with future measurements will be helpful in pinpointing NP effects from those of the SM.

	\item We have extracted the values of the new WCs given in eq.~\ref{eq:heff} from the available data on $R(D)$, $R(D^*)$, $R(\Lambda_c)$, $\mathcal{B}(\Lambda_b\to \Lambda_c\tau^-\bar{\nu})$ and $F_{L}^{D^*}$. This is the first time we are using the most recent measured value of $D^*$ longitudinal polarization asymmetry by LHCb, which is SM consistent. We have done these fits using one operator at a time and in the two-operator scenarios. The detailed phenomenology of each of these scenarios has been discussed in section~\ref{sec:NPtst}.
	
	\item Using the results of the fits, we have obtained the correlations between different observables mentioned earlier. In the case of NP, these correlations between the measured and yet-to-be-measured observables give us an understanding of the expected trend that we should observe in the measurements. This will also help to distinguish the effect of different NP scenarios.
	
	\item Also, we have studied the $q^2$ distributions of the observables in the presence of one or two-operator scenarios to see whether or not it is possible to separate the impact of different NP effects from one another with relatively smaller values of new WCs which are allowed by the current data. The details are given in section~\ref{sec:NPtst}.
	
	\item The $q^2$ distributions of $A^{\ell}_{FB}(q^2)$, $A^{\Lambda_c \ell}_{FB}(q^2)$ and $ P_{\ell}^{(\Lambda_c)}(q^2)$ have zero-crossing points $q_0^2$ (in GeV$^2$). We have estimated the values of $q_0^2$ in the SM. These zero crossing points have unique features for different leptons, which has been thoroughly discussed in section~\ref{sec:SMprd}. Also, we have studied whether or not there are shifts in the values of these zero crossing points in different NP scenarios. Wherever necessary, we have predicted them.
			
	\end{enumerate}

\section{Form Factor Shape} \label{sec:BGLparamfitres}
	
We have noted in the last section that the helicity or transversality amplitudes of the hadronic current depend on the non-perturbative form factors and their $q^2$ shapes. Hence, to get the shape of the decay rate distribution, one needs to know the shape of the corresponding form factors in the whole $q^2$ region. The lattice results on the relevant form factors are available in ref.~\cite{Detmold:2015aaa}. To obtain the $q^2$ shapes of the form factors, in their analysis, the lattice collaboration has used the parametrization proposed by Bourrely-Lellouch-Caprini (BCL) \cite{BCL}. The details of the relevant inputs used in the BCL parameterization can be seen from the ref.~\cite{Detmold:2015aaa}, which we have incorporated into our analysis. We use their fit results for the BCL coefficients to obtain the shapes of the form factors. Using these shapes, we have obtained the decay rate distributions and predicted many other related observables, which we will discuss in the following sections. 
	%%%%%%%%%%%%%%%%%%%%%%%%%%%%%%%%%%%%%%%%%%%%%%%%%%%%%	
	\begin{table}[h]
		\renewcommand{\arraystretch}{1.5}
		\centering
		\setlength\tabcolsep{30pt}
		\begin{tabular}{*{4}{c}}
			\hline\hline
			\text{$\bf 0^-$}/GeV &\text{$\bf 0^+$}/GeV &\text{$\bf1^-$}/GeV & \text{$\bf 1^+$}/GeV\\
			\hline
			\text{6.2749}&\text{6.6925}&\text{6.3290}&\text{6.7305}\\
			\text{6.8710}&\text{7.1045}&\text{6.8975}&\text{6.7385}\\
			&             &\text{7.0065}&\text{7.1355}\\
			&             &             &\text{7.1435}\\
			\hline
		\end{tabular}
		\caption{Pole masses used in the BGL parametrization.}
		\label{tab:BGLpolmass}
	\end{table}
	
	\begin{table}[h]
		\begin{center}
			\begin{tabular}{|c|c|}
				\hline
				\textbf{observables}& 
				\textbf{Values}
				\\
				\hline
				$\tilde{\chi}_{0^-}^L(0)$ & 19.412$\times 10^{-3}$\\
				$\chi_{0^+}^L(0)$ & 6.204$\times 10^{-3}$\\
				$\chi_{1^-}^T(0)$ & 5.131$\times 10^{-4}$ \text{$ GeV^{-2}$}\\
				$\chi_{1^+}^T(0)$ & 3.894$\times 10^{-4}$ \text{$ GeV^{-2}$}\\
				\hline
			\end{tabular}
			\caption{$\chi^i,\tilde{\chi}$ values that are entering into the form factor through the outer function $\phi(t)$.}
			\label{tab:chifunctions}
		\end{center}
	\end{table}
	%%%%%%%%%%%%%%%%%%%%%%%%%%%%%%%%%%%%%%%%%%
Note that the BCL expansion is a simplified series expansion. Another more general series expansion of the form factors was proposed by Boyd, Grinstein and Lebed (BGL) in ref.~\cite{Boyd:1997kz}. There are differences between the BCL and BGL parametrisation of the form factors; for detail, see the discussion in ref.~\cite{BCL}. Given the kinematics of the $\Lambda_{b} \to \Lambda_c^+\ell^- \bar{\nu}_{\ell}$ decay, one can use BGL parametrization to get the $q^2$ shape of the form factors. Following the BGL, we can obtain the shape of the form factors as given below 
\begin{equation}
	\label{eq:BGLexpn}
	F_i (q^2) = \frac{1}{P_i (q^2) \phi_i (q^2,t_0)} \sum_{j=0}^{n} a_{j}^i z^j.
\end{equation}
The conformal map from $q^2$ to $z \equiv z[q^2; t_0]$ is given by :
\begin{equation}\label{eq:z}
	z(q^2,t_0) = \frac{\sqrt{t_+-q^2}-\sqrt{t_+-t_0}}{\sqrt{t_+-q^2}+\sqrt{t_+-t_0}}\,,  
\end{equation}
where
$t_\pm \equiv (m_{\Lambda_b} \pm m_{\Lambda_c})^2$ and $t_0\equiv t_+(1-\sqrt{1-t_-/t_+})$. Here, $t_0$ is a free parameter to customise the value of $z$. The coefficients $a_{j}^i$ are the unknown parameters which need to be extracted from the available inputs. 

In eq.~\ref{eq:BGLexpn}, the variable $z$ is also related to the recoil variable $w$ ($w$ = $v_{\Lambda_b}.v_F$ with $v_{\Lambda_b}$ and $v_F$ being the four-velocities of the $\Lambda_b$ and the final state baryon respectively) as 
\begin{equation}
	z = \frac{\sqrt{w+1}-\sqrt{2}}{\sqrt{w+1}+\sqrt{2}},
\end{equation}
where $w$ is related to the momentum transferred to the dilepton system ($q^2$) as $q^2 = m_{\Lambda_b}^2 + m_F^2 - 2 m_{\Lambda_B} m_Fw$. The Blaschke factor $P_i (q^2)$ accounts for the poles in $f_i(q^2)$ for the subthreshold resonances with the same spin-parity as the current defining the form factors. Mathematically, the $P_i (q^2)$ is defined as    
\begin{equation}
	P_i(q^2) = \prod_p  z[q^2;t_p] ,
	\label{eq:Blaschke-fact}
\end{equation}
where for a single pole
\begin{equation}
	z[q^2;t_p] = \frac{z-z_p}{1 - z z_p}, \text{with}\ \  z_p \equiv z[t_p; t_0].
\end{equation} 
The role of the $P_i (q^2)$ is to eliminate the poles in $f_i(q^2)$ at the resonance masses below the threshold. Following eq.~\ref{eq:z}, we will get 
\begin{equation}
	z_p = \frac{\sqrt{(m_{\Lambda_b} + m_F)^2 - m_P^2} - \sqrt{4 m_{\Lambda_b} m_F}}{\sqrt{(m_{\Lambda_b} + m_F)^2 - m_P^2} + \sqrt{4 m_{\Lambda_b} m_F}}.
	\label{eq:zp}
\end{equation}
Here, $m_P$ denotes the pole masses and the relevant inputs with appropriate spin are presented in table~\ref{tab:BGLpolmass}. The detailed calculation of the outer functions $\phi_i (q^2;t_0)$ are available in \cite{Boyd:1997kz}, which we have presented in eqs.~\ref{eq:outerfunc}. The numerical inputs for the perturbatively calculable $\chi$-functions which appear in the outer functions are shown in table \ref{tab:chifunctions}, which we have extracted using the results in \cite{Boyd:1997kz}.

The BGL coefficients $a_j^{i}$ satisfy the following weak unitarity constraints \cite{Boyd:1997kz}, 
\begin{align}\label{eq:constraintsbgl}
	\sum_{j=0}^{n} (a^{H_V}_j)^2 +(a^{F_1}_j)^2  \leq 1 ,~
	\sum_{j=0}^{n} (a^{H_A}_j)^2 +(a^{G_1}_j)^2  \leq 1,~
	\text{and},~~\sum_{j=0}^{n} (a^{F_0}_j)^2 \leq 1,~
	~~\sum_{j=0}^{n} (a^{G_0}_j)^2 \leq 1.\\ \nn
\end{align}
In principle, by comparing the representations of the BCL and BGL expansion, one could derive inequalities in terms of BCL coefficients. However, in the BCL parametrisation of the form factors, imposing the unitarity bound is much more difficult. For example, one could see the ref.~\cite{BCL} where such relations were derived for $B\to \pi$ form factors. We could expect that the constraint relations, which include the BCL coefficients, will be more complex than in eq.~\ref{eq:constraintsbgl}, the derivation of which is beyond the scope of this paper. Also, in the ref.~\cite{Detmold:2015aaa} analysis, the authors have not used any such constraint relations.   

To check whether or not the unitary bounds on the BGL coefficients have any impact on the estimated error in the shapes of the form factors, we can utilise the inequalities in eq.~\ref{eq:constraintsbgl} as additional information in the fit to the lattice inputs on the form factors. To do so, we have created synthetic data points for the form factors at a few $q^2$ values using the fit results and the correlations given in ref. \cite{Detmold:2015aaa}. Table \ref{tab:syndat} shows the corresponding synthetic data points at values $q^2=0, 5, 10$ \text{$GeV^2$}. These synthetic data points and their respective correlations are used to extract the BGL coefficients following a $\chi^2$ minimisation. Note that in the BGL expansion, we have presented our results considering terms up to order $j=2$. We have checked that the current lattice inputs are insensitive to the higher order ($j >2$) BGL coefficients. The $\chi^2$ function is defined as
\begin{equation}
	\chi^{2}=\sum_{i,j} \big(O_i^{theory}-O_i^{lattice}\big) V_{ij}^{-1}\big(O_j^{theory}-O_j^{lattice}\big).
\end{equation}
Here, $O_i^{theory}$s are the expressions of the form factors at $q^2=0, 5,$ and $10$ \text{$GeV^2$}, respectively, parametrized in terms of the BGL coefficients. The $O_i^{lattice}$s are the respective inputs on the form factors given in table \ref{tab:syndat} and $V_{ij}$ is the covariance matrix comprising the information of the correlation and error value of the corresponding lattice inputs. Our fit results for the BGL coefficients are shown in table \ref{tab:BGLfitres}. Note that we have obtained a p-value of the fit around 58\%, which could be considered as a good fit. Using these fit results, we are able to reproduce the respective shapes of all the form factors obtained in \cite{Detmold:2015aaa}, which we have not shown here. All the fits and subsequent analyses have been carried out using a Mathematica® package\cite{OptEx}.
	%%%%%%%%%%%%%%%%%%%%%%%%
	\begin{table}[t!]
		\begin{center}
			\resizebox{0.45\textwidth}{!}{
				\begin{tabular}{|*{4}{c|}}
					\hline
					\text{Form Factor}&\multicolumn{3}{c|}{$q^2~(GeV^2)$ \bf }\\
					\cline{2-4}
					&$0$&$ 5 $ & $ 10 $   \\
					\hline\hline
					$ f_{+}$ &  $\text{0.431(49)}$& $\text{0.669(34)}$ &  $\text{1.020(32)}$\\ 
					$ f_{0}$  &  $\text{0.431(49)}$ &$\text{0.625(32)}$ &  $\text{0.902(25)}$\\
					$ f_{\perp}$ & $\text{0.588(91)}$& $\text{0.904(58)}$ &  $\text{1.377(49)}$ \\
					$ g_{+}$ &  $\text{0.387(37)}$& $\text{0.572(25)}$ &$\text{0.832(21)}$  \\
					$ g_{0}$  &  $\text{0.387(37)}$ &  $\text{0.607(28)}$&  $\text{0.936(30)}$ \\
					$ g_{\perp}$  &  $\text{0.372(39)}$ &  $\text{0.561(26)}$ &   $\text{0.830(21)}$ \\ 
					\hline
				\end{tabular}
			}
			\caption{Synthetic data for the $\Lambda_{b}\to \Lambda_{c}$ form factors generated using lattice information, BCL parameter along with their correlations.}
			\label{tab:syndat}
		\end{center}
	\end{table}	
	%%%%%%%%%%%%%%%%%%%%%%%%	
	%\paragraph{\bf Extractions of the BGL coefficients:} 
	%%%%%%%%%%%%%%%%%%%%%%%%%%%%%%%%%%%%%5
	\begin{table}[t!]
		\begin{center}
			\resizebox{0.45\textwidth}{!}{
				\begin{tabular}{|*{4}{c|}}
					\hline
					$\text{BGL Coeff.}$ &\text{Fit Results}&$\text{BGL Coeff.}$ &\text{Fit Results}\\
					\hline\hline				
					$ a_{F_0}^0$ &$\text{0.048(1)}$ & $ a_{F_1}^0$ &$\text{0.040(2)}$\\
					$a_{F_0}^1$  &$\text{-0.49(9)}$ & $ a_{F_1}^1$  &$\text{-0.4(1)}$\\
					$ a_{G_0}^0$ &$\text{0.067(2)}$ & $ a_{F_1}^2$ &$\text{0.8(11)}$\\
					$ a_{G_0}^1$ &$\text{-0.7(1)}$&$a_{H_V}^0$&$\text{0.0111(4)}$\\
					$ a_{G_1}^0$  &$\text{0.0161(4)}$&$ a_{H_V}^1$&$\text{-0.13(2)}$\\
					$ a_{G_1}^1$ &$\text{-0.15(2)}$& $ a_{H_V}^2$&$\text{0.3(3)}$\\
					$ a_{G_1}^2$  &$\text{0.10(3)}$&& \\
					$ a_{H_A}^1$ &$\text{-0.024(3)}$&&\\
					$ a_{H_A}^2$ &$\text{0.03(3)}$&&\\
					\hline
					$\text{DOF}$ &$\text{2}$&$\text{p-Value}$&$\text{0.6}$  \\	     
					\hline
				\end{tabular}
			}
			\caption{Fit results for the BGL coefficients $a^n_i$ corresponding to the $\Lambda_{b}\to\Lambda_{c}$ transition form factors. We took $n=2$ for form factor BGL parameterization, exploiting the QCD constraints at zero and large recoil.}
			\label{tab:BGLfitres}
		\end{center}
	\end{table} 
	
As an alternative approach to get the shapes of the form factors, one could see the analysis of ref. \cite{Bernlochner:2023jkp}. Using the lattice results, the $\Lambda_b \to \Lambda_c$ form factors have been estimated in the framework of heavy quark effective theory (HQET). The HQET form factors are obtained in terms of heavy quark symmetry (HQS) functions following residual chiral expansion (RCE)-based parametrizations at order $\mathcal{O}(\alpha_s^2, \alpha_s/m_{c,b}, 1/m_{c,b}^2)$. The free parameters of the model were extracted from the lattice inputs on the form factors and the LHCb data on the measurement of the normalized differential spectrum $(1/\Gamma)\times d\Gamma(\Lambda_b\to \Lambda_c \mu\nu)/dq^2$.

\section{Observables: Standard Model Predictions}\label{sec:SMprd}
	%%%%%%%%%%%%%%%%%%%%%%%%%%%%%%%%%%%%%%%%%%%%%%%5
	\begin{figure}[h!]
		\centering
		\includegraphics[scale=.39]{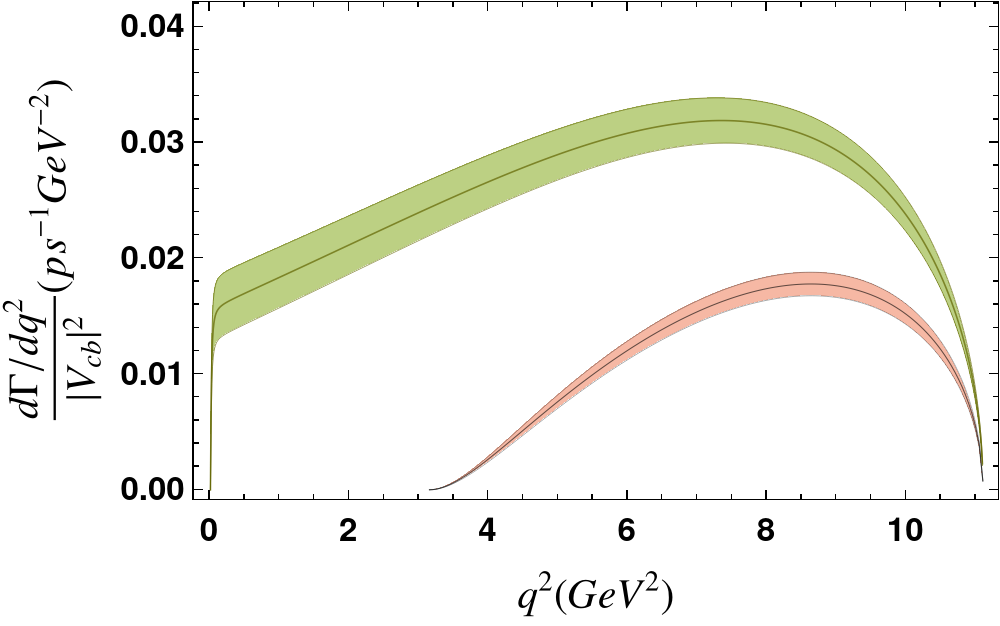}~~
		\includegraphics[scale=.39]{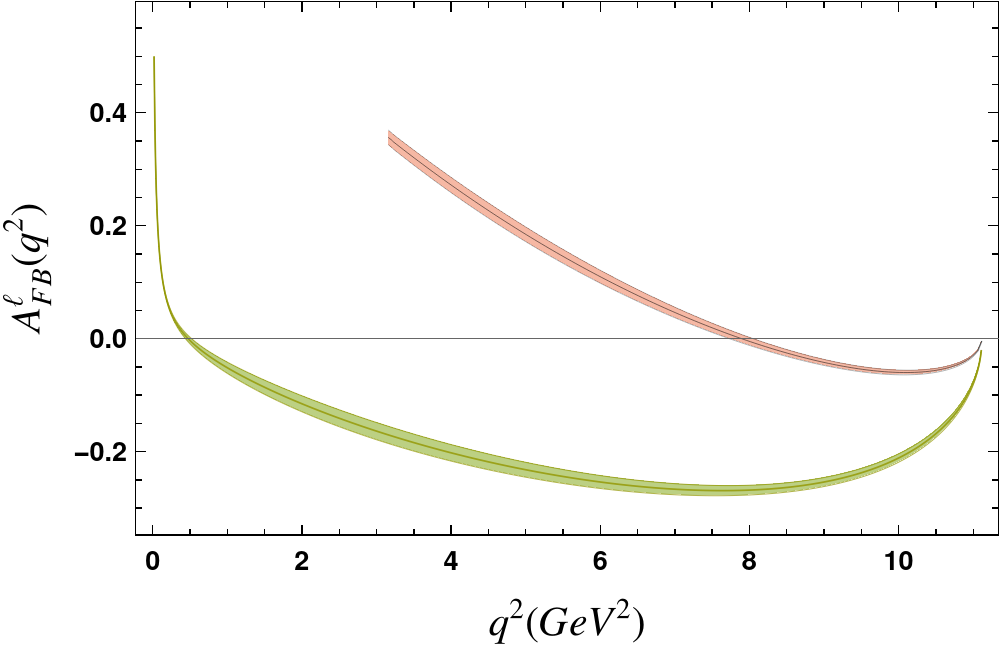}~~
		\includegraphics[scale=.39]{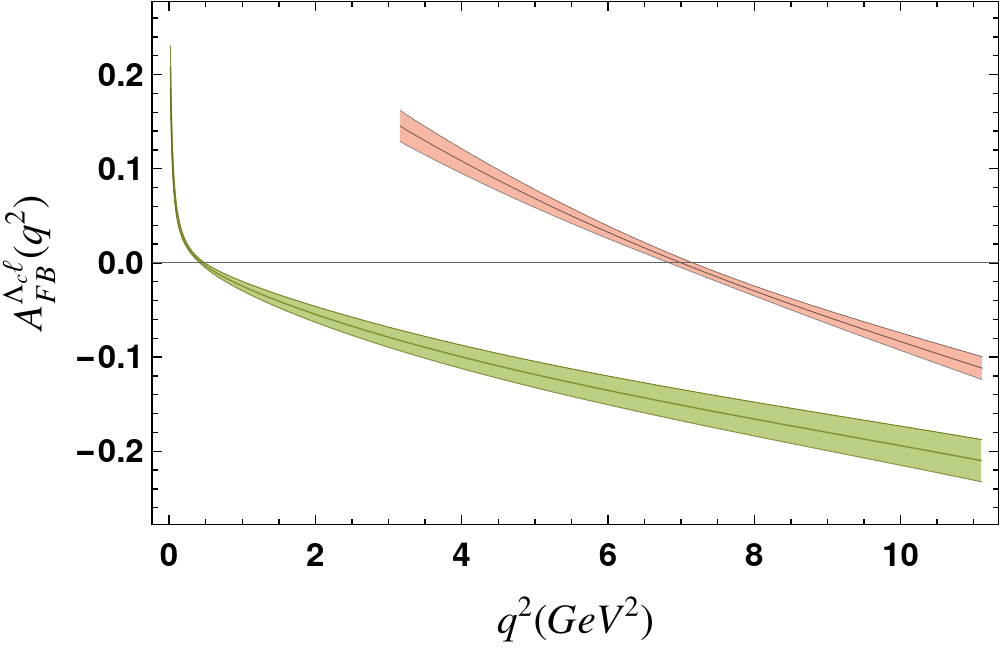}\\
		\includegraphics[scale=.39]{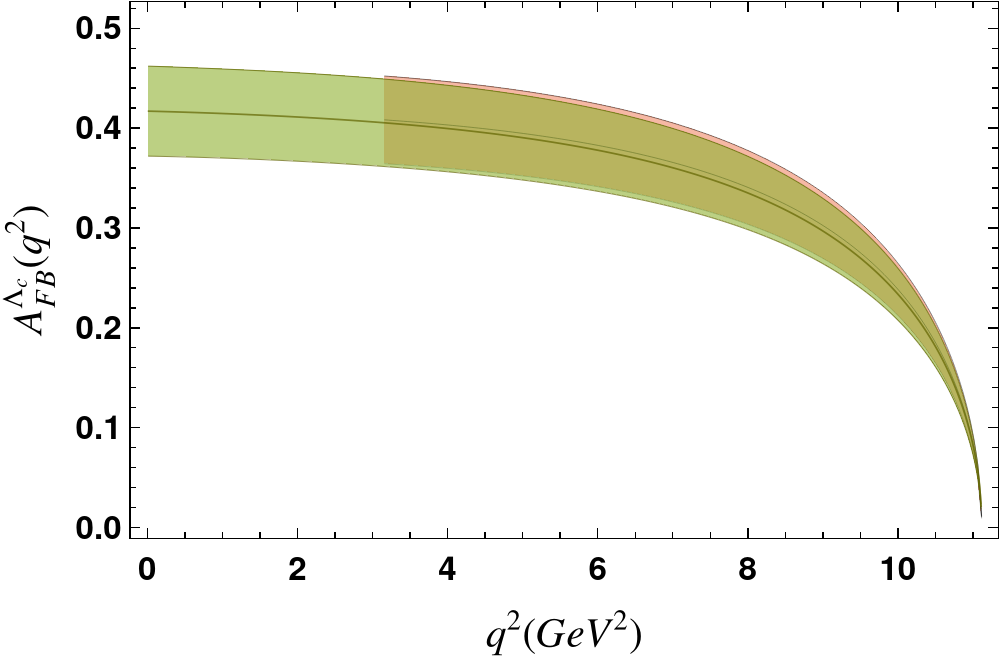}~~
		\includegraphics[scale=.39]{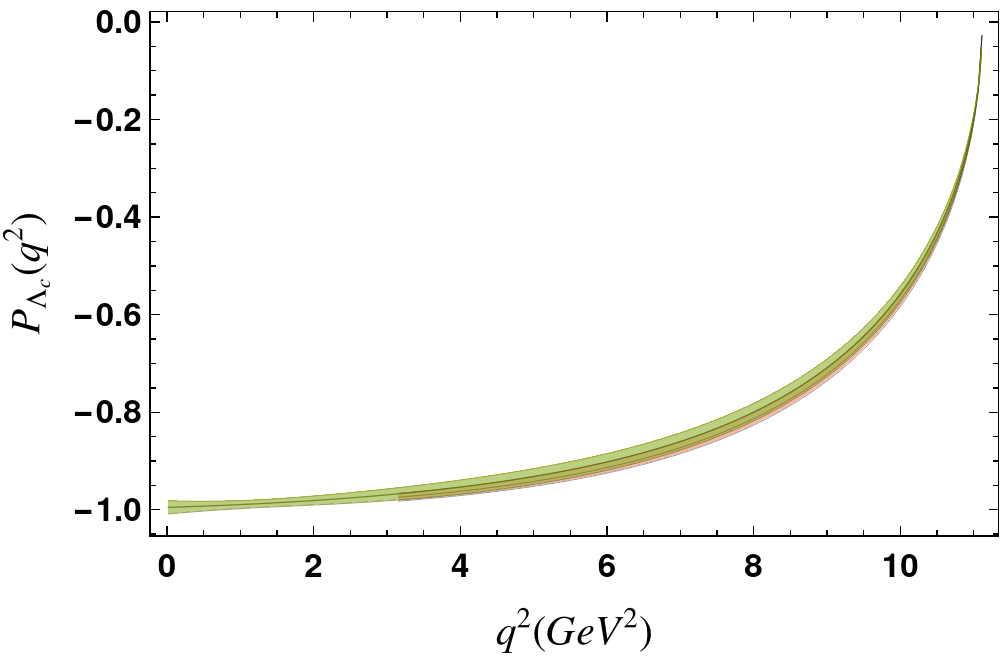}~~
		\includegraphics[scale=.39]{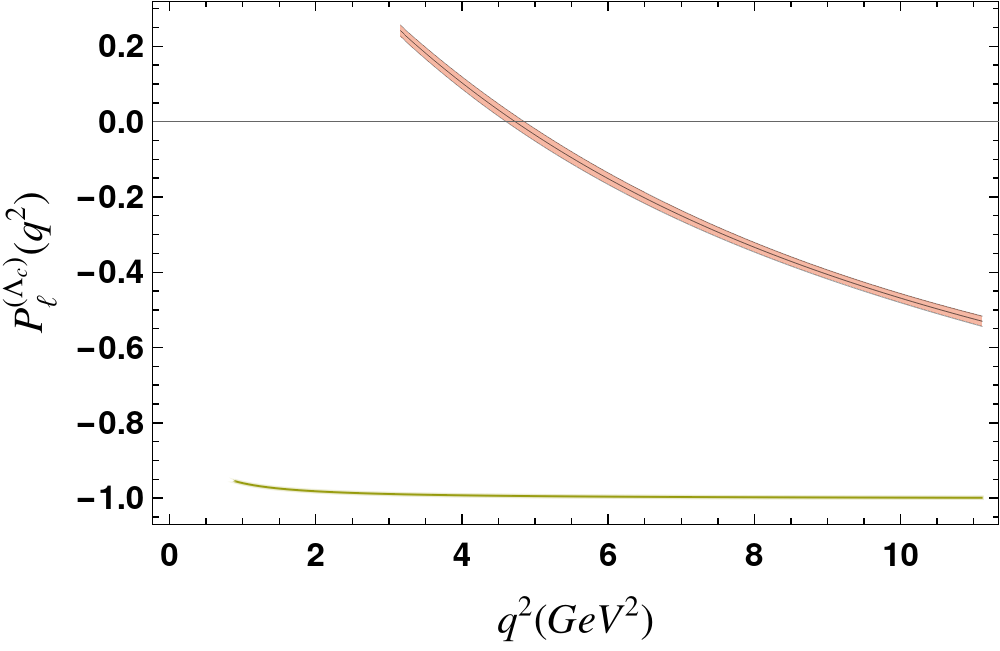}~~\\
		\includegraphics[scale=.45]{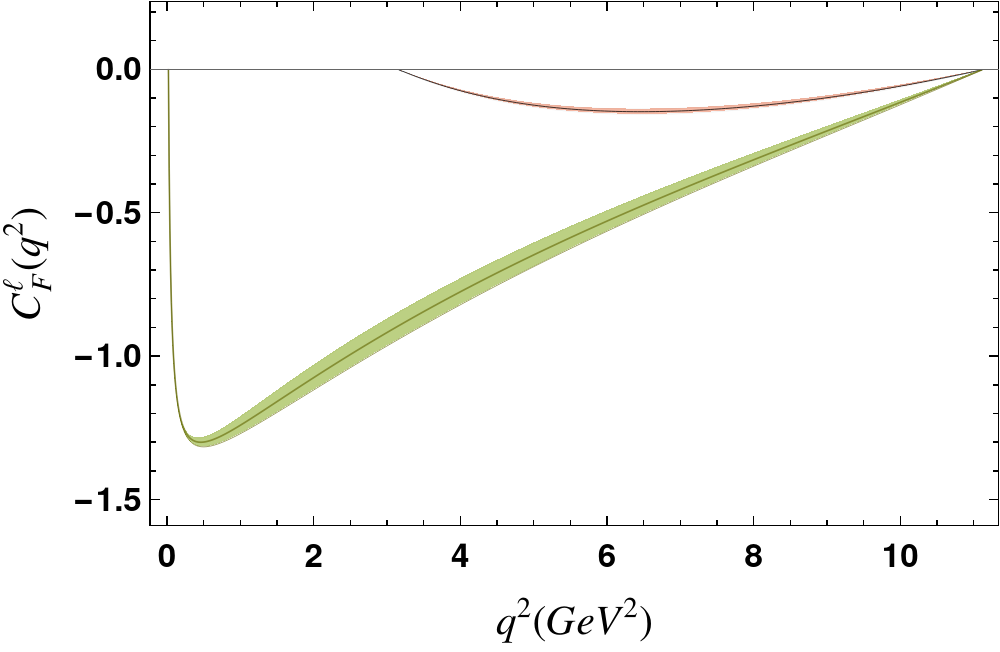}
		\caption{Our estimates for the differential decay distribution, forward-backward asymmetry, and polarization observables for lepton, as well as baryon and convexity observables respectively within BGL parameterization for a tauon (brown) and a muon (green) in the final state in the SM, and the color bands correspond to the 68\% confidence interval.}
		\label{fig:SMobsplots}
	\end{figure}
%%%%%%%%%%%%%%%%%%%%%%%%%%%%%%%%
Using the $q^2$-shapes of the form factors obtained in BCL and BGL parametrisations, we have analysed the $q^2$ distributions of the decay rates and the other angular observables. In both the parametrisations, we have obtained identical results that are difficult to distinguish. Hence, we have not presented these distributions separately in both the parametrisations. In fig.~\ref{fig:SMobsplots}, we have shown the results obtained from the analysis with the BGL parametrisation. The respective $q^2$ distributions are shown for the final state lepton $\ell = \mu$ and $\tau$. The bands indicate the respective 1$\sigma$ error bars.

Furthermore, we have predicted the $q^2$ integrated decay rates and the angular observables in the SM and in the presence of NP. We normalise each observable with the decay rate, which is $2K_{1ss}+K_{1cc}$. This makes the observables independent of $|V_{cb}|$. The observables are then dependent only on the hadronic helicity amplitudes, which are parametrised by the relevant form factors. Here, we have compared the results in both parametrisations, which will help us understand whether or not the predictions are independent of the respective parametrisation of the form factors. The numerical estimates in different small $q^2$-bins are presented in table \ref{tab:binpredmmuSM} for the final state with a muon and in table \ref{tab:binpredmtauSM} for a tauon in the final state, respectively. Note that we have presented the $q^2$ integrated values in small bins. Also, we have shown the SM predictions integrated in the full $q^2$ region. We have shown the SM predictions for the rates $\Gamma ( \Lambda_b\to \Lambda_c^+(\to \Lambda \pi^+)\ell^-\bar{\nu}_{\ell})$ normalized by the CKM element $|V_{cb}|^2$. One can extract the respective number depending on the value of $|V_{cb}|$. Following are a few comments based on the $q^2$ distributions or the predictions of the observables in small $q^2$ bins.
	
\begin{itemize}
		\item $\frac{\rm d\Gamma~}{\rm dq^2~}$: In fig. \ref{fig:SMobsplots}, we have shown the $q^2$ dependence of the differential decay rate for $\Lambda_{b}\to \Lambda_{c}^+(\to \Lambda \pi^+)\ell^- \bar{\nu}_{\ell}$ decay within the standard model. The differential ratio shows a step-like behaviour for the muon mode when $q^2=m_{\mu}^2$. And at the zero recoil, $q^2_{\text{max}}=(m_{\Lambda_{b}}-m_{\Lambda_{c}})^2$, the differential decay rate for both the modes, $\mu$ and $\tau$ converge to zero because of the unavailability of the phase-space.
		
		\item $A_{FB}^{\ell}(q^2)$: At the zero recoil, the forward-backward asymmetries of the leptonic side, $A_{FB}^{\ell}(q^2)$, approaches zero for both the leptonic modes $\mu$ and $\tau$. However, at $q^2=m_{\ell}^2$, the asymmetry takes the value	
		\begin{equation}
			A_{FB}^{\mu}(q^2)\big\vert_{q^2=m^2_{\mu}} = 0.4994(1), \ \  A_{FB}^{\tau}(q^2)\big\vert_{q^2=m^2_{\tau}} = 0.36(1).
		\end{equation}
		Furthermore, $A_{FB}^{\tau}(q^2)$ is positive for most of the kinematic range, and zero crossing occurs at $\bf q^2_0=7.92(17)GeV^2$. It has a small negative value at the high $q^2$ region. Whereas, $A_{FB}^{\mu}(q^2)$ is negative for all of the kinematic range and became positive at large recoil, and zero crossing occurs at  $\bf q^2_0=0.48(5) GeV^2$. 
		
		\item $A_{FB}^{\Lambda_c}(q^2)$: Both the leptonic final states ($\mu$ and $\tau$) have identical $q^2$ distributions, which we can not distinguish. This is due to the absence of any threshold effects with respect to the dilepton invariant mass square $q^2$. Hence, any deviation from the above would exhibit large NP effects, which we will discuss later.
						
		\item $A_{FB}^{\Lambda_c \ell}(q^2)$: For both the leptonic final states, these asymmetries have zero crossings. For taunic mode, the zero crossing occurs at $\bf q^2_0=7.01(17)~GeV^2$ whereas $A_{FB}^{\Lambda_c \mu}(q^2)$ has a negative value for most of the kinematic range and zero crossing occurs at $\bf q^2_0=0.45(4)~GeV^2 $.
		
		\item $P_{\Lambda_c}(q^2)$: We could see that the $q^2$ distributions of the $\Lambda_{c}$ polarization asymmetries for both the leptonic channels overlap. This longitudinal hadronic polarization is independent of the lepton flavour (as expected). The corresponding asymmetries have zero values near the zero recoil, while they approach -1 in the large recoil regions. 
		%%%%%%%%%%%%%%%%%%%%%%%%%%%%%%
		\begin{table}[t]
			\begin{center}
				\resizebox{0.8\textwidth}{!}{
					\begin{tabular}{c|c|ccccccc}
						\hline\hline
						\multirow{1}{*}{Obs.}& \multirow{2}{*}{Parametrization}&\multicolumn{6}{c}{$q^2$ Bin~(in \textbf{$GeV^2$})}\\
						\cline{3-8}
						&&~~$\textbf{\text{$\bf q^2_{min.}$}-3}$~~~&~~~~\text{$\bf 3-5$}~~~~ &~~~~\text{$\bf 5-7$}~~& ~~~~\text{$\bf 7-9$}~~~~&~~\text{$\bf 9-q^2_{max.}$}~~&~~\text{$\bf q^2_{min.}-q^2_{max.}$}~~\\
						\hline
						\multirow{2}{*}{\bf $\langle \frac{\Gamma}{ |V_{cb}|^2} \rangle$}&
						\textbf{BCL} & $\text{0.059(8)}$  &  $\text{0.053(5)}$  &  $\text{0.061(4)}$  &  $\text{0.063(4)}$  &  $\text{0.046(3)}$  &  $\text{0.281(20)}$  \\
						\cline{2-8}
						&\textbf{BGL} & $\text{0.059(8)}$  &  $\text{0.053(5)}$  &  $\text{0.061(4)}$  &  $\text{0.062(4)}$  &  $\text{0.045(3)}$  &  $\text{0.281(20)}$  \\
						\hline
						\multirow{2}{*}{\bf $ A_{FB}^{\mu}=\langle A_{FB}^{\mu}(q^2) \rangle$}
						&\textbf{BCL}  &$\text{0.149(8)}$  &  $\text{-0.199(15)}$  &  $\text{-0.251(11)}$  \
						&  $\text{-0.264(9)}$  &  $\text{-0.189(7)}$  & tab. \ref{tab:belowabovezero} \\
						\cline{2-8}
						&\textbf{BGL} &  $\text{0.149(7)}$  &  $\text{-0.199(15)}$  &  $\text{-0.251(11)}$  \
						&  $\text{-0.264(9)}$  &  $\text{-0.189(7)}$  &  - \\	
						\hline
						\multirow{2}{*}{\bf $A_{FB}^{\Lambda_c \mu}=\langle A_{FB}^{\Lambda_c \mu}(q^2)\rangle$} &\textbf{BCL}   & $\text{0.060(7)}$  &  $\text{-0.098(12)}$  &  $\text{-0.135(15)}$  \
						&  $\text{-0.165(18)}$  &  $\text{-0.195(21)}$  &  tab. \ref{tab:belowabovezero}\\
						\cline{2-8}
						&\textbf{BGL} &$\text{0.060(7)}$  &  $\text{-0.098(12)}$  &  $\text{-0.135(15)}$  \
						&  $\text{-0.165(18)}$  &  $\text{-0.195(21)}$  & - \\
						\hline
						\multirow{2}{*}{\bf $A_{FB}^{\Lambda_c}=\langle A_{FB}^{\Lambda_c}(q^2)\rangle$} &\textbf{BCL}   & $\text{0.416(45)}$  &  $\text{0.400(43)}$  &  $\text{0.377(41)}$  \
						&  $\text{0.333(36)}$  &  $\text{0.212(24)}$  &  $\text{0.387(42)}$\\
						\cline{2-8}
						&\textbf{BGL} & $\text{0.416(45)}$  &  $\text{0.400(43)}$  &  $\text{0.377(41)}$  &  $\text{0.333(36)}$  &  $\text{0.212(23)}$  \
						&  $\text{0.387(42)}$\\
						\hline
						\multirow{2}{*}{\bf $P_{\Lambda_{c}}=\langle P_{\Lambda_{c}}(q^2)\rangle $}& \textbf{BCL}  & $\text{-0.982(8)}$  &  $\text{-0.950(14)}$  &  $\text{-0.896(17)}$  \
						&  $\text{-0.789(18)}$  &  $\text{-0.495(15)}$  &  $\text{-0.757(17)}$\\
						\cline{2-8}
						&\textbf{BGL} & $\text{-0.982(8)}$  &  $\text{-0.950(14)}$  &  $\text{-0.897(17)}$  &  $\text{-0.790(18)}$  &  $\text{-0.495(15)}$  \
						&  $\text{-0.758(17)}$  \\
						\hline
						\multirow{2}{*}{\bf $P_{\mu}^{(\Lambda_{c})}=\langle P_{\mu}^{(\Lambda_{c})}(q^2)\rangle$}& \textbf{BCL}  & $\text{-0.937(3)}$  &  $\text{-0.9914(2)}$  &  $\text{-0.9948(2)}$  &  $\text{-0.9965(1)}$  &  $\text{-0.9975(1)}$  &  $\text{-0.988(1)}$ \\
						\cline{2-8}
						& \textbf{BGL}  & $\text{-0.937(2)}$  &  $\text{-0.9913(3)}$  &  $\text{-0.9948(2)}$  &  $\text{-0.9965(1)}$  &  $\text{-0.9975(1)}$  \
						&  $\text{-0.988(1)}$ \\
						\hline
						\multirow{2}{*}{\bf $C_F^{\mu}=\langle C_F^{\mu}(q^2)\rangle $} &\textbf{BCL}& $\text{-0.854(10)}$  &  $\text{-0.780(45)}$  &  $\text{-0.529(36)}$  \
						&  $\text{-0.313(23)}$  &  $\text{-0.107(9)}$  &  $\text{-0.711(24)}$ \\
						\cline{2-8}
						&\textbf{BGL}& $\text{-0.854(10)}$  &  $\text{-0.779(45)}$  &  $\text{-0.529(35)}$  \
						&  $\text{-0.313(22)}$  &  $\text{-0.107(8)}$  &  $\text{-0.711(24)}$  \\
						\hline\hline
					\end{tabular}
				}
				\caption{Bin-wise observables prediction for$\Lambda_b \to\Lambda_c^+(\to \Lambda \pi^+) \mu \bar{\nu}_{\mu}$ mode.}
				\label{tab:binpredmmuSM}
			\end{center}
		\end{table}
	%%%%%%%%%%%%%%%%%%%%%%%%%%%%%%%%%%%%%%%%%%%%%%%
		\item $P_{\ell}^{(\Lambda_{c})}(q^2)$: For the longitudinal lepton polarization of the charged lepton, it is found that for muon mode, this observable has the least variation and has $P_{\mu}^{(\Lambda_{c})}(q^2) \simeq -1$ almost constant value in the whole region. Meanwhile, the behaviour in the tauonic mode is quite different from that in the former. $P_{\tau}^{(\Lambda_{c})}(q^2)= -0.529(13)$ at the zero recoil, and $P_{\tau}^{(\Lambda_{c})}(q^2)=0.241(15)$ at the $q^2=m^2_{\tau}$, and zero crossing occurs at $\bf q^2_0= 4.73(13)$ \text{$\bf GeV^2$}.
		
		\item $C_F^{\ell}(q^2)$: The convexity parameter has zero value for both the decay modes at zero recoil point. At the large recoil range at $q^2=0.45$ \text{$GeV^2$}, $C_F^{\mu}(q^2)$ reaches its maximum value $-1.297(16)$, and due to the lepton mass effect, it becomes zero suddenly at $q^2$=$m_{\mu}^2$. Whereas $C_F^{\tau}(q^2)$ has a very low value throughout the $q^2$ region and reaches its maximum value, $-0.146(9)$ at $q^2=6.45$ \text{$GeV^2$}.
		
		\item Note that the predictions shown in tables \ref{tab:binpredmmuSM} and  \ref{tab:binpredmtauSM}, respectively, are consistent in both the parametrisation of the form factors. In $q^2$-bins, we predict the normalised decay rates and the angular observables in both the leptonic modes with a 1$\sigma$ error $\lesssim 10$\%. Also, for most observables, the predictions obtained after integration in the full $q^2$ regions have an error $\approx 10\%$ at their 1$\sigma$ C.I.   
		
		\item Among the angular observables, we could predict, for both the lepton final states, the $\Lambda_c$ polarization asymmetry with a 1$\sigma$ error $\lesssim 2$\%. 
		
		\item We obtain the muon polarization asymmetry $P_{\mu}^{(\Lambda_{c})}(q^2)$ with a 1$\sigma$ error of 0.1\% while the tauon polarization asymmetry has an error 6\% at 1$\sigma$.   
		
		\item The $q^2$ distributions of $A_{FB}^{\mu}(q^2)$ and $A_{FB}^{\Lambda_c \mu}(q^2)$ in fig.~\ref{fig:SMobsplots} show that for $q^2 \to m_{\mu}^2$ the corresponding values suddenly increases. This region of $q^2$ is close to the zero crossing of $A_{FB}^{\mu}(q^2)$ and $A_{FB}^{\Lambda_c \mu}(q^2)$, respectively. Hence, for these observables, instead of integrating over the full $q^2$ regions, we have predicted them for $q^2 > q_0^2$ and $q^2 < q_0^2$ which are shown in table \ref{tab:belowabovezero}. In the same table, we have also given the predictions $A_{FB}^{\tau}(q^2)$ and $A_{FB}^{\Lambda_c \tau}(q^2)$ for $q^2 > q_0^2$ and $q^2 < q_0^2$ where $q_0^2$ are the corresponding values of $q^2$ at the zero crossings.      
		
	\end{itemize}

	\begin{table}[t]
		\renewcommand{\arraystretch}{1.4}
		\begin{center}
			\resizebox{0.75\textwidth}{!}{
				\begin{tabular}{c|c|cccccc}
					\hline\hline
					\multirow{1}{*}{Obs.}& \multirow{2}{*}{Parametrization}& \multicolumn{5}{|c}{$q^2$ bin~(in \text{$GeV^2$})}\\
					\cline{3-7}
					&&~~~~\textbf{$\bf q^2_{min.}$-5}~~~~&~~~~\textbf{$\bf 5-7$}~~&~~~~\textbf{\bf 7-9}~~~~&~~\text{\bf 9-\text{$\bf q^2_{max.}$}}~~&~~\text{$\bf q^2_{min.}$}-\text{$\bf q^2_{max.}$}~~\\
					\hline
					\multirow{2}{*}{\bf $\langle \frac{\Gamma}{|V_{cb}|^2} \rangle$}
					&\textbf{BCL}  &  $\text{0.0059(5)}$  &  $\text{0.024(2)}$  &  $\text{0.034(2)}$  &  $\text{0.029(2)}$  &  $\text{0.092(5)}$  \\
					\cline{2-7}
					&\textbf{BGL} &  $\text{0.0060(5)}$  &  $\text{0.024(2)}$  &  $\text{0.034(2)}$  &  $\text{0.029(2)}$  &  $\text{0.093(5)}$    \\
					\hline
					\multirow{2}{*}{\bf $ A_{FB}^{\tau}=\langle A_{FB}^{\tau}(q^2)\rangle$}
					&\textbf{BCL} &  $\text{0.272(14)}$  &  $\text{0.113(12)}$  &  $\text{-0.002(7)}$  \
					&  $\text{-0.050(4)}$  &  tab. \ref{tab:belowabovezero}\\	
					\cline{2-7}
					&\textbf{BGL}  &  $\text{0.273(14)}$  &  $\text{0.115(12)}$  &  $\text{-0.001(7)}$  &  $\text{-0.050(4)}$  &  - \\
					\hline
					\multirow{2}{*}{\bf $A_{FB}^{\Lambda_c \tau}= \langle A_{FB}^{\Lambda_c \tau}(q^2)\rangle$} &\textbf{BCL}   &  $\text{0.108(13)}$  &  $\text{0.034(7)}$  &  $\text{-0.029(5)}$  \
					&  $\text{-0.084(10)}$  &  tab. \ref{tab:belowabovezero}\\
					\cline{2-7}
					&\textbf{BGL}  &  $\text{0.109(13)}$  &  $\text{0.035(7)}$  &  $\text{-0.029(5)}$  \
					&  $\text{-0.084(10)}$  &  -  \\
					\hline
					\multirow{2}{*}{\bf $A_{FB}^{\Lambda_c}= \langle A_{FB}^{\Lambda_c}(q^2)\rangle$} & \textbf{BCL}   &  $\text{0.403(43)}$  &  $\text{0.382(41)}$  &  $\text{0.339(37)}$  \
					&  $\text{0.218(24)}$  &  $\text{0.344(37)}$ \\
					\cline{2-7}
					&\textbf{BGL}  &  $\text{0.403(43)}$  &  $\text{0.383(41)}$  &  $\text{0.339(37)}$  &  $\text{0.218(24)}$  &  $\text{0.345(37)}$ \\
					\hline
					\multirow{2}{*}{\bf $P_{\Lambda_{c}}= \langle P_{\Lambda_{c}}(q^2)\rangle$}& \textbf{BCL}    &  $\text{-0.958(9)}$  &  $\text{-0.908(11)}$  &  $\text{-0.804(13)}$  \
					&  $\text{-0.509(12)}$  &  $\text{-0.761(12)}$  \\
					\cline{2-7}
					&\textbf{BGL}  &  $\text{-0.958(9)}$  &  $\text{-0.908(11)}$  &  $\text{-0.804(13)}$  &  $\text{-0.509(12)}$  &  $\text{-0.761(12)}$\\
					\hline
					\multirow{2}{*}{\bf $P_{\tau}^{(\Lambda_{c})}= \langle P_{\tau}^{(\Lambda_{c})}(q^2)\rangle$}&\textbf{BCL}& $\text{0.088(17)}$  &  $\text{-0.156(16)}$  &  $\text{-0.336(13)}$  \
					&  $\text{-0.472(12)}$  &  $\text{-0.258(16)}$  \\
					\cline{2-7}
					&\textbf{BGL}  &  $\text{0.095(17)}$  &  $\text{-0.150(16)}$  &  $\text{-0.333(13)}$  \
					&  $\text{-0.471(12)}$  &  $\text{-0.254(16)}$  \\
					\hline
					\multirow{2}{*}{\bf $C_F^{\tau}=\langle C_F^{\tau}(q^2)\rangle$} &\textbf{BCL}& $\text{-0.069(4)}$  &  $\text{-0.141(9)}$  &  $\text{-0.125(9)}$  \
					&  $\text{-0.053(4)}$  &  $\text{-0.096(6)}$  \\
					\cline{2-7}
					&\textbf{BGL}& $\text{-0.069(3)}$  &  $\text{-0.140(9)}$  &  $\text{-0.125(8)}$  \
					&  $\text{-0.053(4)}$  &  $\text{-0.096(6)}$    \\
					\hline\hline
				\end{tabular}
			}
			\caption{Bin-wise observables prediction for $\Lambda_b \to\Lambda_c^+(\to \Lambda \pi^+) \tau \bar{\nu}_{\tau}$ mode}
			\label{tab:binpredmtauSM}
		\end{center}
	\end{table}
	%%%%%%%%%%
	%%%%%%%%%%%%%%%%%%%
	\begin{table}[h!]
		\renewcommand{\arraystretch}{2.}
		\begin{center}
			\resizebox{0.35\textwidth}{!}{
				\begin{tabular}{c|c|c|c}
					\hline
					\text{Modes}& \text{$q^2$ Range}&~~\textbf{$\langle A_{FB}^{\ell}(q^2)\rangle$}~~&~~$\langle A_{FB}^{\Lambda_c \ell}(q^2)\rangle$\\
					\hline
					\multirow{2}{*}{$\tau$}
					&\text{[$q^2_{min}- q^2_0$]} &  $\text{0.174(13)}$& $\text{0.075(10)}$\\
					&\text{[$q^2_0-q^2_{max}$]}& $\text{-0.041(4)}$& $\text{-0.057(7)}$\\
					\hline
					\multirow{2}{*}{$\mu$} &\text{[$q^2_{min}- q^2_0$]}   &$\text{0.246(2)}$  & $\text{0.104(11)}$\\
					&\text{[$q^2_0-q^2_{max}$]} & $\text{-0.170(11)}$ &$\text{-0.106(13)}$ \\
					\hline
			\end{tabular}}
			\caption{We provide predictions for the asymmetric observables, $\langle A_{FB}^{\ell}(q^2)\rangle$ and $\langle A_{FB}^{\Lambda_c \ell}(q^2)\rangle$, on both sides of their respective zero crossing values ($q_0$) in $\tau$ and $\mu$ modes. Refer to the text for the observables' zero crossing value ($q_0$).}
			\label{tab:belowabovezero}
		\end{center}
	\end{table}

As mentioned earlier, there are a couple of other angular observables, like $K_{3s}(q^2)$, $K_{3sc}(q^2)$, $K_{2ss}(q^2)$, $K_{2cc}(q^2)$, $K_{2c}(q^2)$, $K_{1ss}(q^2)$, $K_{1cc}(q^2)$, $K_{1c}(q^2)$, which could be extracted from the given angular distribution in eq.~\ref{eq:ang_dist}. The SM predictions of all these observables are given in the table~\ref{tab:predangular_oneopr}, \ref{tab:predangular_twoopr}.    
	
	Apart from the predictions of decay rates and angular observables, we have predicted the ratio of the decay rates
	\begin{equation}
		R(\Lambda_c) = \frac{ \mathcal{B}(\Lambda_{b}\to \Lambda_{c}\tau \overline{\nu}_{\tau})}{ \mathcal{B}(\Lambda_{b}\to \Lambda_{c}\mu \overline{\nu}_{\mu})}. 
	\end{equation}
	
	In the SM, with the BGL parametrization of the form factors, our prediction is given by 
	\begin{equation}
		R(\Lambda_{c}) = 0.330\pm 0.010. 
	\end{equation}
	which is consistent with the respective value obtained in \cite{Detmold:2015aaa} where BCL parametrization had been used. Therefore, the unitarity relations described earlier for the BGL coefficients have negligible impact on all the predictions. Furthermore, we have estimated the following $q^2$ integrated branching fractions:
	\begin{equation}
		\frac{1}{|V_{cb}|^2} \times \mathcal{B}(\Lambda_{b}\to \Lambda_{c}(\to \Lambda \pi)\tau \bar{\nu}_{\tau})_{\text{SM}}= 0.136(8) , ~~~~~~
		\frac{1}{|V_{cb}|^2} \times \mathcal{B}(\Lambda_{b}\to \Lambda_{c}(\to \Lambda \pi)\mu \bar{\nu}_{\mu})_{\text{SM}}=0.413(30).
	\end{equation}
	The above estimates with $|V_{cb}|=40.3(5)\times10^{-3}$ as in ref~\cite{Ray:2023xjn} and $\tau_{\Lambda_{b}}^{\text{HFLAV}}=1.471(9)~\text{ps}~$\cite{HFLAV:2022esi} are given by 
	\begin{equation}
		\mathcal{B}(\Lambda_{b}\to \Lambda_{c}(\to \Lambda \pi)\tau \bar{\nu}_{\tau})_{\text{SM}}= 2.21(14)\times 10^{-4}~~~\text{and},~~~
		\mathcal{B}(\Lambda_{b}\to \Lambda_{c}(\to \Lambda \pi)\mu \bar{\nu}_{\mu})_{\text{SM}}=6.70(52)\times 10^{-4}.
	\end{equation}
	%%%%%%%%%%%%%%%%%%%%%%
	
	\section{Test of New Physics with existing observations: One and two Operator scenario}\label{sec:NPtst}
	
One of our important goals in this analysis is to test the new physics sensitivities of the different observables related to $\Lambda_b \to\Lambda_c^+ \tau^- \bar{\nu}$ decay. We need benchmark value(s) for the new WCs defined in eq. \ref{eq:heff}, which will contribute to this decay. Instead of randomly choosing the benchmarks, we can use the available inputs on the other related channels with similar NP effects to get a reasonable estimate of the relevant WCs. As we have mentioned earlier, a few other lepton flavor violating ratios, like $R(D^{(*)})=\Gamma(B \to D^{(*)} \tau^- \bar{\nu})/\Gamma(B \to D^{(*)} \ell^- \bar{\nu})$ and $R(J/\psi)=\Gamma(B_c \to J/\psi\tau^- \bar{\nu})/\Gamma(B_c\to J/\psi\ell^- \bar{\nu})$ (with $\ell^-=\mu^-, e^-$) are measured by the experimental collaborations BaBar, Belle, and LHCb, respectively \cite{BaBar:2012obs, BaBar:2013mob, Belle:2015qfa, Belle:2016ure, Belle:2016dyj, Belle:2017ilt, Belle:2019rba, LHCb:2015gmp, LHCb:2017smo, LHCb:2017rln, LHCb:2023zxo}\footnote{In the experimental analysis,  the semileptonic rates or ratios like $R(D^*)$ are obtained via a fit to large resource-intensive Monte Carlo (MC) samples, which incorporate detailed simulations of detector responses and physics backgrounds. Therefore, the extracted parameters may be susceptible to the underlying theoretical models used in the MC generation. All existing measurements of $R(D^*)$ rely heavily on the MC fit template and, hence, the underlying theory model, the SM. Usually, in the literature, the measured values of $R(D^{(*)})$ are used to fit the new WCs, which in practice could introduce bias since the experimental analysis of the template model is the SM.   The Hammer software tool developed in ref. \cite{Bernlochner:2020tfi} enables efficient reweighting of MC samples to arbitrary NP scenarios or to any hadronic matrix elements. In the near future, experimental collaboration could use this tool to extract new physics WCs. Note that the probability of bias could be more important when we look for large NP effects. However, for small NP effects, the impact may not be that high, which could also be seen from the results of the toy analysis in \cite{Bernlochner:2020tfi}. As we will see later, the current data do not allow significant NP effects.}. Following are the respective averages estimated by HFLAV \cite{HeavyFlavorAveragingGroup:2022wzx} 
	\begin{align}
		R(D)&=0.357\pm 0.029,\\
		R(D^*)&=0.284\pm 0.012.
	\end{align}
	Based on the most recent lattice inputs, the corresponding SM predictions are given by \cite{Ray:2023xjn}
	\begin{align}
		R(D)_{SM}&=0.304\pm 0.003,\\
		R(D^*)_{SM}&=0.258 \pm 0.012.
	\end{align}
	At the moment, the measurements have relatively large errors, and the SM prediction of $R(D^*)$ has a relatively large error as compared to $R(D)$. In the present scenario, the data deviates from the respective SM predictions at $\approx 2 \sigma$. Note that the measurements on $R(D)$ and $R(D^*)$ also have some correlation due to which there will be little more discrepancies than $2\sigma$ \cite{HeavyFlavorAveragingGroup:2022wzx}. 
	
Recently, LHCb collaboration has measured the branching fraction $\mathcal{B}(\Lambda_{b}\to \Lambda_{c} \tau \bar{\nu})$ and finds \cite{LHCb:2022piu},
	\begin{equation}
		R(\Lambda_{c})=0.242\pm 0.026\pm 0.040 \pm 0.059,
	\end{equation} 
	where the first uncertainty is statistical, the second one is systematic and the third is due to some other inputs used in the analysis. In comparison to our SM prediction, 
	\begin{equation}
		R(\Lambda_{c})_{SM}=0.330\pm 0.010.
	\end{equation}
	It points towards a downward shift from the SM and the data is consistent with the SM prediction at 1.15$\sigma$ level. 
	
	Note that in the measurement for $R(\Lambda_c)$, to normalize $\Lambda_b \to \Lambda_c \tau \nu$ decay rate, LHCb have used $\mathcal{B}(\Lambda_b \to \Lambda_c \mu \nu)_{\text{DELPHI}}$= $6.2(1.4)\% $ \cite{DELPHI:2003qft}. Using this value we have obtained an estimate for $\vert V_{cb}\vert$ which is as given below 
	\begin{equation}
		\vert V_{cb}\vert = (37.9 \pm 4.5)\times 10^{-3}. 
	\end{equation}
	This estimate has a large error and is consistent with those obtained from $B\to D^{(*)}(\mu, e)\nu$ modes \cite{Ray:2023xjn}. Similarly, using the measurement of the branching fraction $\mathcal{B}(\Lambda_b \to \Lambda_c \tau \nu)=(1.5 \pm 0.16 \pm 0.25 \pm 0.23)\%$~\cite{LHCb:2022piu}, we have obtained 
	\begin{equation}
		\vert V_{cb}\vert = (44.0 \pm 5.0)\times 10^{-3}. 
	\end{equation}
	Both of these estimates are consistent within their 0.91$\sigma$ uncertainties, though they have large errors, and there is a gap of about 16\% between the two best-fit values.  
	
It is important to note that $\Lambda_b \to \Lambda_c \tau \nu$ could be potentially sensitive that can contribute beyond the SM. Many model-independent NP analysis have been performed to explain either the deviation observed in $R(D)$ and $R(D^*)$\cite{Bhattacharya:2016zcw, Bhattacharya:2018kig, Huang:2018nnq, Murgui:2019czp,Iguro:2022yzr, Becirevic:2019tpx, Ray:2023xjn} and with $R(\Lambda_{c})$ \cite{DiSalvo:2018ngq, Ray:2018hrx, Penalva:2019rgt, Ferrillo:2019owd, Mu:2019bin, Becirevic:2022bev,Fedele:2022iib} along with NP effects that appears through only tauonic interaction to the theory. However, a simultaneous explanation of these three LFUV ratios is mandatory because the three decay modes are correlated through the same charge current interaction. Therefore, to critically scrutinise the data compatibility in the presence of NP,  we have done a combined $\chi^2$ analysis using these observables. 

Like $R(D^{(*)})$, $R(\Lambda_c)$ we define $R(J/\psi)$ for the $B_c\to J/\psi \ell^-\bar{\nu}$ decays. Measurements are available for $R(J/\psi)$ and the measured values are the following: 
\begin{equation}
	R(J/\psi)|_{LHCb}=0.71\pm 0.17\pm 0.18,
\end{equation}
by LHCb \cite{LHCb:2017vlu} and
\begin{equation}
	R(J/\psi)|_{CMS}=0.17\pm 0.33 
\end{equation}
 by the CMS collaboration \cite{CMS:2023vgr}.
Note that both the measurements have large errors, and they agree with each other at their $1.3\sigma$ error. The corresponding SM prediction for comparison is provided by \cite{Harrison:2020nrv} from the HPQCD lattice collaboration
\begin{equation}
	R(J/\psi)=0.258\pm 0.004.
\end{equation} 
The SM prediction is compatible with LHCb at the $1.8 \sigma$ level, while the measured value by CMS is consistent with the SM predictions within their $1\sigma$ error. We want to wait for more precise data to incorporate these inputs in the analysis of NP. Another essential point is that for this mode, the inputs on the vector form factors are available, but the inputs on the tensor form factors are not available. Later, we will see that the contributions from the tensor current operator will play an essential role in explaining the other available data. In the following subsections, we will discuss the results of our analysis regarding the extraction of new physics information from different fits to the available data.
	
\subsection{New Physics analysis: One operator scenario}

The new physics effective operators relevant to $b\to c\tau^-\bar{\nu}$ transitions are defined in eq.\ref{eq:heff}. The relevant data on $B\to D^{(*)}\mu^- \bar{\nu}$ decays suggest that the allowed new physics contributions in $b\to c\mu^-\bar{\nu}$ transitions are negligibly small \cite{Ray:2023xjn}, which is as per the expectations. To constrain the new WCs, we fit the available data, considering the contributions from one operator at a time. We perform a combine analysis of $R(D)$, $R(D^*)$, $R(\Lambda_c)$, $F_L^{D^*}$ and $\mathcal{B}(\Lambda_b \to \Lambda_c \tau \nu)$ observables. We have used the Heavy Flavor Averaging Group (HFLAV) averages for $R(D)$ and $R(D^*)$ along with their respective correlation \cite{HeavyFlavorAveragingGroup:2022wzx}. Furthermore, we have included the latest LHCb measurement on $\mathcal{B}(\Lambda_b \to \Lambda_{c} \tau \nu)$, $R(\Lambda_{c})$~\cite{LHCb:2022piu} and on $F_L^{D^*}$ \cite{LHCb:2023ssl}. In table \ref{tab:expt_dat}, we have presented the data used in the fit. 

	\begin{table}[t]
		\begin{center}
			\resizebox{0.8\textwidth}{!}{
				\begin{tabular}{|*{6}{c|}}
					\hline
					\text{$R(D)$}~\cite{HeavyFlavorAveragingGroup:2022wzx}&  \text{$ R(D^*)$}~\cite{HeavyFlavorAveragingGroup:2022wzx}& \text{correlation}  \
					~\cite{HeavyFlavorAveragingGroup:2022wzx}& \text{$ R(\Lambda_{c})$}~\cite{LHCb:2022piu}& $\mathcal{B}(\Lambda_{b} \to \Lambda_{c}\tau \bar{\nu}_{\tau})$~\cite{LHCb:2022piu}& \text{$ F_L^{D^*}$}~\cite{LHCb:2023ssl}\\
					\hline
					$0.357(29)$ & $0.284(12)$&$-0.37$&$0.242(76)$&$0.015(4)$&$0.43(7)$\\
					\hline
				\end{tabular}
			}
			\caption{Experimental data used in the $\chi^2$ analysis.}
			\label{tab:expt_dat}
		\end{center}
	\end{table}
	
	We closely follow the treatment in ref~\cite{Ray:2023xjn} for the observables $R(D)$, $R(D^*)$ and $F_L^{D^*}$. For our purposes, we have taken the analytic expressions for the observables $R(D)$ and $R(D^*)$ as well as $F_L^{D^*}$ from that reference. We provide the relevant expressions in the appendix~\ref{Appndx:LFUVtheoexp}. The respective expressions show the SM predictions with the respective 1$\sigma$ error as the overall normalization. The new WCs are the only free parameters in these expressions. In the analysis in \cite{Ray:2023xjn}, the shape of the form factors was obtained using only the lattice inputs \cite{MILC:2015uhg,Na:2015kha,FermilabLattice:2021cdg,Aoki:2023qpa}, and the estimated errors in SM are solely due to the form factors. In the NP scenarios, there will be additional errors due to the uncertainties in the fitted values of the new WCs. For $\Lambda_b \to \Lambda_c\tau^-\bar{\nu}$ decay, we have obtained the shape of all the form factors relevant in NP scenarios from lattice \cite{Detmold:2015aaa,Datta:2017aue}. 	
	For the analysis, we define the following $\chi^2$ function 
	\begin{equation}\label{equchi2}
		\chi^2 = \chi^2_{observables} + \chi^2_{lattice},
	\end{equation}
	with 
	\begin{equation}
		\chi^2_{observables} = \sum_{i,j}\big[\mathcal{O}_i^{th}-\mathcal{O}_i^{exp}\big] (V^{exp}+V^{th})^{-1}_{i,j}\big[\mathcal{O}_j^{th} - \mathcal{O}_j^{exp}\big]
	\end{equation}	
and
	\begin{equation}
	\chi^2_{lattice}= \sum_{i,j}(F_i^{BCL}-F_i^{lattice}) V^{-1}_{ij} (F_j^{BCL}-F_j^{lattice}).
	\end{equation}
We have minimized this $\chi^2$ function defined in eq. \ref{equchi2}. Here, $\mathcal{O}_{i/j}^{exp}$ and $\mathcal{O}_{i/j}^{th}$ are the respective measured values and the theory expressions of the observables used in the fit. The $\mathcal{O}_{i/j}^{th}$ are the functions of the BCL coefficients and the WCs $C_{k}$'s defined in the effective Hamiltonian in eq.~\ref{eq:heff}. The $F_i^{BCL}$  corresponds to the form factor expressed in terms of BCL coefficients at respective $q^2$ values of 4, 8, 10 \text{$GeV^2$} and $F_i^{lattice}$ are the respective lattice inputs with the covariance matrix $V^{-1}_{ij}$ accounts for the correlations between the form factors at respective $q^2$ values. The $V^{exp}_{i,j}$ and $V^{th}_{i,j}$ are, respectively, the covariance matrices of the measured values and the theory predictions of the observables used in the fit. This covariance matrix will take care of the measured uncertainties and their correlation. Here, the theory correlation will be only between $R(D^*)$ and $F_L^{D^*}$.

	\begin{table}[t]
		\begin{center}
			\resizebox{0.65\textwidth}{!}{
				\begin{tabular}{|*{7}{c|}}
					\hline
					\text{Parameter}&\multicolumn{3}{c|}{One Parameter fit scenario}&\multicolumn{3}{c|}{ $\sigma_{dev}$ (in $\sigma$)}\\
					\cline{2-7}
					&  $\text{Fit values}$ &\text{$\chi^2_{min.}/\text{DOF}$}& \text{P-Value} & $ R(D)$ & $ R(D^*)$ & $R(\Lambda_c)$\\
					\hline
					$ Re[C_{S_1}]$  &  $\text{0.104(45)}$  &$4.463/4$& \
					$0 .215$  &  $0.151$  &  $1 .355$  &  $1.372$  \\
					$ Re[C_{S_2}]$ &  $\text{0.101(47)}$ &$5.187/4$& \
					$0.159$  &  $0.098$  &  $1 .709$  &  $1.297$   \\
					$ Re[C_{V_1}]$ &   $\text{0.050(22)}$  &$4.001/4$&  \
					$0 .261$  &  $0 .683$  &  $0.048$  &  $1.524$   \\
					$Re[C_{V_2}]$  &  $\text{-0.004(34)}$ &$9.176/4$& \
					$0 .027$  &  $1.564$  &  $1 .029$  &  $1.128$ \\
					$Re[C_{T}]$  &  $\text{-0.022(18)}$ & $7.903/4$& \
					$0 .048$  &  $1.971$  &  $0.343$  &  $1.385$  \\
					\hline
				\end{tabular}
			}
			\caption{The simultaneous fit of the new physics WCs considers a one-operator scenario. The inputs are $R(D)$, $R(D^*)$, $R(\Lambda_{c})$, $F_L^{D^*}$, $\mathcal{B}(\Lambda_{b} \to \Lambda_{c} \tau\nu)$. Note that our fit results for the form factor parameters are consistent with the lattice results; therefore, we refrain from presenting their fit values. These fit consider new physics contributions solely in the $\tau$ lepton final state. Additionally, we quantify the tension between the predicted observables and the corresponding experimental data for each scenario, presenting these results in units of $\sigma$ in the last three columns.}
			\label{tab:1parmscnr}
		\end{center}
	\end{table}
	\begin{table}[t]
		\renewcommand{\arraystretch}{1.8}
		\centering
		\setlength\tabcolsep{8 pt}
		\begin{center}
			\resizebox{0.8\textwidth}{!}{
				\begin{tabular}{|*{7}{c|}}
					\hline
					Observables  & \multicolumn{5}{c|}{\text{Observables Prediction (One operator scenario)}} &  $\text{Expt. Measurement}$  \\
					\cline{2-6}
					& $ Re[C_{S_1}]$  & $\text{$ Re[C_{S_2}]$}$ &  $\text{$ Re[C_{V_1}]$}$  &  $\text{$ Re[C_{V_2}]$}$  &  $\text{$ Re[C_{T}]$}$  &    \\
					\hline
					\text{$R(D)$} &  $\text{0.363(27)}$  &  $\text{0.361(29)}$  &  $\text{0.335(14)}$  \
					&  $\text{0.301(21)}$  &  $\text{0.299(5)}$  &  $\text{0.357(29)}$ \cite{HeavyFlavorAveragingGroup:2022wzx} \\
					% &\bf Range&$0.390-0.335$&$0.390-0.332$&$0.349-0.321$&$0.322-0.280$&$0.304-0.293$&$0.386-0.328$\\
					\hline
					\text{$ R(D^*)$}  &   $\text{0.261(12)}$  &  $\text{0.255(12)}$  &  $\text{0.285(17)}$  \
					&  $\text{0.260(20)}$  &  $\text{0.276(20)}$  &  $\text{0.284(12)}$ \cite{HeavyFlavorAveragingGroup:2022wzx}\\
					% &\bf Range&$0.273-0.249$&$0.268-0.243$&$0.301-0.268$&$0.280-0.240$&$0.296-0.256$&$0.296-0.272$\\
					\hline
					\text{R($ \Lambda_c$)} &   $\text{0.348(14)}$  &  $\text{0.342(13)}$  &  $\text{0.361(18)}$  \
					&  $\text{0.329(13)}$  &  $\text{0.352(23)}$  &  $\text{0.242(76)}$\cite{LHCb:2022piu}  \\
					%&\bf Range&$0.361-0.334$&$0.355-0.329$&$0.380-0.343$&$0.342-0.316$&$0.375-0.328$&$0.318-0.116$\\
					\hline
					\text{\bf $ F^{D^*}_L$}   &  $\text{0.433(3)}$  &  $\text{0.421(3)}$  &  $\text{0.427(9)}$  \
					&  $\text{0.427(3)}$  &  $\text{0.421(6)}$  &  $\text{0.430(70)}$ \cite{LHCb:2023ssl}\\
					%&\bf Range&$0.436-0.430$&$0.424-0.418$&$0.436-0.418$&$0.430-0.424$&$0.427-0.415$&$0.50-0.36$\\
					\hline\hline
					$\text{$ P^{\tau}(D^*)$}$  &  $\text{-0.502(10)}$  &  $\text{-0.535(9)}$  &  $\text{-0.519(7)}$  \
					&  $\text{-0.519(7)}$  &  $\text{-0.505(14)}$  &  $\text{-0.38(54)}$ \cite{Belle:2017ilt} \\ 
					$\text{$ P^{\tau}(D)$}$ &  $\text{0.433(42)}$  &  $\text{0.431(45)}$  &  $\text{0.324(3)}$  \
					&  $\text{0.324(3)}$  &  $\text{0.336(10)}$  &  N.A.  \\
					\hline
				\end{tabular}
			}
			\caption{Predictions of $R(D)$, $R(D^*)$, $R(\Lambda_{c})$ and $F_L^{D^*}$ using the corresponding fit results from table~\ref{tab:1parmscnr} while considering new physics only in $\tau$ channel.}
			\label{tab:1parmscnrprd}
		\end{center}
	\end{table}				
	%%%%%%%%%%%%%%%%%%%%%%%%%%%%%%%%%%%%%%%%%%%%%%
	\begin{figure}[h!]
		\begin{center}
			\includegraphics[scale=0.37]{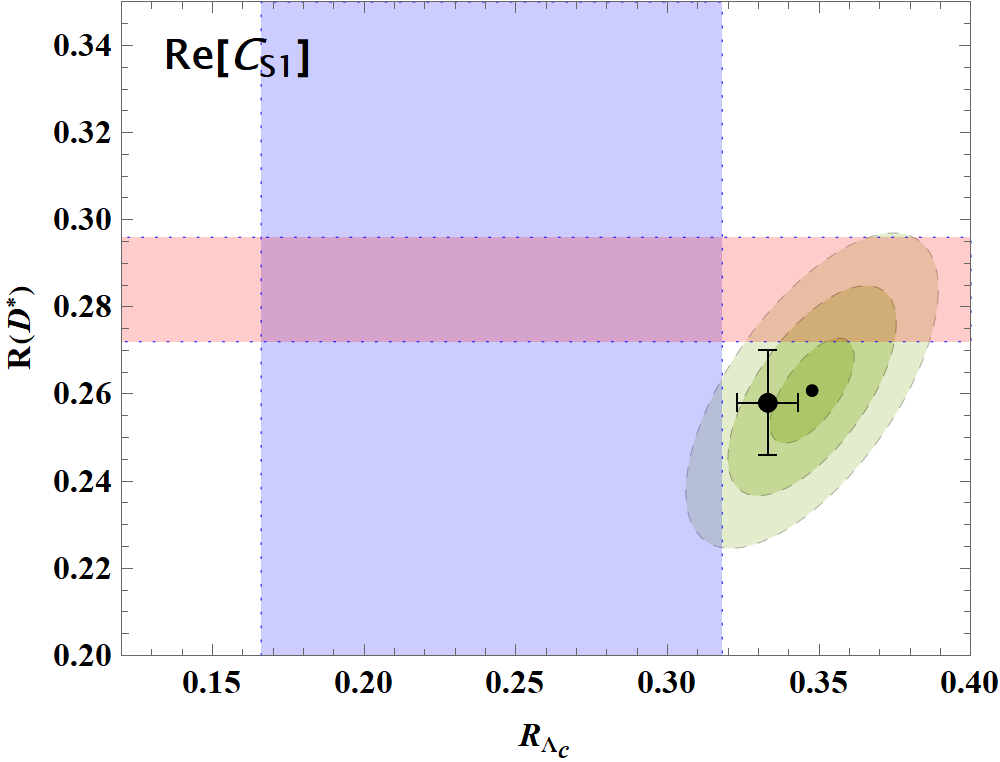}~~
			\includegraphics[scale=0.37]{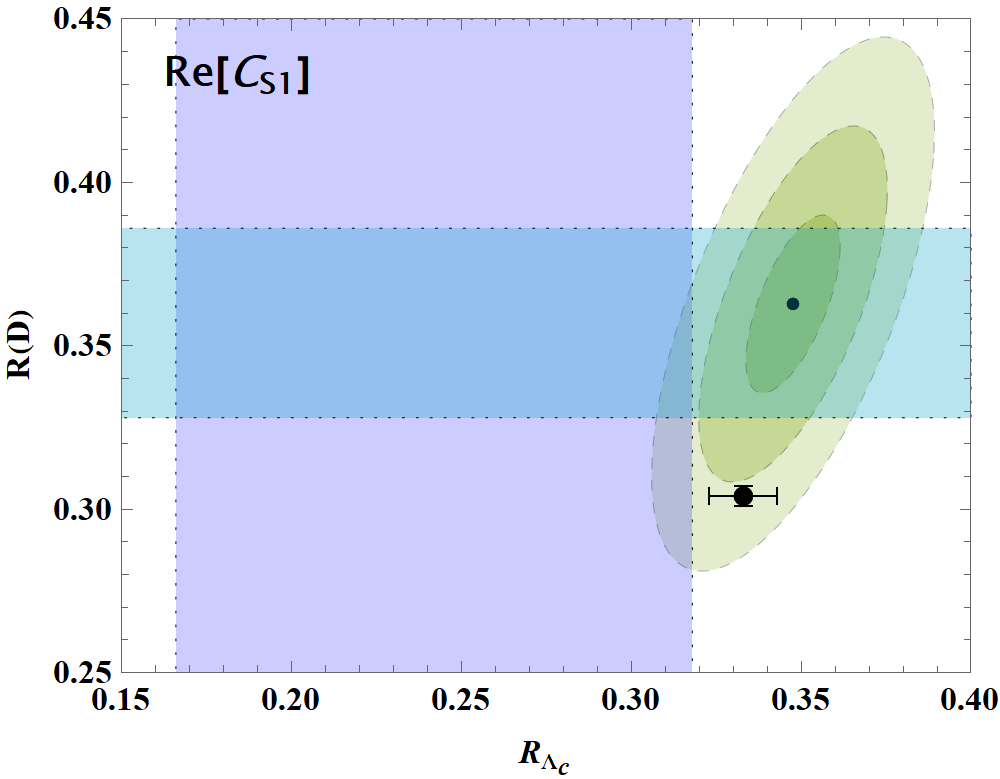}~~	
			\includegraphics[scale=0.37]{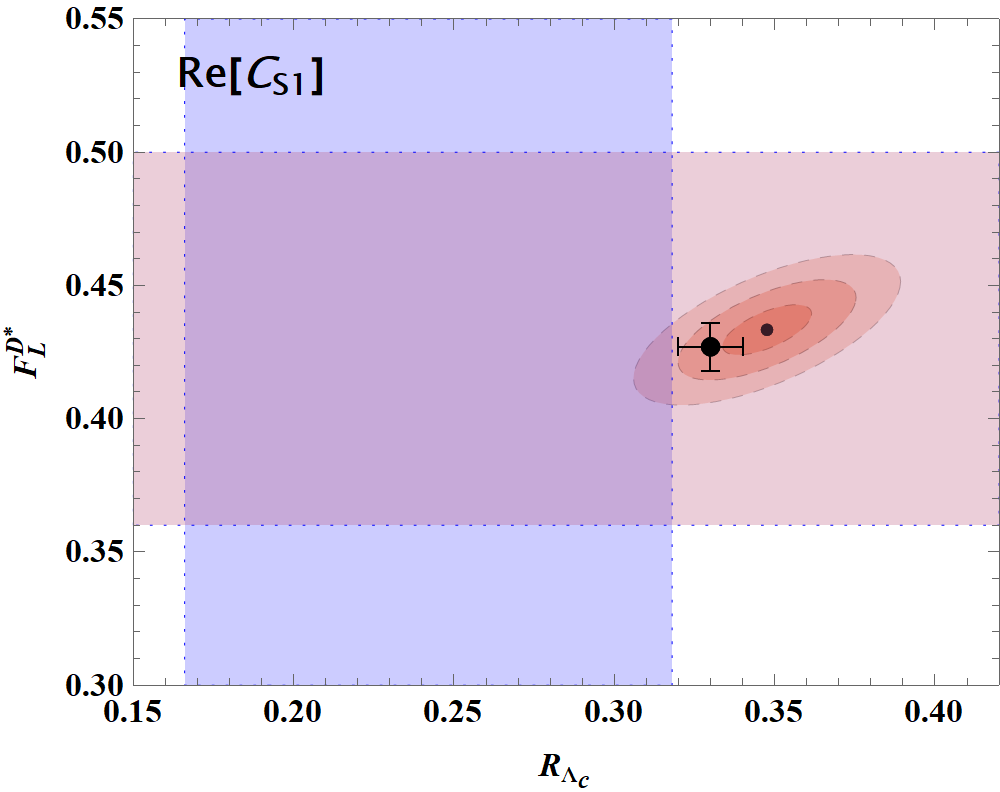}\\	
			\includegraphics[scale=0.37]{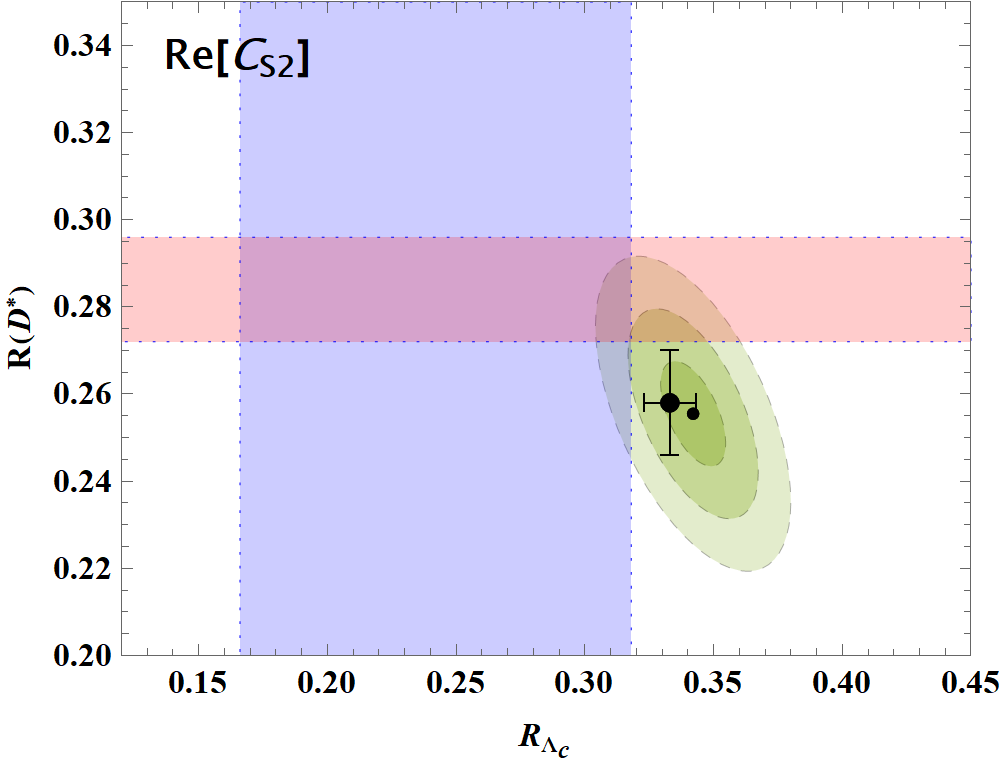}~~
			\includegraphics[scale=0.37]{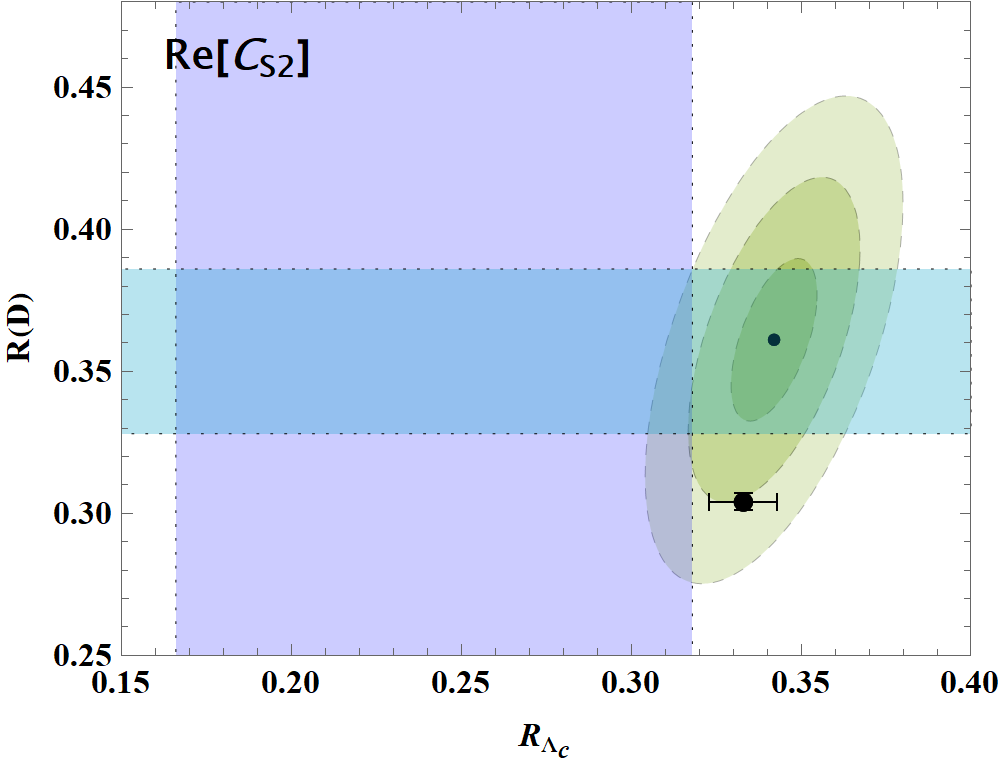}~~
			\includegraphics[scale=0.37]{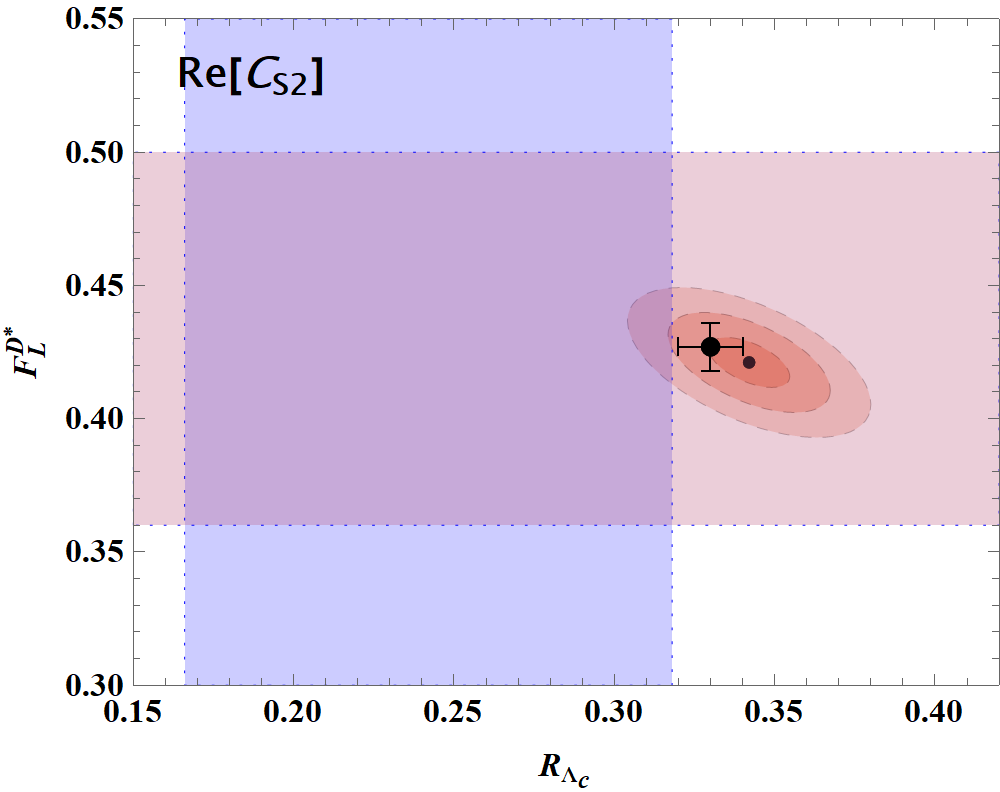}\\	
			\includegraphics[scale=0.37]{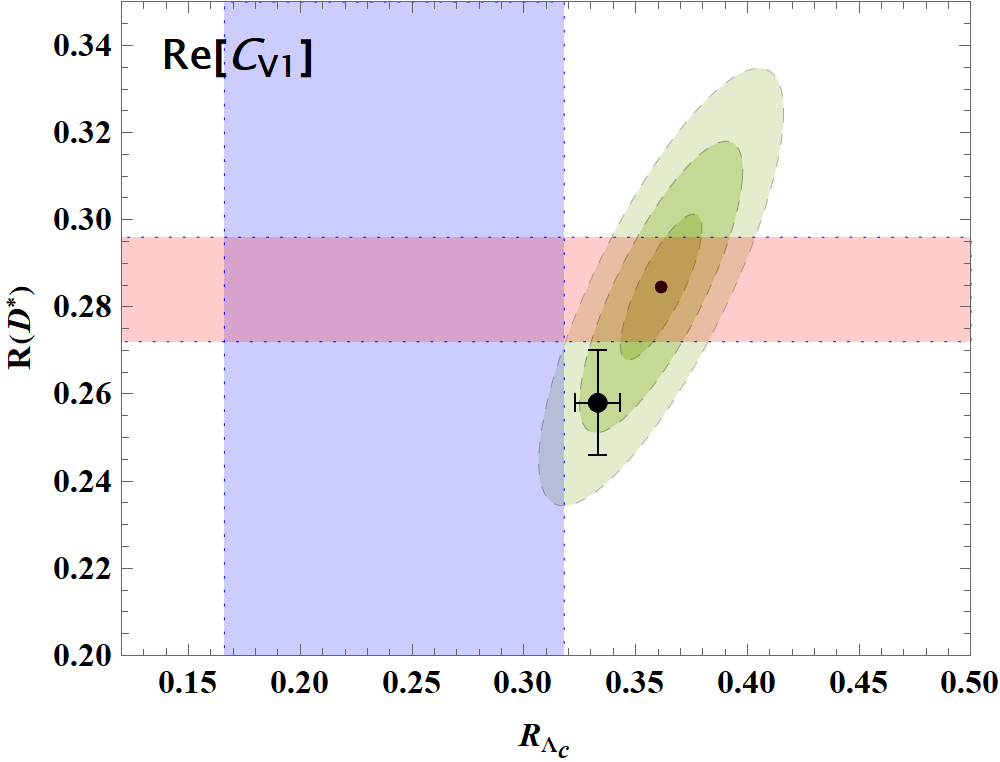}~~
			\includegraphics[scale=0.37]{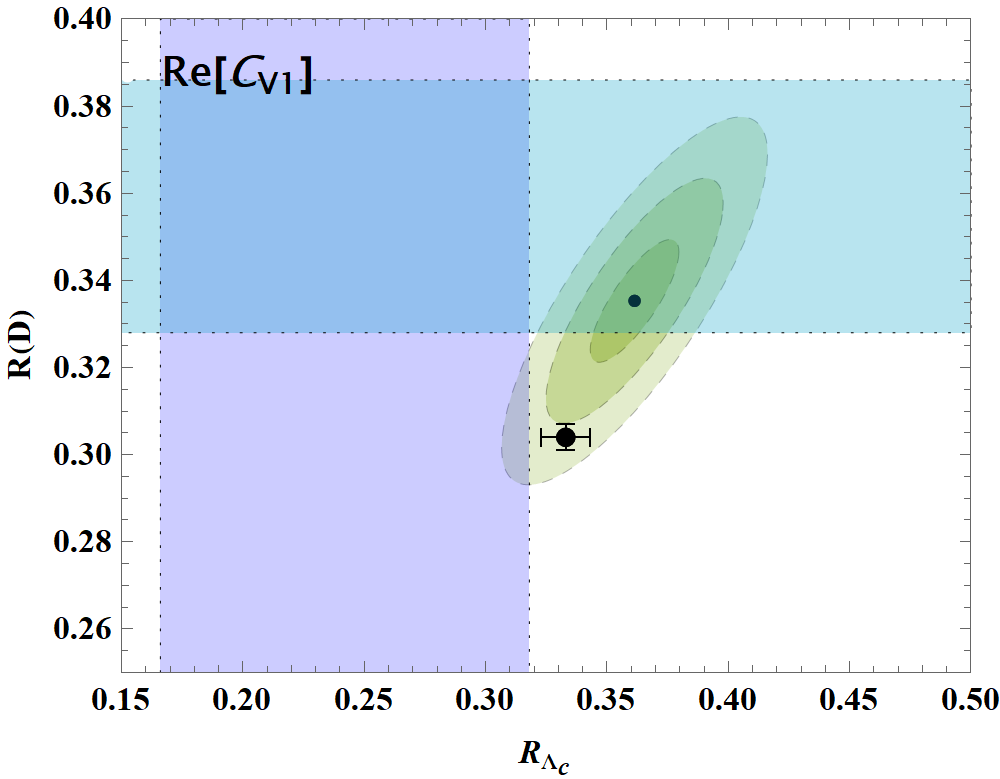}~~
			\includegraphics[scale=0.37]{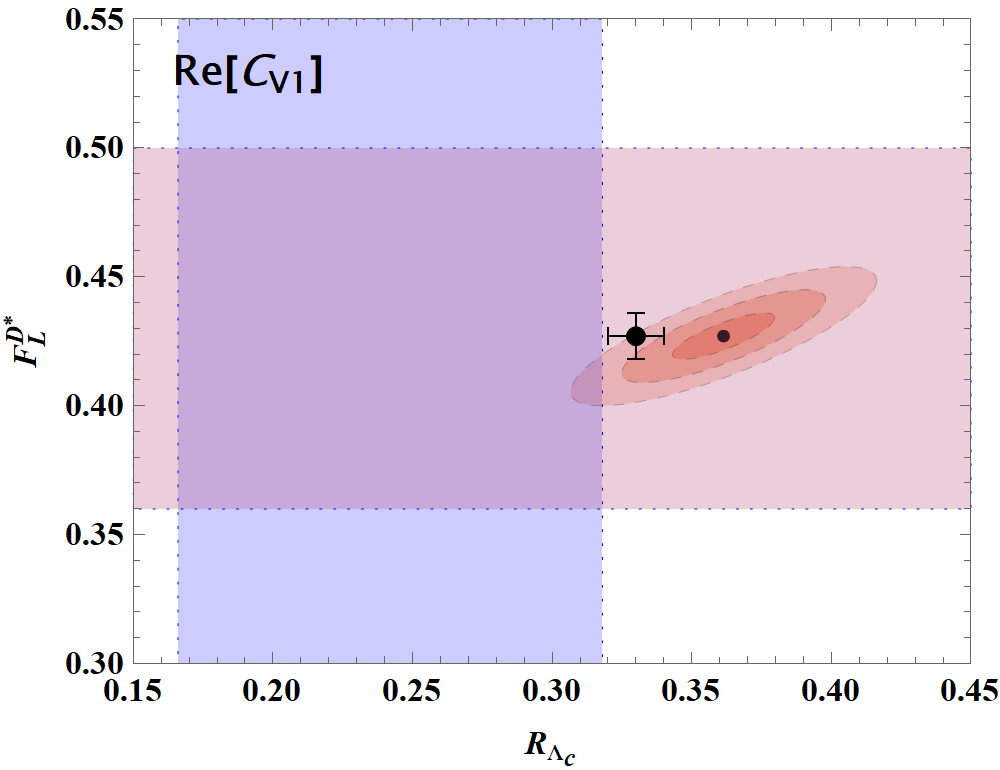}\\
			\includegraphics[scale=0.37]{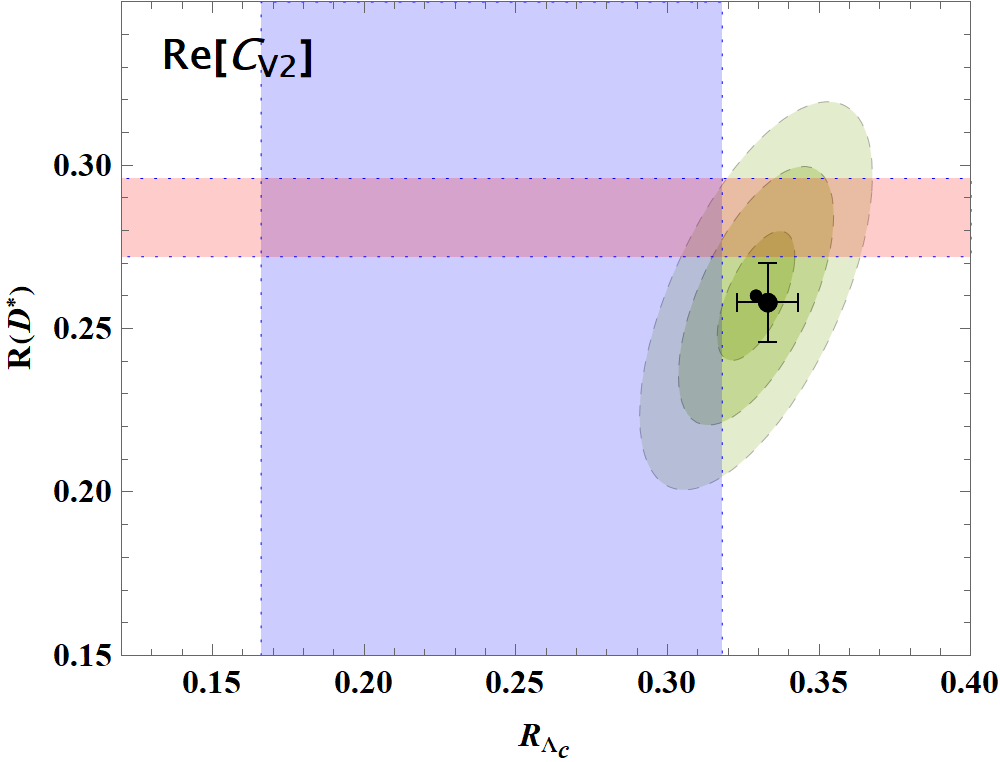}~~
			\includegraphics[scale=0.37]{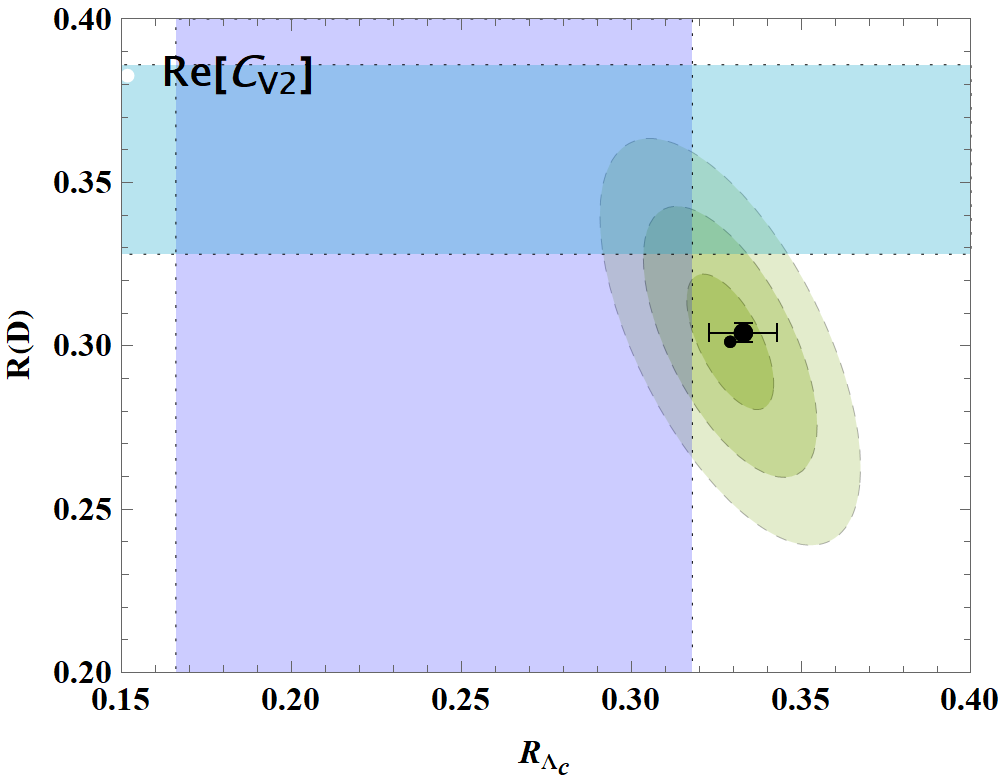}~~		
			\includegraphics[scale=0.37]{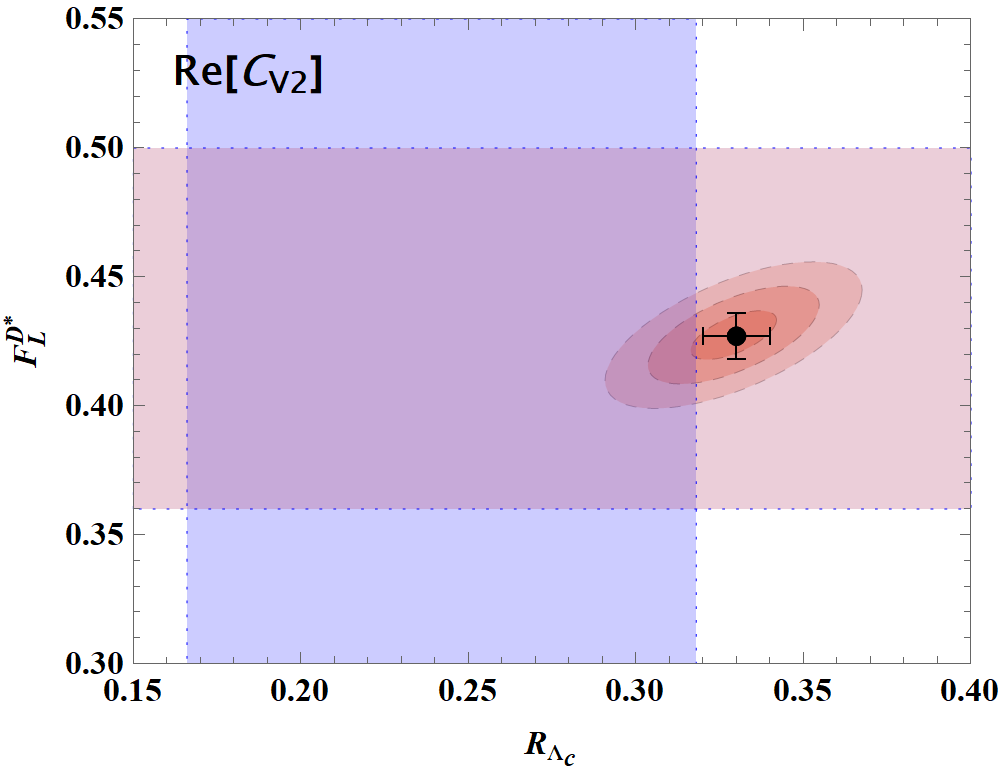}\\
			\includegraphics[scale=0.37]{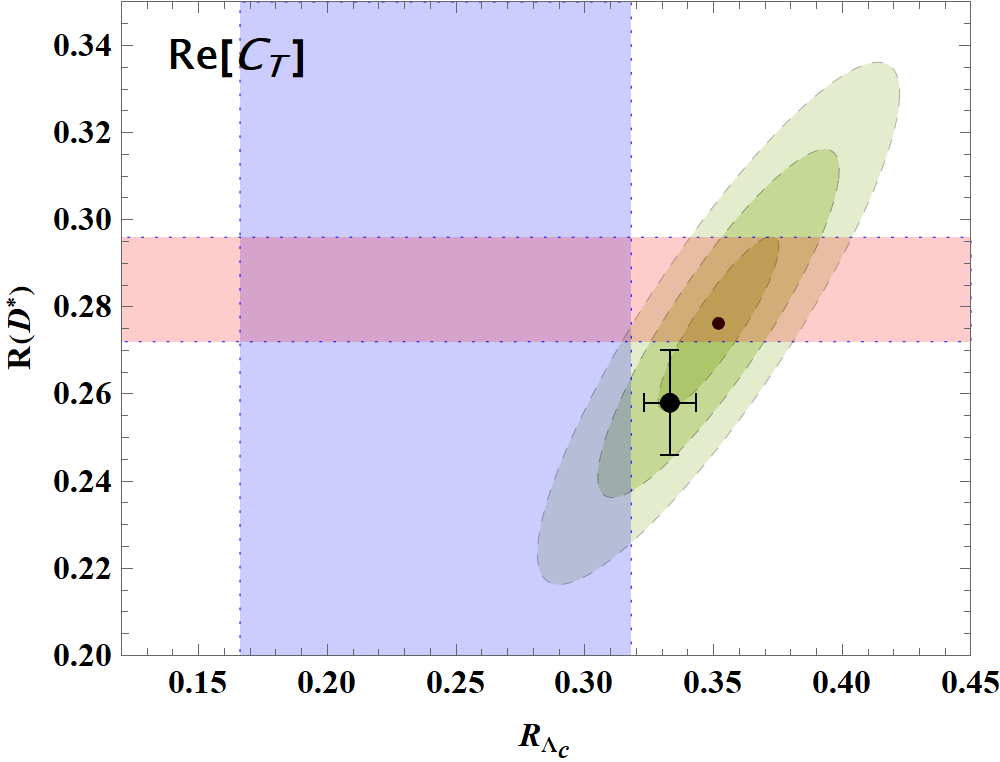}~~
			\includegraphics[scale=0.37]{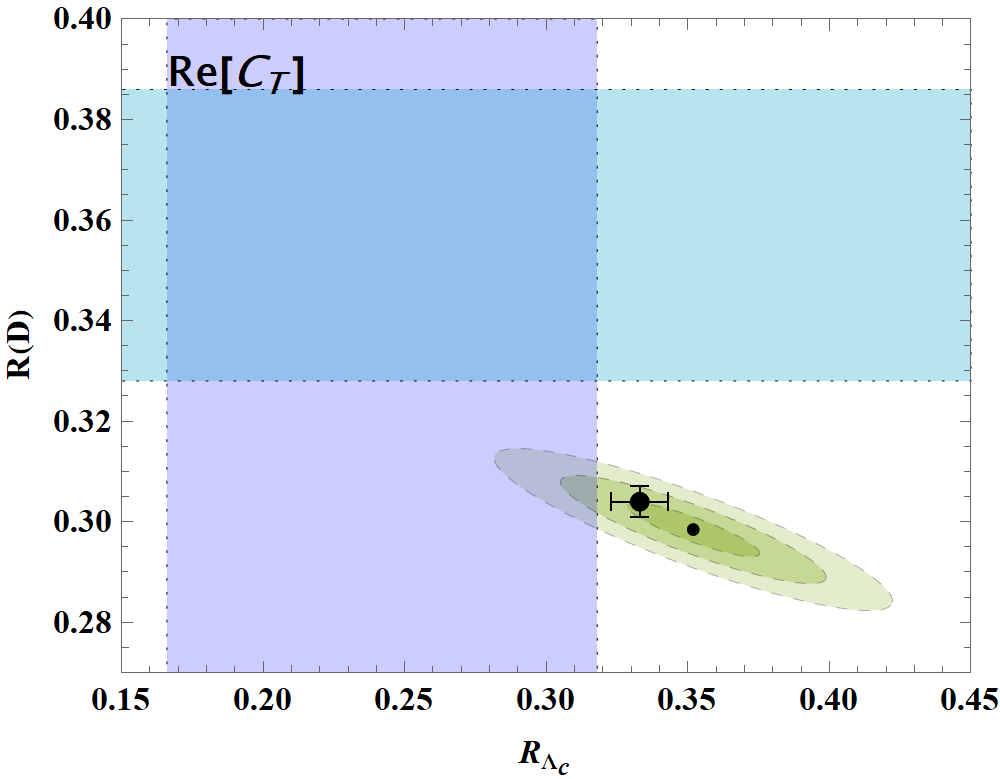}~~
			\includegraphics[scale=0.37]{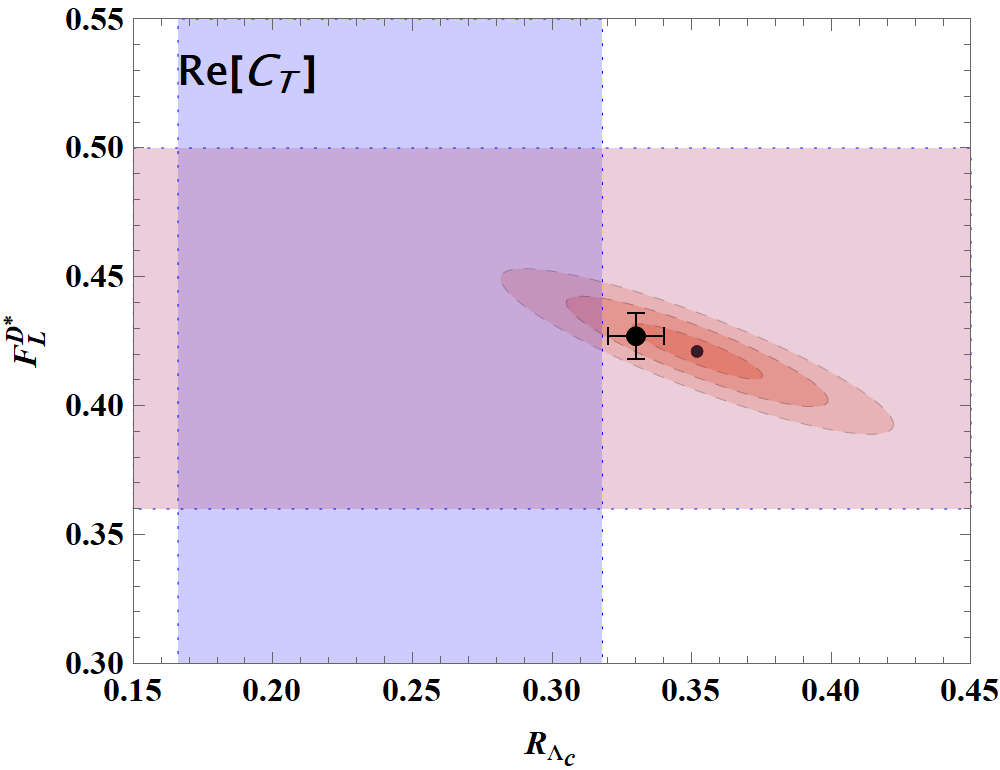}	
		\end{center}
		\caption{The correlations between $R(D)$, $R(D^*)$, $F_L^{D^*}$ with $R(\Lambda_{c})$ in various one-parameter scenarios. The horizontal and vertical color bands indicate the respective experimental results within their 1$\sigma$ CI. The correlations among the observables are represented by ellipses. We display the $1\sigma$, $2\sigma$, and $3\sigma$ contour error bands around the best estimate point (shown in black) using the best-fit values from table~\ref{tab:1parmscnr}. Additionally, the SM estimate is represented by a black point, accompanied by vertical and horizontal $1\sigma$ error bars for the corresponding observables.}
		\label{fig:corrplot1prm}
	\end{figure}
	
	%%%%%%%%%%%%%%%%%%%%%%%%%%%%%%%%%%%%%%%%%%%%%%%%%%%%%%%%
	
The fit results for each one-operator scenario are listed in table~\ref{tab:1parmscnr}. For each of them, we quote the p-value to understand how probable a scenario could be to explain the current data. It is a quantitative measure of compatibility between the hypothesis and measurement. It is calculated using the following formula:
	\begin{equation}\label{pvalue}
		\text{p-value}=1-\text{CDF}_{d.o.f}(\chi^2_{min.})
	\end{equation}
The $\text{CDF}_n$ is the cumulative distribution function of a random variable that follows the $\chi^2$ distribution with degrees of freedom $n$ at the value of the variable $\chi^2_{min}$. Note that in all the cases, we have allowed fit. However, the scenario with $\mathcal{O}_{V_2}$ has a relatively low p-value. To understand the reason for this observation, we predict the values of these observables in all new physics scenarios, as presented in table \ref{tab:1parmscnrprd}. Additionally, we estimate the tension between the experimental data and our new physics predictions by defining the quantity   
	\begin{equation}\label{eq:tsndef}
		\sigma_{dev}=\Big|\frac{\mathcal{O}_i^{exp}-\mathcal{O}_i^{NP}}{\sqrt{\sigma_i^{2}|_{exp}+\sigma_i^{2}|_{NP}}}\Big|.
	\end{equation}
 It is a quantitative measure of the deviation between observed data and the prediction of the observables in the presence of NP (see tab.~\ref{tab:1parmscnr}). One can infer from these estimates that none of the one-operator scenarios could conveniently explain all three LFUV data simultaneously if we take the data and the respective NP predictions within their 1$\sigma$ uncertainties. The situation is even more constrained in the case of $\mathcal{O}_{V_2}$. None of the three data we could explain within their $1\sigma$ confidence interval (CI). In table \ref{tab:1parmscnrprd}, along with $R(D)$, $R(D^*)$, $R(\Lambda_{c})$ and $F_L^{D^*}$ we have also estimated the $\tau$ polarisation asymmetries $P^{\tau}(D^*)$ and $P^{\tau}(D)$. Measurement is also available on $P^{\tau}(D^*)$, which we have shown in the same table. We have not included it in our analysis since the error in the measurement is too large. Also, in all the NP scenarios, the predictions are negative and consistent with the measured value.    
	%%%%%%%%%%%%%%%%%%%%%%%%%%%%%%5	
	\begin{figure}[t!]
		\begin{center}
			\includegraphics[scale=0.37]{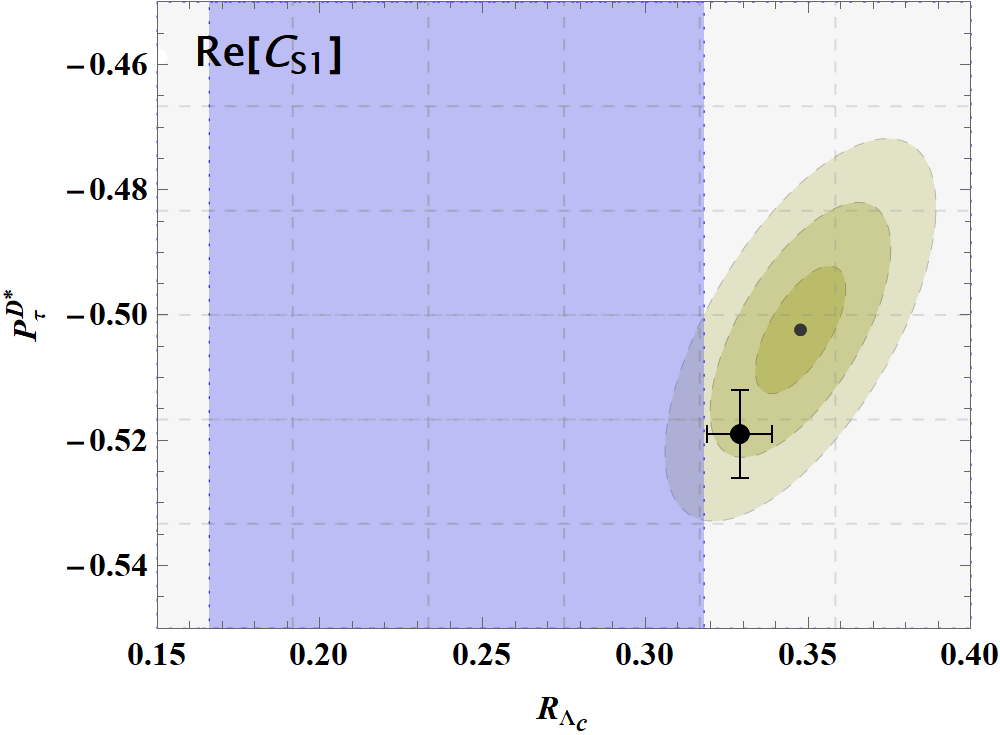}
			\includegraphics[scale=0.37]{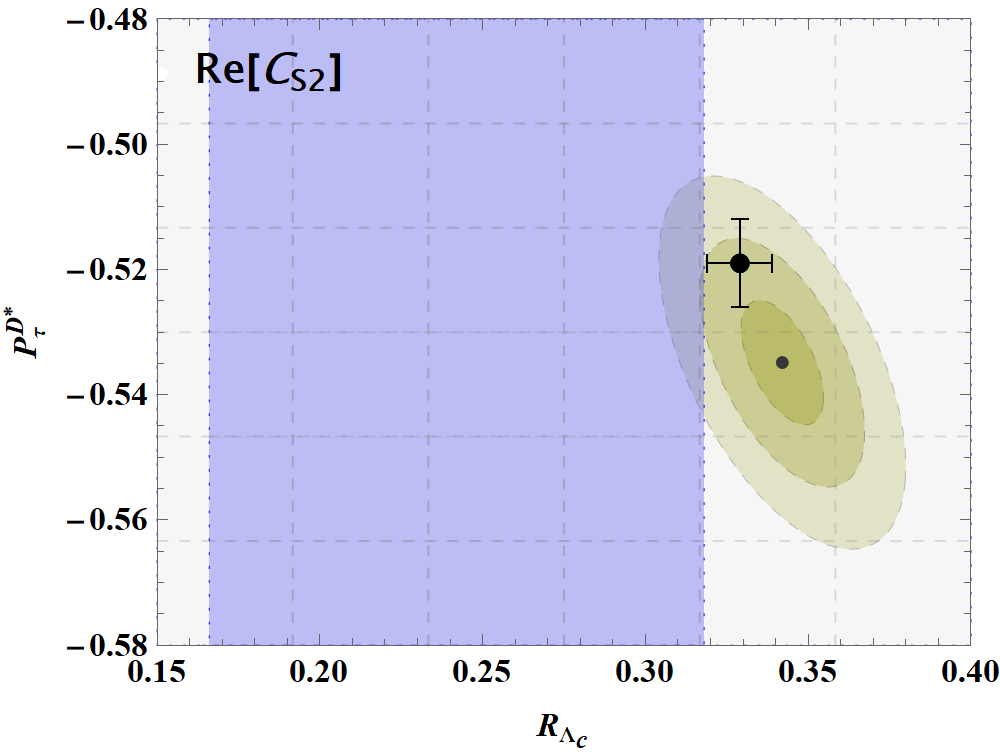}
			\includegraphics[scale=0.37]{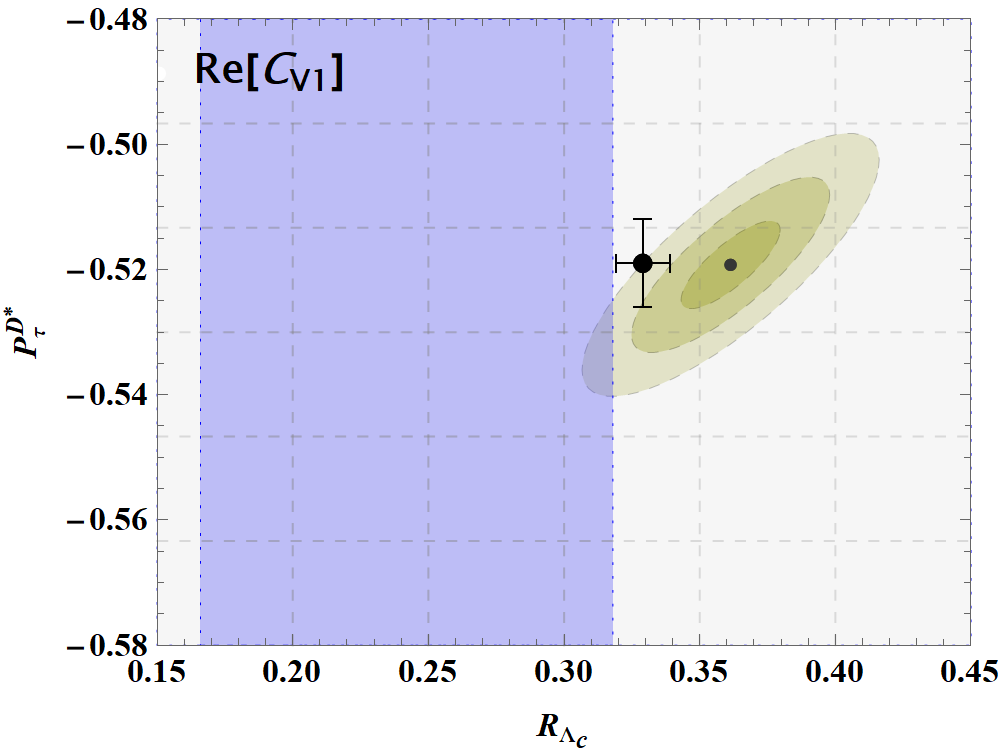}\\
			\includegraphics[scale=0.37]{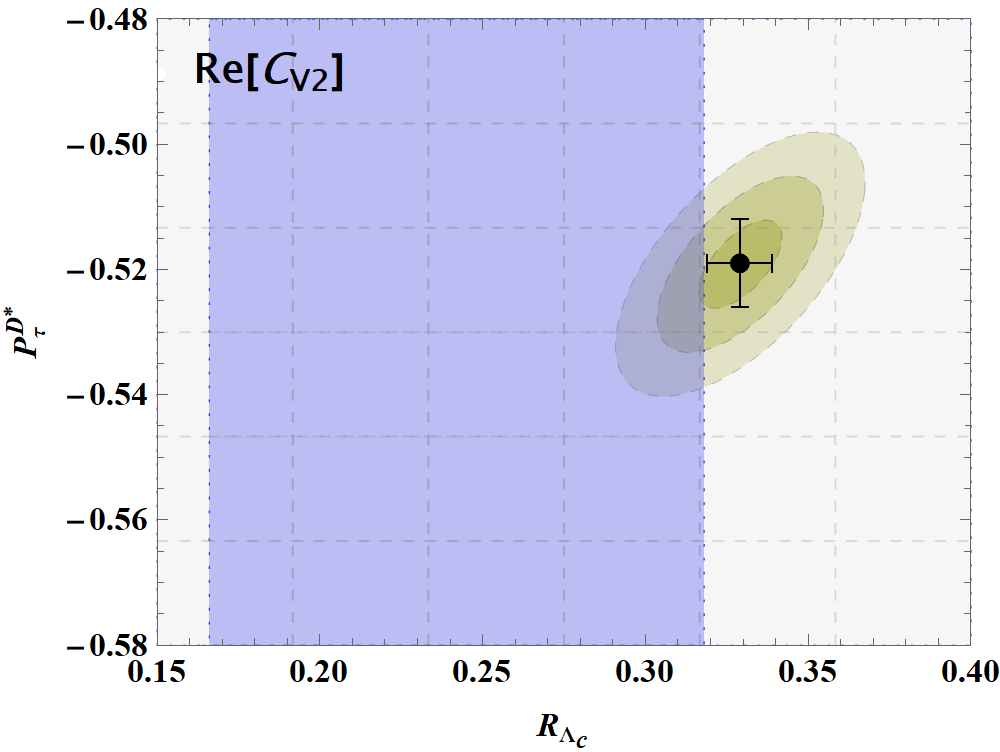}~~~
			\includegraphics[scale=0.37]{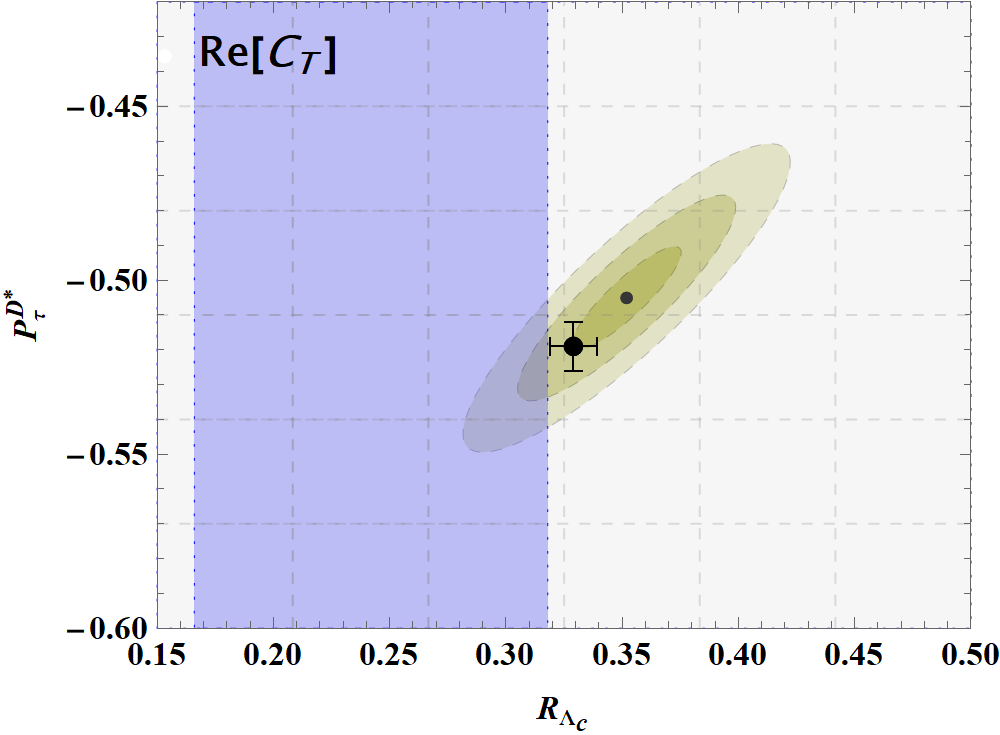}
		\end{center}
		\caption{Correlation plots between $P_{\tau}^{D^*}$ and $R(\Lambda_c)$ in one operator scenarios. The vertical color bands indicate the experimental constraint for $R(\Lambda_{c})$ within 1$\sigma$ confidence level, while the full parameter space is allowed for $P_{\tau}^{D^*}$ in each NP scenario. The correlations between the observables are depicted by ellipses. We include $1\sigma$, $2\sigma$, and $3\sigma$ contour error bands around the best estimate point (shown in black) based on the best-fit values provided in table~\ref{tab:1parmscnr}. The SM estimate is also represented by a black point, along with vertical and horizontal $1\sigma$ error bars for the respective observables.}
		\label{fig:corrPtaudst1parm}
	\end{figure}
%%%%%%%%%%%%%%%%%%%%%%%%%%%%%%%%%%%%%%%%	
Note that the deviations shown in table~\ref{tab:1parmscnr} represent the deviations of each of the individual observables with the respective data. However, as we know, all the observables will be correlated due to similar types of new physics contributions. We have worked out these correlations in different one-operator scenarios, which are shown in fig.~\ref{fig:corrplot1prm}. These correlations will help us understand the pattern of NP effects in different observables, enabling us to determine whether these patterns follow the current data and what we can anticipate from future measurements. Also, we have shown the correlations between the $P^{\tau}(D^*)$ and $R(\Lambda_c)$ in fig.~\ref{fig:corrPtaudst1parm} in each of the one-operator scenarios. In the following items, we will analyse and interpret the information presented in these plots.  
	
	\begin{itemize}
		\item In all the scenarios, we can explain the data on $F_L^{D^*}$ and $P^{\tau}(D^*)$, the 1$\sigma$ contours are well within the current experimental limits. 
		
		\item The scenario with $\mathcal{O}_{V_1}$ could explain quite comfortably both the data on $R(D)$ and $R(D^*)$ even when considering their respective predictions within 1$\sigma$ ranges. This is not as straightforward in other new physics scenarios when looking at predictions within 1$\sigma$. However, if we increase our error margins to 3$\sigma$, we can simultaneously explain the data on $R(D)$ and $R(D^*)$ in all the scenarios except the scenario with $\mathcal{O}_T$.    
		
		\item In all the scenarios, independently, we could explain a small portion (relatively higher values) of the current experimental limit on $R(\Lambda_c)$ if we consider the respective 3$\sigma$ contours. We can do it more comfortably with $\mathcal{O}_T$ and with $\mathcal{O}_{V_2}$. 
		
		\item  The predictions of table~\ref{tab:1parmscnrprd} and the correlation plots in fig.~\ref{fig:corrplot1prm} and \ref{fig:corrPtaudst1parm} suggest that if we consider the errors within their 1$\sigma$ CI, none of the one operator scenarios can explain the current data on $R(D)$, $R(D^*)$, $F_L^{D^*}$, $P^{\tau}(D^*)$ and $R(\Lambda_c)$ simultaneously. However, if we consider the 3$\sigma$ contours of our predictions, then only the scenarios with the operator $\mathcal{O}_{S_2}$ or $\mathcal{O}_{V_2}$ could explain these five data simultaneously. Again, we can explain only a tiny portion of the current experimental limit on $R(\Lambda_c)$. 
	\end{itemize}
	
	The correlation plots are also helpful for understanding how our conclusions might change with more precise data in the near future. For example, if the future data shows only deviations in $R(D)$ but not in $R(D^*)$, this will indicate scalar-pseudoscalar type interactions. On the other hand, if we see the deviations in both these measurements, a new $(V \pm A)$ type of interaction will provide the more probable solutions. Of course, more precise measurements of the other observables will further shed light on a particular type of new interaction in case of deviations, which is also evident from these correlation plots. Furthermore, using the fit values in table \ref{tab:1parmscnr}, we have estimated the values of all the other angular observables (integrated over the full $q^2$) related to $\Lambda_b \to \Lambda_c^+(\to \Lambda \pi^+)\tau^-\bar{\nu}_{\tau}$ in all these one operator scenarios. We have presented the corresponding results in table \ref{tab:predangular_oneopr}, which one can compare with the respective SM predictions and with future measurements. Given the errors in the predictions, it would be hard to distinguish the NP effects in those observables from the respective SM predictions. To improve the situation further, we need more precise inputs from lattice.	
	\begin{table}[t]
		\begin{center}
			\resizebox{0.7\textwidth}{!}{
				\begin{tabular}{|*{7}{c|}}
					\hline
					\text{Observables} &  \text{$Re[C_{S_1}]$}  &  \text{$Re[C_{S_2}]$}  &  \text{$Re[C_{V_1}]$}  &  \text{$Re[C_{V_2}]$}  &  \text{$Re[C_T]$}&  \text{SM}\\
					\hline\hline
					 $\hat{K}_{1cc} = \langle \hat{K}_{1cc}(q^2) \rangle$ &  $\text{0.313(1)}$  &  $\text{0.313(1)}$  &  $\text{0.312(1)}$  \
					&  $\text{0.312(1)}$  &  $\text{0.313(1)}$& $\text{0.312(2)}$\\
					$\text{$ \hat{K}_{1ss}= \langle \hat{K}_{1ss}(q^2) \rangle$}$  & $\text{0.343(1)}$  &  $\text{0.344(1)}$  &  $\text{0.344(1)}$  &  $\text{0.344(1)}$  &  $\text{0.344(1)}$ &  $\text{0.344(1)}$\\
					\text{$\hat{K}_{2cc}=\langle \hat{K}_{2cc}(q^2)\rangle$} & $\text{0.208(22)}$  &  $\text{0.203(22)}$  &  $\text{0.208(22)}$  \
					&  $\text{0.207(23)}$  &  $\text{0.194(25)}$ &  $\text{0.208(22)}$\\
					\text{$ \hat{K}_{2ss}=\langle \hat{K}_{2ss}(q^2)\rangle$} & $\text{0.239(26)}$  &  $\text{0.235(26)}$  &  $\text{0.241(26)}$  \
					&  $\text{0.240(26)}$  &  $\text{0.224(28)}$ &  $\text{0.240(26)}$\\
					\text{$\hat{K}_{3sc}= \langle \hat{K}_{3sc}(q^2)\rangle$}  & $\text{0.016(2)}$  &  $\text{0.017(2)}$  &  $\text{0.017(2)}$  \
					&  $\text{0.017(2)}$  &  $\text{0.016(2)}$&  $\text{0.017(2)}$\\
					\text{$\hat{K}_{3s}= \langle \hat{K}_{3s}(q^2)\rangle$}& $\text{0.034(8)}$  &  $\text{0.032(9)}$  &  $\text{0.043(8)}$  \
					&  $\text{0.045(13)}$  &  $\text{0.047(9)}$  &  $\text{0.044(8)}$\\
					\text{$ A_{FB}^{\tau}=\langle A_{FB}^{\tau}(q^2)\rangle $}& $\text{0.118(12)}$  &  $\text{0.108(11)}$  &  $\text{0.104(11)}$  \
					&  $\text{0.104(11)}$  &  $\text{0.108(11)}$ &  $\text{0.103(11)}$\\
					\text{$ A_{FB}^{\Lambda_{c} \tau}
						=\langle A_{FB}^{\Lambda_{c} \tau}(q^2)\rangle$}  &  $\text{0.028(7)}$  &  $\text{0.024(7)}$  &  $\text{0.019(6)}$  \
					&  $\text{0.018(8)}$  &  $\text{0.014(7)}$ &  $\text{0.018(6)}$\\
					\text{$A_{FB}^{\Lambda_{c}}= \langle A_{FB}^{\Lambda_{c}}(q^2)\rangle$} &$\text{0.343(37)}$  &  $\text{0.336(37)}$  &  $\text{0.345(37)}$  \
					&  $\text{0.344(38)}$  &  $\text{0.321(41)}$ &  $\text{0.344(37)}$\\
					\text{$ P_{\Lambda_c}= \langle P_{\Lambda_c}(q^2)\rangle $} & $\text{-0.758(12)}$  &  $\text{-0.741(16)}$  &  $\text{-0.763(12)}$  \
					&  $\text{-0.760(21)}$  &  $\text{-0.705(51)}$ & $\text{-0.761(12)}$ \
					\\
					\text{$P_{\tau}^{(\Lambda_{c})}= \langle P_{\tau}^{(\Lambda_{c})}(q^2)\rangle$} &$\text{-0.187(35)}$  &  $\text{-0.207(30)}$  &  $    \text{-0.257(16)}$  \
					&  $\text{-0.260(26)}$  &  $\text{-0.260(15)}$  &$\text{-0.258(16)}$  \\
					\text{$ C^{\tau}_F =\langle C^{\tau}_F (q^2) \rangle$}  & \text{-0.091(6)} &  $\text{-0.093(6)}$  &  $\text{-0.096(6)}$  &  $\text{-0.096(6)}$  \
					&  $\text{-0.091(7)}$  &  $\text{-0.096(6)}$  \\
					\hline
				\end{tabular}
			}
			\caption{Predictions for the angular observables with the fit results given in table~\ref{tab:1parmscnr} while considering NP in the $\tau$ lepton final state one at a time.}
			\label{tab:predangular_oneopr}
		\end{center}
	\end{table}
	\begin{figure*}[ht!]
		\centering
		\includegraphics[scale=0.4]{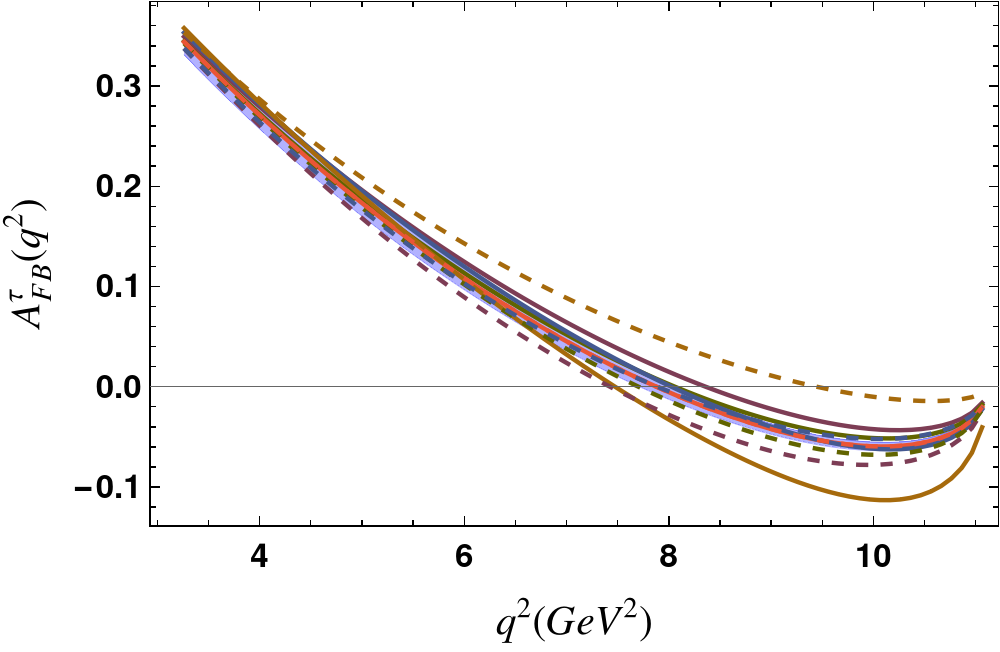}~~
		\includegraphics[scale=0.4]{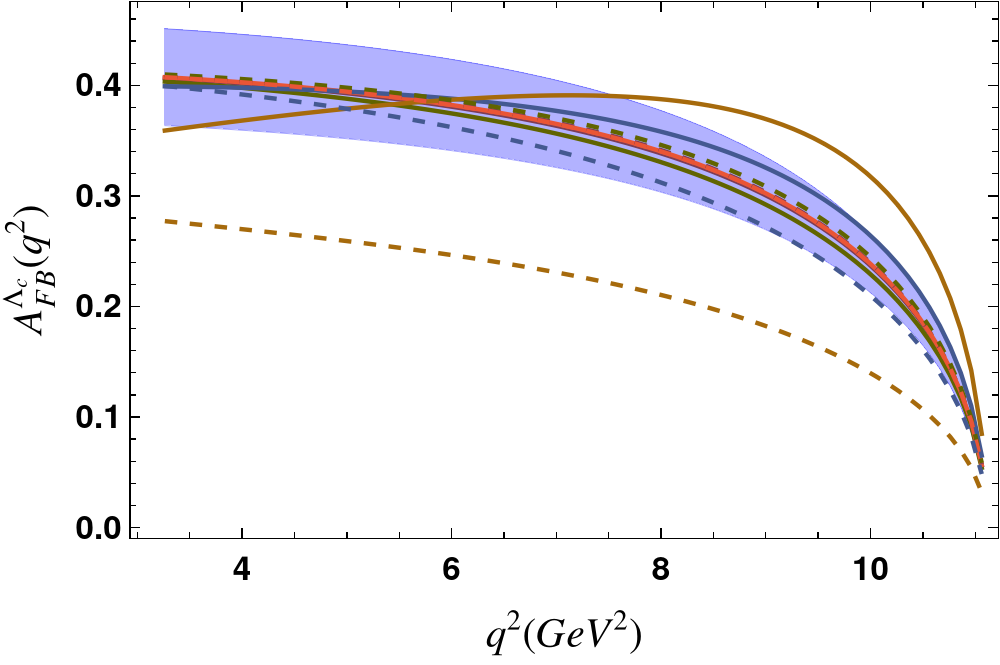}~~
		\includegraphics[scale=0.4]{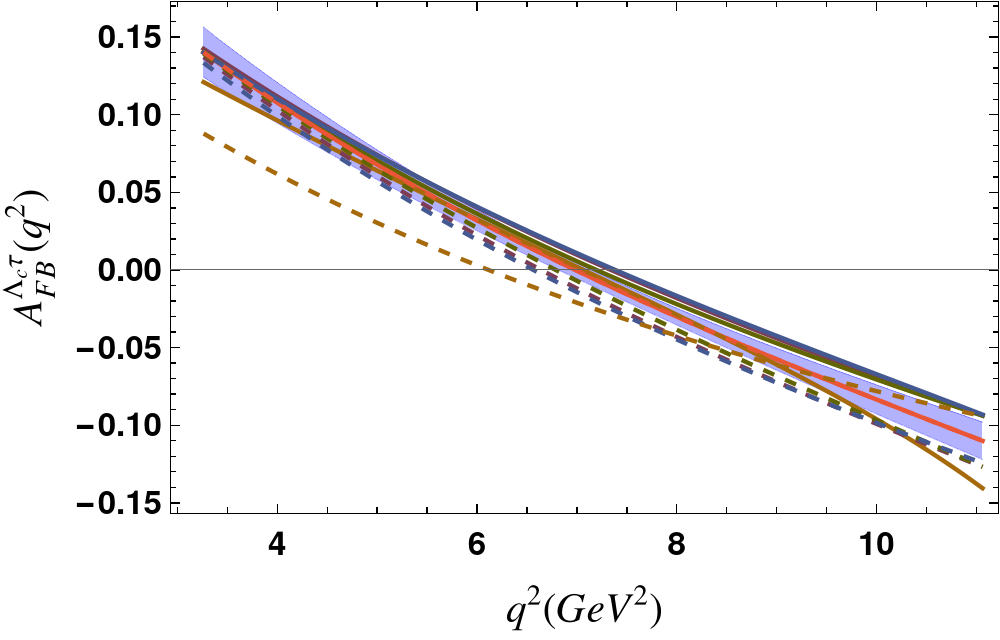}\\
		\includegraphics[scale=0.4]{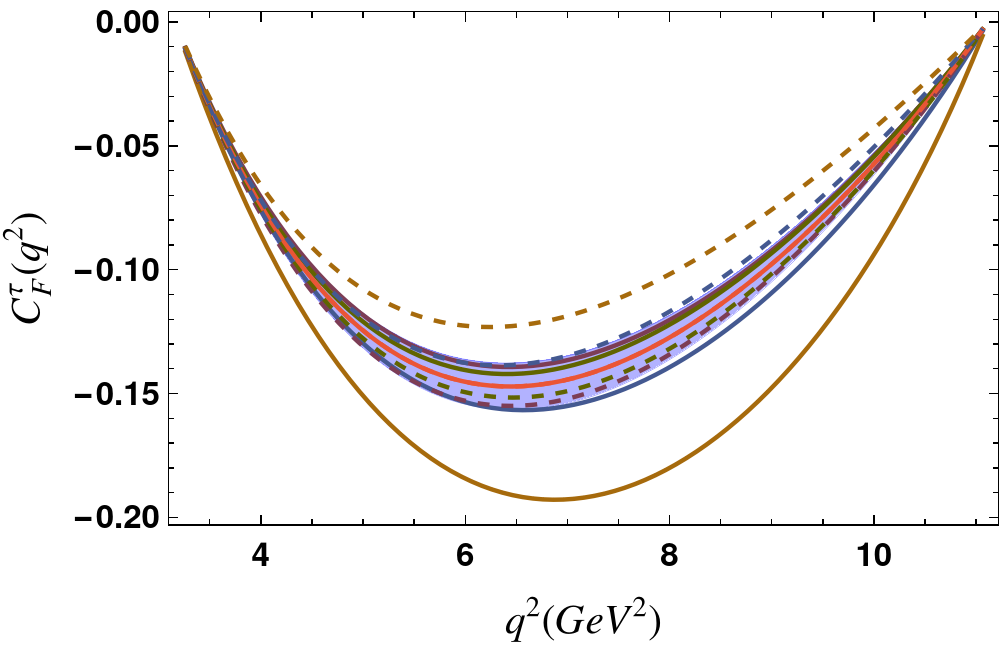}~~
		\includegraphics[scale=0.4]{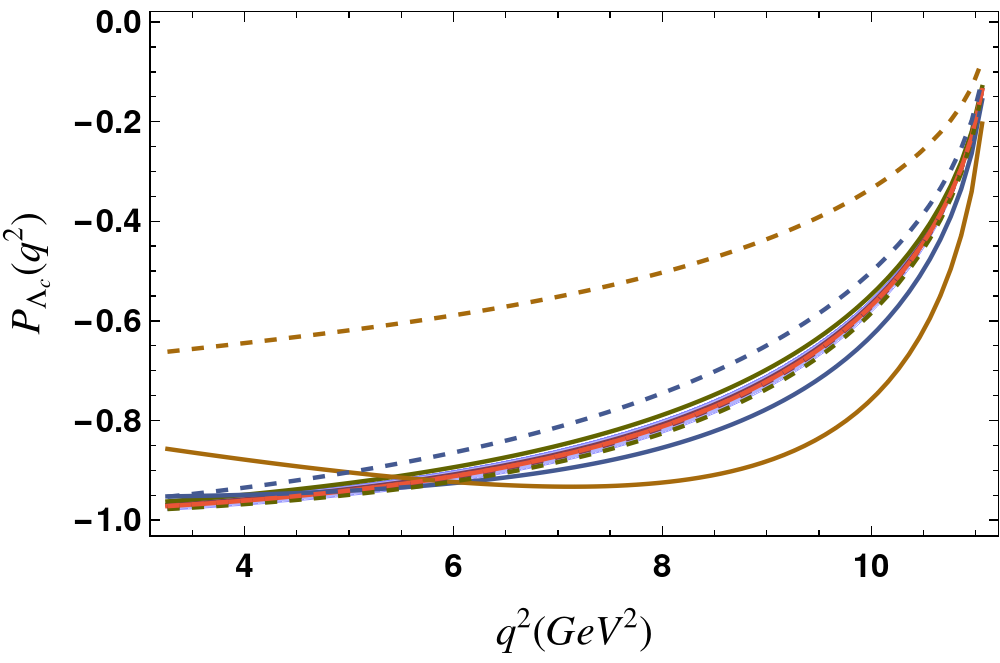}~~
		\includegraphics[scale=0.4]{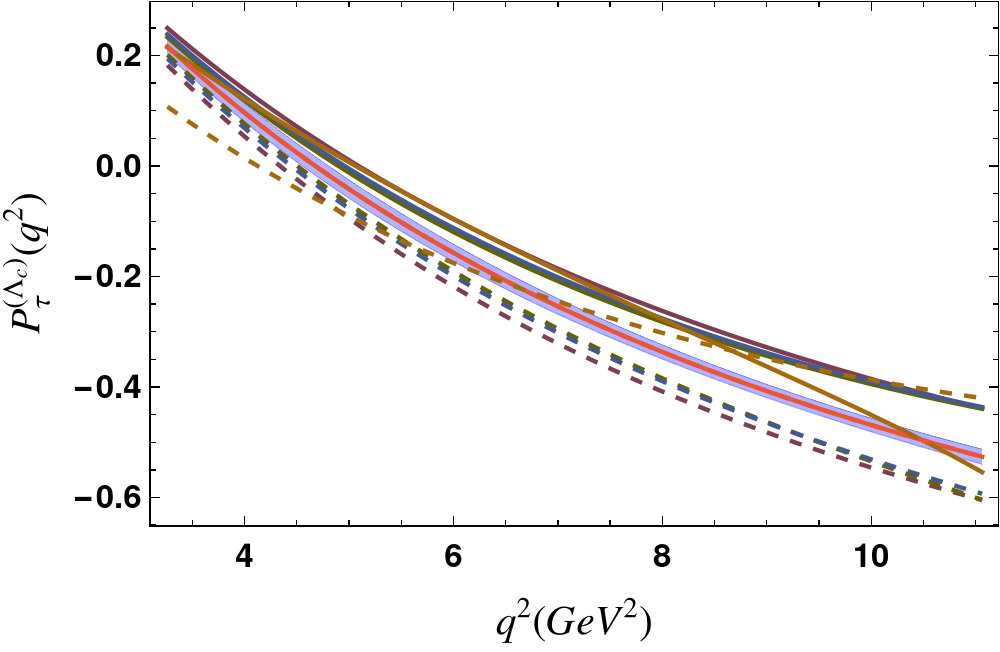}\\
		\includegraphics[scale=0.4]{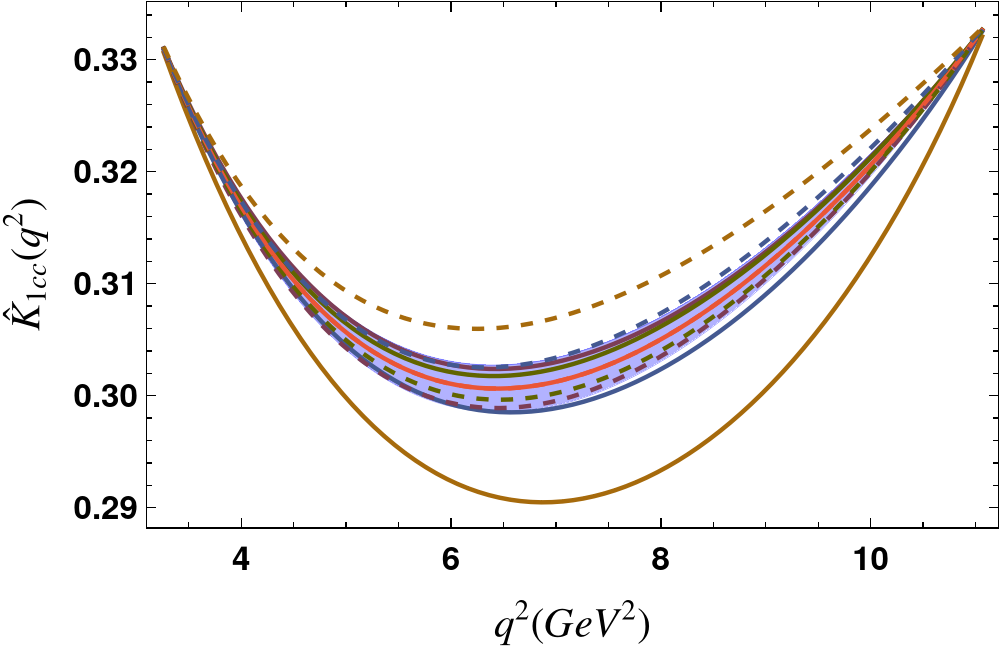}~~
		\includegraphics[scale=0.4]{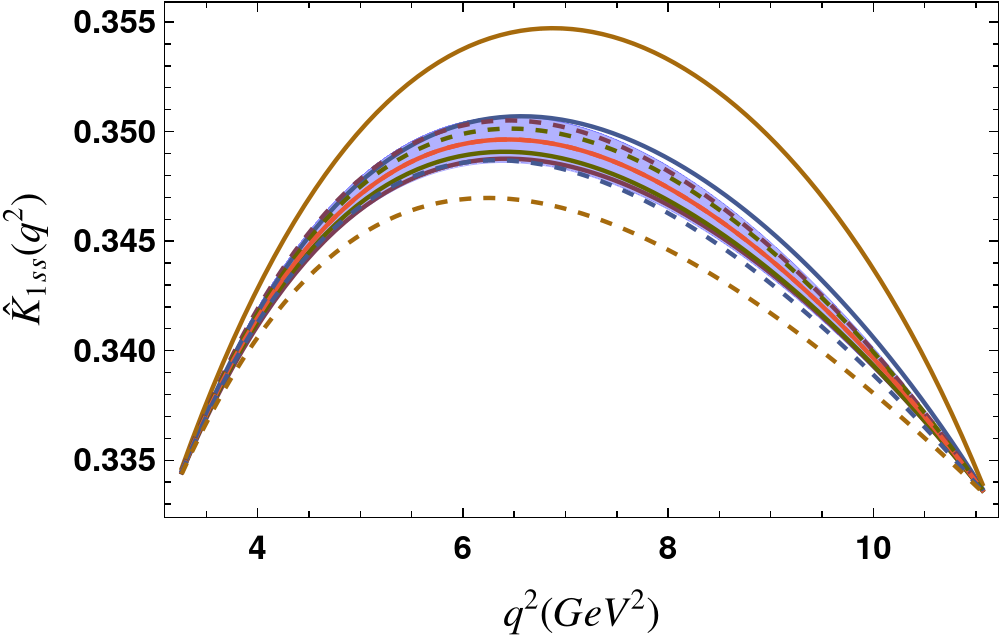}~~	
		\includegraphics[scale=0.4]{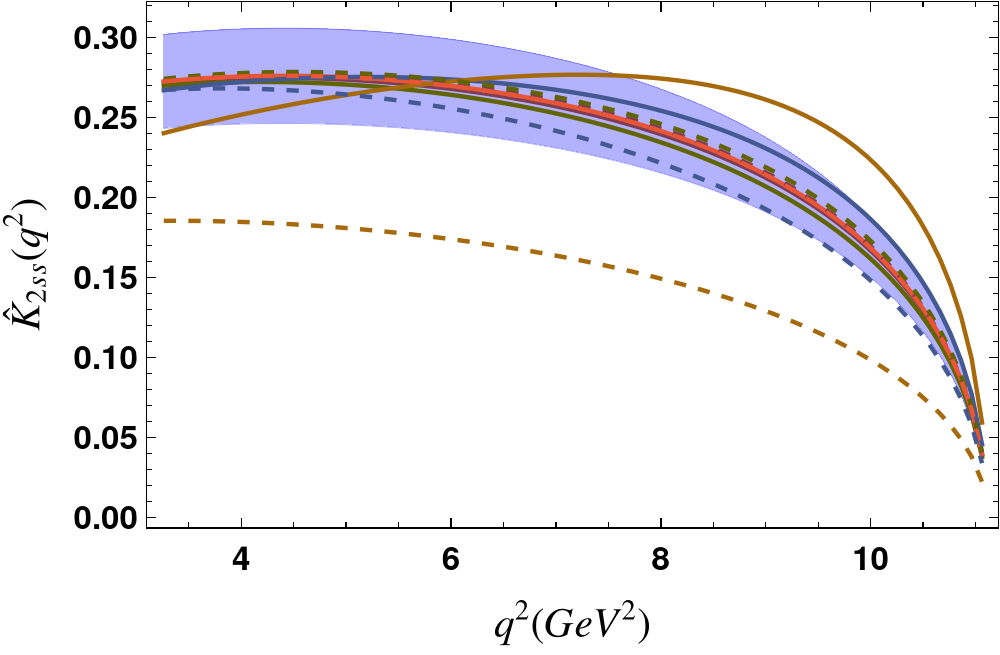}\\
		\includegraphics[scale=0.4]{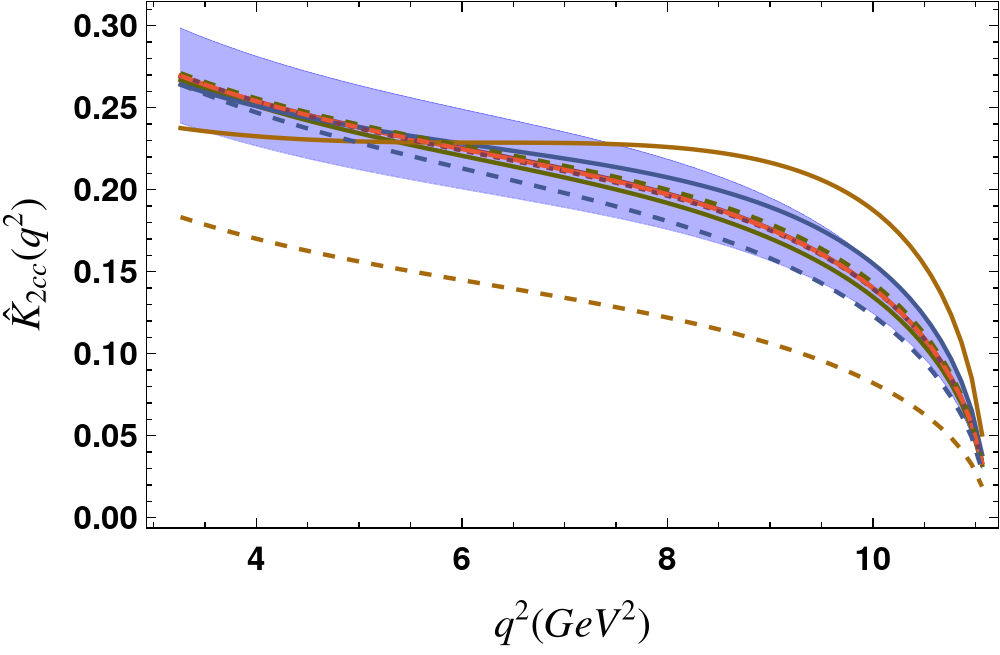}~~
		\includegraphics[scale=0.4]{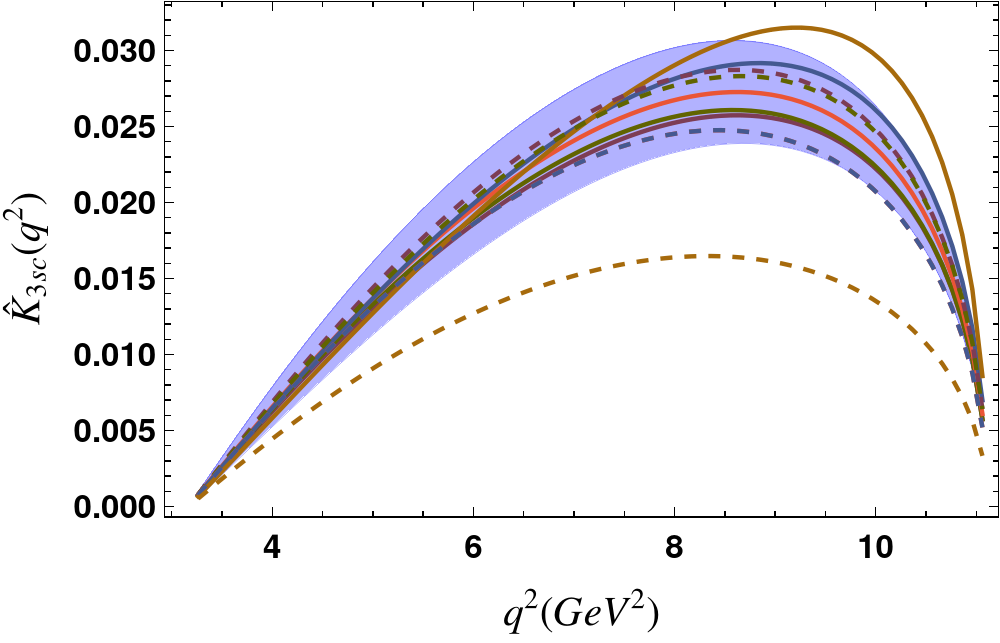}~~
		\includegraphics[scale=0.4]{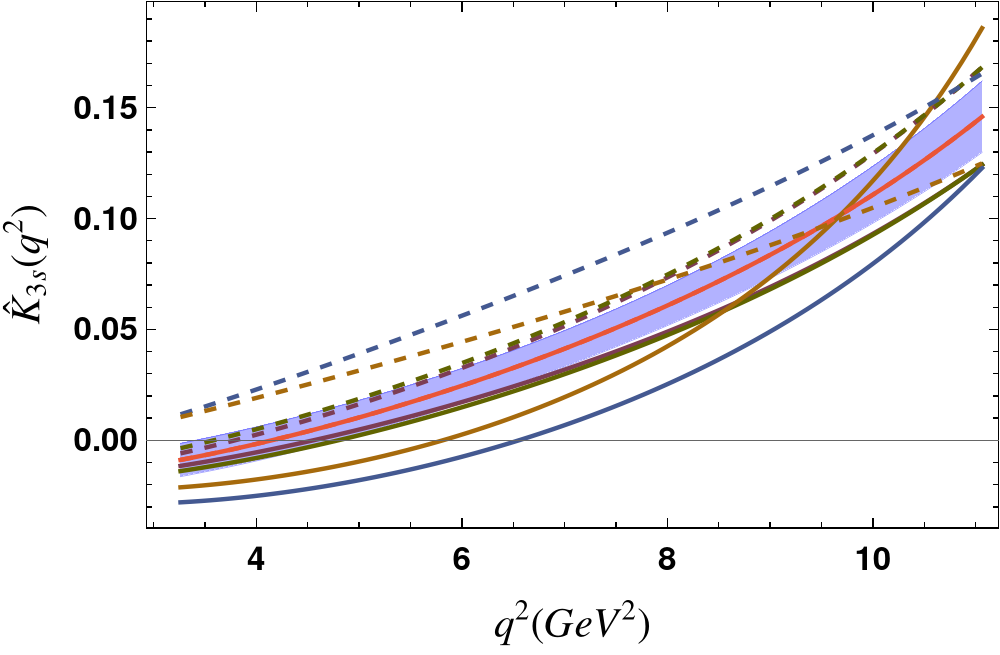}\\~~~
		\includegraphics[scale=0.6]{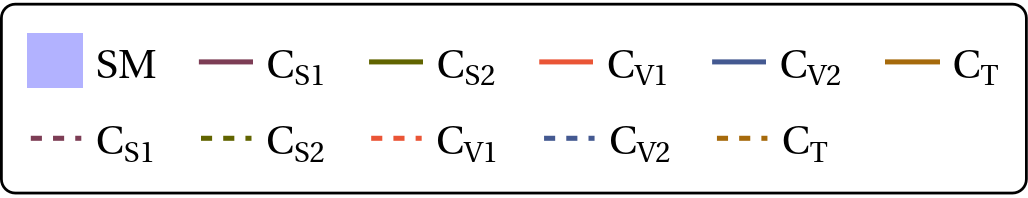}	
		\caption{The $q^2$ dependence of the angular observables for $\Lambda_{b}\to \Lambda^+_{c}(\to \Lambda \pi^+) \tau \bar{\nu}_{\tau} $ decay in the one operator scenarios with $C_i = 0.1$ (solid line) and $C_i=-0.1$ (dashed line).}
		\label{fig:q2shapesobs}
	\end{figure*}

From eq.~\ref{eq:ang_dist}, we see that unpolarized $\Lambda_{b}$ baryon decay gives ten observables with different helicity combinations. We expect that with the next runs, LHCb can measure all these observables. In this work, we also analysed the new physics sensitivities of these observables across the allowed $q^2$ regions. The corresponding results are presented in fig.~\ref{fig:q2shapesobs}. 
To make the analysis more general, in this part of the analysis we have not used the fitted values of the WCs. Rather, we have presented the results (in fig.~\ref{fig:q2shapesobs}) for $Re(C_i) = 0.1$ (solid lines) and $Re(C_i) = - 0.1$ (dashed line), respectively, which is guided by the fitted values of the new physics WCs in table~\ref{tab:1parmscnr}\footnote{Given the errors of the fitted values of the new physics WCs in table~\ref{tab:1parmscnr}, the magnitudes of these coefficients could be as large as $0.1$. Guided by this observation, we restrict the magnitudes of the new WCs to relatively small values to generate these plots. We have not chosen large values for the new WCs to test the NP sensitivities of the angular observables. Hence, we have presented our results while considering small NP effects and assumed $-0.1 \le Re(C_i) \le 0.1$. From the predictions and the respective $q^2$ distributions, one will get an idea of the range of values of these angular and asymmetric observables for the variation of the WCs in the range $[-0.1,0.1]$. If we notice deviations with relatively small values of the WCs, in that case, we could expect a much more significant effect for larger values of the WCs and expect them to detect relatively easily.}. 
 
In the $q^2$ distributions of the observables, we have not shown the errors in the NP cases. However, it is essential to check whether or not we can distinguish the contribution of NP from that of the respective SM within the given errors. Hence, for a better understanding, we have predicted the values of all these angular and asymmetric observables in all the one-operator scenarios in small $q^2$ bins, which we have shown separately in the tables from \ref{tab:obsbinprdall1} to \ref{tab:obsbinprdall12} in the appendix. We have marked in bold the predicted values that are inconsistent with their respective SM predictions at their respective 3$\sigma$ error bars. All these predictions follow the trend we could see in the respective $q^2$ distributions in fig.~\ref{fig:q2shapesobs}. We note that the one-operator scenario with $\mathcal{O}_T$ might show sizeable deviations from the respective SM predictions in a couple of those observables. Other than operator $\mathcal{O}_{T}$, the $\tau$-polarisation asymmetry $P_{\tau}^{\Lambda_c}(q^2)$ also shows sensitivity to the operators $\mathcal{O}_{S_1}$ and $\mathcal{O}_{V_2}$. In the bins within the range $ 7 \le q^2 \le q^2_{max}$ (in GeV$^2$) the predictions for $P_{\tau}^{\Lambda_c}(q^2)$ reveal distinguishable deviations from the SM predictions.

We have noted earlier that the $q^2$ distributions of the observables $A_{FB}^{\tau}(q^2)$ and $A_{FB}^{\Lambda_c \tau}(q^2)$ have zero crossing at $q_0^2$ (in GeV$^2$) in the SM. From the $q^2$ distributions presented in fig.~\ref{fig:q2shapesobs}, it is evident that, aside from the scenario $\mathcal{O}_T$, the other one-operator scenarios do not show any significant shifts in the values of  $q_0^2$ compared to the respective SM predictions. In the scenario $\mathcal{O}_T$, the value of $q^2$ at which the $A_{FB}^{\tau}(q^2) \approx 0$ is given by 
\begin{equation}
	\bf q_0^2\big|_{\mathcal{O}_T} = 9.47 (21)\ \  \text{GeV$^2$},
\end{equation}
which is 5$\sigma$ away from the respective SM prediction. For $A_{FB}^{\Lambda_c \tau}(q^2)$, there is no one-operator scenario that we observe has any considerable shift in the value of $q_0^2$. This is an interesting observation to test in the experiment. In the following subsection we will present the results of the two-operator analysis.

\subsection{New physics analysis: Two operators scenario}

Many UV complete NP models have contributions in $b\to c\tau^-\bar{\nu}$ decays \cite{Freytsis:2015qca}. Among them, there are models that contribute to more than one effective operator, as defined in eq.~\ref{eq:heff}. For example, one could see the analysis in the models like minimal supersymmetric standard model (MSSM) \cite{Boubaa:2016mgn}, (B-L) extension of the MSSM \cite{Boubaa:2022xsk}, nonminimal universal extra dimension (NMUED) \cite{Biswas:2017vhc}, two Higgs doublet model (2HDM) with or without extension \cite{Crivellin:2012ye,Chen:2021vzk}, scalar and vector-leptoquark (LQ) model \cite{Dorsner:2013tla,Freytsis:2015qca,Fedele:2022iib}, extension with a general $W^{\prime}$ boson \cite{Gomez:2019xfw}. All the five new operators in eq.~\ref{eq:heff} might get contributions in the MSSM framework discussed in \cite{Boubaa:2016mgn,Boubaa:2022xsk}. Though their relative sizes may be different. The NMUED model contributes to $C_{S_1}$ and $C_{V_1}$ while in the 2HDM there will be contributions in $C_{S_1}$ and $C_{S_2}$ via the tree level charged Higgs effects \cite{Crivellin:2012ye}. However, in addition, if we consider the neutral components of the 2HDM, which may contribute to $C_{V_1}$ and $C_{V_2}$ via loop effects \cite{Biswas:2021pic,Kolay:2024wns}. There are different types of LQ models, like the scalar or vector type LQs; among them, there are subclasses \cite{Dorsner:2016wpm}. These different LQ models may contribute to different sets of operators in eq.~\ref{eq:heff}. For example, $SU(2)_L$-singlet scalar LQ $S_1$ contributes to $C_{V_1}$ and $C_{S_2}/C_T$ while the $SU(2)_L$-singlet vector LQ $U_1$ contributes to $C_{V_1}$ and $C_{S_1}$. Furthermore, a scalar LQ with the SM quantum numbers $(3, 2)_{7/6}$ could contribute to $C_{S_1}$ and $C_T$. In the extension with a $W^{\prime}$ there are contributions in $C_{V_1}$ and $C_{V_2}$. There may be many other models one could write to describe multi-operator contributions in $b \to c\tau^-\bar{\nu}$, which is beyond the scope of this paper.   

	%%%%%%%%%%%%%%%%%%%%%%%%%%%%%%%%%%%
\begin{table}[t]
	\renewcommand{\arraystretch}{1.7}
	\centering
	\setlength\tabcolsep{8 pt}
	\begin{center}
		\resizebox{0.7\textwidth}{!}{
			\begin{tabular}{|*{9}{c|}}
				\hline
				\text{2 Operator}  &\multicolumn{4}{c|}{Two operator scenarios fit results}&\multicolumn{4}{c|}{$\sigma_{dev}$ (in $\sigma$)}\\
				\cline{2-9}
				Scenario &\multicolumn{2}{c|}{WC fit results}&\text{$\chi_{
						\min}^2$}/\text{DOF}  &  \text{P-Value}&$\text{$
					R(D)$}$&$\text{$ R(D^*)$}$&$\text{$R(\Lambda_c)$}$&$\text{$F_L^{D^*}$}$\\
				\hline
				\multirow{2}{*}{$ C_{S_1}$, $ C_{S_2}$} &$ Re[C_{S_1}]$&  
				$\text{-2.268(207)}$& \multirow{2}{*}{$3.432$/$3$} &\multirow{2}{*}{$0.330$}&\multirow{2}{*}{$0.067$}&\multirow{2}{*}{$0.409$}&\multirow{2}{*}{$1.437$}&\multirow{2}{*}{$0.237$} \\
				%\cline{2-3}
				& $Re[C_{S_2}]$&$\text{0.904(220)}$&&&&&&\\
				\hline
				\multirow{2}{*}{$ C_{S_1}$, $C_{T}$}&$Re[C_{S_1}]$&  
				$\text{0.098(46)}$& \multirow{2}{*}{$3.978$/$3$} &\multirow{2}{*}{$0.264$}&\multirow{2}{*}{$0.038$}&\multirow{2}{*}{$0.526$}&\multirow{2}{*}{$1.494$}&\multirow{2}{*}{$0.010$} \\
				%\cline{2-3}
				& $Re[C_{T}]$&$\text{-0.014(19)}$&&&&&&\\
				\hline
				\multirow{2}{*}{$ C_{S_2}$, $C_{T}$}&$ Re[C_{S_2}]$&  
				$\textbf{-1.255(64)}$&\multirow{2}{*}{$\bf 1.553/3$}&\multirow{2}{*}{$\bf 0.670$}&
				\multirow{2}{*}{$\bf 0.039$}&\multirow{2}{*}{$\bf 0.391$}&\multirow{2}{*}{$\bf 0.963$}&\multirow{2}{*}{$\bf 0.278$}\\
				% \cline{2-3}
				&$Re[C_{T}]$&$\textbf{0.226(32)}$&&&&&&\\
				\hline
				\multirow{2}{*}{$ C_{V_1}$, $C_{V_2}$} &$ Re[C_{V_1}]$&  $\text{-0.978(32)}$   
				&\multirow{2}{*}{$3.557$/$3$}&\multirow{2}{*}{$0.313$}
				&\multirow{2}{*}{$0 .113$}&\multirow{2}{*}{$0 .303$}&\multirow{2}{*}{$1.503$}&\multirow{2}{*}{$0.013$}\\
				% \cline{2-3}
				&$ Re[C_{V_2}]$&  $\text{1.055(23)}$&&&&&&\\
				\hline
				\multirow{2}{*}{$  C_{V_1}$, $C_{T}$}&$Re[C_{V_1}]$& 
				$\text{0.077(31)}$&\multirow{2}{*}{$2.827$/$3$}&\multirow{2}{*}{$0.419$}&
				\multirow{2}{*}{$0.148$}&\multirow{2}{*}{$0 .466$}&\multirow{2}{*}{$1.296$}&\multirow{2}{*}{$0.006$}\\
				%\cline{2-3}
				&$Re[C_{T}]$&$\text{0.037(37)}$&&&&&&\\
				\hline
				\multirow{2}{*}{$ C_{V_2}$, $C_{T}$}  &$ Re[C_{V_2}]$& $\text{0.080(53)}$&\multirow{2}{*}{$5.829$/$3$}&\multirow{2}{*}{$0.120$}&
				\multirow{2}{*}{$0.435$}&\multirow{2}{*}{$0 .560$}&\multirow{2}{*}{$1.634$}&\multirow{2}{*}{$0.389$}\\
				% \cline{2-3}
				&$Re[C_{T}]$& $\text{-0.059(28)}$&&&&&&\\
				\hline
				\multirow{2}{*}{$ C_{S_1}$, $ C_{V_1}$}  &$ Re[C_{S_1}]$&  $\text{0.051(73)}$&  \multirow{2}{*}{$3.534$/$3$}&\multirow{2}{*}{$0.316$}&
				\multirow{2}{*}{$0.102$}&\multirow{2}{*}{$0 .311$}&\multirow{2}{*}{$1.498$}&\multirow{2}{*}{$0.0004$} \\
				%\cline{2-3}
				&$Re[C_{V_1}]$&$\text{0.033(34)}$&&&&&&\\
				\hline
				\multirow{2}{*}{$ C_{S_1}$, $ C_{V_2}$}  &$ Re[C_{S_1}]$&  
				$\text{0.123(48)}$&\multirow{2}{*}{$3.511$/$3$}&\multirow{2}{*}{$0.319$}&
				\multirow{2}{*}{$0 .106$}&\multirow{2}{*}{$0.301$}&\multirow{2}{*}{$1.494$}&\multirow{2}{*}{$0.098$}\\
				% \cline{2-3}
				&$ Re[C_{V_2}]$&$\text{-0.033(33)}$&&&&&&\\
				\hline
				\multirow{2}{*}{$ C_{S_2}$, $ C_{V_1}$}  &$ Re[C_{S_2}]$&  $\text{0.045(65)}$&\multirow{2}{*}{$3.533$/$3$}&\multirow{2}{*}{$0.316$}&
				\multirow{2}{*}{$0 .101$}&\multirow{2}{*}{$0 .311$}&\multirow{2}{*}{$1.496$}&\multirow{2}{*}{$0.079$}\\
				%\cline{2-3}
				&$ Re[C_{V_1}]$ &$\text{0.038(29)}$ &&&&&&\\
				\hline
				\multirow{2}{*}{$ C_{S_2}$, $ C_{V_2}$}  &$Re[C_{S_2}]$&  
				$\text{0.139(54)}$&\multirow{2}{*}{$3.474$/$3$}&\multirow{2}{*}{$0.324$}&
				\multirow{2}{*}{$0.104$}&\multirow{2}{*}{$0.292$}&\multirow{2}{*}{$1.489$}&\multirow{2}{*}{$0.098$}\\
				%\cline{2-3}
				&$Re[C_{V_2}]$&$\text{-0.048(36)}$&&&&&&\\
				\hline
			\end{tabular}
		}
		\caption{ Fit results for the simultaneous fitting of the new physics Wilson coefficients in a two-operator scenario. The inputs used for the fit include $R(D)$, $R(D^*)$, $R(\Lambda_{c})$, $F_L^{D^*}$ and $\mathcal{B}(\Lambda_{b}\to\Lambda_{c} \tau \bar{\nu}_{\tau})$, where we only consider new physics in the $\tau$ final state. Additionally, we quantify the tension between the predicted observables and the corresponding experimental data for each scenario, presenting these results in units of $\sigma$ in the last four columns.}
		\label{tab:2parmfitres}
	\end{center}
\end{table}
%%%%%%%%%%%%%%%%%%%%%%%%%%%%%%%%%%
	\begin{figure}[t!]
	\begin{center}
		\includegraphics[scale=0.45]{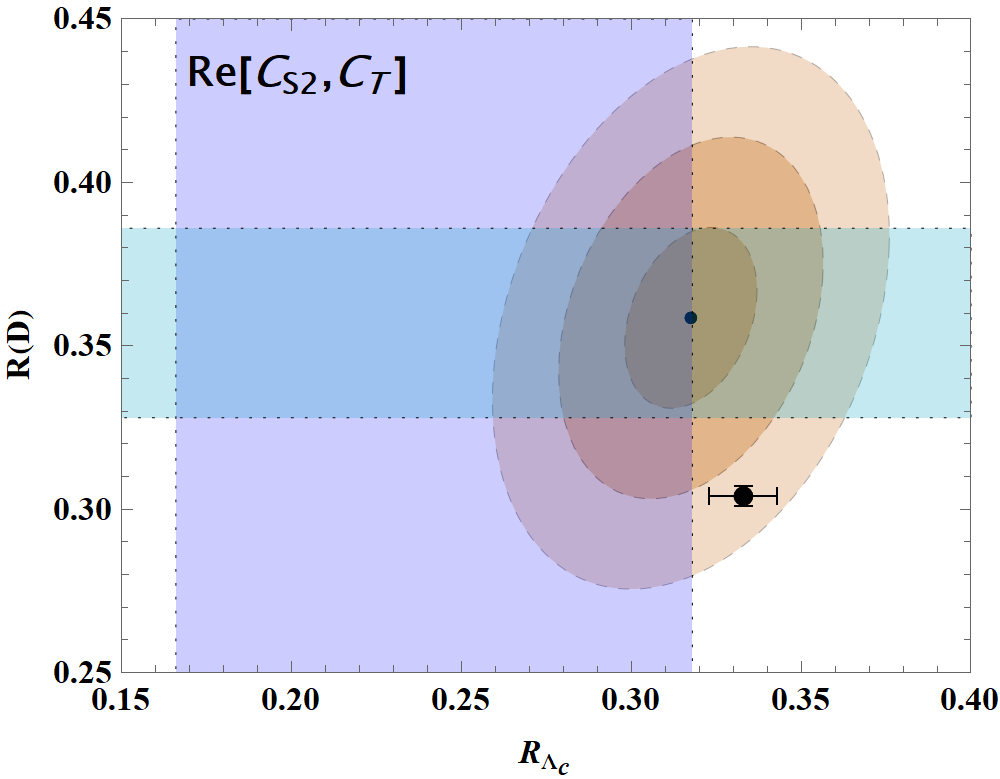}~~~
		\includegraphics[scale=0.45]{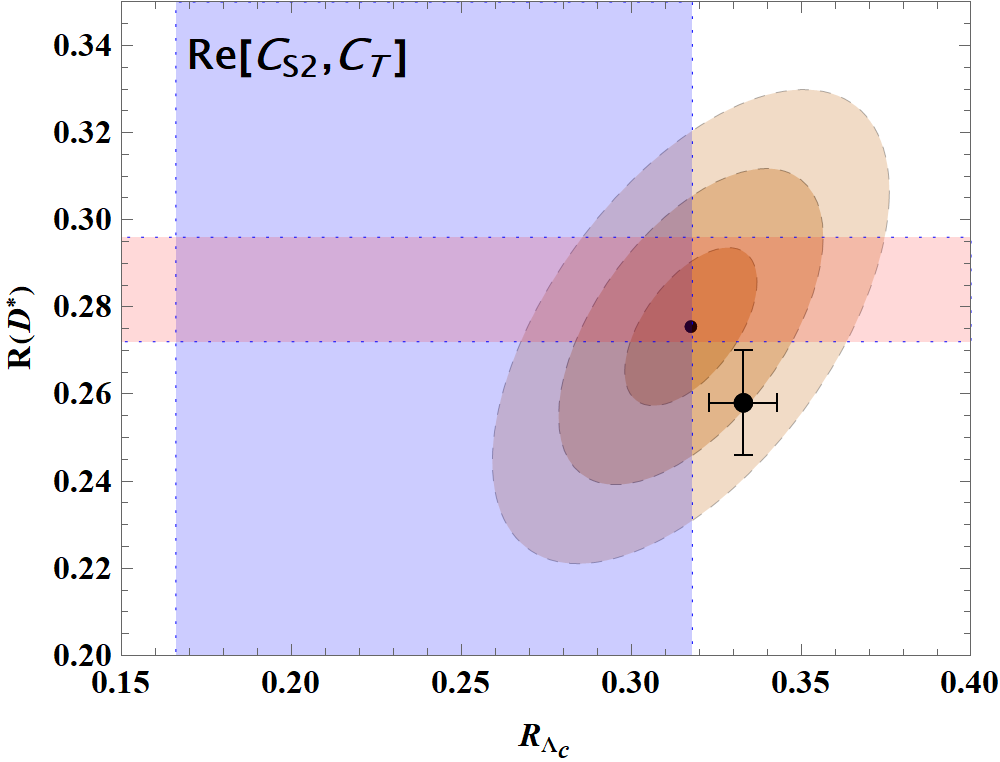}\\
		\includegraphics[scale=0.45]{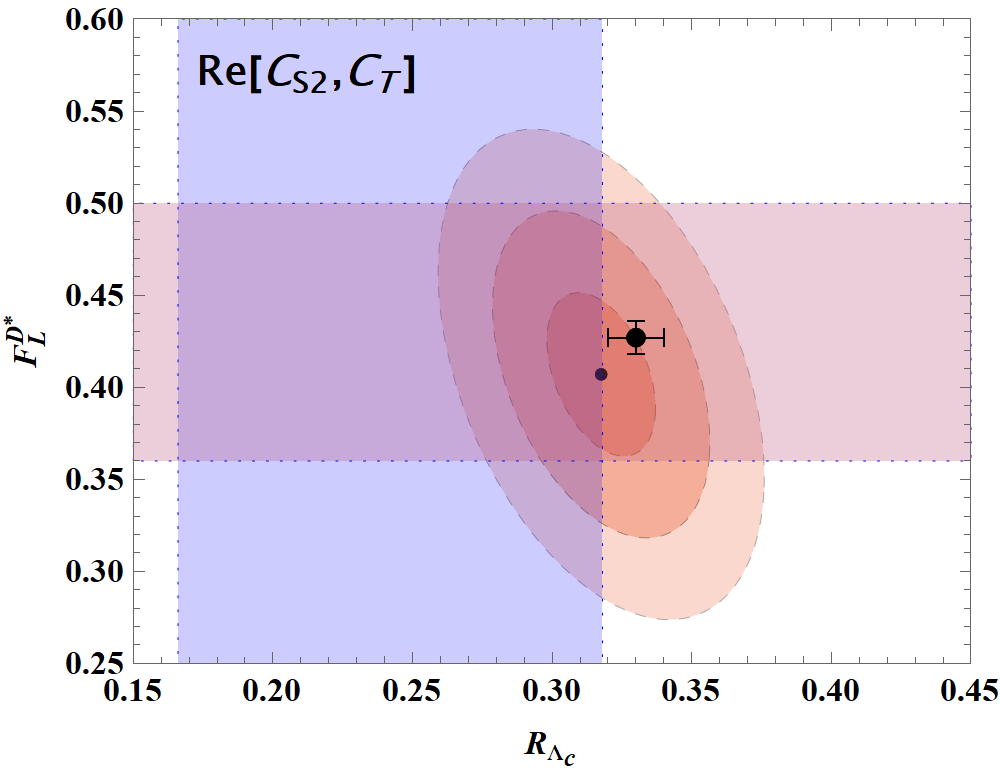}~~~
		\includegraphics[scale=0.45]{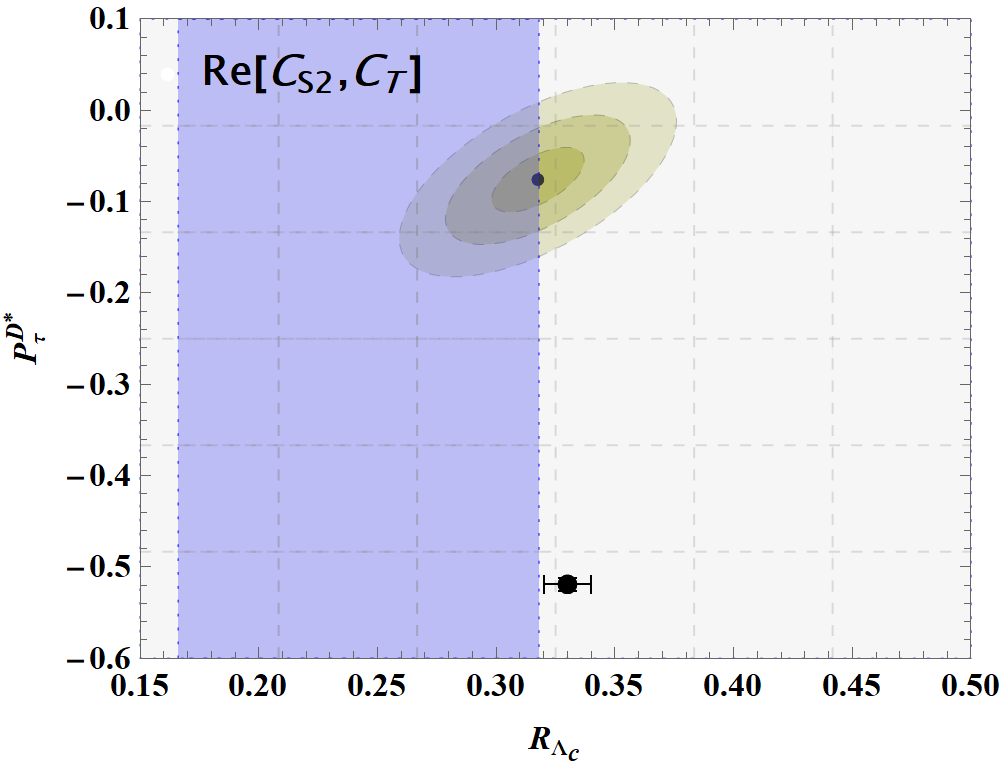}
	\end{center}
	\caption{The plot illustrates the correlation between $R(D^{(*)})$, $F_L^{D^*}$, $P_\tau^{D^*}$ with $R(\Lambda_{c})$ in the $Re[C_{S2}, C_T]$ two-operator NP scenario (best-fit scenario). The horizontal and vertical color bands represent the experimental constraints for respective observable, at a 1$\sigma$ confidence level. Correlations between the observables are depicted by an ellipse. Additionally, we show the  $1\sigma$, $2\sigma$, and $3\sigma$ contour error bands around the best estimate point (in black) based on the best-fit values provided in Table~\ref{tab:2parmfitres}. The Standard Model estimate is also indicated by a black point, accompanied by vertical and horizontal $1\sigma$ error bars for the respective observables.}
	\label{fig:2paracorrcs2ct}
\end{figure}

\begin{table}[t]
	\renewcommand{\arraystretch}{2.}
	\centering
	\setlength\tabcolsep{8 pt}
	\begin{center}
		\resizebox{1.03\textwidth}{!}{
			\begin{tabular}{|*{12}{c|}}
				\hline
				\text{ Obs.}  &  \text{$ [C_{S_1}, C_{S_2}]$}& \text{$ [C_{S_1}, C_T]$}  &  \text{$ [C_{S_2}, C_T]$}  &  \text{$ [C_{V_1}, C_{V_2}]$}  &  \text{$ [C_{V_1},C_T]$}  &  \text{$ [C_{V_2}, C_T]$}  &  \
				\text{$ [C_{S_1}, C_{V_1}]$}  &  \text{$ [C_{S_1}, C_{V_2}]$}  &  \text{$ [C_{S_2},C_{V_1}]$}  &  \text{$ [C_{S_2}, C_{V_2}]$} &  \text{SM} \\
				\hline\hline
				\text{$\hat{K}_{1cc} = \langle \hat{K}_{1cc}(q^2)\rangle$}  & \text{$0.311(2)$} & $\text{0.314(1)}$  &  $\text{0.300(2)}$  &  $\text{0.312(1)}$  \
				&  $\text{0.310(3)}$  &  $\text{0.313(1)}$  &  $\text{0.312(1)}$  &  $\text{0.314(1)}$  &  $\text{0.312(1)}$  \
				&  $\text{0.314(1)}$ & $\text{0.312(2)}$\\
				\hline
				\text{$\hat{K}_{1ss}= \langle  \hat{K}_{1ss} (q^2)\rangle$} &\text{$0.345(1)$}&  $\text{0.343(1)}$  &  $\text{0.350(1)}$  &  $\text{0.344(1)}$ \
				&  $\text{0.345(1)}$  &  $\text{0.343(1)}$  &  $\text{0.344(1)}$  &  $\text{0.343(1)}$  &  $\text{0.344(1)}$  \
				&  $\text{0.343(1)}$ & $\text{0.344(1)}$\\
				\hline
				\text{$ \hat{K}_{2cc}= \langle \hat{K}_{2cc}(q^2) \rangle$}  &\text{$0.202(22)$}& $\text{0.200(25)}$  &  $\text{-0.018(29)}$  &  $\text{-0.210(23)}$  \
				&  $\text{0.223(26)}$  &  $\text{0.178(25)}$  &  $\text{0.208(22)}$  \
				&  $\text{0.205(22)}$  &  $\text{0.206(22)}$  &  $\text{0.197(22)}$ &$\text{0.208(22)}$\\
				\hline
				\text{$ \hat{K}_{2ss}=\langle  \hat{K}_{2ss}(q^2) \rangle$}&\text{$0.237(26)$}&  $\text{0.229(28)}$  &  $\text{-0.008(34)}$  &  $\text{-0.243(26)}$  \
				&  $\text{0.258(30)}$  &  $\text{0.205(29)}$  &  $\text{0.240(26)}$  \
				&  $\text{0.235(26)}$  &  $\text{0.238(26)}$  &  $\text{0.227(25)}$ &$\text{0.240(26)}$\\
				\hline
				\text{$\hat{K}_{3sc}= \langle \hat{K}_{3sc}(q^2) \rangle$} &\text{$0.018(2)$}&   $\text{0.016(2)}$  &  $\text{0.005(4)}$  &  $\text{-0.017(2)}$  \
				&  $\text{0.018(3)}$  &  $\text{0.014(3)}$  &  $\text{0.017(2)}$  &  $\text{0.016(2)}$  &  $\text{0.017(2)}$  \
				&  $\text{0.016(2)}$ &$\text{0.017(2)}$\\
				\hline
				\text{$ \hat{K}_{3s}=\langle \hat{K}_{3s}(q^2) \rangle$} &\text{$0.131(19)$}& $\text{0.036(9)}$  &  $\text{-0.010(20)}$  &  $\text{0.132(18)}$  \
				&  $\text{0.037(10)}$  &  $\text{0.030(14)}$  &  $\text{0.039(10)}$  \
				&  $\text{0.041(12)}$  &  $\text{0.038(10)}$  &  $\text{0.042(12)}$ &$\text{0.044(8)}$\\
				\hline
				\text{$ A_{FB}^{\tau}= \langle  A_{FB}^{\tau}(q^2) \rangle$}  &\text{$-0.303(17)$}& $\text{0.119(12)}$  &  $\text{-0.081(28)}$  &  $\text{0.377(9)}$  \
				&  $\text{0.101(12)}$  &  $\text{0.121(13)}$  &  $\text{0.111(15)}$  \
				&  $\text{0.118(12)}$ & $\text{0.106(11)}$&  $\text{0.104(11)}$ & $\text{0.103(11)}$\\
				\hline
				\text{$A_{FB}^{\Lambda_{c} \tau}=\langle  A_{FB}^{\Lambda_{c} \tau}(q^2) \rangle$}  &\text{$-0.169(18)$}&  $\text{0.025(9)}$  &  $\text{0.020(8)}$  &  $\text{-0.195(21)}$  \
				&  $\text{0.024(7)}$  &  $\text{0.017(8)}$  &  $\text{0.023(9)}$  &  $\text{0.025(8)}$  &  $\text{0.021(7)}$  \
				&  $\text{0.020(7)}$ &$\text{0.018(6)}$\\
				\hline
				\text{$A_{FB}^{\Lambda_{c}}=\langle A_{FB}^{\Lambda_{c}}(q^2) \rangle$}  &\text{$0.338(37)$}&   $\text{0.329(41)}$  &  $\text{-0.017(49)}$  &  $\text{-0.348(38)}$  \
				&  $\text{0.370(43)}$  &  $\text{0.294(42)}$  &  $\text{0.344(37)}$  \
				&  $\text{0.337(37)}$  &  $\text{0.341(37)}$  &  $\text{0.326(36)}$ & $\text{0.344(37)}$\\
				\hline
				\text{$P_{\Lambda_c}=\langle P_{\Lambda_c}(q^2) \rangle$}  &\text{$-0.757(22)$}& $\text{-0.725(49)}$  &  $\text{0.065(109)}$  &  $\text{0.772(18)}$  \
				&  $\text{-0.830(51)}$  &  $\text{-0.647(62)}$  &  $\text{-0.761(12)}$  &  $\text{-0.742(21)}$  &  $\text{-0.754(18)}$  \
				&  $\text{-0.712(27)}$ &$\text{-0.761(12)}$\\
				\hline
				\text{$P_{\tau }^{(\Lambda_{c})}= \langle P_{\tau }^{(\Lambda_{c})}(q^2) \rangle$}  &\text{$-0.218(47)$}&  $\text{-0.196(36)}$  &  $\text{0.098(37)}$  &  $\text{-0.244(25)}$  \
				&  $\text{-0.245(24)}$  &  $\text{-0.214(35)}$  &  $\text{-0.224(51)}$  &  $\text{-0.196(35)}$  &  $\text{-0.236(35)}$  \
				&  $\text{-0.220(31)}$  &$\text{-0.258(16)}$\\
				\hline
				\text{$C^{\tau}_F= \langle C^{\tau}_F(q^2) \rangle$}  &\text{-0.102(8)}&  $\text{-0.088(7)}$  &  $\text{-0.151(11)}$  &  $\text{-0.098(6)}$  \
				&  $\text{-0.106(12)}$  &  $\text{-0.091(6)}$  &  $\text{-0.094(7)}$  &  $\text{-0.088(7)}$  &  $\text{-0.095(6)}$  \
				&  $\text{-0.089(7)}$  &$\text{-0.096(6)}$\\
				\hline
			\end{tabular}
		}
		\caption{Predictions of angular observables with the two operator scenarios fit results considering different new physics in $\tau$ channel.}
		\label{tab:predangular_twoopr}
	\end{center}
\end{table}	

%%%%%%%%%%%%%%%%%%%%%%%%%%%%%%%%%%%%%%%%%%%%%%%%%%%%%%%%

Following the discussion in the last subsection and the above paragraph, there is enough motivation to look for the impact of data on the two operator scenarios. We have separately studied the impact of the two-operator scenarios. Moreover, while analysing the contributions to the helicity amplitudes due to the BSM operators, in several of them, we notice terms proportional to $m_{\tau}$ which are due to the interference of $(V \pm A)$ with either $(S \pm P)$ or the tensor operators. We will miss these contributions and the relevant phenomenology when analysing one-operator scenarios. Hence, studying the two-operator scenarios may be useful in understanding the impact of these mass-dependent interference terms, which will otherwise be missing in the one-operator scenarios. Also, for simplicity, we have considered only the real WCs.

	%%%%%%%%%%%%%%%%%%%%%%%%%%%%%%%%%%%

In table~\ref{tab:2parmfitres}, we have presented the fit results for each two-operator scenario. Using these fit results, we first predict the values of the observables used in the fit and the $\tau$-polarisation asymmetries $P_{\tau}^{D}$ and $P_{\tau}^{D*}$ and we have presented them in table \ref{tab:predmain_twoopr} in the appendix. In all the two-operator scenarios, we can comfortably explain the observed data on $R(D^{(*)})$, $F_L^{D*}$ and $P_{\tau}^{D*}$, the respective data and the predicted values are consistent within their 1$\sigma$ CI. However, apart from the scenario $[\mathcal{O}_{S_2},\mathcal{O}_T]$, in all the other scenarios, the predicted values of $R(\Lambda_c)$ have slight deviation with respect to the measured value. It is only the scenario $[\mathcal{O}_{S_2},\mathcal{O}_T]$ which could accommodate the data on $R(\Lambda_c)$ alongside $R(D)$, $R(D^*)$, $F_L^{D^*}$ within their 1$\sigma$ uncertainties. This is also the best-fit scenario, with the largest p-value of 67\% among all others. In each of these cases, using the formula given in eq.~\ref{eq:tsndef}, we have estimated the deviations ($\sigma_{dev}$) between each NP prediction and the respective measured values. The respective estimates of $\sigma_{dev}$ are shown in the last four columns of table~\ref{tab:2parmfitres}. In all the other scenarios, the estimated $\sigma_{dev}$ in $R(\Lambda_c)$ are little more than 1$\sigma$, which are not large deviations. Therefore, if we consider the predictions within their 2$\sigma$ CI, each of the data independently explains the respective measured values. However, note that in the estimate of $\sigma_{dev}$, we have not included the correlations between the observables. Like the one-operator scenarios, we have studied the correlations between the observables in different two-operator scenarios. We have shown such correlations between the observables in the scenario $[\mathcal{O}_{S_2},\mathcal{O}_T]$ in fig. \ref{fig:2paracorrcs2ct}, which indicates that in this scenario, we can comfortably explain all the four data used in the analysis along with $P_{\tau}^{D^*}$. Furthermore, we have analysed the correlation between these observables in all the other two operator scenarios, which we have not shown, and noted that it is possible to simultaneously explain the data on $R(\Lambda_c)$ with $R(D)$, $F_L^{D^*}$ and $P_{\tau}^{D^*}$, respectively if we take the 3$\sigma$ contours of the respective predictions. However, while it is coming to a simultaneous explanation of $R(\Lambda_c)$ and $R(D^*)$, most scenarios fail to do so even if we consider the 3$\sigma$ contours of the respective predictions. There are too many such plots; we have not shown them here separately.

	\begin{table}[t!]\centering
		\renewcommand{\arraystretch}{1.8}
		\resizebox{0.9\textwidth}{!}{
			\begin{tabular}{|*{13}{c|}}
				\hline
				& \multicolumn{12}{c|}{Deviations w.r.t. SM predictions (in $\sigma$ level)} \\
				\cline{2-13}
				~~$\text{Scenario}$~~  &  \text{$ \hat{K}_{1cc}$}  &  \text{$\hat{K}_{1ss}$}  &  \text{$ \hat{K}_{2cc}$}  \
				&  \text{$\hat{K}_{2ss}$}  &  \text{$\hat{K}_{3sc}$}  &  \text{$\hat{K}_{3s}$}  &  \text{$ A_{FB}^{\tau}$} \
				&  \text{$ A_{FB}^{\Lambda_c \tau}$}  &  \text{$A_{FB}^{\Lambda_c}$}  &  \text{$ P_{\Lambda_c}$}  &  \text{$P_{\tau }^{(\Lambda_{c})}$} & \text{$C^{\tau}_F$}  \\
				\hline
				\text{$ [C_{S_1}, ~C_{S_2}]$} &  $0.35$  &  $0.82$  &  $0.19$  &  $0.08$  \
				&  $0.28$  &  $\bf 4.22$  &  $\bf 20.05$  &  $\bf 9.86$  &  $0.16$  & \
				$0.16$  &  $0.81$ & $0.60$\\
				\text{$ [C_{S_1}, ~C_{T}]$}  &  $0.89$  &  $0.82$  &  $0.24$  &  $0.29$  & \
				$0.35$  &  $0.66$  &  $0.98$  &  $0.65$  &  $0.36$  &  \
				$0.71$  &  $ 1.57$ & $0.87$ \\
				\text{$ [C_{S_2}, ~C_{T}]$}  &  $\bf 3.33$  &  $\bf 4.91$  &  $\bf 6.21$  &  $\bf 5.79$  \
				&  $\bf 2.68$  &  $ 1.34$  &  $\bf 6.12$  &  $0.20$  &  $\bf 7.34$  &  \
				$\bf 7.53$  &  $\bf 8.83$  &$\bf 4.39$\\
				\text{$ [C_{V_1}, C_{V_2}]$}  &  $0.0$  &  $0.0$  &  $\bf 13.13$  &  $\bf 13.14$  &  \
				$\bf 12.02$  &  $\bf 4.47$  &  $\bf 19.28$  &  $\bf 9.75$  &  $\bf 18.11$  &  \
				$\bf 70.86$  &  $0.47$ &$0.24$ \\
				\text{$ [C_{V_1}, ~C_{T}]$}  &  $0.55$  &  $0.82$  &  $0.44$  &  $0.45$  &  \
				$0.28$  &  $0.55$  &  $0.12$  &  $0.65$  &  $0.60$  &  \
				$1.32$  &  $0.45$ & $0.74$\\
				\text{$ [C_{V_2}, ~C_{T}]$}  &  $0.45$  &  $0.82$  &  $0.90$  &  $0.90$  \
				&  $0.83$  &  $0.87$  &  $1.06$  &  $0.11$  &  $1.19$  &  \
				$1.81$  &  $1.14$ & $0.59$\\
				\text{$ [C_{S_1}, ~C_{V_1}]$}  &  $0.35$  &  $0.0$  &  $0.0$  &  $0.0$  &  $0.0$  &  \
				$0.39$  &  $0.43$  &  $0.46$  &  $0.0$  &  $0.0$  &  $0.64$ & $0.22$ \\
				\text{$ [C_{S_1}, ~C_{V_2}]$}  &  $0.89$  &  $0.82$  &  $0.10$  &  $0.14$  \
				&  $0.35$  &  $0.21$  &  $0.92$  &  $0.70$  &  $0.19$  &  \
				$0.79$  &  $1.61$ & $0.87$ \\
				\text{$ [C_{S_2}, ~C_{V_1}]$}  &  $0.0$  &  $0.$  &  $0.06$  &  $0.05$  &  \
				$0.0$  &  $0.47$  &  $0.19$  &  $0.32$  &  $0.08$  &  $0.32$  & \
				$0.57$ & $0.12$\\
				\text{$[C_{S_2}, ~C_{V_2}]$}  &  $0.89$  &  $0.82$  &  $0.35$  &  $0.36$  \
				&  $0.35$  &  $0.14$  &  $0.06$  &  $0.22$  &  $0.50$  &  \
				$1.66$  &  $1.09$ & $0.76$ \\
				\hline
		\end{tabular}}
		\caption{The discrepancies/agreement between the predictions in the NP two-operator scenarios and in the SM. The respective NP predictions for the observables are shown in table \ref{tab:predangular_twoopr}, which are obtained using the fit results of table \ref{tab:2parmfitres}.}
		\label{tab:discrepancy_twoopr}
	\end{table}

In table \ref{tab:predangular_twoopr}, the predictions of the full $q^2$ integrated asymmetric and angular observables in $\Lambda_b \to \Lambda_c^+(\to\Lambda\pi^+)\tau^-\bar{\nu}_{\tau}$ decays in all the two-operator scenarios are presented which we have obtained using the fit results in table \ref{tab:2parmfitres}. Also, we have estimated the $\sigma$ level discrepancies between the prediction in the NP scenarios and the SM. We present the corresponding results in table \ref{tab:discrepancy_twoopr}, and the scenarios showing discrepancies of more than 2.5$\sigma$ level are pointed in bold font. We observe discrepancies in a couple of observables only in the scenarios:  $[\mathcal{O}_{S_1},\mathcal{O}_{S_2}]$, $[\mathcal{O}_{S_2},\mathcal{O}_T]$ and  $[\mathcal{O}_{V_1},\mathcal{O}_{V_2}]$. In the rest of the two operator scenarios, we do not observe any significant deviations in these observables with respect to the SM\footnote{We should note from the results of table \ref{tab:2parmfitres} that in the scenarios  $[\mathcal{O}_{S_1},\mathcal{O}_{S_2}]$, $[\mathcal{O}_{S_2},\mathcal{O}_T]$ and  $[\mathcal{O}_{V_1},\mathcal{O}_{V_2}]$, the available data allows values of the WCs of order $\approx 1$. On the other hand, in the rest of the two-operator scenarios, within the error bars, the allowed values of the WCs could be at most of order $10^{-1}$. The observation of large effects only in the above three scenarios is partly due to the relatively large allowed values of the WCs. Hence, we can state that our observations are associated with the current data set, and any major changes in the data might change the allowed ranges of the WCs, hence the outcome of the analysis.}. In the scenario $[\mathcal{O}_{S_2},\mathcal{O}_T]$, apart from $A^{\Lambda_c \tau}_{FB}$ and $\hat{K}_{3s}$, all the other observables will show discrepancies for the corresponding SM predictions. As we can see from table \ref{tab:predangular_twoopr}, in this scenario, the predicted values of a couple of observables will be lower than the respective SM predictions, while in a few, they are larger than the SM. For the scenario $[\mathcal{O}_{S_1},\mathcal{O}_{S_2}]$, we observe sizeable deviations in $\hat{K}_{3s}$, $A^{\tau}_{FB}$ and $A^{\Lambda_c \tau}_{FB}$ which is significant in $A^{\tau}_{FB}$. The predictions of $A^{\tau}_{FB}$ and $A^{\Lambda_c \tau}_{FB}$ are lower than the SM and have opposite signs, while the prediction of $\hat{K}_{3s}$ is larger than the SM. In the scenario $[\mathcal{O}_{V_1},\mathcal{O}_{V_2}]$, apart from $\hat{K}_{1cc}$, $\hat{K}_{1ss}$, $C_F^{\tau}$ and  $P^{(\Lambda_c)}_\tau$ the predictions of the rest of the observables have considerable deviations with respect the SM values. Also, in a couple of observables, we see the predicted values are lower than and have opposite signs, and the rest are higher than SM. Therefore, once we have precise measurements of all these observables, a comparative study will be helpful to distinguish these three types of scenarios.

	\begin{figure*}[t]
		\centering
		\includegraphics[scale=0.4]{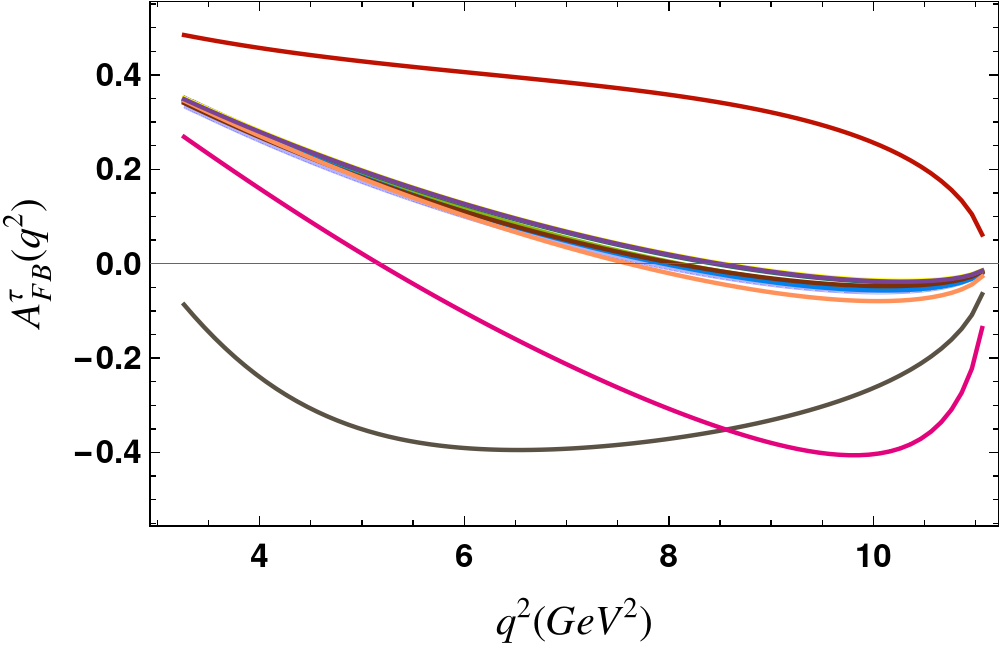}~~
		\includegraphics[scale=0.4]{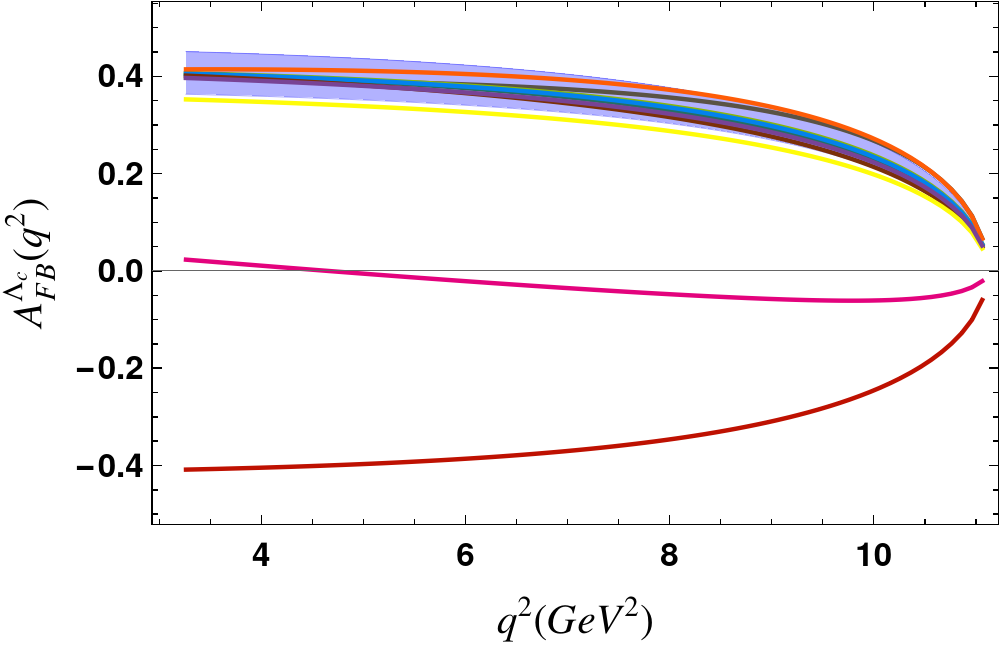}~~
		\includegraphics[scale=0.4]{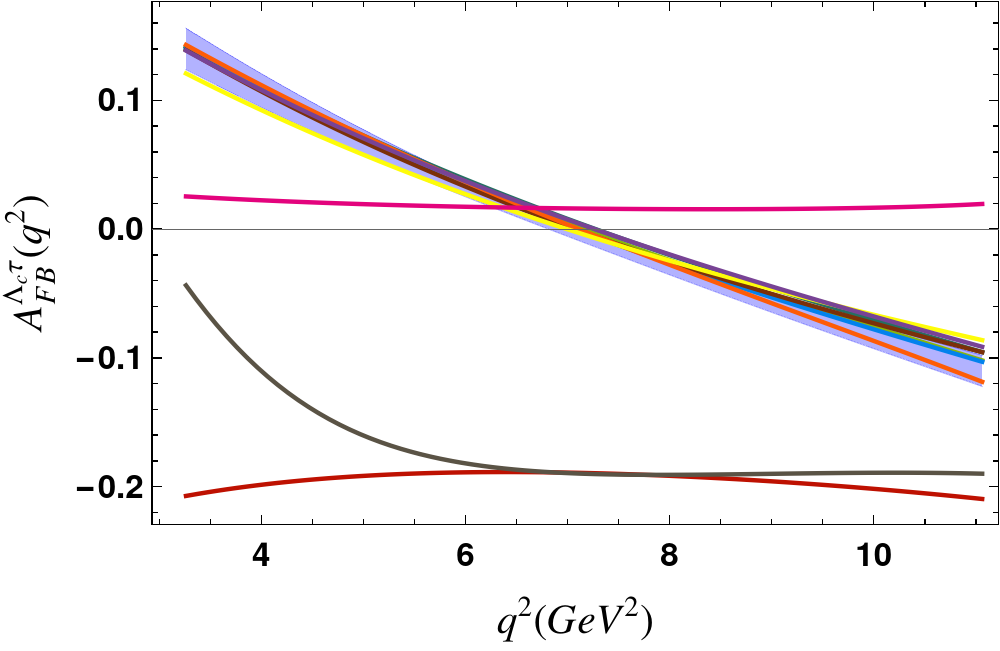}\\
		\includegraphics[scale=0.4]{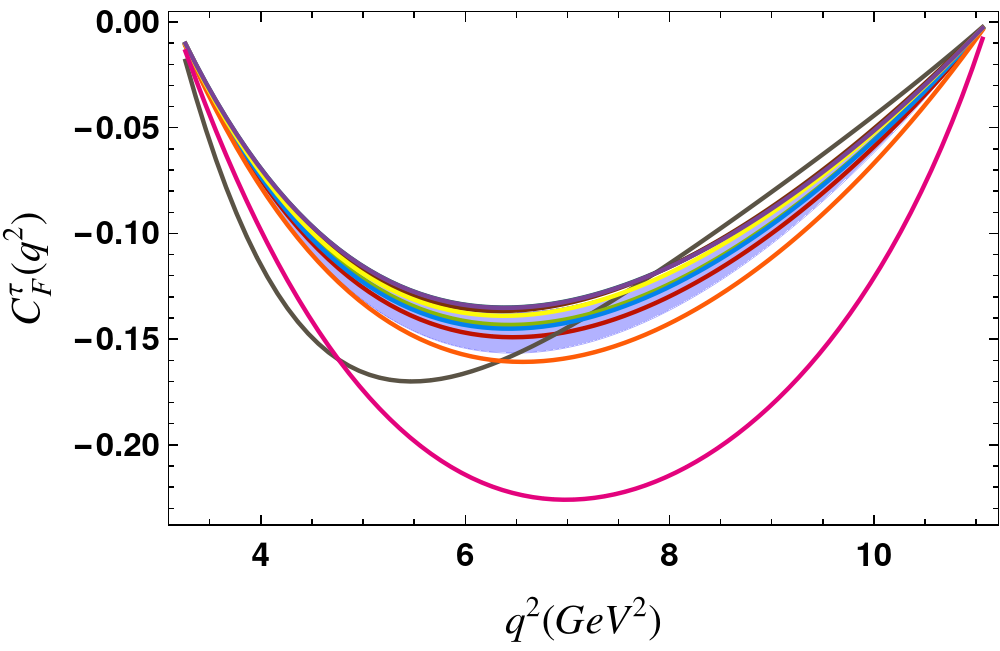}~~
		\includegraphics[scale=0.4]{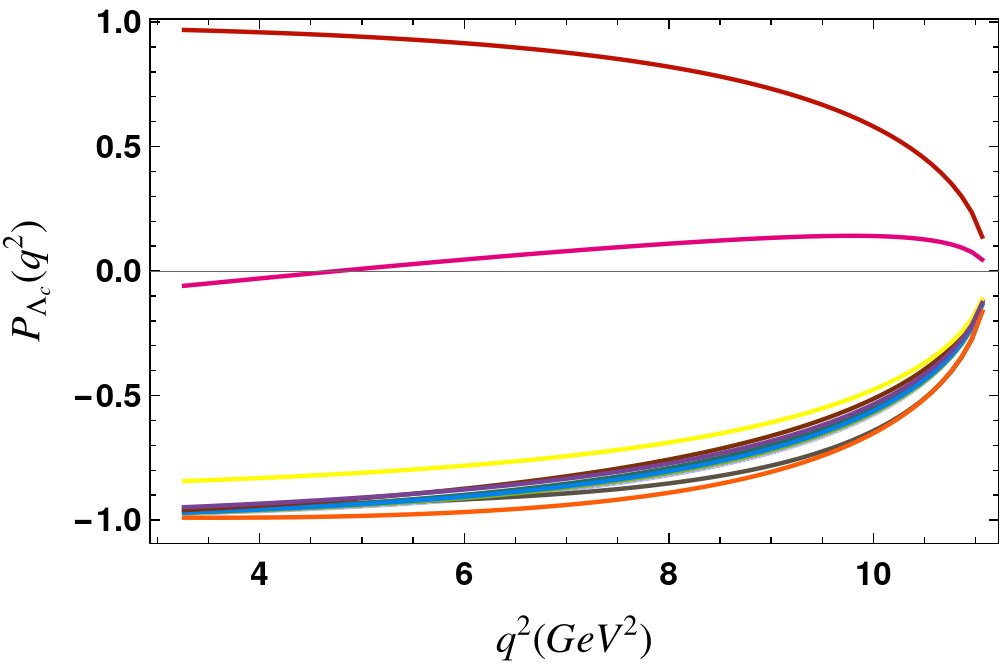}~~
		\includegraphics[scale=0.4]{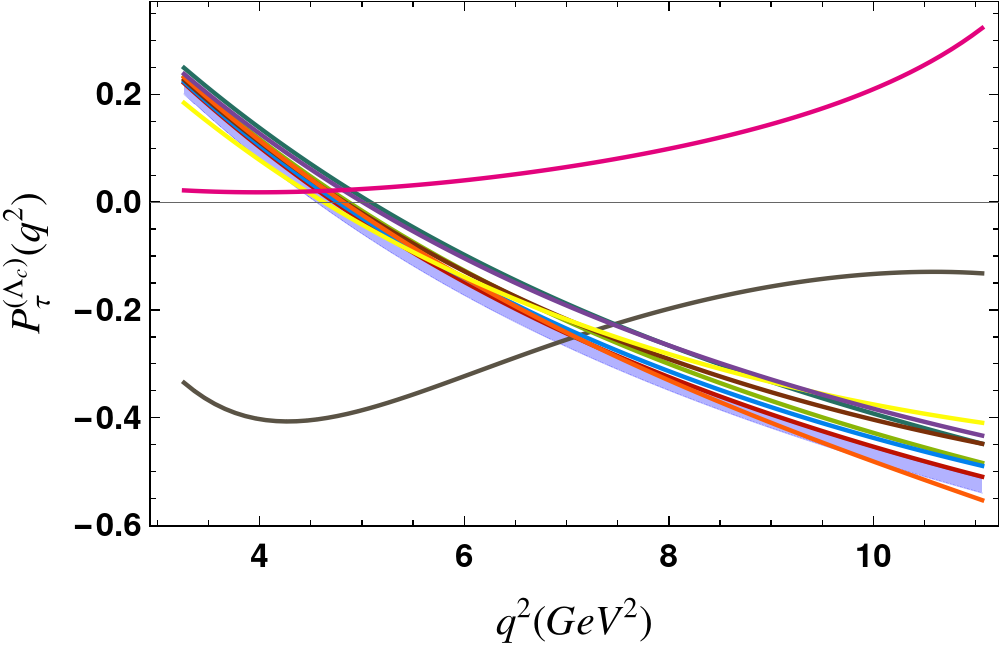}\\
		\includegraphics[scale=0.4]{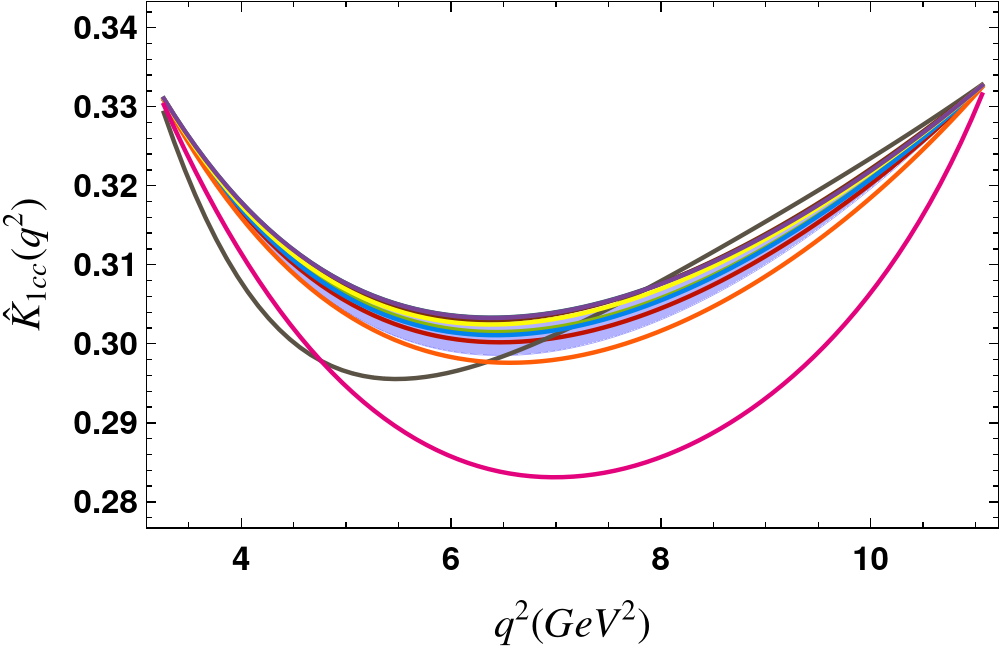}~~
		\includegraphics[scale=0.4]{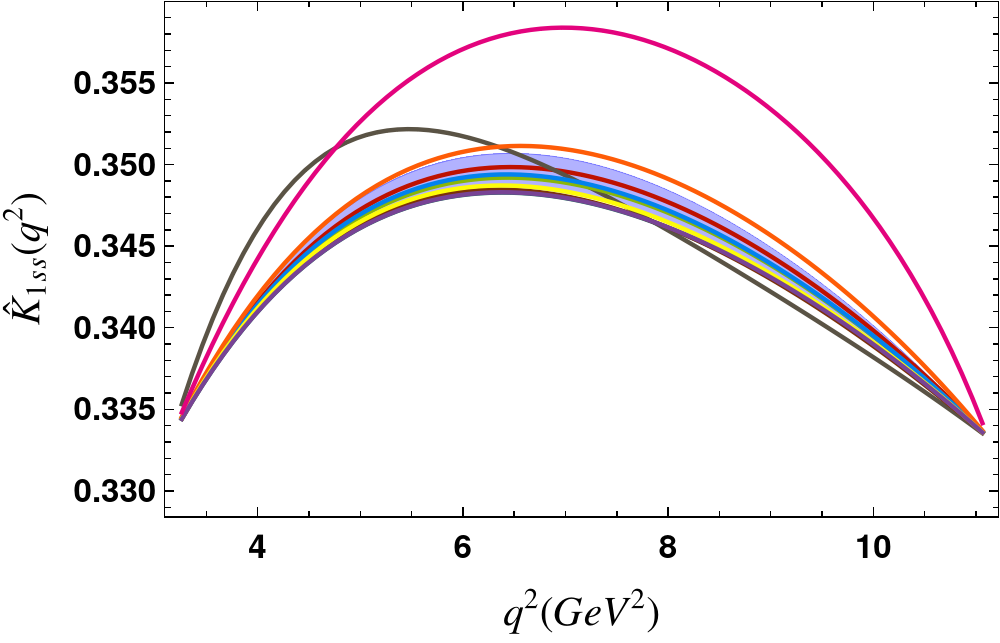}~~	
		\includegraphics[scale=0.4]{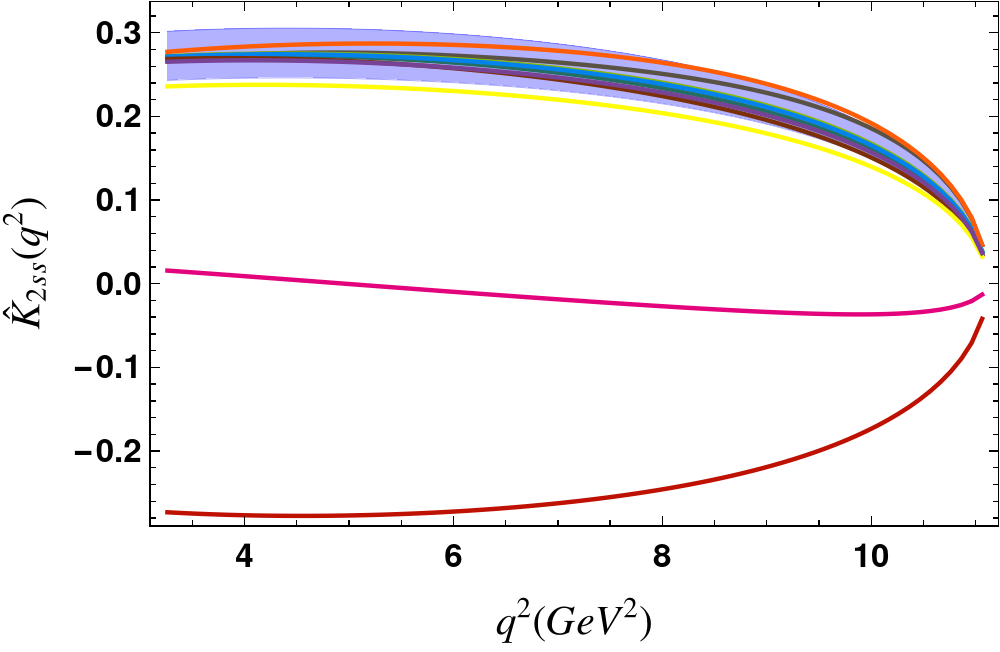}\\
		\includegraphics[scale=0.4]{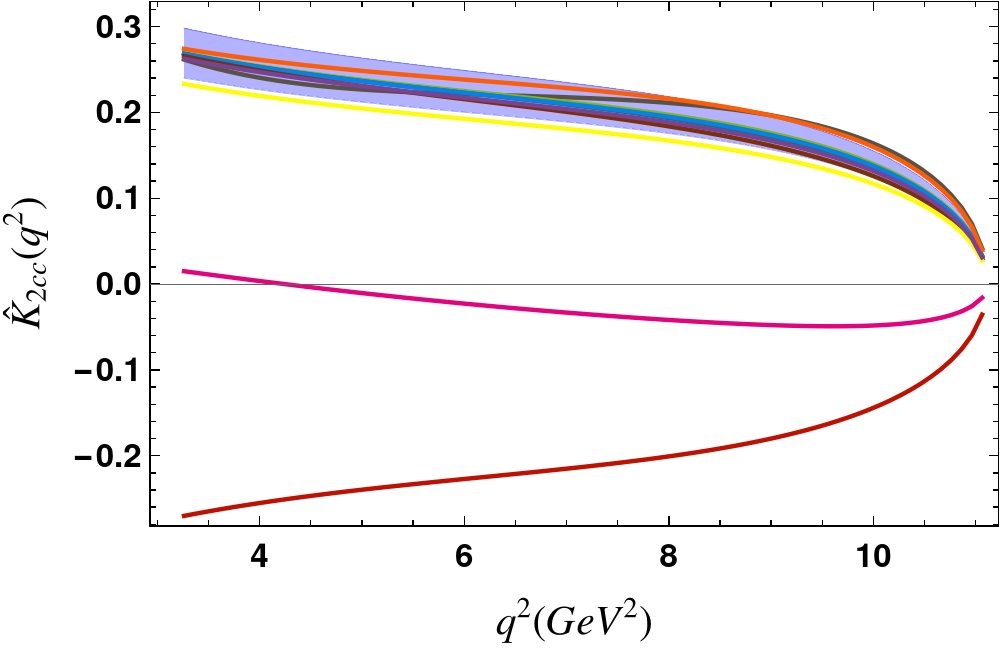}~~	
		\includegraphics[scale=0.4]{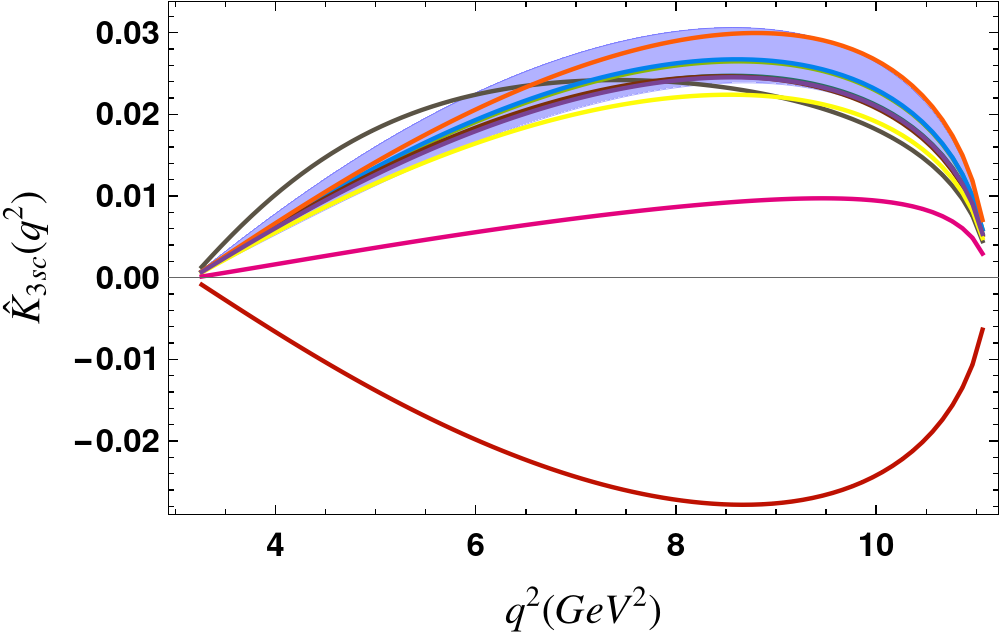}~~
		\includegraphics[scale=0.4]{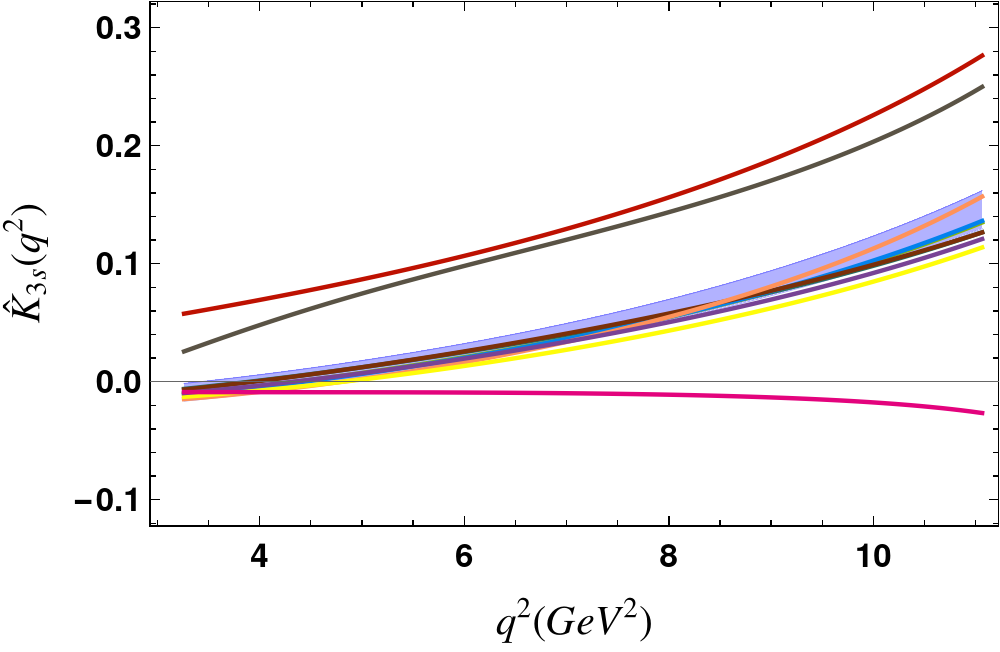}\\
		\includegraphics[scale=0.5]{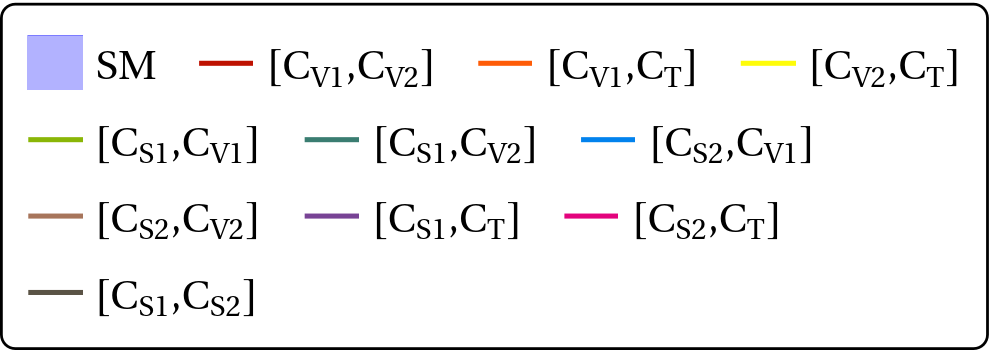}	
		\caption{The $q^2$ dependence of the angular observables for $\Lambda_{b}\to \Lambda_{c}^+(\to \Lambda \pi^+) \tau^- \bar{\nu}_{\tau} $ decay. The variations have been shown for different NP two parameter scenarios.}
		\label{fig:q2shapesobs2opr}
	\end{figure*}
	
In addition, we have studied the $q^2$ distributions of all the asymmetric and angular observables in the two-operator scenarios and presented them in fig. \ref{fig:q2shapesobs2opr}. These distributions might be helpful to test the NP sensitivities specific to the $q^2$ regions, which may be otherwise missing in the full $q^2$ integrated observables. We have taken the numerical values of the WCs from our fit results of two-operator scenarios given in table \ref{tab:2parmfitres} and the respective correlations. Unlike one-operator scenarios, we did not choose the benchmark values of WCs randomly here. If we choose them randomly, the choices of the benchmark values will be more critical and ambiguous for two-operator scenarios. For simplicity of the presentation, we have presented the results in two-operator scenarios by directly using the fit results.

Like one-operator scenarios, also in the two-operator scenarios, we have predicted all the observables with the respective 1$\sigma$ errors in small bins of $q^2$ which we have shown in the appendix in the tables from \ref{tab:obsbinprdall1} to \ref{tab:obsbinprdall12}. In those tables, we have marked (in bold font) the bins and the scenarios in which the predicted values are either inconsistent or marginally consistent with the SM within their respective 3$\sigma$ error bars. Like the full $q^2$ integrated observables, the $q^2$ distributions or the bin predictions of the observables show sizeable NP effects only for the scenarios $[\mathcal{O}_{S_1},\mathcal{O}_{S_2}]$, $[\mathcal{O}_{S_2},\mathcal{O}_T]$ and  $[\mathcal{O}_{V_1},\mathcal{O}_{V_2}]$, respectively. Based on the results we have mentioned above, we will point out a few important observations from the items below. 
	\begin{itemize}
		\item Apart from $\hat{K}_{3s}$, all the observables listed above are sensitive to the scenario $[\mathcal{O}_{S_2},\mathcal{O}_T]$. Interestingly, the $q^2$ integrated $A^{\Lambda_c \tau}_{FB}$ are consistent with the SM. However, we can see from the respective distribution in figure \ref{fig:q2shapesobs2opr} and from table \ref{tab:obsbinprdall8} that the corresponding NP predictions have discrepancies with SM in both the high and low $q^2$ regions. The predicted values throughout the $q^2$ regions are minimal. The SM predictions are negative in the high-$q^2$ regions and positive in the low-$q^2$ regions. Hence, due to a relative cancellation, the $q^2$ integrated value becomes very small and consistent with the respective NP prediction. 
		
		\item In the scenario $[\mathcal{O}_{S_2},\mathcal{O}_T]$, the predictions for the full $q^2$ integrated observables $A^{\Lambda_c}_{FB}$, $P_{\Lambda_c}$, $\hat{K}_{2ss}$, $\hat{K}_{2cc}$, $\hat{K}_{3s}$ and $\hat{K}_{3sc}$ are very small and consistent with zero. However, the respective SM predictions are either positive or negative and deviate from the NP predictions. Though the distribution of $\hat{K}_{3s}(q^2)$ at the large $q^2$ bin shows a shift in values from the respective SM prediction, they are consistent within their 3$\sigma$ error uncertainties. The predictions for the other observables, like $C_F^{\ell}$, $\hat{K}_{1cc}$ and $\hat{K}_{1ss}$, are different from zero but largely deviated from the respective SM predictions in most of the $q^2$ regions.   
		
		\item In the scenario $[\mathcal{O}_{V_1},\mathcal{O}_{V_2}]$, apart from the observables $\hat{K}_{1ss}$, $\hat{K}_{1cc}$, $C_F^{\tau}$ and $P^{(\Lambda_c)}_{\tau}$ the predictions of all the other observables have significant deviations from the respective SM predictions. Furthermore, the predictions for $A^{\Lambda_c}_{FB}$, $P_{\Lambda_c}$, $\hat{K}_{2ss}$, $\hat{K}_{2cc}$ and $\hat{K}_{3sc}$ have opposite sign of the respective SM predictions. A comparison of these observations with those in item-2 indicates that the effects the scenario $[\mathcal{O}_{V_1},\mathcal{O}_{V_2}]$ can be clearly distinguished from the effects of the $[\mathcal{O}_{S_2},\mathcal{O}_T]$. 
		
		\item The observables $A^{\tau}_{FB}(q^2)$, $A^{\Lambda_c}_{FB}(q^2)$, $A^{\Lambda_c \tau}_{FB}(q^2)$  $P_{\Lambda_c}(q^2)$, $\hat{K}_{2ss}(q^2)$, $\hat{K}_{2cc}(q^2)$ and $\hat{K}_{3sc}(q^2)$ are sensitive to both the scenarios $[\mathcal{O}_{V_1},\mathcal{O}_{V_2}]$ and $[\mathcal{O}_{S_2},\mathcal{O}_T]$. However, the respective $q^2$ distributions or the bin predictions show that their effects are clearly distinguishable.  
		
	  \item In most of the $q^2$ regions, the operator $[\mathcal{O}_{S_1},\mathcal{O}_{S_2}]$ have significant contributions in $A^{\Lambda_c \tau}_{FB}$, $A^{\tau}_{FB}$, $P^{(\Lambda_c)}_{\tau}$ and $\hat{K}_{3s}$ which may be difficult to separate from the effect of either $[\mathcal{O}_{S_2},\mathcal{O}_T]$ or $[\mathcal{O}_{V_1},\mathcal{O}_{V_2}]$. However, the scenarios $[\mathcal{O}_{S_2},\mathcal{O}_T]$ or $[\mathcal{O}_{V_1},\mathcal{O}_{V_2}]$ have significant contributions in other observables which are distinguishable from each other and from $[\mathcal{O}_{S_1},\mathcal{O}_{S_2}]$. Also, $[\mathcal{O}_{S_1},\mathcal{O}_{S_2}]$ contributes to $\hat{K}_{1ss}$, $\hat{K}_{1cc}$, $C_F^{\tau}$ only in the low-$q^2$ bin which are clearly distinguishable from the contribution of $[\mathcal{O}_{S_2},\mathcal{O}_T]$.       
		
	%	\item The predictions of $A^{\Lambda_c}_{FB}$ in the scenarios $[\mathcal{O}_{V_1},\mathcal{O}_{V_2}]$ and $[\mathcal{O}_{S_2},\mathcal{O}_T]$ have opposite sign. Also, it is possible to distinguish the effects of these operators in $A^{\Lambda_c \tau}_{FB}$ and $\hat{K}_{3s}$.
		
	%	\item If the measurements show deviations only in $C^{\tau}_{F}$, $\hat{K}_{1cc}$ and $\hat{K}_{1ss}$ this could be an indication for contributions from $[\mathcal{O}_{S_2},\mathcal{O}_T]$. Also, in this scenario in the high-$q^2$ regions, the prediction of $P^{(\Lambda_c)}_{\tau}$ has an opposite sign than SM, which is not the case in any other scenario.  

		\item  In the observable $P^{(\Lambda_c)}_{\tau}(q^2)$, for the scenario $[\mathcal{O}_{S_1},\mathcal{O}_{S_2}]$, we see sizeable contribution in the low and high-$q^2$ regions which are distinguishable from the SM. However, this information is missing in the $q^2$ integrated prediction since in the high-$q^2$, the predicted value is lower than the SM, and in the low-$q^2$ regions, the prediction is higher than the SM.    
		
		\item In the SM, we observe zero crossing in the $q^2$ distributions of $A_{FB}^{\Lambda_c \tau}(q^2)$ and $A_{FB}^{\tau}(q^2)$. For both these observables, in the scenarios $[\mathcal{O}_{V_1},\mathcal{O}_{V_2}]$ and $[\mathcal{O}_{S_1},\mathcal{O}_{S_2}]$ the $q^2$ distributions do not have any zero crossing points. However, for $A_{FB}^{\tau}(q^2)$ we notice the zero crossing point 
		\begin{equation}
		\bf q_0^2\big|_{[\mathcal{O}_{S_2},\mathcal{O}_T]} = 5.17 (13) \ \ \text{GeV$^2$}
		\end{equation}
		in the scenario $[\mathcal{O}_{S_2},\mathcal{O}_T]$, which is much lower than the respective SM value and very easily detectable. In the rest of the two-operator scenarios, it would be hard to distinguish the zero crossing points from that of the SM. This pattern will be helpful to distinguished the impact of $[\mathcal{O}_{S_2},\mathcal{O}_T]$ from the other operators.       
		
	\end{itemize}
	
	Following the above discussions, we can argue that the information available in table  \ref{tab:discrepancy_twoopr} and the tables from \ref{tab:obsbinprdall1} to \ref{tab:obsbinprdall12} and in fig. \ref{fig:q2shapesobs2opr} clearly show that the effects of the two operator scenarios are distinguishable from each other once we have measurements of all the observables mentioned above. Also, the observed pattern of the $q^2$ distributions of the observables are very different in the two-operator scenarios than those observed in the one-operator scenarios. Hence, if we find discrepancies in a few observables, the pattern of the discrepancies will help distinguish the effects of the one-operator scenario from those of two-operator scenarios.  
	
	Finally, we would like to comment on the observables $\hat{K}_{4s}$ and $\hat{K}_{4sc}$. We can see from the respective expressions in eqs. \ref{eq:k4s} and \ref{eq:k4sc} in the appendix that these observables are sensitive to the imaginary components of the helicity amplitudes, and the complex WCs will be the source of these imaginary components. The values of $\hat{K}_{4s}$ and $\hat{K}_{4sc}$ will be zero if the imaginary part of the relevant WC is zero. Also, we can obtain non-zero contributions in both these observables only in the one operator scenarios $\mathcal{O}_{V_1}$, $\mathcal{O}_{V_2}$ and $\mathcal{O}_{T}$, respectively, with complex WCs. The contributions from the scalar or pseudoscalar operators will always appear in combinations with the other operators. Therefore, these operators with complex WCs in combination with vector, axial-vector and tensor operators can contribute to $\hat{K}_{4s}(q^2)$ and $\hat{K}_{4sc}(q^2)$. In this analysis, we are not considering complex WCs; hence, we have not presented any numerical analysis. We are leaving it for dedicated future work.

	\section{Summary}\label{sec:summary}

In this study, we have analysed the four-fold angular distribution of the semileptonic $\Lambda_b\to \Lambda_c^+(\to \Lambda \pi^+) \mu^-\bar{\nu}$ and $\Lambda_b\to \Lambda_c^+(\to \Lambda \pi^+) \tau^-\bar{\nu}$ decays in the SM and the presence of model-independent BSM effective operators in $b\to c\ell^-\bar{\nu}$ transition. Considering the polarisations of the $\Lambda_{c}$ baryons, we have worked out the analytical expressions for various asymmetric and angular observables for the above cascade decays. Wherever available, we have compared our analytical expressions with the literature. In the framework of SM, using the available lattice inputs on the form factor shapes, we have predicted all these asymmetric and angular observables alongside the decay rates and LFU ratio $R(\Lambda_c)$ integrated over the whole $q^2$ regions and small $q^2$ bins. In addition, we have predicted the values $q_0^2$ GeV$^2$ for which the forward-backward asymmetries $A_{FB}^{\ell}(q^2)$, $A_{FB}^{\Lambda_c\ell}(q^2)$ and the lepton-polarisation asymmetry $P_\ell^{(\Lambda_c)}(q^2)$ have zero crossing in the respective $q^2$ distributions in the SM. 

Using the available data on $R(D^{(*)})$, $F_{L}^{D^*}$, $\mathcal{B}(\Lambda_b\to \Lambda_c^+\tau^-\bar{\nu})$ and $R(\Lambda_c)$, we have extracted the WCs of the BSM operators contribute to $b\to c\tau^-\bar{\nu}$ transitions. First, we have analysed the data using the contributions of one operator at a time, and given the error bars, the magnitudes of extracted WC could be $\mathcal{O}(10^{-1})$. Using the extracted values of the WC, we have predicted the observables used in the fits and $P^{\tau}(D^{(*)})$ and looked into the observable by-observable correlations in each scenario. We have found that not all the scenarios could comfortably explain all the data simultaneously. If we consider the predictions of the observables in their 3$\sigma$ ranges, then only the scenarios $\mathcal{O}_{V_2}$ and $\mathcal{O}_{S_2}$, independently, could accommodate all the five data used in the fit. Also, we have predicted all the other observables in $\Lambda_b\to \Lambda_c^+(\to \Lambda \pi^+) \tau^-\bar{\nu}$ decays in small $q^2$-bins and those integrated over the entire allowed $q^2$ interval. In addition, we have presented the $q^2$ distributions of all these observables in all the one-operator scenarios. After analysing these results, we find that most of the asymmetric and angular observables in $\Lambda_b\to \Lambda_c^+(\to \Lambda \pi^+) \tau^-\bar{\nu}$ decays are sensitive to the contribution from $\mathcal{O}_T$, it will be hard to distinguish the contributions of other one-operator scenarios from the respective SM predictions. Here, we would like to point out that our observation is based on the values of WCs of order $\mathcal{O}(10^{-1})$. If we allow these values to be ordered one, we might see deviations in the scenario $\mathcal{O}_{V_2}$ in a couple of observables.

We repeat the exercise mentioned in the above paragraph for the scenarios where we have considered the contributions of two operators at a time. This study is relevant since many UV complete models exist where multiple operators may contribute to $b\to c\tau^-\bar{\nu}$ transitions. Also, the findings of the one-operator scenarios motivate us to look for the impacts of two-operator scenarios. In these studies, we have found that in all the two operator scenarios, we could explain all the data in $B\to D^{(*)}\tau^-\bar{\nu}$ decays within their 1$\sigma$ error bars. However, not all these scenarios could comfortably explain $R(\Lambda_c)$ even if we take the respective predictions at their 3$\sigma$ CI. It is only the scenario $[\mathcal{O}_{S_2},\mathcal{O}_T]$ that could accommodate comfortably all the measured data simultaneously even if we take the respective predictions within their 1$\sigma$ CI. In the other two operator scenarios, apart from $R(D^*)$, we are able to explain all the other data simultaneously if we take the uncertainties of our predictions at the 3$\sigma$ level.  

 We have presented the $q^2$ distributions of each of the angular and asymmetric observables in $\Lambda_b\to \Lambda_c^+(\to \Lambda \pi^+) \tau^-\bar{\nu}$ decays in all the two operator scenarios. Furthermore, in each scenario, we have predicted these observables' full $q^2$ integrated values and their values in small $q^2$ bins. From the analysis of these results, we note that many of these observables show distinguishable sensitivity to the operators $[\mathcal{O}_{S_2},\mathcal{O}_T]$, $[\mathcal{O}_{V_1}$, $\mathcal{O}_{V_2}]$ and $[\mathcal{O}_{S_1}$, $\mathcal{O}_{S_2}]$. For example, all the twelve observables which we have discussed are sensitive to $[\mathcal{O}_{S_2},\mathcal{O}_T]$. In comparison, out of this, only eight observables are sensitive to $[\mathcal{O}_{V_1}$, $\mathcal{O}_{V_2}]$. The observables like $\hat{K}_{1cc}$, $\hat{K}_{1ss}$, $P_\tau^{(\Lambda_c)}$ and $C_F^{\tau}$ are not sensitive to $[\mathcal{O}_{V_1}$, $\mathcal{O}_{V_2}]$. In the rest of the eight observables, the effects of $[\mathcal{O}_{S_2},\mathcal{O}_T]$ are distinguishable from those of $[\mathcal{O}_{V_1}$, $\mathcal{O}_{V_2}]$. Similarly, the scenario $[\mathcal{O}_{S_1}$, $\mathcal{O}_{S_2}]$ have impact only on $\hat{K}_{3s}$, $A_{FB}^{\tau}$, $A_{FB}^{\Lambda_c \tau}$ and $P_{\tau}^{\Lambda_c}$. In the scenario $[\mathcal{O}_{S_2},\mathcal{O}_T]$, we note a sizable shift in the value of zero crossing point $q_0^2$ GeV$^2$ in the $q^2$ distribution of $A_{FB}^{\tau}(q^2)$. However, in the other two scenarios, we do not observe any zero crossing points in the $q^2$ distributions of $A_{FB}^{\tau}(q^2)$ and in $A_{FB}^{\Lambda_c\tau}(q^2)$, which is in sharp contrast with the SM. Similarly, for $[\mathcal{O}_{S_1}$, $\mathcal{O}_{S_2}]$ we do not observe any zero crossing point in $P_\tau^{(\Lambda_c)}(q^2)$, and in the other two scenarios the shift in the zero crossing point are not distinguishable from that in the SM. In all these observables, for the other two-operator scenarios, it would be hard to distinguish their contributions from the SM. Furthermore, a comparison of the respective $q^2$ distributions of these observables suggests that the effects of one-operator scenarios are distinguishable from those of two-operator scenarios.

 %%%%%%%%%%%%%%%%%%%%%%%%%%%%%%%%%%%%%%%%%%%%%%%%%%%%5
	\section{Acknowledgements}
SS acknowledges the Council of Scientific and Industrial Research (CSIR), Govt. of India for JRF fellowship grant with File No. 09/731(0173)/2019-EMR-I. RS acknowledges the financial support from the Science and Engineering Research Board (SERB) for the National PostDoctoral Fellowship (file no. NPDF/PDF/2021/003328).

\section{Appendix}
In the appendix, we provide explicit expressions for the spinors of baryons and leptons, as well as the polarization vectors used to calculate the helicity amplitudes for the decay $\Lambda_b \to \Lambda_c \ell \bar{\nu}_{\ell}$. We also discuss the secondary decay $\Lambda^+_c \to \Lambda \pi^+$ and provide explicit expressions for angular observables in the $\Lambda_b \to \Lambda^+_c(\to \Lambda \pi^+) \ell \bar{\nu}_{\ell}$ decay.
\subsection{Transition Matrix elements and Kinematics} \label{app:hadronicmatrix}
\begin{eqnarray}
	\bra{\lc}\bar{c}\gamma^\mu b\ket{\lb}&=&\bar{u}_{\lc}\Big[ f_0 (q^2)(m_{\lb} - m_{\lc})\frac{q^\mu}{q^2} 
	+f_+ (q^2)\frac{m_{\lb} + m_{\lc}}{Q_+}(p_{\lb}^{\mu} +p_{\lc}^{\mu}-(m_{\lb} ^2 - m_{\lc} ^2)\frac{q^\mu}{q^2}) \nonumber\\
	&&+f_\perp (q^2)(\gamma^\mu - \frac{2m_{\lc}}{Q_+}p_{\lb}^{\mu} - \frac{2m_{\lb}}{Q_+}p_{\lc}^{\mu})\Big]u_{\lb}, \label{eq:VFF} \\
	\bra{\lc}\bar{c}\gamma^\mu \gamma_5 b\ket{\lb}&=&-\bar{u}_{\lc}\gamma_5\Big[ g_0 (q^2)(m_{\lb} + m_{\lc})\frac{q^\mu}{q^2} \nonumber\\
	&&+g_+ (q^2)\frac{m_{\lb} - m_{\lc}}{Q_-}(p_{\lb}^{\mu} +p_{\lc}^{\mu}-(m_{\lb} ^2 - m_{\lc} ^2)\frac{q^\mu}{q^2})\nonumber\\
	&&+g_\perp (q^2)(\gamma^\mu + \frac{2m_{\lc}}{Q_-}p_{\lb}^{\mu} - \frac{2m_{\lb}}{Q_-}p_{\lc}^{\mu})\Big]u_{\lb} \label{eq:AFF}.
\end{eqnarray}

The matrix elements of the scalar and pseudo-scalar currents can be obtained from the vector and axial vector matrix elements using the equations of motion:
\begin{align}
	\nonumber \bra{\lc}\bar{c} b\ket{\lb} =& \frac{q_\mu}{m_b-m_c}\bra{\lc}\bar{c}\gamma^\mu b\ket{\lb} \\
	=& f_0(q^2)  \frac{m_{\lb} - m_{\lc}}{m_b-m_c} \bar{u}_{\lc}u_{\lb}, \\
	\nonumber \bra{\lc}\bar{c}\gamma_5 b\ket{\lb} =& \frac{q_\mu}{m_b+m_c}\bra{\lc}\bar{c}\gamma^\mu\gamma_5 b\ket{\lb} \\
	=& g_0(q^2)  \frac{m_{\lb} + m_{\lc}}{m_b+m_c} \bar{u}_{\lc}\gamma_5 u_{\lb}.
\end{align}
In our numerical analysis, we use $m_b = 4.18(4)$ GeV, $m_c = 1.27(3)$ GeV \cite{Workman:2022ynf}. The matrix element of the tensor current can be written in terms of four form factors $h_+$, $h_\perp$, $\widetilde{h}_+$, $\widetilde{h}_\perp$,
\begin{align}
	\bra{\lc}\bar{c}i\sigma^{\mu\nu} b\ket{\lb}=&\bar{u}_{\lc}\Big[2h_+(q^2)\frac{p_{\lb}^\mu p_{\lc}^{ \nu}-p_{\lb}^\nu p_{\lc}^{\mu}}{Q_+} \nonumber\\
	&+h_\perp (q^2)\Big(\frac{m_{\lb}+m_{\lc}}{q^2}(q^\mu \gamma^\nu -q^\nu \gamma^\mu)-2(\frac{1}{q^2}+\frac{1}{Q_+})(p_{\lb}^\mu p_{\lc}^{\nu}-p_{\lb}^\nu p_{\lc}^{\mu}) \Big) \nonumber\\
	&+\widetilde{h}_+ (q^2)\Big(i\sigma^{\mu \nu}-\frac{2}{Q_-}(m_{\lb}(p_{\lc}^{\mu}\gamma^\nu -p_{\lc}^{\nu}\gamma^\mu)\nonumber\\
	&-m_{\lc}(p_{\lb}^\mu \gamma^\nu -p_{\lb}^\nu \gamma^\mu)+p_{\lb}^\mu p_{\lc}^{\nu}-p_{\lb}^\nu p_{\lc}^{\mu}) \Big) \nonumber\\
	&+\widetilde{h}_\perp(q^2) \frac{m_{\lb}-m_{\lc}}{q^2 Q_-}\Big((m_{\lb}^2-m_{\lc}^2-q^2)(\gamma^\mu p_{\lb}^\nu - \gamma^\nu p_{\lb}^\mu)\nonumber\\
	&-(m_{\lb}^2-m_{\lc}^2+q^2)(\gamma^\mu p_{\lc}^{\nu}-\gamma^\nu p_{\lc}^{\mu})+2(m_{\lb}-m_{\lc})(p_{\lb}^\mu p_{\lc}^{\nu}-p_{\lb}^\nu p_{\lc}^{\mu}) \Big)
	\Big]u_{\lb}. \nonumber \\ \label{eq:TFF}
\end{align}
The matrix elements of the pseudo-tensor current $\bar{c}i\sigma^{\mu\nu}\gamma_5 b$ can be obtained from the above equation by using the identity
\begin{align}
	\sigma^{\mu \nu}\gamma_{5}=\frac{i}{2}\epsilon^{\mu \nu \alpha \beta}\sigma_{\alpha \beta}.
\end{align}
Here, we adopt $\epsilon^{0123}=+1$ for evaluating the pseudo-tensor matrix elements in this work.
\subsubsection{Kinematics in $\Lambda_b$ rest frame and baryonic spinors}
To calculate the hadronic helicity amplitudes, we work in the $\Lambda_b$ rest frame, positioning the three-momentum of the $\Lambda_c$ along the +z direction and the three-momentum of the virtual vector boson along the -z direction. The baryon spinors are then expressed as follows:
%%%%%%%%%%%%%%%%%%%%%%%%%%%%%%%%%%%%%%%%%%%%%%%%%%%%%%%%%% 
\begin{eqnarray}
	u_\Lb\Big(\pm \frac{1}{2}, p_\Lb \Big) &=& \sqrt{2m_\Lb} 
	\left(
	\begin{array}{l}
		\chi_\pm \\
		0 \\
	\end{array}
	\right) \,, \nonumber\\
	\bar u_{\Lambda_c}\Big( \pm \frac{1}{2}, p_{\Lambda_c}\Big) &=& \sqrt{E_{\Lambda_c} + m_{\Lambda_c}}
	\Big( \chi_\pm^\dagger, \frac{\mp |{\bf p_{\Lambda_c}}|}{E_{\Lambda_c} + m_{\Lambda_c}}  
	\chi_\pm^\dagger \Big)
\end{eqnarray}
where $\chi_+ = \left(
\begin{array}{l}
	1 \\
	0 \\
\end{array} \right)$ 
and $\chi_- = \left(
\begin{array}{l}
	0 \\
	1 \\
\end{array} \right)$ are two-component Pauli spinors.    
%%%%%%%%%%%%%%%%%%%%%%%%%%%%%%%%%%%%%%%%%%%%%%%%%%%
In the $\Lambda_b$ rest frame, the four-momenta of $\Lambda_b(p_{\Lb})$, $\Lambda_c(p_{\Lambda_c})$ and q are
\begin{eqnarray}
	p_{\Lb}^{\mu} &=& \left(m_\Lb, 0, 0, 0\right)\,, \nonumber \\
	p_{\Lambda_c}^{\mu} &=& \left(E_{\Lambda_c}, 0, 0, |\mathbf q|\right) \,,\nonumber \\
	q^{\mu} &=& \left(q_0, 0, 0, -|\mathbf q|\right)\,.
\end{eqnarray}
%%%%%%%%%%%%%%%%%%%%%%%%%%%%%%%%%%%%%%%%%%%%%%%%%%%%%%%%%%%%%%
The polarization vectors of virtual vector boson($\epsilon$) are defined in the $\Lambda_b$ rest frame as follows:
\begin{eqnarray}
	\label{}
	\epsilon^{\mu *}(t) = \frac{1}{\sqrt{q^2}}\left(q_0,0,0,-|\mathbf q|\right) ,\, 
	\epsilon^{\mu *}(0) = \frac{1}{\sqrt{q^2}} \left(|\mathbf q|, 0, 0, -q_0\right) , \, 
	\epsilon^{\mu *}(\pm 1) = \frac{1}{\sqrt{2}}\left(0, \pm 1, i, 0\right)
\end{eqnarray}
where,
\begin{equation}
	|\mathbf P_{\Lambda_c}|=|\mathbf q|=\frac{\sqrt{Q_+ Q_-}}{2 m_\Lb},\qquad
	E_{\Lambda_c}=\frac{m_\Lb^2+m_{\Lambda_c}^2-q^2}{2m_\Lb},\qquad 
	q_0= \frac{m_\Lb^2-m_{\Lambda_c}^2+q^2}{2m_\Lb}. 
\end{equation}
The vectors satisfy the following orthonormality and completeness relations
\begin{align}
	&\epsilon^{*\mu}(n) \epsilon_{\mu}(n^{\prime})=g_{n n^{\prime}}, \, \, \sum_{n, n^{\prime}}\epsilon^{*\mu}(n) \epsilon^{\nu}(n^{\prime}) g_{n n^{\prime}}=g^{\mu \nu}, \, \,~~\text{with}~~~~ \, n,n^{\prime}=t, \pm ,0 \nn \\
	&\epsilon^{\mu}(\lambda) \cdot q_{\mu}=0  \, \,~~\text{with}~~~~ \, \lambda= \pm ,0.
\end{align}
where, $g_{n n^{\prime}}=\text{diag}(+1,-1,-1,-1)$ and our choice of the metric tensor is $g^{\mu \nu}=\text{diag}(+1,-1,-1,-1)$.
%%%%%%%%%%%%%%%%%%%%%%%%%%%%%%%%%%%%%%%%%%%%%%%%%%
\subsubsection{Kinematics in dilepton rest frame and lepton spinors}
\label{app:leptonichel}
The polarization vectors of the virtual vector boson in this frame are
\begin{eqnarray}\label{eq:polWst}
	\label{polvec2}
	\epsilon^{\mu }(t) &=& \left(1; 0, 0, 0\right)\,, \nonumber \\
	\epsilon^{\mu}(\pm 1) &=& \frac{1}{\sqrt{2}}\left(0;\pm 1, -i, 0\right) \,,\nonumber \\
	\epsilon^{\mu}(0) &=& \left(0; 0, 0, -1\right)\,.
\end{eqnarray}
The three-momentum and energy of the $\ell$-lepton in this $\ell-\nu$ rest frame can be written as
\begin{eqnarray}
	|\mathbf{p}_\ell| &=& \sqrt{q^2}\, v^2/2,\nonumber\\
	E_\ell &=& |\mathbf{p}_\ell| + m_\ell^2/\sqrt{q^2},
\end{eqnarray}
where, $v=\sqrt{1-\frac{m_\ell^2}{q^2}}$
The lepton spinors for $\mathbf{p}_\ell$ pointing in the $+z$ direction and $\mathbf{p}_{\bar{\nu}_\ell}$ pointing in the $-z$ direction are

The Leptonic three momentum $p_{\ell}$ along any arbitrary direction $p_{\ell}(\theta, \phi=0)$ with polar angle $\theta$, and azimuthal angle $\phi$ calculated as follow 
\begin{eqnarray}\label{eq:spinors}
	\label{spinor3}
	\bar{u}_\ell(+{\textstyle \frac{1}{2}},p_\ell)&=& \sqrt{E_\ell+m_\ell}\left( \cos (\theta_\ell/2),\sin (\theta_\ell/2),
	\frac{-|\mathbf{p}_\ell|}{E_\ell+m_\ell}\cos (\theta_\ell/2),\frac{-|\mathbf{p}_\ell|}{E_\ell+m_\ell}\sin (\theta_\ell/2) \right) \,,\nonumber \\
	\bar{u}_\ell(-{\textstyle \frac{1}{2}},p_\ell)&=& \sqrt{E_\ell+m_\ell}\left( -\sin (\theta_\ell/2),\cos (\theta_\ell/2),
	\frac{-|\mathbf{p}_\ell|}{E_\tau+m_\ell}\sin (\theta_\ell/2),\frac{|\mathbf{p}_\ell|}{E_\tau+m_\ell}\cos (\theta_\ell/2) \right) \,,\nonumber \\
	v_{\bar{\nu}_\ell}({\textstyle \frac{1}{2}},p_{\bar{\nu}_\ell}) &=& \sqrt{E_\nu}
	\left(\begin{array}{c}\cos (\theta_\ell/2) ~~ \sin (\theta_\ell/2) ~~ -\cos (\theta_\ell/2) ~~ -\sin (\theta_\ell/2)  \end{array}\right)^T \, .
\end{eqnarray}
%%%%%%%%%%%%%%%%%%%%%%%%%%%%%%%%%
\subsection{Hadronic Couplings in $\Lambda_c^+\to \Lambda\pi^+$}\label{subsec:lambdactolambda}

In the SM the decay $\Lambda_c^+  \to\Lambda \pi^+$ is described by the effective Hamiltonian
\begin{equation}\label{eq:heffds1}
	H^\text{eff}_{\Delta S = 1} = \frac{4 \gfermi}{\sqrt{2}} \, V_{ud}^* V_{us}
	\left[\bar{d} \gamma_\mu P_L u\right]\left[\bar{s} \gamma^\mu P_L c\right] \,.
\end{equation}
The hadronic matrix element which determines the $\Lambda_c \to \Lambda \pi$ decay
can be parametrized as 
\begin{align}
	&
	\frac{4 \gfermi}{\sqrt{2}} \, V_{ud}^* V_{us} \bra{\Lambda(k_1, \lambda_{\Lambda}) \pi^+(k_2)} \left[\bar{d} \gamma_\mu P_L u\right]\left[\bar{s} \gamma^\mu P_L c\right]
	\ket{\Lambda(k, \lambda_{\Lambda_c})} \cr
	& = N_2 \big[\bar u(k_1, \lambda_\Lambda) \big(\xi \,\gamma_5 + \omega\big) u(k, \lambda_{\Lambda_c})\big] \equiv 
	\mathcal{A}^{\lambda_{\Lambda_c}}_{\lambda_\Lambda} \,
\end{align}
with $N_2=\frac{4 \gfermi}{\sqrt{2}} \, V_{ud}^* V_{us}$. As a consequence of the equations of motion, only two independent hadronic parameters
appear which we have denoted as $\xi$ and $\omega$.
They can be extracted from the $\Lambda_c^+  \to\Lambda \pi^+$ decay width and polarization measurements.

In terms of the kinematic variables introduced above, the helicity
amplitudes for the secondary decay can be written as
\begin{align}
	\mathcal{A}^{+1/2}_{+1/2} &= N_2  \left(\sqrt{r_+} \, \omega - \sqrt{r_-} \, \xi \right) \cos\frac{\theta_\Lambda}{2} \,, 
	\cr
	\mathcal{A}^{+1/2}_{-1/2} &= N_2  \left(\sqrt{r_+} \, \omega + \sqrt{r_-} \, \xi \right) \sin\frac{\theta_\Lambda}{2}\, 
	e^{i \phi}  \,, \cr
	\mathcal{A}^{-1/2}_{+1/2} &= N_2  \left(-\sqrt{r_+} \, \omega + \sqrt{r_-} \, \xi \right) \sin\frac{\theta_\Lambda}{2}\,
	e^{-i \phi} \,, \cr
	\mathcal{A}^{-1/2}_{-1/2} &= N_2  \left(\sqrt{r_+}\, \omega + \sqrt{r_-} \,\xi \right) \cos\frac{\theta_\Lambda}{2} \,.
\end{align}
where we abbreviate
\begin{equation}
	r_{\pm} \equiv (m_{\Lambda_c} \pm m_{\Lambda})^2 - m_{\pi}^2\,.
\end{equation}
The corresponding helicity contributions to the decay width can be defined as
\begin{equation}
	\Gamma_2(\lambda_{\Lambda_c}^{(a)},\lambda_{\Lambda_c}^{(b)}) = \frac{\sqrt{r_+ r_-}}{16 \pi m_{\Lambda}^3} \,
	\sum_{\lambda_\Lambda} \mathcal{A}^{\lambda_{\Lambda_c}^{(a)}}_{\lambda_\Lambda} \, \bigl(\mathcal{A}^{\lambda_{\Lambda_c}^{(b)}}_{\lambda_\Lambda}\bigr)^* \,,
\end{equation}
which yield
\begin{align}
	\Gamma_2(+1/2, +1/2) &=  (1 + \alpha  \, \cos\theta_\Lambda) \, \Gamma_\Lambda\,,\qquad
	\Gamma_2(+1/2, -1/2) = -\alpha \, \sin\theta_\Lambda \, e^{i \phi} \, \Gamma_\Lambda \,,\cr
	\Gamma_2(-1/2, -1/2) &= (1 - \alpha \, \cos\theta_\Lambda) \, \Gamma_\Lambda \,,\qquad
	\Gamma_2(-1/2, +1/2) = -\alpha \, \sin\theta_\Lambda e^{-i \phi}\, \Gamma_\Lambda \,.
\end{align}
Here the $\Lambda^+_c \to \Lambda \pi^+$ decay width is given as
\begin{align}
	\Gamma_\Lambda
	= \frac{|N_2|^2 \sqrt{r_+ r_-}}{16 \pi m_\Lambda^3} \left(
	r_- \,|\xi|^2 + r_+ \, |\omega|^2 \right)\,.
\end{align}
%\textcolor{red}{insert the decay width measurement from PDG}
and the parity-violating decay parameter $\alpha_\Lambda$ reads
\begin{equation}
	\alpha_\Lambda = \frac{-2 \,re[\omega \,\xi]}{\sqrt{\frac{r_-}{r_+}} \, |\xi|^2 + \sqrt{\frac{r_+}{r_-}} \, |\omega|^2} = +\alpha^{\text{exp}}_\Lambda\,.
\end{equation}
%%%%%%%%%%%%%%%%%%%%%%%%%%%%%%%%%%%%%%%%%%%%%%%%%%%%%%%
\subsection {Explicit expression for Angular Observables }\label{sec:anay_expr}
	%%%%%%%%%%%%%%%%%%%%%%%%%%%%%%
	%%%%%%%%%%%%%%%%%%%%%%%%%%%%%%
	\begin{align}
		K_{1c}=&N\biggl[2\biggl(\bigl|H_{\frac{1}{2},+1}^{\text{VA}}\bigr|^2
		-\bigl|H_{-\frac{1}{2},-1}^{\text{VA}}\bigr|^2\biggr)
		+8~\text{Re}\biggl[H_{\frac{1}{2},0}^{\text{SP}*} H_{\frac{1}{2},t,0}^{T,\frac{1}{2}}
		+H_{-\frac{1}{2},0}^{\text{SP}*} H_{-\frac{1}{2},t,0}^{T,-\frac{1}{2}}
		-H_{-\frac{1}{2},0}^{\text{SP}*}H_{-\frac{1}{2},+1,-1}^{T,-\frac{1}{2}}
		-H_{\frac{1}{2},0}^{\text{SP}*}H_{\frac{1}{2},+1,-1}^{T,\frac{1}{2}}
		\biggr]\nn \\& + \frac{\text{$m_{\ell}$}}{\sqrt{q^2}}\biggl\{
		\text{Re}\biggl[4 H_{-\frac{1}{2},0}^{\text{SP}*} H_{-\frac{1}{2},0}^{\text{VA}}+4
		H_{\frac{1}{2},0}^{\text{SP}*} H_{\frac{1}{2},0}^{\text{VA}} +8\biggl( H_{\frac{1}{2},+1}^{\text{VA}*} H_{\frac{1}{2},t,1}^{T,-\frac{1}{2}}+ H_{-\frac{1}{2},t,0}^{T,-\frac{1}{2}}H_{-\frac{1}{2},t}^{\text{VA}*} + H_{\frac{1}{2},t,0}^{T,\frac{1}{2}} H_{\frac{1}{2},t}^{\text{VA}*}\nn \\& -
		H_{-\frac{1}{2},-1,0}^{T,\frac{1}{2}}
		H_{-\frac{1}{2},-1}^{\text{VA}*}- H_{\frac{1}{2},+1,0}^{T,-\frac{1}{2}} H_{\frac{1}{2},+1}^{\text{VA}*}
		- H_{-\frac{1}{2},+1,-1}^{T,-\frac{1}{2}} H_{-\frac{1}{2},t}^{\text{VA}*}-
		H_{\frac{1}{2},+1,-1}^{T,\frac{1}{2}}
		H_{\frac{1}{2},t}^{\text{VA}*}- H_{-\frac{1}{2},-1}^{\text{VA}*}
		H_{-\frac{1}{2},t,-1}^{T,\frac{1}{2}}                                    
		\biggr)\biggr] \biggr\} \nn \\  
		&+ \frac{\text{$m^2_{\ell}$}}{q^2} \biggl\{
		8\biggl(\bigl|H_{\frac{1}{2},+1,0}^{T,-\frac{1}{2}}\bigr|^2
		-\bigl|H_{-\frac{1}{2},-1,0}^{T,\frac{1}{2}}\bigr|^2 -
		\bigl|H_{-\frac{1}{2},t,-1}^{T,\frac{1}{2}}\bigr|^2 +\bigl|H_{\frac{1}{2},t,+1}^{T,-\frac{1}{2}}\bigr|^2 \biggr)
		\nn\\&
		+\text{Re}  
		\biggl[4 H_{-\frac{1}{2},0}^{\text{VA}*}H_{-\frac{1}{2},t}^{\text{VA}}+ 4H_{\frac{1}{2},0}^{\text{VA}*}H_{\frac{1}{2},t}^{\text{VA}}
		-16 H_{-\frac{1}{2},-1,0}^{T,\frac{1}{2}*} H_{-\frac{1}{2},t,-1}^{T,\frac{1}{2}} 
		-16 H_{\frac{1}{2},+1,0}^{T,-\frac{1}{2}*} H_{\frac{1}{2},t,1}^{T,-\frac{1}{2}}\biggr] \biggr\}\biggr]
	\end{align}
	%%%%%%%%%%%%%%%%%%%%%%%%%%%%%%%%%%%%%%%%%%%%%%%%%%%%		
	\begin{align}
		K_{1cc}=&N \biggl[2\biggl(\bigl|H_{-\frac{1}{2},0}^{\text{SP}}\bigr|^2+
		\bigl|H_{\frac{1}{2},0}^{\text{SP}}\bigr|^2+
		\bigl|H_{-\frac{1}{2},-1}^{\text{VA}}\bigr|^2 
		+\bigl|H_{\frac{1}{2},+1}^{\text{VA}}\bigr|^2\biggr)\nn \\&
		+8 \bigg(\bigl|H_{-\frac{1}{2},+1,-1}^{T,-\frac{1}{2}}\bigr|^2 + \bigl|H_{\frac{1}{2},+1,-1}^{T,\frac{1}{2}}\bigr|^2 +
		\bigl|H_{-\frac{1}{2},t,0}^{T,-\frac{1}{2}}\bigr|^2 + \bigl|H_{\frac{1}{2},t,0}^{T,\frac{1}{2}}\bigr|^2 \bigg)\nn \\&
		-16\text{Re}\biggl[H_{-\frac{1}{2},+1,-1}^{T,-\frac{1}{2}*} H_{-\frac{1}{2},t,0}^{T,-\frac{1}{2}}+
		H_{\frac{1}{2},+1,-1}^{T,\frac{1}{2}*} H_{\frac{1}{2},t,0}^{T,\frac{1}{2}}\biggr] \nn \\ & +\frac{\text{$m_{\ell}$}}{\sqrt{q^2}}\biggl\{\text{Re}\bigg[
		4\biggl(H_{-\frac{1}{2},0}^{\text{SP}*} H_{-\frac{1}{2},t}^{\text{VA}}
		+ H_{\frac{1}{2},0}^{\text{SP}*} H_{\frac{1}{2},t}^{\text{VA}}\biggr)\nn \\&
		+8 \biggl(H_{-\frac{1}{2},0}^{\text{VA}*}H_{-\frac{1}{2},t,0}^{T,-\frac{1}{2}}
		-H_{-\frac{1}{2},0}^{\text{VA}*} H_{-\frac{1}{2},+1,-1}^{T,-\frac{1}{2}} -H_{\frac{1}{2},0}^{\text{VA}*}H_{\frac{1}{2},+1,-1}^{T,\frac{1}{2}}
		+H_{\frac{1}{2},0}^{\text{VA}*} H_{\frac{1}{2},t,0}^{T,\frac{1}{2}}\nn \\&
		+H_{-\frac{1}{2},-1,0}^{T,\frac{1}{2}}H_{-\frac{1}{2},-1}^{\text{VA}*}
		-H_{\frac{1}{2},+1,0}^{T,-\frac{1}{2}} H_{\frac{1}{2},+1}^{\text{VA}*}+
		H_{-\frac{1}{2},-1}^{\text{VA}*} H_{-\frac{1}{2},t,-1}^{T,\frac{1}{2}}+ H_{\frac{1}{2},+1}^{\text{VA}*} H_{\frac{1}{2},t,1}^{T,-\frac{1}{2}}\biggr)\bigg]\biggr\}\nn\\&
		+\frac{\text{$m^2_{\ell}$}}{q^2} \biggl\{
		2\biggl(
		\bigl|H_{-\frac{1}{2},0}^{\text{VA}}\bigr|^2
		+\bigl|H_{\frac{1}{2},0}^{\text{VA}}\bigr|^2
		+\bigl|H_{-\frac{1}{2},t}^{\text{VA}}\bigr|^2 
		+\bigl|H_{\frac{1}{2},t}^{\text{VA}}\bigr|^2
		\biggr)\nn \\&
		+8\biggl(
		\bigl|H_{-\frac{1}{2},-1,0}^{T,\frac{1}{2}}\bigr|^2+ \bigl|H_{\frac{1}{2},+1,0}^{T,-\frac{1}{2}}\bigr|^2+ \bigl|H_{-\frac{1}{2},t,-1}^{T,\frac{1}{2}}\bigr|^2 + \bigl|H_{\frac{1}{2},t,+1}^{T,-\frac{1}{2}}\bigr|^2
		\biggr)\nn\\& 
		+16 Re\biggl[H_{-\frac{1}{2},-1,0}^{T,\frac{1}{2}*}H_{-\frac{1}{2},t,-1}^{T,\frac{1}{2}}
		-H_{\frac{1}{2},+1,0}^{T,-\frac{1}{2}*} H_{\frac{1}{2},t,1}^{T,-\frac{1}{2}}\biggr] \biggr\} \biggr]
	\end{align}
	%%%%%%%%%%%%%%%%%%%%%%%%%%%%%%%%%%%%%%%
	\begin{align}
		K_{1ss}=&N\Biggl[
		\bigl|H_{-\frac{1}{2},-1}^{\text{VA}}\bigr|^2
		+\bigl|H_{\frac{1}{2},+1}^{\text{VA}}\bigr|^2
		+2\biggl(\bigl|H_{\frac{1}{2},0}^{\text{SP}}\bigr|^2
		+\bigl|H_{-\frac{1}{2},0}^{\text{SP}}\bigr|^2
		+\bigl|H_{-\frac{1}{2},0}^{\text{VA}}\bigr|^2
		+\bigl|H_{\frac{1}{2},0}^{\text{VA}}\bigr|^2 \biggr)\nn\\&
		+4\biggl(\bigl|H_{\frac{1}{2},+1,0}^{T,-\frac{1}{2}}\bigr|^2
		+\bigl|H_{-\frac{1}{2},-1,0}^{T,\frac{1}{2}}\bigr|^2
		+\bigl|H_{-\frac{1}{2},t,-1}^{T,\frac{1}{2}}\bigr|^2
		+\bigl|H_{\frac{1}{2},t,+1}^{T,-\frac{1}{2}}\bigr|^2\biggr) +8\text{Re}\biggl[H_{-\frac{1}{2},-1,0}^{T,\frac{1}{2}*} H_{-\frac{1}{2},t,-1}^{T,\frac{1}{2}}
		-H_{\frac{1}{2},+1,0}^{T,-\frac{1}{2}*} H_{\frac{1}{2},t,1}^{T,-\frac{1}{2}}\biggr]\nn \\&
		+\frac{\text{$m_{\ell}$}}{\sqrt{q^2}} \biggl\{4\text{Re}\biggl[H_{-\frac{1}{2},0}^{\text{SP}*} H_{-\frac{1}{2},t}^{\text{VA}}+
		H_{\frac{1}{2},0}^{\text{SP}*} H_{\frac{1}{2},t}^{\text{VA}}\biggr]\nn \\&
		+8 \text{Re}\biggl[H_{-\frac{1}{2},0}^{\text{VA}*} H_{-\frac{1}{2},t,0}^{T,-\frac{1}{2}}
		-H_{-\frac{1}{2},0}^{\text{VA}*} H_{-\frac{1}{2},+1,-1}^{T,-\frac{1}{2}} -H_{\frac{1}{2},0}^{\text{VA}*} H_{\frac{1}{2},+1,-1}^{T,\frac{1}{2}}\nn \\&
		+H_{\frac{1}{2},0}^{\text{VA}*} H_{\frac{1}{2},t,0}^{T,\frac{1}{2}}+ H_{-\frac{1}{2},-1,0}^{T,\frac{1}{2}}
		H_{-\frac{1}{2},-1}^{\text{VA}*}- H_{\frac{1}{2},+1,0}^{T,-\frac{1}{2}} H_{\frac{1}{2},+1}^{\text{VA}*}\nn \\&+H_{-\frac{1}{2},-1}^{\text{VA}*} H_{-\frac{1}{2},t,-1}^{T,\frac{1}{2}} + H_{\frac{1}{2},+1}^{\text{VA}*} H_{\frac{1}{2},t,1}^{T,-\frac{1}{2}}\biggr]\biggr\}\nn\\
		&+\frac{\text{$m^2_{\ell}$}}{q^2}\biggl\{
		\bigl|H_{-\frac{1}{2},-1}^{\text{VA}}\bigr|^2
		+\bigl|H_{\frac{1}{2},+1}^{\text{VA}}\bigr|^2
		+2 \biggl(\bigl|H_{-\frac{1}{2},t}^{\text{VA}}\bigr|^2
		+\bigl|H_{\frac{1}{2},t}^{\text{VA}}\bigr|^2
		\biggr)\nn \\&
		+4\biggl(
		\bigl|H_{-\frac{1}{2},-1,0}^{T,\frac{1}{2}}\bigr|^2
		+\bigl|H_{\frac{1}{2},+1,0}^{T,-\frac{1}{2}}\bigr|^2
		+\bigl|H_{-\frac{1}{2},t,-1}^{T,\frac{1}{2}}\bigr|^2
		+\bigl|H_{\frac{1}{2},t,+1}^{T,-\frac{1}{2}}\bigr|^2\biggr)\nn \\ & 
		+8\biggl(\bigl|H_{-\frac{1}{2},+1,-1}^{T,-\frac{1}{2}}\bigr|^2
		+\bigl|H_{\frac{1}{2},+1,-1}^{T,\frac{1}{2}}\bigr|^2
		+\bigl|H_{-\frac{1}{2},t,0}^{T,-\frac{1}{2}}\bigr|^2
		+\bigl|H_{\frac{1}{2},t,0}^{T,\frac{1}{2}}\bigr|^2 \biggr)
		\nn\\& +8~\text{Re}\biggl[H_{-\frac{1}{2},-1,0}^{T,\frac{1}{2}*}H_{-\frac{1}{2},t,-1}^{T,\frac{1}{2}}
		-H_{\frac{1}{2},+1,0}^{T,-\frac{1}{2}*} H_{\frac{1}{2},t,1}^{T,-\frac{1}{2}}
		-2H_{-\frac{1}{2},+1,-1}^{T,-\frac{1}{2}*} H_{-\frac{1}{2},t,0}^{T,-\frac{1}{2}}
		-2 H_{\frac{1}{2},+1,-1}^{T,\frac{1}{2}*} H_{\frac{1}{2},t,0}^{T,\frac{1}{2}}\biggr]\biggr\}\Biggr]
	\end{align}
	%%%%%%%%%%%%%%%%%%%%%%%%%%%%%%%%%
	\begin{align}
		K_{2c}=&N \alpha_{\Lambda} \Biggl[
		2\biggl\{\bigl|H_{-\frac{1}{2},-1}^{\text{VA}}\bigr|^2
		+ \bigl|H_{\frac{1}{2},+1}^{\text{VA}}\bigr|^2 
		+4 \text{Re}\biggl[H_{-\frac{1}{2},0}^{\text{SP}*} H_{-\frac{1}{2},+1,-1}^{T,-\frac{1}{2}}
		- H_{\frac{1}{2},0}^{\text{SP}*}H_{\frac{1}{2},+1,-1}^{T,\frac{1}{2}}
		- H_{-\frac{1}{2},0}^{\text{SP}*} H_{-\frac{1}{2},t,0}^{T,-\frac{1}{2}} 
		+H_{\frac{1}{2},0}^{\text{SP}*} H_{\frac{1}{2},t,0}^{T,\frac{1}{2}}\bigg]\biggr\}\nn \\&
		-\frac{4 \text{$m_{\ell}$}}{\sqrt{q^2}}
		\biggl\{Re\biggl[H_{-\frac{1}{2},0}^{\text{SP}*}
		H_{-\frac{1}{2},0}^{\text{VA}}- H_{\frac{1}{2},0}^{\text{SP}*}
		H_{\frac{1}{2},0}^{\text{VA}}-2 H_{-\frac{1}{2},-1,0}^{T,\frac{1}{2}} H_{-\frac{1}{2},-1}^{\text{VA}*}+2 H_{\frac{1}{2},+1,0}^{T,-\frac{1}{2}} H_{\frac{1}{2},+1}^{\text{VA}*} -2 H_{-\frac{1}{2},+1,-1}^{T,-\frac{1}{2}} H_{-\frac{1}{2},t}^{\text{VA}*}\nn \\&+2 H_{\frac{1}{2},+1,-1}^{T,\frac{1}{2}} H_{\frac{1}{2},t}^{\text{VA}*}-2 H_{-\frac{1}{2},-1}^{\text{VA}*} H_{-\frac{1}{2},t,-1}^{T,\frac{1}{2}}-2 H_{\frac{1}{2},+1}^{\text{VA}*} H_{\frac{1}{2},t,1}^{T,-\frac{1}{2}}+2 H_{-\frac{1}{2},t,0}^{T,-\frac{1}{2}} H_{-\frac{1}{2},t}^{\text{VA}*}-2 H_{\frac{1}{2},t,0}^{T,\frac{1}{2}} H_{\frac{1}{2},t}^{\text{VA}*}\biggr] \biggr\} \nn \\
		& +\frac{4\text{$m^2_{\ell}$}}{q^2} 
		\biggl\{2 
		\bigl|H_{-\frac{1}{2},-1,0}^{T,\frac{1}{2}}\bigr|^2
		+2 \bigl|H_{\frac{1}{2},+1,0}^{T,-\frac{1}{2}}\bigr|^2
		+2\bigl|H_{-\frac{1}{2},t,-1}^{T,\frac{1}{2}}\bigr|^2
		+2 \bigl|H_{\frac{1}{2},t,+1}^{T,-\frac{1}{2}}\bigr|^2 \nn\\ & 
		+\text{Re}\biggl[H_{\frac{1}{2},0}^{\text{VA}*} H_{\frac{1}{2},t}^{\text{VA}}
		-H_{-\frac{1}{2},0}^{\text{VA}*} H_{-\frac{1}{2},t}^{\text{VA}}
		+4 H_{-\frac{1}{2},-1,0}^{T,\frac{1}{2}*} H_{-\frac{1}{2},t,-1}^{T,\frac{1}{2}}
		-4 H_{\frac{1}{2},+1,0}^{T,-\frac{1}{2}*} H_{\frac{1}{2},t,1}^{T,-\frac{1}{2}}\biggr]\biggr\} \Biggr]
	\end{align}
	%%%%%%%%%%%%%%%%%%%%%%%%%%%%%%%%%%%%%%%%%%%%%%%%%%
	\begin{align}
		K_{2cc}=&-\alpha_{\Lambda}N\biggl[2 \biggl\{
		\bigl|H_{-\frac{1}{2},0}^{\text{SP}}\bigr|^2
		-\bigl|H_{\frac{1}{2},0}^{\text{SP}}\bigr|^2
		+\bigl|H_{-\frac{1}{2},-1}^{\text{VA}}\bigr|^2 
		-\bigl|H_{\frac{1}{2},+1}^{\text{VA}}\bigr|^2 \nn\\&
		+4 \bigl|H_{-\frac{1}{2},+1,-1}^{T,-\frac{1}{2}}\bigr|^2
		-4\bigl|H_{\frac{1}{2},+1,-1}^{T,\frac{1}{2}}\bigr|^2
		+4 \bigl|H_{-\frac{1}{2},t,0}^{T,-\frac{1}{2}}\bigr|^2
		-4\bigl|H_{\frac{1}{2},t,0}^{T,\frac{1}{2}}\bigr|^2 \nn\\&
		-8~\text{Re}\biggl[H_{-\frac{1}{2},+1,-1}^{T,-\frac{1}{2}*} H_{-\frac{1}{2},t,0}^{T,-\frac{1}{2}}
		- H_{\frac{1}{2},+1,-1}^{T,\frac{1}{2}*} H_{\frac{1}{2},t,0}^{T,\frac{1}{2}}\biggr]\biggr\} \nn\\&
		-\frac{4  \text{$m_{\ell}$}}{\sqrt{q^2}}
		\biggl\{\text{Re}\biggl[H_{-\frac{1}{2},0}^{\text{SP}*}
		H_{-\frac{1}{2},t}^{\text{VA}}- H_{\frac{1}{2},0}^{\text{SP}*}
		H_{\frac{1}{2},t}^{\text{VA}}-2 H_{-\frac{1}{2},0}^{\text{VA}*}
		H_{-\frac{1}{2},+1,-1}^{T,-\frac{1}{2}} +2 H_{-\frac{1}{2},0}^{\text{VA}*} H_{-\frac{1}{2},t,0}^{T,-\frac{1}{2}} \nn\\&
		+2 H_{\frac{1}{2},0}^{\text{VA}*} H_{\frac{1}{2},+1,-1}^{T,\frac{1}{2}}-2 H_{\frac{1}{2},0}^{\text{VA}*}
		H_{\frac{1}{2},t,0}^{T,\frac{1}{2}}\nn \\&
		+2 H_{-\frac{1}{2},-1,0}^{T,\frac{1}{2}} H_{-\frac{1}{2},-1}^{\text{VA}*}+2
		H_{\frac{1}{2},+1,0}^{T,-\frac{1}{2}}
		H_{\frac{1}{2},+1}^{\text{VA}*}
		+2 H_{-\frac{1}{2},-1}^{\text{VA}*}
		H_{-\frac{1}{2},t,-1}^{T,\frac{1}{2}}-2
		H_{\frac{1}{2},+1}^{\text{VA}*} H_{\frac{1}{2},t,+1}^{T,-\frac{1}{2}}\biggr]\biggr\}\nn\\
		&-\frac{2  \text{$m^2_{\ell}$}}{q^2}\biggl\{
		\bigl|H_{-\frac{1}{2},0}^{\text{VA}}\bigr|^2
		-\bigl|H_{\frac{1}{2},0}^{\text{VA}}\bigr|^2 
		+\bigl|H_{-\frac{1}{2},t}^{\text{VA}}\bigr|^2
		-\bigl|H_{\frac{1}{2},t}^{\text{VA}}\bigr|^2 +4 \bigl|H_{-\frac{1}{2},-1,0}^{T,\frac{1}{2}}\bigr|^2 \nn\\ &
		-4 \bigl|H_{\frac{1}{2},+1,0}^{T,-\frac{1}{2}}\bigr|^2
		+4 \bigl|H_{-\frac{1}{2},t,-1}^{T,\frac{1}{2}}\bigr|^2 
		-4 \bigl|H_{\frac{1}{2},t,+1}^{T,-\frac{1}{2}}\bigr|^2  +8 \text{Re}\biggl[H_{-\frac{1}{2},-1,0}^{T,\frac{1}{2}*}
		H_{-\frac{1}{2},t,-1}^{T,\frac{1}{2}}
		+ H_{\frac{1}{2},+1,0}^{T,-\frac{1}{2}*} H_{\frac{1}{2},t,+1}^{T,-\frac{1}{2}}\biggr]\biggr\}
		\biggr]
	\end{align}
	%%%%%%%%%%%%%%%%%%%%%%%%%%%%%%%%%%%%%%%%%%%%%%
	\begin{align}
		K_{2ss}=&-\alpha_{\Lambda}N\Biggl[\biggl\{\biggl(\bigl|H_{-\frac{1}{2},-1}^{\text{VA}}\bigr|^2
		-\bigl|H_{\frac{1}{2},+1}^{\text{VA}}\bigr|^2\biggr)
		+2\biggl(\bigl|H_{-\frac{1}{2},0}^{\text{SP}}\bigr|^2
		+ \bigl|H_{-\frac{1}{2},0}^{\text{VA}}\bigr|^2
		-\bigl|H_{\frac{1}{2},0}^{\text{SP}}\bigr|^2 
		-\bigl|H_{\frac{1}{2},0}^{\text{VA}}\bigr|^2 \biggr)\nn \\&
		+4 \biggl(\bigl|H_{-\frac{1}{2},-1,0}^{T,\frac{1}{2}}\bigr|^2
		+\bigl|H_{-\frac{1}{2}, t, -1}^{T,\frac{1}{2}}\bigr|^2 
		-\bigl|H_{\frac{1}{2},+1,0}^{T,-\frac{1}{2}}\bigr|^2 
		-\bigl|H_{\frac{1}{2},t,+1}^{T,-\frac{1}{2}}\bigr|^2\biggr)
		\nn\\& 
		+8 \text{Re}\biggl[H_{-\frac{1}{2},-1,0}^{T,\frac{1}{2}*} H_{-\frac{1}{2},t,-1}^{T,\frac{1}{2}}+8 H_{\frac{1}{2},+1,0}^{T,-\frac{1}{2}*} H_{\frac{1}{2},t,+1}^{T,-\frac{1}{2}}\biggr]\biggr\}\nn\\
		& +\frac{4 \text{$m_{\ell}$}}{\sqrt{q^2}} 
		\biggl\{\text{Re}\biggl[H_{-\frac{1}{2},0}^{\text{SP}*}
		H_{-\frac{1}{2},t}^{\text{VA}}- H_{\frac{1}{2},0}^{\text{SP}*}
		H_{\frac{1}{2},t}^{\text{VA}}-2 H_{-\frac{1}{2},0}^{\text{VA}*}
		H_{-\frac{1}{2},+1, -1}^{T,-\frac{1}{2}}+2
		H_{-\frac{1}{2},0}^{\text{VA}*} H_{-\frac{1}{2},t,0}^{T,-\frac{1}{2}}+2 H_{\frac{1}{2},0}^{\text{VA}*} H_{\frac{1}{2},+1,-1}^{T,\frac{1}{2}} 
		\nn \\& 
		-2 H_{\frac{1}{2},0}^{\text{VA}*} H_{\frac{1}{2},t,0}^{T,\frac{1}{2}} 
		+2 H_{-\frac{1}{2},-1,0}^{T,\frac{1}{2}} H_{-\frac{1}{2},-1}^{\text{VA}*}
		+2 H_{\frac{1}{2},+1,0}^{T,-\frac{1}{2}}
		H_{\frac{1}{2},+1}^{\text{VA}*} +2 H_{-\frac{1}{2},-1}^{\text{VA}*}
		H_{-\frac{1}{2},t,-1}^{T,\frac{1}{2}}-2 H_{\frac{1}{2},+1}^{\text{VA}*} H_{\frac{1}{2},t,+1}^{T,-\frac{1}{2}} \biggr]\biggr\}\nn\\&
		+\frac{ \text{$m^2_{\ell}$}}{q^2} \biggl\{
		\bigl|H_{-\frac{1}{2},-1}^{\text{VA}}\bigr|^2
		-\bigl|H_{\frac{1}{2},+1}^{\text{VA}}\bigr|^2
		+2\bigl|H_{-\frac{1}{2},t}^{\text{VA}}\bigr|^2
		-2 \bigl|H_{\frac{1}{2},t}^{\text{VA}}\bigr|^2 \nn\\&
		+4\bigl|H_{-\frac{1}{2},t,-1}^{T,\frac{1}{2}}\bigr|^2
		-4\bigl|H_{\frac{1}{2},t,+1}^{T,-\frac{1}{2}}\bigr|^2+
		4\bigl|H_{-\frac{1}{2},-1,0}^{T,\frac{1}{2}}\bigr|^2-
		4\bigl|H_{\frac{1}{2},+1,0}^{T,-\frac{1}{2}}\bigr|^2 \nn\\&	
		+8\bigl|H_{-\frac{1}{2},+1,-1}^{T,-\frac{1}{2}}\bigr|^2
		-8\bigl|H_{\frac{1}{2},+1,-1}^{T,\frac{1}{2}}\bigr|^2   
		+8\bigl|H_{-\frac{1}{2},t,0}^{T,-\frac{1}{2}}\bigr|^2
		-8\bigl|H_{\frac{1}{2},t,0}^{T,\frac{1}{2}}\bigr|^2 \nn \\& 
		+8 \text{Re}\biggl[H_{-\frac{1}{2},-1, 0}^{T,\frac{1}{2}*}
		H_{-\frac{1}{2},t,-1}^{T,\frac{1}{2}}
		+8 H_{\frac{1}{2},+1,0}^{T,-\frac{1}{2}*} H_{\frac{1}{2},t,+1}^{T,-\frac{1}{2}}
		-16 H_{-\frac{1}{2},+1,-1}^{T,-\frac{1}{2}*} H_{-\frac{1}{2},t,0}^{T,-\frac{1}{2}}
		+16 H_{\frac{1}{2},+1,-1}^{T,\frac{1}{2}*} H_{\frac{1}{2},t,0}^{T,\frac{1}{2}}\biggr]\biggr\}\Biggr]
	\end{align}
	%%%%%%%%%%%%%%%%%%%%%%%%%%%%%%%%%%%%%%%%%
	\begin{align}
		K_{3sc}=&-2 \sqrt{2}\alpha_{\Lambda}N\biggl(1-\frac{\text{$m^2_{\ell}$}}{q^2}\biggr)\times \text{Re}\biggl[
		\biggl(H_{-\frac{1}{2},0}^{\text{VA}*} H_{\frac{1}{2},+1}^{\text{VA}}
		-H_{\frac{1}{2},0}^{\text{VA}*} H_{-\frac{1}{2},-1}^{\text{VA}}\biggr)\nn\\&
		+4 \biggl(H_{-\frac{1}{2},-1,0}^{T,\frac{1}{2}*} H_{\frac{1}{2},t,0}^{T,\frac{1}{2}}
		-H_{-\frac{1}{2},-1,0}^{T,\frac{1}{2}} H_{\frac{1}{2},+1,-1}^{T,\frac{1}{2}*}
		-H_{\frac{1}{2},+1,0}^{T,-\frac{1}{2}*} H_{-\frac{1}{2},+1,-1}^{T,-\frac{1}{2}} 
		+H_{\frac{1}{2},+1,0}^{T,-\frac{1}{2}*} H_{-\frac{1}{2},t,0}^{T,-\frac{1}{2}}
		+H_{-\frac{1}{2},+1,-1}^{T,-\frac{1}{2}*} H_{\frac{1}{2},t,1}^{T,-\frac{1}{2}}\nn \\&
		-H_{\frac{1}{2},+1,-1}^{T,\frac{1}{2}*} H_{-\frac{1}{2},t,-1}^{T,\frac{1}{2}}
		-H_{-\frac{1}{2},t,0}^{T,-\frac{1}{2}} H_{\frac{1}{2},t,1}^{T,-\frac{1}{2}*}
		+H_{\frac{1}{2},t,0}^{T,\frac{1}{2}*} H_{-\frac{1}{2},t,-1}^{T,\frac{1}{2}}\biggr)\biggr]
	\end{align}
	%%%%%%%%%%%%%%%%%%%%%%%%%%%%%%%%%%%%%%
	\begin{align}
		K_{3s}=&-2 \sqrt{2}\alpha_{\Lambda}N\times\nn\\& 
		\Biggl[
		\biggl\{ \text{Re}\biggl[
		H_{-\frac{1}{2},0}^{\text{VA}*}H_{\frac{1}{2},+1}^{\text{VA}}
		+H_{\frac{1}{2},0}^{\text{VA}*}H_{-\frac{1}{2},-1}^{\text{VA}}\nn\\&  +2\biggl(H_{-\frac{1}{2},0}^{\text{SP}*} H_{\frac{1}{2},+1,0}^{T,-\frac{1}{2}}
		-H_{-\frac{1}{2},0}^{\text{SP}*} H_{\frac{1}{2},t,1}^{T,-\frac{1}{2}}
		+H_{\frac{1}{2},0}^{\text{SP}*} H_{-\frac{1}{2},-1,0}^{T,\frac{1}{2}} 
		+H_{\frac{1}{2},0}^{\text{SP}*}H_{-\frac{1}{2},t,-1}^{T,\frac{1}{2}}\biggr)
		\biggr]\biggr\}\nn\\
		&-\frac{\text{$m_{\ell}$}}{\sqrt{q^2}}
		\biggl\{\text{Re}\biggl[H_{-\frac{1}{2},0}^{\text{SP}*}
		H_{\frac{1}{2},+1}^{\text{VA}}-H_{\frac{1}{2},0}^{\text{SP}*}
		H_{-\frac{1}{2},-1}^{\text{VA}}
		+2\biggl(H_{-\frac{1}{2},0}^{\text{VA}*}H_{\frac{1}{2},+1,0}^{T,-\frac{1}{2}}
		-H_{-\frac{1}{2},0}^{\text{VA}*} H_{\frac{1}{2},t,1}^{T,-\frac{1}{2}}
		- H_{\frac{1}{2},0}^{\text{VA}*} H_{-\frac{1}{2},-1,0}^{T,\frac{1}{2}} \nn \\&
		- H_{\frac{1}{2},0}^{\text{VA}*}H_{-\frac{1}{2},t,-1}^{T,\frac{1}{2}}
		- H_{-\frac{1}{2},-1,0}^{T,\frac{1}{2}} H_{\frac{1}{2},t}^{\text{VA}*}
		-H_{\frac{1}{2},+1,0}^{T,-\frac{1}{2}}H_{-\frac{1}{2},t}^{\text{VA}*}
		+ H_{-\frac{1}{2},+1,-1}^{T,-\frac{1}{2}} H_{\frac{1}{2},+1}^{\text{VA}*}
		+H_{\frac{1}{2},+1,-1}^{T,\frac{1}{2}} H_{-\frac{1}{2},-1}^{\text{VA}*}\nn \\&
		-H_{-\frac{1}{2},-1}^{\text{VA}*} H_{\frac{1}{2},t,0}^{T,\frac{1}{2}}
		- H_{\frac{1}{2},+1}^{\text{VA}*} H_{-\frac{1}{2},t, 0}^{T,-\frac{1}{2}}
		- H_{-\frac{1}{2},t,-1}^{T,\frac{1}{2}} H_{\frac{1}{2},t}^{\text{VA}*}
		+ H_{\frac{1}{2},t,1}^{T,-\frac{1}{2}} H_{-\frac{1}{2},t}^{\text{VA}*}\biggr)\biggr]\biggr\}\nn\\
		& +\frac{\text{$m^2_{\ell}$}}{q^2}\biggl\{Re\biggl[
		H_{-\frac{1}{2},-1}^{\text{VA}}H_{\frac{1}{2},t}^{\text{VA}*}
		-H_{\frac{1}{2},+1}^{\text{VA}}H_{-\frac{1}{2},t}^{\text{VA}*}\nn \\&
		+4\biggl(H_{-\frac{1}{2},-1,0}^{T,\frac{1}{2}*}H_{\frac{1}{2},t,0}^{T,\frac{1}{2}}
		-H_{-\frac{1}{2},-1,0}^{T,\frac{1}{2}} H_{\frac{1}{2},+1,-1}^{T,\frac{1}{2}*}
		+ H_{\frac{1}{2},+1,0}^{T,-\frac{1}{2}*} H_{-\frac{1}{2},+1,-1}^{T,-\frac{1}{2}}
		-H_{\frac{1}{2},+1,0}^{T,-\frac{1}{2}*} H_{-\frac{1}{2},t,0}^{T,-\frac{1}{2}}
		- H_{-\frac{1}{2}, +1,-1}^{T,-\frac{1}{2}*} H_{\frac{1}{2}, t, 1}^{T,-\frac{1}{2}}\nn \\&
		-H_{\frac{1}{2},+1,-1}^{T,\frac{1}{2}*} H_{-\frac{1}{2},t,-1}^{T,\frac{1}{2}}
		+H_{-\frac{1}{2},t,0}^{T,-\frac{1}{2}} H_{\frac{1}{2},t,1}^{T,-\frac{1}{2}*}
		+H_{\frac{1}{2},t,0}^{T,\frac{1}{2}*} H_{-\frac{1}{2},t,-1}^{T,\frac{1}{2}}\biggr)
		\biggr]\biggr\}
		\Biggr]
	\end{align}
	%%%%%%%%%%%%%%%%%%%%%%%%%%%%%%%%%%%%%%%%%%%%%%
	\begin{align}\label{eq:k4s}
		K_{4s}=&2\sqrt{2}\alpha_{\Lambda}N\times\nn\\& \Biggl[
		\biggl\{\text{Im}\biggl[H_{-\frac{1}{2},0}^{\text{VA}*}H_{\frac{1}{2},+1}^{\text{VA}}
		-H_{\frac{1}{2},0}^{\text{VA}*}H_{-\frac{1}{2},-1}^{\text{VA}}\nn\\& +
		2\biggl(H_{-\frac{1}{2},0}^{\text{SP}*} H_{\frac{1}{2},+1,0}^{T,-\frac{1}{2}}
		-H_{-\frac{1}{2},0}^{\text{SP}*} H_{\frac{1}{2},t,1}^{T,-\frac{1}{2}}
		-H_{\frac{1}{2},0}^{\text{SP}*}H_{-\frac{1}{2},-1,0}^{T,\frac{1}{2}}
		-H_{\frac{1}{2},0}^{\text{SP}*} H_{-\frac{1}{2},t,-1}^{T,\frac{1}{2}}\biggr)
		\biggr]\biggr\}\nn\\&
		-\frac{\text{$m_{\ell}$}}{\sqrt{q^2}} 
		\biggl\{\text{Im}\biggl[
		H_{-\frac{1}{2},0}^{\text{SP}*}H_{\frac{1}{2},+1}^{\text{VA}}
		+ H_{\frac{1}{2},0}^{\text{SP}*}H_{-\frac{1}{2},-1}^{\text{VA}}
		+2\biggl(H_{-\frac{1}{2},0}^{\text{VA}*}H_{\frac{1}{2},+1,0}^{T,-\frac{1}{2}}
		-H_{-\frac{1}{2},0}^{\text{VA}*} H_{\frac{1}{2},t,1}^{T,-\frac{1}{2}}
		+H_{\frac{1}{2},0}^{\text{VA}*} H_{-\frac{1}{2},-1,0}^{T,\frac{1}{2}}\nn \\& 
		+H_{\frac{1}{2},0}^{\text{VA}*}H_{-\frac{1}{2},t,-1}^{T,\frac{1}{2}}
		+H_{-\frac{1}{2},-1,0}^{T,\frac{1}{2}} H_{\frac{1}{2},t}^{\text{VA}*}
		-H_{\frac{1}{2},+1,0}^{T,-\frac{1}{2}}H_{-\frac{1}{2},t}^{\text{VA}*}
		-H_{-\frac{1}{2}, +1,-1}^{T,-\frac{1}{2}} H_{\frac{1}{2},+1}^{\text{VA}*}
		+H_{\frac{1}{2},+1,-1}^{T,\frac{1}{2}} H_{-\frac{1}{2},-1}^{\text{VA}*}\nn \\&
		-H_{-\frac{1}{2},-1}^{\text{VA}*} H_{\frac{1}{2},t,0}^{T,\frac{1}{2}}
		+H_{\frac{1}{2},+1}^{\text{VA}*} H_{-\frac{1}{2},t,0}^{T,-\frac{1}{2}}
		+H_{-\frac{1}{2},t,-1}^{T,\frac{1}{2}} H_{\frac{1}{2},t}^{\text{VA}*}
		+H_{\frac{1}{2},t,1}^{T,-\frac{1}{2}} H_{-\frac{1}{2},t}^{\text{VA}*}\biggr)\biggr]
		\biggr\}\nn\\&
		+\frac{\text{$m^2_{\ell}$}}{q^2} \biggl\{\text{Im}\biggl[
		-H_{-\frac{1}{2},-1}^{\text{VA}}H_{\frac{1}{2},t}^{\text{VA}*}
		-H_{\frac{1}{2},+1}^{\text{VA}}H_{-\frac{1}{2},t}^{\text{VA}*}\nn \\& 
		+4\Bigl(H_{-\frac{1}{2},-1,0}^{T,\frac{1}{2}*} H_{\frac{1}{2},t,0}^{T,\frac{1}{2}}
		+H_{-\frac{1}{2},-1,0}^{T,\frac{1}{2}} H_{\frac{1}{2},+1,-1}^{T,\frac{1}{2}*}
		-H_{\frac{1}{2},+1,0}^{T,-\frac{1}{2}*} H_{-\frac{1}{2},+1, -1}^{T,-\frac{1}{2}}
		+H_{\frac{1}{2},+1,0}^{T,-\frac{1}{2}*} H_{-\frac{1}{2},t,0}^{T,-\frac{1}{2}}
		-H_{-\frac{1}{2},+1,-1}^{T,-\frac{1}{2}*} H_{\frac{1}{2},t, 1}^{T,-\frac{1}{2}}\nn \\&
		+H_{\frac{1}{2},+1,-1}^{T,\frac{1}{2}*} H_{-\frac{1}{2},t,-1}^{T,\frac{1}{2}}
		-H_{-\frac{1}{2},t,0}^{T,-\frac{1}{2}} H_{\frac{1}{2},t,1}^{T,-\frac{1}{2}*}
		-H_{\frac{1}{2},t,0}^{T,\frac{1}{2}*} H_{-\frac{1}{2},t,-1}^{T,\frac{1}{2}}\Bigr)
		\biggr]\biggr\}\Biggr]
	\end{align}
	%%%%%%%%%%%%%%%%%%%%%%%%%%%%%%%%%%%%%%%%%%%%%%%%%%
	
	\begin{align}\label{eq:k4sc}
		K_{4sc}=&2\sqrt{2}\alpha_{\Lambda} N\bigl(1-\frac{\text{$m^2_{\ell}$}}{q^2}\bigr)\times\nn \\&
		\text{Im}\biggl[
		H_{-\frac{1}{2},0}^{\text{VA}*} H_{\frac{1}{2},+1}^{\text{VA}}
		+H_{\frac{1}{2},0}^{\text{VA}*} H_{-\frac{1}{2},-1}^{\text{VA}} 
		+4\biggl(H_{-\frac{1}{2},-1,0}^{T,\frac{1}{2}*} H_{\frac{1}{2},t,0}^{T,\frac{1}{2}}
		+H_{-\frac{1}{2},-1, 0}^{T,\frac{1}{2}} H_{\frac{1}{2},+1,-1}^{T,\frac{1}{2}*}
		+H_{\frac{1}{2}, +1,0}^{T,-\frac{1}{2}*} H_{-\frac{1}{2},+1,-1}^{T,-\frac{1}{2}}\nn \\& 
		-H_{\frac{1}{2},+1,0}^{T,-\frac{1}{2}*} H_{-\frac{1}{2},t,0}^{T,-\frac{1}{2}}
		+H_{-\frac{1}{2},+1,-1}^{T,-\frac{1}{2}*} H_{\frac{1}{2},t,1}^{T,-\frac{1}{2}} 
		+H_{\frac{1}{2},+1,-1}^{T,\frac{1}{2}*} H_{-\frac{1}{2},t,-1}^{T,\frac{1}{2}} 
		+H_{-\frac{1}{2},t,0}^{T,-\frac{1}{2}} H_{\frac{1}{2},t, 1}^{T,-\frac{1}{2}*}
		-H_{\frac{1}{2},t,0}^{T,\frac{1}{2}*} H_{-\frac{1}{2},t,-1}^{T,\frac{1}{2}}\biggr)\biggr]
	\end{align}
	%%%%%%%%%%%%%%%%%%%%%%%%%%%%%
	\begin{align}
		\frac{d\Gamma}{dq^2}^{\lambda_{\ell}=+1/2}=&
		\frac{3 N}{2}\biggl[4\bigl|H_{-\frac{1}{2},0}^{\text{SP}}\bigr|^2
		+4\bigl|H_{\frac{1}{2},0}^{\text{SP}}\bigr|^2\nn\\&
		+\frac{16}{3}\biggl(\bigl|H_{-\frac{1}{2},-1,0}^{T,\frac{1}{2}}\bigr|^2 
		+\bigl|H_{\frac{1}{2},+1,0}^{T,-\frac{1}{2}}\bigr|^2\nn
		+ \bigl|H_{-\frac{1}{2},+1,-1}^{T,-\frac{1}{2}}\bigr|^2
		+ \bigl|H_{\frac{1}{2},+1,-1}^{T,\frac{1}{2}}\bigr|^2\nn\\&
		+ \bigl|H_{-\frac{1}{2},t,0}^{T,-\frac{1}{2}}\bigr|^2
		+\bigl|H_{\frac{1}{2},t,0}^{T,\frac{1}{2}}\bigr|^2
		+ \bigl|H_{-\frac{1}{2},t,-1}^{T,\frac{1}{2}}\bigr|^2
		+\bigl|H_{\frac{1}{2},t,+1}^{T,-\frac{1}{2}}\bigr|^2\biggr)\nn\\&
		+\frac{32}{3}\text{Re}\bigl[H_{-\frac{1}{2},-1,0}^{T,\frac{1}{2}*} H_{-\frac{1}{2},t,-1}^{T,\frac{1}{2}}
		- H_{\frac{1}{2},+1,0}^{T,-\frac{1}{2}*}H_{\frac{1}{2},t,+1}^{T,-\frac{1}{2}}
		- H_{-\frac{1}{2},+1,-1}^{T,-\frac{1}{2}*}H_{-\frac{1}{2},t,0}^{T,-\frac{1}{2}}
		-H_{\frac{1}{2},+1,-1}^{T,\frac{1}{2}*} H_{\frac{1}{2},t,0}^{T,\frac{1}{2}}\bigr]\nn\\&
		+\frac{\text{$m_{\ell}$}}{\sqrt{\text{$q^2$}}} \biggl\{ 8 \text{Re}\bigl[H_{-\frac{1}{2},0}^{\text{SP}*}
		H_{-\frac{1}{2},t}^{\text{VA}}+H_{\frac{1}{2},0}^{\text{SP}*} H_{\frac{1}{2},t}^{\text{VA}}\bigr]
		\nn\\
		&+\frac{16}{3}
		\text{Re}\biggl[
		H_{-\frac{1}{2},0}^{\text{VA}*}H_{-\frac{1}{2},t,0}^{T,-\frac{1}{2}}
		-H_{-\frac{1}{2},0}^{\text{VA}*} H_{-\frac{1}{2},+1,-1}^{T,-\frac{1}{2}}	
		-H_{\frac{1}{2},0}^{\text{VA}*}H_{\frac{1}{2},+1,-1}^{T,\frac{1}{2}}
		+ H_{\frac{1}{2},0}^{\text{VA}*} H_{\frac{1}{2},t,0}^{T,\frac{1}{2}} 
		+H_{-\frac{1}{2},-1,0}^{T,\frac{1}{2}} H_{-\frac{1}{2},-1}^{\text{VA}*}\nn\\ &
		-H_{\frac{1}{2},+1,0}^{T,-\frac{1}{2}} H_{\frac{1}{2},+1}^{\text{VA}*}
		+H_{-\frac{1}{2},-1}^{\text{VA}*} H_{-\frac{1}{2},t,-1}^{T,\frac{1}{2}} 
		+H_{\frac{1}{2},+1}^{\text{VA}*}H_{\frac{1}{2},t,+1}^{T,-\frac{1}{2}}\biggr]\biggr\}\nn\\&
		+\frac{\text{$m^2_{\ell}$}}{\text{$q^2$}} \biggl\{ 
		\frac{4}{3}\biggl(\bigl|H_{-\frac{1}{2},0}^{\text{VA}}\bigr|^2
		+ \bigl|H_{\frac{1}{2},0}^{\text{VA}}\bigr|^2
		+  \bigl|H_{-\frac{1}{2},-1}^{\text{VA}}\bigr|^2
		+  \bigl|H_{\frac{1}{2},+1}^{\text{VA}}\bigr|^2\biggr) 
		+4 \bigl|H_{-\frac{1}{2},t}^{\text{VA}}\bigr|^2
		+4 \bigl|H_{\frac{1}{2},t}^{\text{VA}}\bigr|^2\biggr\}\biggr]
	\end{align}
	%%%%%%%%%%%%%%%%%%%%%%%%%%%%%%%%%%%%%%%%%
	\begin{align}
		\frac{d\Gamma}{dq^2}^{\lambda_{\ell}=-1/2}=&
		\frac{3 N}{2}\biggl[\frac{8}{3}\biggl(\bigl|H_{-\frac{1}{2},0}^{\text{VA}}\bigr|^2
		+ \bigl|H_{\frac{1}{2},0}^{\text{VA}}\bigr|^2
		+\bigl|H_{-\frac{1}{2},-1}^{\text{VA}}\bigr|^2
		+\bigl|H_{\frac{1}{2},+1}^{\text{VA}}\bigr|^2\biggr) \nn\\&
		+\frac{32}{3} \frac{\text{$m_{\ell}$}}{\sqrt{\text{$q^2$}}}
		\text{Re}\biggl[
		H_{-\frac{1}{2},0}^{\text{VA}*} H_{-\frac{1}{2},t,0}^{T,-\frac{1}{2}}
		-H_{-\frac{1}{2},0}^{\text{VA}*}H_{-\frac{1}{2},+1,-1}^{T,-\frac{1}{2}}		 
		-H_{\frac{1}{2},0}^{\text{VA}*} H_{\frac{1}{2},+1,-1}^{T,\frac{1}{2}}
		+H_{\frac{1}{2},0}^{\text{VA}*} H_{\frac{1}{2},t,0}^{T,\frac{1}{2}}
		+H_{-\frac{1}{2},-1,0}^{T,\frac{1}{2}} H_{-\frac{1}{2},-1}^{\text{VA}*} \nn\\& 
		-H_{\frac{1}{2},+1,0}^{T,-\frac{1}{2}}H_{\frac{1}{2},+1}^{\text{VA}*}
		+H_{-\frac{1}{2},-1}^{\text{VA}*} H_{-\frac{1}{2},t,-1}^{T,\frac{1}{2}}
		+H_{\frac{1}{2},+1}^{\text{VA}*} H_{\frac{1}{2},t,+1}^{T,-\frac{1}{2}}
		\biggr] \nn\\&
		+\frac{32}{3}\frac{\text{$m^2_{\ell}$}}{\text{$q^2$}}\biggl\{
		2\text{Re}\biggl[
		H_{-\frac{1}{2},-1,0}^{T,\frac{1}{2}*}H_{-\frac{1}{2},t,-1}^{T,\frac{1}{2}}
		- H_{\frac{1}{2},+1,0}^{T,-\frac{1}{2}*} H_{\frac{1}{2},t,+1}^{T,-\frac{1}{2}}
		-H_{-\frac{1}{2},+1,-1}^{T,-\frac{1}{2}*} H_{-\frac{1}{2},t,0}^{T,-\frac{1}{2}}
		-H_{\frac{1}{2},+1,-1}^{T,\frac{1}{2}*} H_{\frac{1}{2},t,0}^{T,\frac{1}{2}}\biggr] \nn\\&
		+\biggl(\bigl|H_{-\frac{1}{2},-1,0}^{T,\frac{1}{2}}\bigr|^2
		+\bigl|H_{\frac{1}{2},+1,0}^{T,-\frac{1}{2}}\bigr|^2
		+ \bigl|H_{-\frac{1}{2},+1,-1}^{T,-\frac{1}{2}}\bigr|^2
		+\bigl|H_{\frac{1}{2},+1,-1}^{T,\frac{1}{2}}\bigr|^2 \nn\\&
		+ \bigl|H_{-\frac{1}{2},t,0}^{T,-\frac{1}{2}}\bigr|^2
		+\bigl|H_{\frac{1}{2},t,0}^{T,\frac{1}{2}}\bigr|^2
		+ \bigl|H_{-\frac{1}{2},t,-1}^{T,\frac{1}{2}}\bigr|^2
		+\bigl|H_{\frac{1}{2},t,+1}^{T,-\frac{1}{2}}\bigr|^2 
		\biggr)\biggr\}\biggr]
	\end{align}
	%%%%%%%%%%%%%%%%%%%%%%%%%%%%%%%%%%%%%%%%%
	\begin{align}
		\frac{d\Gamma}{dq^2}^{\lambda_{\Lambda_c}=+1/2}=&
		\frac{3 N}{2}\biggl[4\bigl|H_{\frac{1}{2},0}^{\text{SP}}\bigr|^2
		+\frac{8}{3}\biggl(\bigl|H_{\frac{1}{2},0}^{\text{VA}}\bigr|^2
		+\bigl|H_{\frac{1}{2},+1}^{\text{VA}}\bigr|^2\biggr)
		+\frac{16}{3}
		\biggl( \bigl|H_{\frac{1}{2},+1,0}^{T,-\frac{1}{2}}\bigr|^2
		+\bigl|H_{\frac{1}{2},+1,-1}^{T,\frac{1}{2}}\bigr|^2
		+\bigl|H_{\frac{1}{2},t,0}^{T,\frac{1}{2}}\bigr|^2
		+ \bigl|H_{\frac{1}{2},t,+1}^{T,-\frac{1}{2}}\bigr|^2 \biggr)
		\nn\\&
		-\frac{32}{3}\text{Re}\biggl[
		H_{\frac{1}{2},+1,0}^{T,-\frac{1}{2}*} H_{\frac{1}{2},t,+1}^{T,-\frac{1}{2}}
	+H_{\frac{1}{2},+1,-1}^{T,\frac{1}{2}*} H_{\frac{1}{2},t,0}^{T,\frac{1}{2}}\biggr]
		\nn\\&
		+\frac{\text{$m_{\ell}$}}{\sqrt{\text{$q^2$}}}
		\biggl\{8\text{Re}\bigl[H_{\frac{1}{2},0}^{\text{SP}*}
		H_{\frac{1}{2},t}^{\text{VA}}\bigr]
		+16 \text{Re}\bigl[H_{\frac{1}{2},0}^{\text{VA}*}H_{\frac{1}{2},t,0}^{T,\frac{1}{2}}
		+H_{\frac{1}{2},+1}^{\text{VA}*}
		H_{\frac{1}{2},t,+1}^{T,-\frac{1}{2}}
		- H_{\frac{1}{2},0}^{\text{VA}*} H_{\frac{1}{2},+1,-1}^{T,\frac{1}{2}}
		- H_{\frac{1}{2},+1,0}^{T,-\frac{1}{2}}
		H_{\frac{1}{2},+1}^{\text{VA}*}\bigr]
		\biggr\}\nn\\&
		+\frac{\text{$m^2_{\ell}$}}{\text{$q^2$}}\biggl\{4\bigl|H_{\frac{1}{2},t}^{\text{VA}}\bigr|^2
		+\frac{4}{3} \bigl(\bigl|H_{\frac{1}{2},0}^{\text{VA}}\bigr|^2
		+\bigl|H_{\frac{1}{2},+1}^{\text{VA}}\bigr|^2 \bigr)\nn\\&
		+\frac{32}{3}
		\bigl(\bigl|H_{\frac{1}{2},+1,0}^{T,-\frac{1}{2}}\bigr|^2
		+\bigl|H_{\frac{1}{2},+1,-1}^{T,\frac{1}{2}}\bigr|^2
		+\bigl|H_{\frac{1}{2},t,0}^{T,\frac{1}{2}}\bigr|^2
		+\bigl|H_{\frac{1}{2},t,+1}^{T,-\frac{1}{2}}\bigr|^2 \bigr)\nn\\&
		-\frac{64}{3}\text{Re}\bigl[
		H_{\frac{1}{2},+1,0}^{T,-\frac{1}{2}*} H_{\frac{1}{2},t,+1}^{T,-\frac{1}{2}} 
	+H_{\frac{1}{2},+1,-1}^{T,\frac{1}{2}*}H_{\frac{1}{2},t,0}^{T,\frac{1}{2}}\bigr]\biggr\}\biggr]
	\end{align}
	%%%%%%%%%%%%%%%%%%%%%%%%%%%%%
	\begin{align}
		\frac{d\Gamma}{dq^2}^{\lambda_{\Lambda_c}=-1/2}=& 
		\frac{3 N}{2}\biggl[4 \bigl|H_{-\frac{1}{2},0}^{\text{SP}}\bigr|^2
		+\frac{8}{3}\biggl(\bigl|H_{-\frac{1}{2},0}^{\text{VA}}\bigr|^2
		+ \bigl|H_{-\frac{1}{2},-1}^{\text{VA}}\bigr|^2 \biggr)
		\nn\\&
		+\frac{16}{3}\biggl(
		\bigl|H_{-\frac{1}{2},-1,0}^{T,\frac{1}{2}}\bigr|^2 
		+\bigl|H_{-\frac{1}{2},+1,-1}^{T,-\frac{1}{2}}\bigr|^2
		+\bigl|H_{-\frac{1}{2},t,0}^{T,-\frac{1}{2}}\bigr|^2
		+ \bigl|H_{-\frac{1}{2},t,-1}^{T,\frac{1}{2}}\bigr|^2 
		\biggr)
		\nn\\&
		+\frac{32}{3}\text{Re}\biggl[(H_{-\frac{1}{2},-1,0}^{T,\frac{1}{2}*}
		H_{-\frac{1}{2},t,-1}^{T,\frac{1}{2}}
		- H_{-\frac{1}{2},+1,-1}^{T,-\frac{1}{2}*}
		H_{-\frac{1}{2},t,0}^{T,-\frac{1}{2}}\biggr] \nn\\&
		+\frac{\text{$m_{\ell}$}}{\sqrt{\text{q2}}} \biggl\{8 \text{Re}\biggl[H_{-\frac{1}{2},0}^{\text{SP}*}
		H_{-\frac{1}{2},t}^{\text{VA}}\biggr] \nn\\&
		+16
		\text{Re}\biggl[ H_{-\frac{1}{2},0}^{\text{VA}*} H_{-\frac{1}{2},t,0}^{T,-\frac{1}{2}}
		- H_{-\frac{1}{2},0}^{\text{VA}*} H_{-\frac{1}{2},+1,-1}^{T,-\frac{1}{2}}
		+ H_{-\frac{1}{2},-1,0}^{T,\frac{1}{2}}
		H_{-\frac{1}{2},-1}^{\text{VA}*}
		+ H_{-\frac{1}{2},-1}^{\text{VA}*}
		H_{-\frac{1}{2},t,-1}^{T,\frac{1}{2}}\biggr]
		\biggr\}\nn\\&
		+\frac{\text{$m^2_{\ell}$}}{\text{$q^2$}}\biggl\{4
		\bigl|H_{-\frac{1}{2},t}^{\text{VA}}\bigr|^2
		+\frac{4}{3}\bigl(\bigl|H_{-\frac{1}{2},0}^{\text{VA}}\bigr|^2
		+\bigl|H_{-\frac{1}{2},-1}^{\text{VA}}\bigr|^2 \bigr)\nn\\&
		+\frac{32}{3}\bigl(\bigl|H_{-\frac{1}{2},-1,0}^{T,\frac{1}{2}}\bigr|^2
		+\bigl|H_{-\frac{1}{2},+1,-1}^{T,-\frac{1}{2}}\bigr|^2
		+\bigl|H_{-\frac{1}{2},t,0}^{T,-\frac{1}{2}}\bigr|^2
		+\bigl|H_{-\frac{1}{2},t,-1}^{T,\frac{1}{2}}\bigr|^2\bigr)\nn\\&
		+\frac{64}{3} \text{Re}\bigl[H_{-\frac{1}{2},-1,0}^{T,\frac{1}{2}*} H_{-\frac{1}{2},t,-1}^{T,\frac{1}{2}}
		- H_{-\frac{1}{2},+1,-1}^{T,-\frac{1}{2}*} H_{-\frac{1}{2},t,0}^{T,-\frac{1}{2}}\bigr] \biggr\}\biggr]
	\end{align}
	%%%%%%%%%%%%%%%%%%%%%%%%%%%%%%%%%%%%%%%%%%
	\begin{align}
		C_F^{\ell}(q^2)=& \biggl(\frac{\rm d \Gamma}{\rm d q^2}\biggr)^{-1}\times \frac{3N}{4} \biggl(1-\frac{\text{$m_{\ell}$}^2}{\text{$q^2$}} \biggr)\biggl\{
		4\biggl(\bigl|H_{-\frac{1}{2},-1}^{\text{VA}}\bigr|^2
		+\bigl|H_{\frac{1}{2},+1}^{\text{VA}}\bigr|^2 \biggr)
		-8\biggl(\bigl|H_{-\frac{1}{2},0}^{\text{VA}}\bigr|^2
		+\bigl|H_{\frac{1}{2},0}^{\text{VA}}\bigr|^2 \biggr)\nn\\&
		-16\biggl(\bigl|H_{-\frac{1}{2},t,-1}^{T,\frac{1}{2}}\bigr|^2
		+\bigl|H_{\frac{1}{2},t,+1}^{T,-\frac{1}{2}}\bigr|^2
		+\bigl|H_{-\frac{1}{2},-1,0}^{T,\frac{1}{2}}\bigr|^2
		+\bigl|H_{\frac{1}{2},+1,0}^{T,-\frac{1}{2}}\bigr|^2
		\biggr)\nn\\&
		+32\biggl(\bigl|H_{-\frac{1}{2},+1,-1}^{T,-\frac{1}{2}}\bigr|^2
		+\bigl|H_{\frac{1}{2},+1,-1}^{T,\frac{1}{2}}\bigr|^2
		+\bigl|H_{-\frac{1}{2},t,0}^{T,-\frac{1}{2}}\bigr|^2
		+\bigl|H_{\frac{1}{2},t,0}^{T,\frac{1}{2}}\bigr|^2 \biggr)\nn\\&
		+32\text{Re}\bigl[H_{\frac{1}{2},+1,0}^{T,-\frac{1}{2}*} H_{\frac{1}{2},t,+1}^{T,-\frac{1}{2}}
		-H_{-\frac{1}{2},-1,0}^{T,\frac{1}{2}*} H_{-\frac{1}{2},t,-1}^{T,\frac{1}{2}}\bigr]
		-64\text{Re}\bigl[H_{-\frac{1}{2},+1,-1}^{T,-\frac{1}{2}*} H_{-\frac{1}{2},t,0}^{T,-\frac{1}{2}}
		+H_{\frac{1}{2},+1,-1}^{T,\frac{1}{2}*}H_{\frac{1}{2},t,0}^{T,\frac{1}{2}}\bigr]
		\biggr\}
	\end{align}
	%%%%%%%%%%%%%%%%%%%%%%%%%%%%%%%%%%%%%%%%%%%
	with $N=\frac{G_{F}^2 |V_{cb}|^2}{1536\pi^3} \,q^2\,\bigl(1-\frac{m_\ell ^2}{q^2}\bigr)^2\frac{\sqrt{Q_+ Q_-}}{m_{\lb}^{3}} \mathcal{B}(\Lambda_c\to \Lambda \pi)$, $Q_{\pm}$=$(m_{\lb} \pm m_{\lc})^2-q^2$ and $\mathcal{B}(\Lambda_c\to \Lambda \pi)=(1.29\pm0.05)\%$~\cite{Workman:2022ynf}
	\subsection{BGL Parametrization: Outer function in HQET basis}
	The outer functions $\phi_i(z)$ has the form-
	\begin{align}\label{eq:outerfunc}
		\phi_{F_0}=&\frac{8}{M}\sqrt{\frac{n_I}{\pi \chi^L(0)}}\frac{r^{3/2} (1+z)^2 (1-z)^2}{\bigg((1+r)(1-z)+2\sqrt{r} (1+z)\bigg)^4} \\\nonumber
		\phi_{F_1}=&\frac{16\sqrt{2}}{M}\sqrt{\frac{n_I}{3 \pi \chi^T(0)}}\frac{r^{3/2 }(1+z) (1-z)^{1/2}}{\bigg((1+r)(1-z)+2\sqrt{r} (1+z)\bigg)^4}\\\nonumber
		\phi_{H_V}=&\frac{16 r^{3/2}}{M^2}\sqrt{\frac{n_I}{3 \pi \chi^T(0)}}\frac{(1+z) (1-z)^{3/2}}{\bigg((1+r)(1-z)+2\sqrt{r} (1+z)\bigg)^5} \\\nonumber
		\phi_{G_0}=&\frac{16 r^{3/2}}{M}\sqrt{\frac{n_I}{\pi \chi^L(0)}}\frac{(1+z) (1-z)^{1/2}}{\bigg((1+r)(1-z)+2\sqrt{r} (1+z)\bigg)^4}\\\nonumber
		\phi_{G_1}=&\frac{8\sqrt{2} r^{3/2}}{M}\sqrt{\frac{n_I}{3 \pi \chi^T(0)}}\frac{(1+z)^2 (1-z)^{1/2}}{\bigg((1+r)(1-z)+2\sqrt{r} (1+z)\bigg)^4}\\\nonumber
		\phi_{H_A}=&\frac{8 r^{3/2}}{M^2}\sqrt{\frac{n_I}{3 \pi \chi^T(0)}}\frac{(1+z)^2 (1-z)^{3/2}}{\bigg((1+r)(1-z)+2\sqrt{r} (1+z)\bigg)^5}
	\end{align}
	where, $r=\frac{m_{\Lambda_{c}}}{m_{\Lambda_{b}}}$, $n_I$, is an isospin Clebsch-Gordan factor, which is 1.0 for $\Lambda_{b} \to \Lambda_{c}$ transition.\\
	%%%%%%%%%%%%%%%%%%%%%%%%%%%%%%%%%%%%%%%%%%%%%%%%%%%%%%%%%
	
	%%%%%%%%%%%%%%%%%%%%%%%%%%%%%%%
	\begin{align}
		F_0=&(M-m) f_0^{lat}\nn\\
		H_V=&(M+m) f_+^{lat}\nn\\
		F_1=&f_{\perp}^{lat}\nn\\
		G_0=&(M+m) g_0^{lat}\nn\\
		H_A=&(M-m) g_+^{lat}\nn\\
		G_1=&g_{\perp}^{lat}
	\end{align}
	Satisfy the kinematic constrain: $H_A(t_{-})=(M-m)G_1(t_{-})$,
	$(M+m) F_0(0)=(M-m)H_V(0)$, ~$(M-m) G_0(0)=(M+m)H_A(0)$.

	\begin{table}[t]
		\renewcommand{\arraystretch}{1.8}
		\centering
		\setlength\tabcolsep{8 pt}
		\begin{center}
			\resizebox{0.9\textwidth}{!}{
				\begin{tabular}{|*{7}{c|}}
					\hline
					Scenarios &\multicolumn{6}{c|}{Observables/Predictions} \\
					\cline{2-7}
					&~~~~$\text{$R(D)$}$~~~~&~~$\text{$ R(D^*)$}$~~&$\text{$ R(\Lambda_c)$}$ &$ F_L^{D^*}$&$ P_{\tau}^{D^*}$&$ P_{\tau}^D$ \\ \hline
					\text{$ Re[C_{S_1}]$,$ Re[C_{S_2}]$} & $\text{0.354(29)}$  &  $\text{0.275(19)}$  &  $\text{0.354(19)}$  \
					&  $\text{0.448(30)}$  &  $\text{-0.463(77)}$  &  $\text{0.421(47)}$  \\
					%&\textbf{Range}&$\{0.383,0.325\}$  &  $\{0.294,0.255\}$\&$\{0.373,0.336\}$  &  $\{0.478,0.418\}$&&\\
					\hline
					\text{$ Re[C_{S_1}]$,$ Re[C_{T}]$} & $\text{0.355(29)}$  &  $\text{0.272(20)}$  &  $\text{0.361(24)}$  \
					&  $\text{0.429(11)}$  &  $\text{-0.496(14)}$  &  $\text{0.435(43)}$  \\
					%&\textbf{Range}&$\{0.385,0.326\}$  &  $\{0.292,0.252\}$\&$\{0.385,0.337\}$  &  $\{0.441,0.418\}$&&\\
					\hline
					\text{$ Re[C_{S_2}]$,$ Re[C_{T}]$}  & $\text{0.359(28)}$  &  $\text{0.275(18)}$  &  $\text{0.318(19)}$  \
					&  $\text{0.407(44)}$  &  $\text{-0.076(38)}$  &  $\text{0.196(81)}$ \\
					%&\textbf{Range}&$\{0.386,0.331\}$  &  $\{0.294,0.257\}$\&$\{0.337,0.298\}$  &  $\{0.451,0.362\}$&&\\
					\hline
					\text{$ Re[C_{V_1}]$,$ Re[C_{T}]$} &  $\text{0.363(29)}$  &  $\text{0.273(20)}$  &  $\text{0.345(24)}$  \
					&  $\text{0.430(9)}$  &  $\text{-0.532(10)}$  &  $\text{0.306(18)}$  \\
					%&\textbf{Range}&$\{0.392,0.334\}$  &  $\{0.293,0.253\}$\&$\{0.369,0.322\}$  &  $\{0.439,0.421\}$&&\\
					\hline
					\text{$ Re[ C_{V_2}]$, $ Re[C_{T}]$}  &   $\text{0.339(29)}$  &  $\text{0.271(20)}$  &  $\text{0.374(27)}$  \
					&  $\text{0.402(17)}$  &  $\text{-0.474(28)}$  &  $\text{0.354(14)}$ \\
					%&\textbf{Range}&$\{0.368,0.310\}$  &  $\{0.291,0.251\}$\&$\{0.401,0.347\}$&$\{0.419,0.384\}$&&\\
					\hline
					\text{$ Re[C_{V_1}]$,$ Re[C_{V_2}]$} &  $\text{0.352(29)}$  &  $\text{0.277(20)}$  &  $\text{0.360(18)}$  \
					&  $\text{0.429(9)}$  &  $\text{-0.520(7)}$  &  $\text{0.324(3)}$\\
					%&\textbf{Range}&$\{0.382,0.323\}$  &  $\{0.297,0.256\}$\&$\{0.378,0.341\}$  &  $\{0.439,0.420\}$&&\\
					\hline
					\text{$ Re[C_{S_1}]$,$ Re[C_{V_1}]$}  &  $\text{0.353(29)}$  &  $\text{0.277(20)}$  &  $\text{0.359(19)}$  \
					&  $\text{0.430(10)}$  &  $\text{-0.511(13)}$  &  $\text{0.378(76)}$  \\
					%&\textbf{Range}&$\{0.382,0.324\}$  &  $\{0.297,0.256\}$\&$\{0.378,0.341\}$  &  $\{0.440,0.420\}$&&\\
					\hline
					\text{$ Re[C_{S_1}]$,$ Re[C_{V_2}]$} & $\text{0.353(29)}$  &  $\text{0.277(20)}$  &  $\text{0.359(18)}$  \
					&  $\text{0.437(10)}$  &  $\text{-0.499(11)}$  &  $\text{0.455(47)}$ \\
					%&\textbf{Range}&$\{0.382,0.323\}$  &  $\{0.297,0.256\}$\&$\{0.377,0.341\}$&$\{0.447,0.427\}$&&\\
					\hline
					\text{$ Re[C_{S_2}]$,$ Re[C_{V_1}]$}  & $\text{0.353(29)}$  &  $\text{0.277(20)}$  &  $\text{0.359(19)}$  \
					&$\text{0.424(10)}$  &  $\text{-0.526(12)}$  &  $\text{0.373(68)}$ \\
					%&\textbf{Range}&$\{0.382,0.324\}$&$\{0.297,0.256\}$\&$\{0.378,0.341\}$&$\{0.434,0.415\}$&&\\
					\hline
					\text{$ Re[C_{S_2}]$,$ Re[C_{V_2}]$} & $\text{0.353(29)}$  &  $\text{0.277(20)}$  &  $\text{0.358(18)}$  \
					&  $\text{0.423(9)}$  &  $\text{-0.539(9)}$  &  $\text{0.472(52)}$ \\
					% &\textbf{Range}&$\{0.382,0.324\}$  &  $\{0.298,0.257\}$&$\{0.377,0.340\}$  &  $\{0.433,0.414\}$&&\\
					\hline
					\hline 
					\text{Measurement}&$\textbf{0.357(29)}$~~~&~~~$\textbf{0.284(13)}$~~~  \
					& $\textbf{0.242(76)}$&$\textbf{0.430(70)}$ &\textbf{-0.38(54)}&\text{N.A}\\ 
					% &\textbf{Range}&$\{0.386,0.328\}$  &  $\{0.296,0.272\}$  &  $\{0.318,0.166\}$\&  $\{0.5,0.36\}$&&\\ 
					\hline 
				\end{tabular}
			}
			\caption{Predictions of $R(D)$, $R(D^*)$, $R(\Lambda_{c})$, $F_L^{D^*}$ in the two operator scenarios considering different new physics in $\tau$ channel.}
			\label{tab:predmain_twoopr}
		\end{center}
	\end{table}		

	%%%%%%%%%%%%%%%%%%%%%%%%%%%%%%%%%%%%%%%%%%%%%%%%%%%%%%%%%%%%%%	
	\subsection{Transformation: Helicity basis to Transversality basis}
 \label{Appndx:litreview}
	As in Ref.~\cite{Boer:2019zmp}, the transformations are not explicitly provided for how the helicity amplitudes are transform into transversality amplitudes, especially for the tensors which are non-trivial. We assume the transformation follows the form:
	\begin{eqnarray}
		A^T_{\perp,0}&=\frac{1}{\sqrt{2}}(H^{T,1/2}_{1/2,t,0} +H^{T,1/2}_{1/2,-1,+1}+H^{T,-1/2}_{-1/2,-1,1} +H^{T,-1/2}_{-1/2,t,0})\\   
		A^T_{\para,0}&=\frac{1}{\sqrt{2}}(H^{T,-1/2}_{-1/2,-1,+1}+H^{T,-1/2}_{-1/2,t,0}-(H^{T,1/2}_{1/2,t,0}+H^{T,1/2}_{1/2,-1,1}))\\
		A^T_{\para,1}&=\frac{1}{\sqrt{2}}(H^{T,1/2}_{-1/2,-1,0}+H^{T,1/2}_{-1/2,t,-1}-(H^{T,-1/2}_{1/2,0,1}+H^{T,-1/2}_{1/2,t,1}))\\
		A^T_{\perp,1}&=\frac{1}{\sqrt{2}}(H^{T,1/2}_{-1/2,-1,0}+H^{T,1/2}_{-1/2,t,-1}+(H^{T,-1/2}_{1/2,0,1}+H^{T,-1/2}_{1/2,t,1}))
	\end{eqnarray} 
	% \begin{eqnarray}
	%A^T_{\perp,0}&=\frac{1}{\sqrt{2}}(H^{T,1/2}_{1/2,-1,+1}+H^{T,1/2}_{1/2,t,0})\\
	%A^T_{\para,0}&=\frac{1}{\sqrt{2}}(H^{T,-1/2}_{-1/2,-1,+1}+H^{T,-1/2}_{-1/2,t,0})\\
	%A^T_{\para,1}&=\frac{1}{\sqrt{2}}(H^{T,1/2}_{-1/2,-1,0}+H^{T,1/2}_{-1/2,t,-1})\\
	%A^T_{\perp,1}&=\frac{1}{\sqrt{2}}(H^{T,-1/2}_{1/2,0,+1}+H^{T,-1/2}_{1/2,t,+1})
	%   \end{eqnarray}
	and for the vector amplitudes the transformation looks like
	\begin{eqnarray}
		A_{\perp,1}&=&\frac{1}{\sqrt{2}}(H^{VA}_{1/2,1}-H^{VA}_{-1/2,-1})\\
		A_{\para,1}&=&\frac{1}{\sqrt{2}}(H^{VA}_{1/2,1}+H^{VA}_{-1/2,-1})\\
		A_{\perp,0}&=&\frac{1}{\sqrt{2}}(H^{VA}_{1/2,0}-H^{VA}_{-1/2,0})\\
		A_{\para,0}&=&\frac{1}{\sqrt{2}}(H^{VA}_{1/2,0}+H^{VA}_{-1/2,0})\\
		A_{\perp,t}&=&\frac{1}{\sqrt{2}}\bigl\{(H^{VA}_{1/2,t}-H^{VA}_{-1/2,t})+\frac{\sqrt{q^2}}{m_l}(H^{SP}_{1/2,0}-H^{SP}_{-1/2,0})\bigr\}\\
		A_{\para,t}&=&\frac{1}{\sqrt{2}}\bigl\{(H^{VA}_{1/2,t}+H^{VA}_{-1/2,t})+\frac{\sqrt{q^2}}{m_l}(H^{SP}_{1/2,0}+H^{SP}_{-1/2,0})\bigr\}.
	\end{eqnarray}
	
	\subsection{Theory Inputs}
	\label{Appndx:LFUVtheoexp}
	These Analytic expressions have been taken from \cite{Ray:2023xjn} for our NP analysis.
	\begin{align}
		R(D)=&(0.304\pm 0.003)\times\big(1+1.35 C_{S_1}^2+C_{S_1}(2.70C_{S_2}+1.72 C_{V_1}+1.72 C_{V_2}+1.72)\nn\\ &+1.35 C_{S_2}^2 + 0.83 C_{T} C_{V_1} + C_{S_2}(1.72 C_{V_1}+1.72 C_{V_2}+1.72)+0.83 C_T C_{V_2}\nn \\ &+(0.49 C_T+ 0.83)C_T +C_{V_1}^2  + 2.00 C_{V_1} C_{V_2} +2.00 C_{V_1}+C_{V_2}^2+ 2.00 C_{V_2}\big)\\
		R(D^*)=&(0.258\pm 0.012)\times\big(1+0.04 C_{S_1}^2+C_{S_1}(-0.07C_{S_2}+0.10 C_{V_1}-0.10 C_{V_2}+0.10)\nn\\ &+0.04 C_{S_2}^2 + C_{S_2}(-0.10 C_{V_1}+ 0.10 C_{V_2}-0.10)-2.94 C_T C_{V_1}+4.79 C_T C_{V_2}\nn\\ &+(10.65 C_T- 2.94)C_T+C_{V_1}^2 -1.79 C_{V_1} C_{V_2}+2.00 C_{V_1}+C_{V_2}^2- 1.79 C_{V_2}\big)\\
		F_L^{D^*}=&(0.427\pm0.009)\times\bigl(27.26+2.34C_{S_1}^2+C_{S_1}(-4.68 C_{S_2}+6.65 C_{V_1}-6.65 C_{V_2}+6.65)\nn\\&+2.34 C_{S_2}^2-66.82 C_T C_{V_1}
		+C_{S_2}(-6.65 C_{V_1}+6.65 C_{V_2}-6.65)+66.82 C_T C_{V_2}\nn\\&+C_T(69.62 C_T-66.82)+27.26 C_{V_1}^2-54.52 C_{V_1} C_{V_2}+54.52 C_{V_1}\nn\\&+27.26 C_{V_2}^2-54.26 C_{V_2}\bigr)/
		\bigl(27.26+C_{S_1}^2+C_{S_1}(-2.00 C_{S_2}+2.84 C_{V_1}-2.84 C_{V_2}+2.84)\nn\\&+ C_{S_2}^2 
		+C_{S_2}(-2.84C_{V_1}+2.84 C_{V_2}-2.84)-80.15 C_T C_{V_1}+130.72 C_T C_{V_2}\nn\\& +C_T(290.34 C_T-80.15)+27.26 C_{V_1}^2-48.81 C_{V_1} C_{V_2}+54.52 C_{V_1}+27.26 C_{V_2}^2-48.81 C_{V_2}\bigr)
	\end{align}
	
%%%%%%%%%%%%%%%%%%%%%%%%%%%%%%%%%%%%%%%%%%%%%%%%%%%%%%%%%%%%%%%%%%%%%
\subsection{Observables bin predictions: SM, NP with both One and Two parameter scenario}\label{sec:Ang_bin_pred}
%%%%%%%%%%%%%%%%%%%%%%%%%%%%%%%%%%%%%%%%%%%%%%%%%%%%%%%%%%%%%%%%%%%%%%%%%%%%%%%%%%%%%%%%%%
\begin{table}
	\centering
	\resizebox{0.78\textwidth}{!}{
		\begin{tabular}{|*{7}{c|}}
			\hline\hline
			\multirow{2}{*}{\bf Observables}  & \multicolumn{2}{|c|}{ \bf Scenario } & \multicolumn{4}{|c|}{\bf $\bf q^2$ bin(in \text{$\bf GeV^2$})} \\
			\cline{4-7}
			& \multicolumn{2}{|c|}{\bf (Real $C_i$'s)} &\textbf{$\bf q^2_{min}$-5}  &  \textbf{5-7}  &  $\textbf{7-9}$  &  \textbf{9- $ \bf q^2_{max}$ } \\
			\hline\hline
			& \multicolumn{2}{c|}{\bf SM ($Re(C_i) = 0$)} &  $\text{0.318(1)}$  &  $\text{0.302(2)}$  &  $\text{0.305(2)}$  &  $\text{0.322(1)}$ \\
			\cline{2-7}
			$\hat{K}_{1cc}$ & \multirow{2}{*}{$C_{S_1}$}&  $\text{[0.1]}$& $\text{0.319(1)}$  &  $\text{0.304(2)}$  &  
			$\text{0.307(2)}$  &  $\text{0.322(1)}$   \\
			\cline{3-7}
			&&  $\text{[-0.1]}$  &  $\text{0.317(1)}$  &  $\text{0.300(2)}$  &  $\text{0.304(2)}$  &  $\text{0.321(1)}$  
			\\
			\cline{2-7}
			&$C_{S_2}$  &  $\text{[0.1]}$  &  $\text{0.318(1)}$  &  $\text{0.303(2)}$  &  $\text{0.307(2)}$  &  
			$\text{0.322(1)}$   \\
			\cline{3-7}
			&&  $\text{[-0.1]}$  &  $\text{0.318(1)}$  &  $\text{0.301(2)}$  &  $\text{0.304(2)}$  &  $\text{0.321(1)}$    \\
			\cline{2-7}
			&$C_{V_1}$  &  $\text{[0.1]}$  &  $\text{0.318(1)}$  &  $\text{0.302(2)}$  &  $\text{0.305(2)}$  &  $\text{0.322(1)}$    \\
			\cline{3-7}
			&&$\text{[-0.1]}$  &  $\text{0.318(1)}$  &  $\text{0.302(2)}$  &  $\text{0.305(2)}$  &  $\text{0.322(1)}$    \\
			\cline{2-7}
			&$C_{V_2}$&  $\text{[0.1]}$  &  $\text{0.318(1)}$  &  $\text{0.300(2)}$  &  $\text{0.303(2)}$  &  $\text{0.320(1)}$   \\
			\cline{3-7}
			&&  $\text{[-0.1]}$  &  $\text{0.318(1)}$  &  $\text{0.304(2)}$  &  $\text{0.308(2)}$  &  $\text{0.323(1)}$   \\
			\cline{2-7}
			&$C_T$  &  $\text{[0.1]}$  &  $\text{0.316(1)}$  &  $\text{0.294(2)}$  &  $\text{0.294(2)}$  &  $\textbf{0.314(1)}$    \\
			\cline{3-7}
			&& $\text{[-0.1]}$  &  $\text{0.320(1)}$  &  $\text{0.307(2)}$  &  $\text{0.311(1)}$  &  $\text{0.325(1)}$    \\
			\cline{2-7}
			&[$C_{S_1},C_{S_2}$]  & $\text{[-2.27(21), 0.90(22)]}$&  $\textbf{0.310(1)}$  &  $\text{0.297(3)}$  &  $\text{0.308(2)}$  &  $\text{0.324(1)}$    \\
			\cline{2-7}
			&[$C_{V_1},C_T$] &$\text{[0.08(3), 0.04(4)]}$ &  $\text{0.317(1)}$  &  $\text{0.299(3)}$  &  $\text{0.302(4)}$  &  $\text{0.320(2)}$   \\
			\cline{2-7}
			&[$C_{V_2},C_T$] &$\text{[0.08(5), -0.06(3)]}$ &  $\text{0.319(1)}$  &  $\text{0.303(2)}$  &  $\text{0.307(2)}$  &  $\text{0.323(1)}$   \\
			\cline{2-7}
			&[$C_{V_1},C_{V_2}$] &$\text{[-0.98(3), 1.05(2)]}$ &  $\text{0.318(1)}$  &  $\text{0.302(2)}$  &  $\text{0.305(2)}$  &  $\text{0.321(1)}$    \\
			\cline{2-7}
			&[$C_{S_1},C_{V_1}$] &$\text{[0.05(7), 0.03(3)]}$  &  $\text{0.318(1)}$  &  $\text{0.303(2)}$  &  $\text{0.306(2)}$  &  $\text{0.322(1)}$    \\
			\cline{2-7}
			&[$C_{S_1},C_{V_2}$] & $\text{[0.12(5), -0.03(3)]}$ &  $\text{0.319(9)}$  &  $\text{0.304(2)}$  &  $\text{0.308(2)}$  &  $\text{0.323(1)}$    \\
			\cline{2-7}
			&[$C_{S_2},C_{V_1}$] & $\text{[0.04(7), 0.04(3)]}$ &  $\text{0.318(1)}$  &  $\text{0.302(2)}$  &  $\text{0.306(2)}$  &  $\text{0.322(1)}$    \\
			\cline{2-7}
			&[$C_{S_2},C_{V_2}$] &$\text{[0.14(5), -0.05(4)]}$ &  $\text{0.319(1)}$  &  $\text{0.304(2)}$  &  $\text{0.308(2)}$  &  $\text{0.323(1)}$    \\
			\cline{2-7}
			&[$C_{S_1},C_T$] &$\text{[0.10(5), -0.01(2)]}$ &  $\text{0.319(1)}$  &  $\text{0.304(2)}$  &  $\text{0.308(2)}$  &  $\text{0.323(1)}$    \\
			\cline{2-7}
			&[$C_{S_2},C_T$] &$\text{[-1.25(6), 0.23(3)]}$ &  $\textbf{0.313(1)}$  &  $\textbf{0.287(3)}$  &  $\textbf{0.286(5)}$  &  $\textbf{0.308(4)}$    \\
			\hline\hline
		\end{tabular}
	}
	\caption{Bin prediction for $\hat{K}_{1cc}$ observable.}
	\label{tab:obsbinprdall1}
\end{table}
%%%%%%%%%%%%%%%%%%%%%%%%%%%%%%%%%%%%%%%%%%%
\begin{table}
	\centering
	\resizebox{0.65\textwidth}{!}{
		\begin{tabular}{|*{7}{c|}}
			\hline\hline
			\multirow{2}{*}{\bf Observables}  & \multicolumn{2}{|c|}{ \bf Scenario } & \multicolumn{4}{|c|}{\bf $\bf q^2$ bin(in \text{$\bf GeV^2$})} \\
			\cline{4-7}
			& \multicolumn{2}{|c|}{\bf (Real $C_i$'s)} &\textbf{$\bf q^2_{min}$-5}  &  \textbf{5-7}  &  $\textbf{7-9}$  &  \textbf{9- $ \bf q^2_{max}$ } \\
			\hline\hline
			&  \multicolumn{2}{c|}{\bf SM ($Re(C_i) = 0$)} &  $\text{0.25(3)}$  &  $\text{0.23(2)}$  &  $\text{0.20(2)}$  &  $\text{0.13(1)}$  \\
			\cline{2-7}
			$\hat{K}_{2cc}$  &  \multirow{2}{*}{$C_{S_1}$} &  $\text{[0.1]}$  &  $\text{0.25(3)}$  &  $\text{0.23(2)}$  &  $\text{0.20(2)}$  &  $\text{0.13(1)}$    \\
			&&  $\text{[-0.1]}$  &  $\text{0.25(3)}$  &  $\text{0.22(2)}$  &  $\text{0.20(2)}$  &  $\text{0.13(1)}$   \\
			\cline{2-7}
			&  \multirow{2}{*}{$C_{S_2}$}  &  $\text{[0.1]}$  &  $\text{0.25(3)}$  &  $\text{0.22(2)}$  &  $\text{0.19(2)}$  &  $\text{0.12(1)}$   \\
			&&  $\text{[-0.1]}$  &  $\text{0.26(3)}$  &  $\text{0.23(2)}$  &  $\text{0.20(2)}$  &  $\text{0.13(1)}$   \\
			\cline{2-7}
			&  \multirow{2}{*}{$C_{V_1}$}  &  $\text{[0.1]}$  &  $\text{0.25(3)}$  &  $\text{0.23(2)}$  &  $\text{0.20(2)}$  &  $\text{0.13(1)}$   \\
			\cline{3-7}
			&&  $\text{[-0.1]}$  &  $\text{0.25(3)}$  &  $\text{0.23(2)}$  &  $\text{0.20(2)}$  &  $\text{0.13(1)}$    \\
			\cline{2-7}
			&  \multirow{2}{*}{$C_{V_2}$}  &  $\text{[0.1]}$  &  $\text{0.25(3)}$  &  $\text{0.23(2)}$  &  $\text{0.21(2)}$  &  $\text{0.14(2)}$   \\
			\cline{3-7}
			&&  $\text{[-0.1]}$  &  $\text{0.25(3)}$  &  $\text{0.21(2)}$  &  $\text{0.18(2)}$  &  $\text{0.11(1)}$   \\
			\cline{2-7}
			&  \multirow{2}{*}{$C_{T}$} &  $\text{[0.1]}$  &  $\text{0.23(3)}$  &  $\text{0.23(3)}$  &  $\text{0.23(2)}$  &  $\text{0.17(2)}$   \\
			&&  $\text{[-0.1]}$  &  $\text{0.17(2)}$  &  $\text{0.15(2)}$  &  $\text{0.12(1)}$  &  $\text{0.08(1)}$    \\
			\cline{2-7}
			&[$C_{S_1},C_{S_2}$]  &$\text{[-2.27(21), 0.90(22)]}$&  $\text{0.24(3)}$  &  $\text{0.22(2)}$  &  $\text{0.21(2)}$  &  $\text{0.15(2)}$    \\
			\cline{2-7}
			&[$C_{V_1},C_T$] &$\text{[0.08(3), 0.04(4)]}$ &  $\text{0.26(3)}$  &  $\text{0.24(3)}$  &  $\text{0.22(3)}$  &  $\text{0.15(2)}$    \\
			\cline{2-7}
			&[$C_{V_2},C_T$] &$\text{[0.08(5), -0.06(3)]}$ &  $\text{0.22(3)}$  &  $\text{0.19(3)}$  &  $\text{0.17(2)}$  &  $\text{0.11(2)}$   \\
			\cline{2-7}
			&[$C_{V_1},C_{V_2}$] &$\text{[-0.98(3), 1.05(2)]}$ &  $\textbf{-0.26(3)}$  &  $\textbf{-0.23(2)}$  &  $\textbf{-0.20(2)}$  &  $\textbf{-0.13(1)}$   \\
			\cline{2-7}
			&[$C_{S_1},C_{V_1}$] &$\text{[0.05(7), 0.03(3)]}$  &  $\text{0.25(3)}$  &  $\text{0.23(2)}$  &  $\text{0.20(2)}$  &  $\text{0.13(1)}$   \\
			\cline{2-7}
			&[$C_{S_1},C_{V_2}$] & $\text{[0.12(5), -0.03(3)]}$ &  $\text{0.25(3)}$  &  $\text{0.22(2)}$  &  $\text{0.19(2)}$  &  $\text{0.12(1)}$    \\
			\cline{2-7}
			&[$C_{S_2},C_{V_1}$] & $\text{[0.04(7), 0.04(3)]}$ &  $\text{0.25(3)}$  &  $\text{0.22(2)}$  &  $\text{0.19(2)}$  &  $\text{0.13(1)}$    \\
			\cline{2-7}
			&[$C_{S_2},C_{V_2}$] &$\text{[0.14(5), -0.05(4)]}$ &  $\text{0.25(3)}$  &  $\text{0.22(2)}$  &  $\text{0.18(2)}$  &  $\text{0.12(1)}$    \\
			\cline{2-7}
			&[$C_{S_1},C_T$] &$\text{[0.10(5), -0.01(2)]}$ &  $\text{0.25(3)}$  &  $\text{0.22(3)}$  &  $\text{0.19(2)}$  &  $\text{0.12(2)}$   \\
			\cline{2-7}
			&[$C_{S_2},C_T$] &$\text{[-1.25(6), 0.23(3)]}$ &  $\textbf{0.01(4)}$  &  $\textbf{-0.02(3)}$  &  $\textbf{-0.04(3)}$  &  $\textbf{-0.04(2)}$    \\
			\hline
		\end{tabular}
	}
	\caption{Bin prediction for $\hat{K}_{2cc}$ observable.}
	\label{tab:obsbinprdall2}
\end{table}
%%%%%%%%%%%%%%%%%%%%%%%%%%%%%%%%%%%%%%%%%%%%%		
%%%%%%%%%%%%%%%%%%%%%%%%%%%%%%%%%%%%%%%%%%%%%		
\begin{table}[t]
	\centering
	\resizebox{0.65\textwidth}{!}{
		\begin{tabular}{|*{7}{c|}}
			\hline\hline
			\multirow{2}{*}{\bf Observables}  & \multicolumn{2}{|c|}{ \bf Scenario } & \multicolumn{4}{|c|}{\bf $\bf q^2$ bin(in \text{$\bf GeV^2$})} \\
			\cline{4-7}
			& \multicolumn{2}{|c|}{\bf (Real $C_i$'s)} &\textbf{$\bf q^2_{min}$-5}  &  \textbf{5-7}  &  $\textbf{7-9}$  &  \textbf{9- $ \bf q^2_{max}$ } \\
			\hline\hline
			& \multicolumn{2}{c|}{\bf SM ($Re(C_i) = 0$)}  &  $\text{0.3410(4)}$  &  $\text{0.3490(10)}$  &  $\text{0.3473(10)}$  \
			&  $\text{0.3392(5)}$ \\
			\cline{2-7}
			$\hat{K}_{1ss}$  &  \multirow{2}{*}{$C_{S_1}$} &  $\text{[0.1]}$  &  $\text{0.3407(4)}$  &  $\text{0.3482(9)}$  &  
			$\text{0.3465(9)}$  \  &  $\text{0.3389(4)}$  \\
			&& $\text{[-0.1]}$  &  $\text{0.3414(4)}$  &  $\text{0.3498(10)}$  &  $\text{0.3480(10)}$  &  $\text{0.3395(5)}$   \\
			\cline{2-7}
			& \multirow{2}{*}{$C_{S_2}$}  &  $\text{[0.1]}$  &  $\text{0.3408(4)}$  &  $\text{0.3485(9)}$  &  $\text{0.3467(9)}$  
			&  $\text{0.3389(4)}$    \\
			&&  $\text{[-0.1]}$  &  $\text{0.3412(4)}$  &  $\text{0.3495(10)}$  &  $\text{0.3478(10)}$  &  $\text{0.3395(5)}$    \\
			\cline{2-7}
			&  \multirow{2}{*}{$C_{V_1}$} &  $\text{[0.1]}$  &  $\text{0.3410(4)}$  &  $\text{0.3490(10)}$  &  $\text{0.3473(10)}$ &  $\text{0.3392(5)}$   \\
			&&  $\text{[-0.1]}$  &  $\text{0.3410(4)}$  &  $\text{0.3490(10)}$  &  $\text{0.3473(10)}$  \
			&  $\text{0.3392(5)}$   \\
			\cline{2-7}
			& \multirow{2}{*}{$C_{V_2}$} &  $\text{[0.1]}$  &  $\text{0.3412(4)}$  &  $\text{0.3499(9)}$  &  $\text{0.3486(9)}$
			&  $\text{0.3401(5)}$   \\
			&&  $\text{[-0.1]}$  &  $\text{0.3408(4)}$  &  $\text{0.3482(10)}$  &  $\text{0.3461(10)}$  &  $\text{0.3385(5)}$   \\
			\cline{2-7}
			&  \multirow{2}{*}{$C_{T}$}  &  $\text{[0.1]}$ &  $\text{0.3422(4)}$  &  $\text{0.3532(11)}$  &  $\text{0.3530(10)}$  &  $\text{0.3430(7)}$   \\
			&&  $\text{[-0.1]}$  &  $\text{0.3402(4)}$  &  $\text{0.3466(8)}$  &  $\text{0.3445(7)}$  \
			&  $\text{0.3377(2)}$    \\
			\cline{2-7}
			&[$C_{S_1},C_{S_2}$]  &$\text{[-2.27(21), 0.90(22)]}$&  $\textbf{0.3452(7)}$  &  $\text{0.3514(14)}$  &  $\text{0.3459(11)}$  &  $\text{0.3379(4)}$    \\
			\cline{2-7}
			&[$C_{V_1},C_T$] &$\text{[0.08(3), 0.04(4)]}$ &  $\text{0.3414(5)}$  &  $\text{0.3504(17)}$  &  $\text{0.3489(21)}$  &  $\text{0.3402(11)}$   \\
			\cline{2-7}
			&[$C_{V_2},C_T$] &$\text{[0.08(5), -0.06(3)]}$ &  $\text{0.3408(4)}$  &  $\text{0.3483(9)}$  &  $\text{0.3464(9)}$  \
			&  $\text{0.3387(4)}$    \\  
			\cline{2-7}
			&[$C_{V_1},C_{V_2}$] & $\text{[-0.98(3), 1.05(2)]}$ &  $\text{0.3411(4)}$  &  $\text{0.3492(10)}$  &  $\text{0.3476(10)}$  \
			&  $\text{0.3394(5)}$    \\
			\cline{2-7}
			&[$C_{S_1},C_{V_1}$] & $\text{[0.05(7), 0.03(3)]}$  &  $\text{0.3408(5)}$  &  $\text{0.3487(11)}$  &  $\text{0.3469(11)}$  &  $\text{0.3391(5)}$    \\
			\cline{2-7}
			&[$C_{S_1},C_{V_2}$] & $\text{[0.12(5), -0.03(3)]}$ &  $\text{0.3405(4)}$  &  $\text{0.3478(10)}$  &  $\text{0.3460(11)}$  &  $\text{0.3386(5)}$   \\
			\cline{2-7}
			&[$C_{S_2},C_{V_1}$] & $\text{[0.04(7), 0.04(3)]}$ &  $\text{0.3409(4)}$  &  $\text{0.3488(10)}$  &  $\text{0.3471(11)}$  &  $\text{0.3391(5)}$    \\
			\cline{2-7}
			&[$C_{S_2},C_{V_2}$] &$\text{[0.14(5), -0.05(4)]}$ &  $\text{0.3407(4)}$  &  $\text{0.3480(10)}$  &  $\text{0.3460(10)}$  &  $\text{0.3385(5)}$    \\
			\cline{2-7}
			&[$C_{S_1},C_T$] &$\text{[0.10(5), -0.01(2)]}$ &  $\text{0.3405(4)}$  &  $\text{0.3478(11)}$  &  $\text{0.3460(11)}$  &  $\text{0.3386(5)}$   \\
			\cline{2-7}
			&[$C_{S_2},C_T$] &$\text{[-1.25(6), 0.23(3)]}$ &  $\textbf{0.3436(5)}$  &  $\textbf{0.3565(15)}$  &  $\textbf{0.3569(23)}$  &  $\textbf{0.3459(20)}$   \\
			\hline
		\end{tabular}
	}
	\caption{Bin prediction for $\hat{K}_{1ss}$ observable.}
	\label{tab:obsbinprdall3}
\end{table}
%%%%%%%%%%%%%%%%%%%%%%%%%%%%%%%%%%%%%%%%%%%
%%%%%%%%%%%%%%%%%%%%%%%%%%%%%%%%%%%%%%%%%%%
\begin{table}
	\centering
	\resizebox{0.65\textwidth}{!}{
		\begin{tabular}{|*{7}{c|}}
			\hline\hline
			\multirow{2}{*}{\bf Observables}  & \multicolumn{2}{|c|}{ \bf Scenario } & \multicolumn{4}{|c|}{\bf $\bf q^2$ bin(in \text{$\bf GeV^2$})} \\
			\cline{4-7}
			& \multicolumn{2}{|c|}{\bf (Real $C_i$'s)} &\textbf{$\bf q^2_{min}$-5}  &  \textbf{5-7}  &  $\textbf{7-9}$  & \textbf{9- $ \bf q^2_{max}$ } \\
			\hline\hline
			&  \multicolumn{2}{c|}{\bf SM ($Re(C_i) = 0$)}  &  $\text{0.28(3)}$  &  $\text{0.27(3)}$  &  $\text{0.24(3)}$  &  $\text{0.15(2)}$   \\
			\cline{2-7}
			$\hat{K}_{2ss}$  &  \multirow{2}{*}{$C_{S_1}$}  &  $\text{[0.1]}$  &  $\text{0.27(3)}$  &  $\text{0.27(3)}$  &  $\text{0.24(3)}$  &  $\text{0.15(2)}$   \\
			&&  $\text{[-0.1]}$  &  $\text{0.28(3)}$  &  $\text{0.27(3)}$  &  $\text{0.24(3)}$  &  $\text{0.15(2)}$    \\
			\cline{2-7}
			& \multirow{2}{*}{$C_{S_2}$}  &  $\text{[0.1]}$  &  $\text{0.27(3)}$  &  $\text{0.26(3)}$  &  $\text{0.23(3)}$  &  $\text{0.15(2)}$    \\
			&&  $\text{[-0.1]}$  &  $\text{0.28(3)}$  &  $\text{0.27(3)}$  &  $\text{0.24(3)}$  &  $\text{0.16(2)}$   \\
			\cline{2-7}
			&  \multirow{2}{*}{$C_{V_1}$} &  $\text{[0.1]}$  &  $\text{0.28(3)}$  &  $\text{0.27(3)}$  &  $\text{0.24(3)}$  &  $\text{0.15(2)}$   \\
			&&  $\text{[-0.1]}$  &  $\text{0.28(3)}$  &  $\text{0.27(3)}$  &  $\text{0.24(3)}$  &  $\text{0.15(2)}$   \\
			\cline{2-7}
			&  \multirow{2}{*}{$C_{V_2}$} &  $\text{[0.1]}$  &  $\text{0.27(3)}$  &  $\text{0.27(3)}$  &  $\text{0.25(3)}$  &  $\text{0.17(2)}$   \\
			&&  $\text{[-0.1]}$  & $\text{0.27(3)}$  &  $\text{0.26(3)}$  &  $\text{0.22(2)}$  &  $\text{0.14(1)}$  \\
			\cline{2-7}
			& \multirow{2}{*}{$C_{T}$} &  $\text{[0.1]}$  &  $\text{0.25(3)}$  &  $\text{0.27(3)}$  &  $\text{0.27(3)}$  &  $\text{0.21(2)}$   \\
			&  &  $\text{[-0.1]}$  &  $\text{0.18(2)}$  &  $\text{0.17(2)}$  &  $\text{0.15(2)}$  &  $\text{0.09(1)}$   \\
			\cline{2-7}
			&[$C_{S_1},C_{S_2}$]  &$\text{[-2.27(21), 0.90(22)]}$&  $\text{0.27(3)}$  &  $\text{0.27(3)}$  &  $\text{0.25(3)}$  &  $\text{0.17(2)}$   \\
			\cline{2-7}
			&[$C_{V_1},C_T$] &$\text{[0.08(3), 0.04(4)]}$ &  $\text{0.28(3)}$  &  $\text{0.29(3)}$  &  $\text{0.26(3)}$  &  $\text{0.18(3)}$   \\
			\cline{2-7}
			&[$C_{V_2},C_T$] &$\text{[0.08(5), -0.06(3)]}$ &  $\text{0.24(3)}$  &  $\text{0.23(3)}$  &  $\text{0.20(3)}$  &  $\text{0.13(2)}$  \\
			\cline{2-7}
			&[$C_{V_1},C_{V_2}$] &$\text{[-0.98(3), 1.05(2)]}$ &  $\textbf{-0.28(3)}$  &  $\textbf{-0.27(3)}$  &  $\textbf{-0.24(3)}$  &  $\textbf{-0.16(2)}$    \\
			\cline{2-7}
			&[$C_{S_1},C_{V_1}$] &$\text{[0.05(7), 0.03(3)]}$  & $\text{0.27(3)}$  &  $\text{0.27(3)}$  &  $\text{0.24(3)}$  &  $\text{0.15(2)}$   \\
			\cline{2-7}
			&[$C_{S_1},C_{V_2}$] & $\text{[0.12(5), -0.03(3)]}$ & $\text{0.27(3)}$  &  $\text{0.26(3)}$  &  $\text{0.23(3)}$  &  $\text{0.15(2)}$    \\
			\cline{2-7}
			&[$C_{S_2},C_{V_1}$] & $\text{[0.04(7), 0.04(3)]}$ &  $\text{0.27(3)}$  &  $\text{0.27(3)}$  &  $\text{0.24(3)}$  &  $\text{0.15(2)}$   \\
			\cline{2-7}
			&[$C_{S_2},C_{V_2}$] &$\text{[0.14(5), -0.05(4)]}$ &  $\text{0.27(3)}$  &  $\text{0.26(3)}$  &  $\text{0.22(3)}$  &  $\text{0.14(2)}$    \\
			\cline{2-7}
			&[$C_{S_1},C_T$] &$\text{[0.10(5), -0.01(2)]}$ &  $\text{0.27(3)}$  &  $\text{0.26(3)}$  &  $\text{0.23(3)}$  &  $\text{0.14(2)}$   \\
			\cline{2-7}
			&[$C_{S_2},C_T$] &$\text{[-1.25(6), 0.23(3)]}$ &  $\textbf{0.01(4)}$  &  $\textbf{-0.01(4)}$  &  $\textbf{-0.03(3)}$  &  $\textbf{-0.03(3)}$   \\
			\hline
		\end{tabular}
	}
	\caption{Bin prediction for $\hat{K}_{2ss}$ observable.}
	\label{tab:obsbinprdall4}
\end{table}
%%%%%%%%%%%%%%%%%%%%%%%%%%%%%%%%%%%%%%%%%%%%%	
\begin{table}
	\centering
	\resizebox{0.65\textwidth}{!}{
		\begin{tabular}{|*{7}{c|}}
			\hline\hline
			\multirow{2}{*}{\bf Observables}  & \multicolumn{2}{|c|}{ \bf Scenario } & \multicolumn{4}{|c|}{\bf $\bf q^2$ bin(in \text{$\bf GeV^2$})}\\
			\cline{4-7}
			& \multicolumn{2}{|c|}{\bf (Real $C_i$'s)} &\textbf{$\bf q^2_{min}$-5}  &  \textbf{5-7}  &  $\textbf{7-9}$  &  \textbf{9- $ \bf q^2_{max}$ }\\
			\hline\hline
			&  \multicolumn{2}{c|}{\bf SM ($Re(C_i) = 0$)}   &  $\text{0.007(1)}$  &  $\text{0.019(3)}$  &  $\text{0.026(3)}$  \
			&  $\text{0.021(2)}$    \\
			\cline{2-7}
			$\hat{K}_{3s}$  &  \multirow{2}{*}{$C_{S_1}$}  &  $\text{[0.1]}$  &  $\text{-0.005(7)}$  &  $\text{0.017(7)}$  &  $\text{0.050(8)}$  &  $\text{0.096(11)}$    \\
			\cline{3-7}
			&&  $\text{[-0.1]}$  &  $\text{0.003(8)}$  &  $\text{0.020(3)}$  &  $\text{0.074(10)}$  &  $\text{0.132(15)}$    \\
			\cline{2-7}
			&  \multirow{2}{*}{$C_{S_2}$}  &  $\text{[0.1]}$  &  $\text{-0.007(7)}$  &  $\text{0.015(7)}$  &  $\text{0.048(8)}$  &  $\text{0.095(11)}$ \\
			\cline{3-7}
			&&  $\text{[-0.1]}$  &  $\text{0.006(8)}$  &  $\text{0.035(8)}$  &  $\text{0.075(10)}$  &  $\text{0.133(15)}$    \\
			\cline{2-7}
			&  \multirow{2}{*}{$C_{V_1}$}  &  $\text{[0.1]}$  &  $\text{-0.001(8)}$  &  $\text{0.025(8)}$  &  $\text{0.061(9)}$ &  $\text{0.114(13)}$  \\
			\cline{3-7}
			&&  $\text{[-0.1]}$  &  $\text{-0.001(8)}$  &  $\text{0.025(8)}$  &  $\text{0.061(9)}$ &  $\text{0.114(1)}$    \\
			\cline{2-7}
			&  \multirow{2}{*}{$C_{V_2}$}  &  $\text{[0.1]}$  &  $\text{-0.024(7)}$  &  $\text{-0.007(7)}$  &  $\text{0.026(7)}$  &  $\text{0.083(10)}$   \\
			\cline{3-7}
			&&  $\text{[-0.1]}$  &  $\text{0.024(8)}$  &  $\text{0.056(9)}$  &  $\text{0.094(12)}$  &  $\text{0.140(15)}$   \\
			\cline{2-7}
			&  \multirow{2}{*}{$C_{T}$} &  $\text{[0.1]}$  &  $\text{-0.017(7)}$  &  $\text{0.003(7)}$  &  $\text{0.043(8)}$  &  $\text{0.027(3)}$   \\
			\cline{3-7}
			&&  $\text{[-0.1]}$  &  $\text{0.020(7)}$  &  $\text{0.044(8)}$  &  $\text{0.073(9)}$  &  $\text{0.107(12)}$    \\
			\cline{2-7}
			&[$C_{S_1},C_{S_2}$]  &$\text{[-2.27(21), 0.90(22)]}$&  $\text{0.049(17)}$  &  $\textbf{0.098(19)}$  &  $\textbf{0.145(20)}$  &  $\textbf{0.208(23)}$    \\
			\cline{2-7}
			&[$C_{V_1},C_T$] &$\text{[0.08(3), 0.04(4)]}$ &  $\text{-0.008(10)}$  &  $\text{0.017(11)}$  &  $\text{0.056(11)}$  &  $\text{0.116(14)}$  \\
			\cline{2-7}
			&[$C_{V_2},C_T$] &$\text{[0.08(5), -0.06(3)]}$ &  $\text{-0.007(10)}$  &  $\text{0.013(13)}$  &  $\text{0.044(16)}$  &  $\text{0.087(19)}$   \\
			\cline{2-7}
			&[$C_{V_1},C_{V_2}$] &$\text{[-0.98(3), 1.05(2)]}$ &  $\textbf{0.070(14)}$  &  $\textbf{0.107(17)}$  &  $\textbf{0.157(20)}$  &  $\textbf{0.230(25)}$    \\
			\cline{2-7}
			&[$C_{S_1},C_{V_1}$] &$\text{[0.05(7), 0.03(3)]}$  &  $\text{-0.003(8)}$  &  $\text{0.021(9)}$  &  $\text{0.055(12)}$  &  $\text{0.105(18)}$    \\
			\cline{2-7}
			&[$C_{S_1},C_{V_2}$] & $\text{[0.12(5), -0.03(3)]}$ &  $\text{0.002(10)}$  &  $\text{0.025(12)}$  &  $\text{0.057(13)}$  &  $\text{0.101(15)}$    \\
			\cline{2-7}
			&[$C_{S_2},C_{V_1}$] & $\text{[0.04(7), 0.04(3)]}$ &  $\text{-0.004(8)}$  &  $\text{0.020(10)}$  &  $\text{0.055(12)}$  &  $\text{0.105(17)}$   \\
			\cline{2-7}
			&[$C_{S_2},C_{V_2}$] &$\text{[0.14(5), -0.05(4)]}$ &  $\text{0.001(10)}$  &  $\text{0.026(12)}$  &  $\text{0.058(13)}$  &  $\text{0.101(15)}$   \\
			\cline{2-7}
			&[$C_{S_1},C_T$] &$\text{[0.10(5), -0.01(2)]}$ &  $\text{-0.003(8)}$  &  $\text{0.020(9)}$  &  $\text{0.051(10)}$  &  $\text{0.095(14)}$ \\
			\cline{2-7}
			&[$C_{S_2},C_T$] &$\text{[-1.25(6), 0.23(3)]}$ &  $\text{-0.008(11)}$  &  $\text{-0.009(17)}$  &  $\text{-0.010(27)}$  \
			&  $\text{-0.018(46)}$    \\
			\hline
		\end{tabular}
	}
	\caption{Bin prediction for $\hat{K}_{3s}$ observable.}
	\label{tab:obsbinprdall5}
\end{table}
%%%%%%%%%%%%%%%%%%%%%%%%%%%%%%%%%%%%%%%%%%
\begin{table}
	\centering
	\resizebox{0.65\textwidth}{!}{
		\begin{tabular}{|*{7}{c|}}
			\hline\hline
			\multirow{2}{*}{\bf Observables}  & \multicolumn{2}{|c|}{ \bf Scenario } & \multicolumn{4}{|c|}{\bf $\bf q^2$ bin(in \text{$\bf GeV^2$})} \\
			\cline{4-7}
			& \multicolumn{2}{|c|}{\bf (Real $C_i$'s)} &\textbf{$\bf q^2_{min}$-5}  &  \textbf{5-7}  &  $\textbf{7-9}$  &  \textbf{9- $ \bf q^2_{max}$ }\\
			\hline\hline
			&\multicolumn{2}{c|}{\bf SM ($Re(C_i) = 0$)}&  $\text{0.007(1)}$  &  $\text{0.019(3)}$  &  $\text{0.026(3)}$  &  $\text{0.021(3)}$    \\
			\cline{2-7}
			$\hat{K}_{3sc}$ &  \multirow{2}{*}{$C_{S_1}$}  &  $\text{[0.1]}$  &  $\text{0.006(1)}$  &  $\text{0.018(3)}$  &  $\text{0.025(3)}$  &  $\text{0.020(2)}$  \\
			&&  $\text{[-0.1]}$  &  $\text{0.007(1)}$  &  $\text{0.020(3)}$  &  $\text{0.028(4)}$  &  $\text{0.022(3)}$    \\
			\cline{2-7}
			&\multirow{2}{*}{$C_{S_2}$} &  $\text{[0.1]}$  &  $\text{0.006(1)}$  &  $\text{0.019(3)}$  &  $\text{0.025(3)}$  &  $\text{0.020(2)}$    \\
			&&  $\text{[-0.1]}$  &  $\text{0.007(1)}$  &  $\text{0.020(3)}$  &  $\text{0.027(4)}$  &  $\text{0.022(3)}$    \\
			\cline{2-7}
			&\multirow{2}{*}{$C_{V_1}$}  &  $\text{[0.1]}$  &  $\text{0.007(1)}$  &  $\text{0.019(3)}$  &  $\text{0.026(3)}$  &  $\text{0.021(3)}$    \\
			&&  $\text{[-0.1]}$  &  $\text{0.007(1)}$  &  $\text{0.019(3)}$  &  $\text{0.026(3)}$  &  $\text{0.021(3)}$   \\
			\cline{2-7}
			&\multirow{2}{*}{$C_{V_2}$}  &  $\text{[0.1]}$  &  $\text{0.007(1)}$  &  $\text{0.020(3)}$  &  $\text{0.028(4)}$  &  $\text{0.024(3)}$   \\
			&&  $\text{[-0.1]}$  &  $\text{0.006(1)}$  &  $\text{0.018(3)}$  &  $\text{0.024(3)}$  &  $\text{0.019(2)}$   \\
			\cline{2-7}
			&\multirow{2}{*}{$C_{T}$} &  $\text{[0.1]}$  &  $\text{0.006(1)}$  &  $\text{0.019(3)}$  &  $\text{0.029(4)}$  &  $\text{0.027(3)}$    \\
			&&$\text{[-0.1]}$  &  $\text{0.005(1)}$  &  $\text{0.012(2)}$  &  $\text{0.016(2)}$  &  $
			\text{0.012(2)}$    \\
			\cline{2-7}
			&[$C_{S_1},C_{S_2}$]  &$\text{[-2.27(21), 0.90(22)]}$&  $\text{0.010(2)}$  &  $\text{0.022(4)}$  &  $\text{0.024(3)}$  &  $\text{0.016(2)}$    \\
			\cline{2-7}
			&[$C_{V_1},C_T$] &$\text{[0.08(3), 0.04(4)]}$ &  $\text{0.007(1)}$  &  $\text{0.020(3)}$  &  $\text{0.029(4)}$  &  $\text{0.024(4)}$    \\
			\cline{2-7}
			&[$C_{V_2},C_T$] &$\text{[0.08(5), -0.06(3)]}$ &  $\text{0.006(1)}$  &  $\text{0.016(3)}$  &  $\text{0.022(4)}$  &  $\text{0.017(3)}$   \\
			\cline{2-7}
			&[$C_{V_1},C_{V_2}$] &$\text{[-0.98(3), 1.05(2)]}$ &  $\textbf{-0.007(1)}$  &  $\textbf{-0.019(3)}$  &  $\textbf{-0.027(3)}$  \
			&  $\textbf{-0.022(3)}$    \\
			\cline{2-7}
			&[$C_{S_1},C_{V_1}$] &$\text{[0.05(7), 0.03(3)]}$  &   $\text{0.006(1)}$  &  $\text{0.019(3)}$  &  $\text{0.026(3)}$  &  $\text{0.021(3)}$    \\
			\cline{2-7}
			&[$C_{S_1},C_{V_2}$] & $\text{[0.12(5), -0.03(3)]}$ &  $\text{0.006(1)}$  &  $\text{0.018(3)}$  &  $\text{0.024(3)}$  &  $\text{0.019(3)}$   \\
			\cline{2-7}
			&[$C_{S_2},C_{V_1}$] & $\text{[0.04(7), 0.04(3)]}$ &  $\text{0.007(1)}$  &  $\text{0.019(3)}$  &  $\text{0.026(3)}$  &  $\text{0.021(3)}$    \\
			\cline{2-7}
			&[$C_{S_2},C_{V_2}$] & $\text{[0.14(5), -0.05(4)]}$ &  $\text{0.006(1)}$  &  $\text{0.018(3)}$  &  $\text{0.024(3)}$  &  $\text{0.019(3)}$    \\
			\cline{2-7}
			&[$C_{S_1},C_T$] &$\text{[0.10(5), -0.01(2)]}$ &  $\text{0.006(1)}$  &  $\text{0.018(3)}$  &  $\text{0.024(3)}$  &  $\text{0.019(3)}$    \\
			\cline{2-7}
			&[$C_{S_2},C_T$] &$\text{[-1.25(6), 0.23(3)]}$ &  $\text{0.002(2)}$  &  $\text{0.006(5)}$  &  $\textbf{0.009(6)}$  &  $\text{0.009(5)}$    \\
			\hline
		\end{tabular}
	}
	\caption{Bin prediction for $\hat{K}_{3sc}$ observable.}
	\label{tab:obsbinprdall6}
\end{table}
%%%%%%%%%%%%%%%%%%%%%%%%%%%%%%%%%%%%%%%%%%%%%%%%%%%%%%%%%%%%%%%	
\begin{table}
	\centering
	\resizebox{0.65\textwidth}{!}{
		\begin{tabular}{|*{7}{c|}}
			\hline\hline
			\multirow{2}{*}{\bf Observables}  & \multicolumn{2}{|c|}{ \bf Scenario } & \multicolumn{4}{|c|}{\bf $\bf q^2$ bin(in \text{$\bf GeV^2$})} \\
			\cline{4-7}
			& \multicolumn{2}{|c|}{\bf (Real $C_i$'s)} &\textbf{$\bf q^2_{min}$-5}  &  \textbf{5-7}  &  $\textbf{7-9}$  &  \textbf{9- $ \bf q^2_{max}$ } \\
			\hline\hline
			&  \multicolumn{2}{c|}{\bf SM ($Re(C_i) = 0$)} &  $\text{0.40(4)}$  &  $\text{0.38(4)}$  &  $\text{0.34(4)}$  &  $\text{0.22(2)}$    \\
			\cline{2-7}
			$A_{FB}^{\Lambda_c}$  &  \multirow{2}{*}{$C_{S_1}$}  &  $\text{[0.1]}$  &  $\text{0.40(4)}$  &  $\text{0.38(4)}$  &  $\text{0.34(4)}$  &  $\text{0.22(2)}$    \\
			\cline{3-7}
			&&  $\text{[-0.1]}$  &  $\text{0.40(4)}$  &  $\text{0.38(4)}$  &  $\text{0.34(4)}$  &  $\text{0.22(2)}$    \\
			\cline{2-7}
			&  \multirow{2}{*}{$C_{S_2}$}  &  $\text{[0.1]}$  &  $\text{0.40(4)}$  &  $\text{0.37(4)}$  &  $\text{0.33(4)}$  &  $\text{0.21(2)}$    \\
			\cline{3-7}
			&&  $\text{[-0.1]}$  &  $\text{0.41(4)}$  &  $\text{0.39(4)}$  &  $\text{0.34(4)}$  &  $\text{0.22(2)}$    \\
			\cline{2-7}
			&  \multirow{2}{*}{$C_{V_1}$} &  $\text{[0.1]}$  &  $\text{0.40(4)}$  &  $\text{0.38(4)}$  &  $\text{0.34(4)}$  &  $\text{0.22(2)}$    \\
			\cline{3-7}
			&&  $\text{[-0.1]}$  &  $\text{0.40(4)}$  &  $\text{0.38(4)}$  &  $\text{0.34(4)}$  &  $\text{0.22(2)}$   \\
			\cline{2-7}
			&  \multirow{2}{*}{$C_{V_2}$}  &  $\text{[0.1]}$  &  $\text{0.40(4)}$  &  $\text{0.39(4)}$  &  $\text{0.36(4)}$  &  $\text{0.24(3)}$    \\
			\cline{3-7}
			&&  $\text{[-0.1]}$  &  $\text{0.39(4)}$  &  $\text{0.36(4)}$  &  $\text{0.31(3)}$  &  $\text{0.19(2)}$    \\
			\cline{2-7}
			&  \multirow{2}{*}{$C_{T}$}  &  $\text{[0.1]}$  &  $\text{0.37(4)}$  &  $\text{0.39(4)}$  &  $\text{0.39(4)}$  &  $\text{0.29(3)}$    \\
			\cline{3-7}
			&&  $\text{[-0.1]}$  &  $\text{0.27(3)}$  &  $\text{0.25(3)}$  &  $\text{0.21(2)}$  &  $\text{0.13(1)}$    \\
			\cline{2-7}
			&[$C_{S_1},C_{S_2}$]  &$\text{[-2.27(21), 0.90(22)]}$&  $\text{0.40(4)}$  &  $\text{0.38(4)}$  &  $\text{0.35(4)}$  &  $\text{0.24(3)}$  \\
			\cline{2-7}
			&[$C_{V_1},C_T$] &$\text{[0.08(3), 0.04(4)]}$ &  $\text{0.41(4)}$  &  $\text{0.40(5)}$  &  $\text{0.37(5)}$  &  $\text{0.25(4)}$    \\
			\cline{2-7}
			&[$C_{V_2},C_T$] &$\text{[0.08(5), -0.06(3)]}$ &  $\text{0.35(5)}$  &  $\text{0.33(5)}$  &  $\text{0.29(4)}$  &  $\text{0.18(3)}$    \\
			\cline{2-7}
			&[$C_{V_1},C_{V_2}$] &$\text{[-0.98(3), 1.05(2)]}$ &  $\textbf{-0.40(4)}$  &  $\textbf{-0.38(4)}$  &  $\textbf{-0.34(4)}$  &  $\textbf{-0.22(3)}$    \\
			\cline{2-7}
			&[$C_{S_1},C_{V_1}$] &$\text{[0.05(7), 0.03(3)]}$  &  $\text{0.40(4)}$  &  $\text{0.38(4)}$  &  $\text{0.34(4)}$  &  $\text{0.22(2)}$    \\
			\cline{2-7}
			&[$C_{S_1},C_{V_2}$] & $\text{[0.12(5), -0.03(3)]}$ &  $\text{0.40(4)}$  &  $\text{0.38(4)}$  &  $\text{0.33(4)}$  &  $\text{0.21(2)}$    \\
			\cline{2-7}
			&[$C_{S_2},C_{V_1}$] & $\text{[0.04(7), 0.04(3)]}$ &  $\text{0.40(4)}$  &  $\text{0.38(4)}$  &  $\text{0.34(4)}$  &  $\text{0.21(2)}$    \\
			\cline{2-7}
			&[$C_{S_2},C_{V_2}$] &$\text{[0.14(5), -0.05(4)]}$ & $\text{0.39(4)}$  &  $\text{0.37(4)}$  &  $\text{0.32(4)}$  &  $\text{0.20(2)}$   \\
			\cline{2-7}
			&[$C_{S_1},C_T$] &$\text{[0.10(5), -0.01(2)]}$ &  $\text{0.39(5)}$  &  $\text{0.37(4)}$  &  $\text{0.32(4)}$  &  $\text{0.20(3)}$   \\
			\cline{2-7}
			&[$C_{S_2},C_T$] &$\text{[-1.25(6), 0.23(3)]}$ &  $\textbf{0.01(6)}$  &  $\textbf{-0.02(5)}$  &  $\textbf{-0.05(5)}$  &  $\textbf{-0.05(4)}$    \\
			\hline\hline
		\end{tabular}
	}
	\caption{Bin prediction for $A_{FB}^{\Lambda_c}$ observable.}
	\label{tab:obsbinprdall7}
\end{table}
%%%%%%%%%%%%%%%%%%%%%%%%%%%%%%%%%%%%%%%%%%%%%%%%%%
\begin{table}
	\centering
	\resizebox{0.65\textwidth}{!}{
		\begin{tabular}{|*{7}{c|}}
			\hline\hline
			\multirow{2}{*}{\bf Observables}  & \multicolumn{2}{|c|}{ \bf Scenario } & \multicolumn{4}{|c|}{\bf $\bf q^2$ bin(in \text{$\bf GeV^2$})}\\
			\cline{4-7}
			& \multicolumn{2}{|c|}{\bf (Real $C_i$'s)} &\textbf{$\bf q^2_{min}$-5}  &  \textbf{5-7}  &  $\textbf{7-9}$  &  \textbf{9- $ \bf q^2_{max}$ }\\
			\hline\hline
			&  \multicolumn{2}{c|}{\bf SM ($Re(C_i) = 0$)} &  $\text{0.11(1)}$  &  $\text{0.03(1)}$  &  $\text{-0.03(1)}$  &  $\text{-0.08(1)}$    \\
			\cline{2-7}
			$A_{FB}^{\Lambda_c \tau}$  &  \multirow{2}{*}{$C_{S_1}$}  &  $\text{[0.1]}$ \
			&  $\text{0.11(1)}$  &  $\text{0.04(1)}$  &  $\text{-0.02(1)}$  &  $\text{-0.07(1)}$   \\
			\cline{3-7}
			&&  $\text{[-0.1]}$  &  $\text{0.10(1)}$  &  $\text{0.02(1)}$  &  $\text{-0.04(1)}$  &  $\text{-0.10(1)}$  \\
			\cline{2-7}
			&  \multirow{2}{*}{$C_{S_2}$}  &  $\text{[0.1]}$  &  $\text{0.11(1)}$  &  $\text{0.04(1)}$  &  $\text{-0.02(1)}$  &  $\text{-0.07(1)}$ \\
			\cline{3-7}
			&&  $\text{[-0.1]}$  &  $\text{0.11(1)}$  &  $\text{0.03(1)}$  &  $\text{-0.04(1)}$  &  $\text{-0.10(1)}$    \\
			\cline{2-7}
			&  \multirow{2}{*}{$C_{V_1}$}  &  $\text{[0.1]}$  &  $\text{0.11(1)}$  &  $\text{0.03(1)}$  &  $\text{-0.03(1)}$  &  $\text{-0.08(1)}$    \\
			\cline{3-7}
			&&  $\text{[-0.1]}$  &  $\text{0.11(1)}$  &  $\text{0.03(1)}$  &  $\text{-0.03(1)}$  &  $\text{-0.08(1)}$    \\
			\cline{2-7}
			&  \multirow{2}{*}{$C_{V_2}$}  &  $\text{[0.1]}$  &  $\text{0.11(1)}$  &  $\text{0.04(1)}$  &  $\text{-0.016(4)}$  &  $\text{-0.07(1)}$   \\
			\cline{3-7}
			&&  $\text{[-0.1]}$  &  $\text{0.10(1)}$  &  $\text{0.02(1)}$  &  $\text{-0.04(1)}$  &  $\text{-0.10(1)}$    \\
			\cline{2-7}
			&  \multirow{2}{*}{$C_{T}$}  &  $\text{[0.1]}$  &  $\text{0.10(1)}$  &  $\text{0.03(1)}$  &  $\text{-0.03(1)}$  &  $\text{-0.10(1)}$    \\
			\cline{3-7}
			&&  $\text{[-0.1]}$  &  $\text{0.06(1)}$  &  $\text{0.004(6)}$  &  $\text{-0.04(1)}$  &  $\text{-0.08(1)}$    \\
			\cline{2-7}
			&[$C_{S_1},C_{S_2}$]  &$\text{[-2.27(21), 0.90(22)]}$&  $\textbf{-0.11(2)}$  &  $\textbf{-0.18(2)}$  &  $\textbf{-0.19(2)}$  &  $\textbf{-0.19(2)}$    \\
			\cline{2-7}
			&[$C_{V_1},C_T$] &$\text{[0.08(3), 0.04(4)]}$ &  $\text{0.11(1)}$  &  $\text{0.04(1)}$  &  $\text{-0.03(1)}$  &  $\text{-0.09(1)}$    \\
			\cline{2-7}
			&[$C_{V_2},C_T$] &$\text{[0.08(5), -0.06(3)]}$ &  $\text{0.09(2)}$  &  $\text{0.03(1)}$  &  $\text{-0.02(1)}$  &  $\text{-0.07(1)}$    \\
			\cline{2-7}
			&[$C_{V_1},C_{V_2}$] &$\text{[-0.98(3), 1.05(2)]}$ & $\textbf{-0.20(2)}$  &  $\textbf{-0.19(2)}$  &  $\textbf{-0.19(2)}$  &  $\textbf{-0.20(2)}$    \\
			\cline{2-7}
			&[$C_{S_1},C_{V_1}$] &$\text{[0.05(7), 0.03(3)]}$  &  $\text{0.11(1)}$  &  $\text{0.04(1)}$  &  $\text{-0.02(1)}$  &  $\text{-0.08(1)}$    \\
			\cline{2-7}
			&[$C_{S_1},C_{V_2}$] & $\text{[0.12(5), -0.03(3)]}$ &   $\text{0.11(1)}$  &  $\text{0.04(1)}$  &  $\text{-0.02(1)}$  &  $\text{-0.07(1)}$    \\
			\cline{2-7}
			&[$C_{S_2},C_{V_1}$] & $\text{[0.04(7), 0.04(3)]}$ &   $\text{0.11(1)}$  &  $\text{0.04(1)}$  &  $\text{-0.02(1)}$  &  $\text{-0.08(1)}$    \\
			\cline{2-7}
			&[$C_{S_2},C_{V_2}$] &$\text{[0.14(5), -0.05(4)]}$ &  $\text{0.11(1)}$  &  $\text{0.03(1)}$  &  $\text{-0.02(1)}$  &  $\text{-0.07(1)}$    \\
			\cline{2-7}
			&[$C_{S_1},C_T$] &$\text{[0.10(5), -0.01(2)]}$ &  $\text{0.11(1)}$  &  $\text{0.04(1)}$  &  $\text{-0.02(1)}$  &  $\text{-0.07(1)}$    \\
			\cline{2-7}
			&[$C_{S_2},C_T$] &$\text{[-1.25(6), 0.23(3)]}$ &   $\textbf{0.02(2)}$  &  $\text{0.02(1)}$  &  $\text{0.02(1)}$  &  $\text{0.02(3)}$    \\
			\hline\hline
		\end{tabular}
	}
	\caption{Bin prediction for $A_{FB}^{\Lambda_c \tau}$ observable.}
	\label{tab:obsbinprdall8}
\end{table}
%%%%%%%%%%%%%%%%%%%%%%%%%%%%%%%%%%%%%%%%%%%%%%%%%%%%%%%%%%%%%%%	
\begin{table}
	\centering
	\resizebox{0.65\textwidth}{!}{
		\begin{tabular}{|*{8}{c|}}
			\hline\hline
			\multirow{2}{*}{\bf Observables}  & \multicolumn{2}{|c|}{ \bf Scenario } & \multicolumn{4}{|c|}{\bf $\bf q^2$ bin(in \text{$\bf GeV^2$})} \\
			\cline{4-7}
			& \multicolumn{2}{|c|}{\bf (Real $C_i$'s)} &\textbf{$\bf q^2_{min}$-5}  &  \textbf{5-7}  &  $\textbf{7-9}$  &  \textbf{9- $ \bf q^2_{max}$ } \\
			\hline\hline
			& \multicolumn{2}{c|}{\bf SM ($Re(C_i) = 0$)}&  $\text{0.272(14)}$  &  $\text{0.113(12)}$  &  $\text{-0.002(7)}$  &  $\text{-0.050(4)}$  \\
			\cline{2-7}
			$A_{FB}^{\tau}$  &  \multirow{2}{*}{$C_{S_1}$} &  $[0.1]$  &  $\text{0.280(13)}$  &  $\text{0.129(11)}$  &  $\text{0.0178(71)}$ &  $\text{-0.035(4)}$   \\
			\cline{3-7}
			&& $[-0.1]$  &  $\text{0.262(14)}$  &  $\text{0.094(12)}$  &  $\text{-0.024(7)}$  \
			&  $\text{-0.067(4)}$   \\
			\cline{2-7}
			& \multirow{2}{*}{$C_{S_2}$}  &  $[0.1]$  &  $\text{0.273(14)}$  &  $\text{0.118(12)}$  &  $\text{0.006(7)}$  &  $\text{-0.043(4)}$   \\
			\cline{3-7}
			&&  $[-0.1]$ &  $\text{0.269(14)}$  &  $\text{0.107(12)}$  &  $\text{-0.011(7)}$  \
			&  $\text{-0.058(4)}$  \\
			\cline{2-7}
			&  \multirow{2}{*}{$C_{V_1}$}  &  $[0.1]$  &  $\text{0.272(14)}$  &  $\text{0.113(12)}$  &  $\text{-0.002(7)}$ &  $\text{-0.050(4)}$  \\
			\cline{3-7}
			&& $[-0.1]$ &  $\text{0.272(14)}$  &  $\text{0.113(12)}$  &  $\text{-0.002(7)}$  \
			&  $\text{-0.050(4)}$    \\
			\cline{2-7}
			&  \multirow{2}{*}{$C_{V_2}$} &  $[0.1]$  &  $\text{0.283(14)}$  &  $\text{0.125(12)}$  &  $\text{0.005(8)}$ &  $\text{-0.052(4)}$   \\
			\cline{3-7}
			&&  $[-0.1]$  &  $\text{0.265(14)}$  &  $\text{0.107(11)}$  &  $\text{-0.002(7)}$  &  $\text{-0.044(3)}$     \\
			\cline{2-7}
			&  \multirow{2}{*}{$C_{T}$} &  $[0.1]$  &  $\text{0.285(17)}$  &  $\text{0.113(15)}$  &  $\text{-0.028(10)}$ &  $\textbf{-0.098(5)}$   \\
			\cline{3-7}
			&&  $[-0.1]$  &  $\text{0.287(14)}$  &  $\text{0.147(11)}$  &  $\textbf{0.047(7)}$  \
			&  $\textbf{-0.006(3)}$    \\
			\cline{2-7}
			&[$C_{S_1},C_{S_2}$]  &$\text{[-2.27(21), 0.90(22)]}$&  $\textbf{-0.231(38)}$       & $\textbf{-0.384(16)}$  &  $\textbf{-0.367(9)}$  \
			&  $\textbf{-0.237(11)}$    \\
			\cline{2-7}
			&[$C_{V_1},C_T$] &$\text{[0.08(3), 0.04(4)]}$ & $\text{0.273(15)}$  &  $\text{0.108(13)}$  &  $\text{-0.016(14)}$ &  $\text{-0.068(17)}$ \\
			\cline{2-7}
			&[$C_{V_2},C_T$] &$\text{[0.08(5), -0.06(3)]}$ &  $\text{0.284(15)}$  &  $\text{0.133(15)}$  &  $\text{0.024(13)}$ &  $\text{-0.028(10)}$     \\
			\cline{2-7}
			&[$C_{V_1},C_{V_2}$] & $\text{[-0.98(3), 1.05(2)]}$ &  $\textbf{0.459(5)}$  &  $\textbf{0.408(8)}$  &  $\textbf{0.357(10)}$ &  $\textbf{0.235(10)}$    \\
			\cline{2-7}
			&[$C_{S_1},C_{V_1}$] &$\text{[0.05(7), 0.03(3)]}$  &  $\text{0.277(15)}$  &  $\text{0.121(16)}$  &  $\text{0.008(16)}$ &  $\text{-0.043(12)}$    \\
			\cline{2-7}
			&[$C_{S_1},C_{V_2}$] & $\text{[0.12(5), -0.03(3)]}$ & $\text{0.280(14)}$ &  $\text{0.130(13)}$  &  $\text{0.021(11)}$ &  $\text{-0.031(8)}$ \\
			\cline{2-7}
			&[$C_{S_2},C_{V_1}$] & $\text{[0.04(7), 0.04(3)]}$ &  $\text{0.273(14)}$  &  $\text{0.115(12)}$  &  $\text{0.001(9)}$ & $\text{-0.047(6)}$ \\
			\cline{2-7}
			&[$C_{S_2},C_{V_2}$] &$\text{[0.14(5), -0.05(4)]}$ &  $\text{0.270(14)}$  &  $\text{0.115(11)}$  &  $\text{0.007(7)}$ &  $\text{-0.039(5)}$    \\
			\cline{2-7}
			&[$C_{S_1},C_T$] &$\text{[0.10(5), -0.01(2)]}$ &  $\text{0.281(13)}$  &  $\text{0.132(13)}$  &  $\text{0.023(12)}$ &  $\text{-0.030(10)}$    \\
			\cline{2-7}
			&[$C_{S_2},C_T$] &$\text{[-1.25(6), 0.23(3)]}$ &  $\text{0.162(30)}$  &  $\textbf{-0.094(29)}$  &  $\textbf{-0.300(14)}$ &  $\textbf{-0.364(23)}$   \\
			\hline\hline
		\end{tabular}
	}
	\caption{Bin prediction for $A_{FB}^{\tau}$ observable.}
	\label{tab:obsbinprdall9}
\end{table}
%%%%%%%%%%%%%%%%%%%%%%%%%%%%%
\begin{table}
	\centering
	\resizebox{0.65\textwidth}{!}{
		\begin{tabular}{|*{7}{c|}}
			\hline\hline
			\multirow{2}{*}{\bf Observables}  & \multicolumn{2}{|c|}{ \bf Scenario } & \multicolumn{4}{|c|}{\bf $\bf q^2$ bin(in \text{$\bf GeV^2$})}\\
			\cline{4-7}
			& \multicolumn{2}{|c|}{\bf (Real $C_i$'s)} &\textbf{$\bf q^2_{min}$-5}  &  \textbf{5-7}  &  $\textbf{7-9}$  & \textbf{9- $ \bf q^2_{max}$ }\\
			\hline\hline
			&\multicolumn{2}{c|}{\bf SM ($Re(C_i) = 0$)}&  $\text{-0.069(3)}$  &  $\text{-0.141(9)}$  &  $\text{-0.125(9)}$  \
			&  $\text{-0.053(4)}$    \\
			\cline{2-7}
			$C_F^{\tau}$  & \multirow{2}{*}{$C_{S_1}$}  &  $[0.1]$  &  $\text{-0.066(3)}$  &  $\text{-0.134(8)}$  &  $\text{-0.118(8)}$  &  $\text{-0.050(4)}$    \\
			&&  $[-0.1]$  &  $\text{-0.072(4)}$  &  $\text{-0.149(9)}$  &  $\text{-0.132(9)}$  \
			&  $\text{-0.056(4)}$   \\
			\cline{2-7}
			&  \multirow{2}{*}{$C_{S_2}$} &  $[0.1]$  &  $\text{-0.067(3)}$  &  $\text{-0.137(8)}$  &  $\text{-0.120(8)}$  \
			&  $\text{-0.050(4)}$    \\
			&& $[-0.1]$  &  $\text{-0.071(4)}$  &  $\text{-0.145(9)}$  &  $\text{-0.130(9)}$  \
			&  $\text{-0.055(4)}$    \\
			\cline{2-7}
			&  \multirow{2}{*}{$C_{V_1}$}  &  $[0.1]$  &  $\text{-0.069(3)}$  &  $\text{-0.141(9)}$  &  $\text{-0.125(9)}$  \
			&  $\text{-0.053(4)}$    \\
			&&  $[-0.1]$  &  $\text{-0.069(3)}$  &  $\text{-0.141(9)}$  &  $\text{-0.125(9)}$  \
			&  $\text{-0.053(4)}$    \\
			\cline{2-7}
			&  \multirow{2}{*}{$C_{V_2}$}  &  $[0.1]$  &  $\text{-0.071(3)}$  &  $\text{-0.149(8)}$  &  $\text{-0.137(8)}$  \
			&  $\text{-0.061(4)}$    \\
			&&  $[-0.1]$  &  $\text{-0.067(4)}$  &  $\text{-0.134(9)}$  &  $\text{-0.115(9)}$  \
			&  $\text{-0.046(4)}$    \\
			\cline{2-7}
			&  \multirow{2}{*}{$C_T$}  &  $[0.1]$ &  $\text{-0.080(4)}$  &  $\text{-0.179(10)}$  &  $\text{-0.177(11)}$  \
			&  $\textbf{-0.087(6)}$    \\
			&&  $[-0.1]$  &  $\text{-0.061(3)}$  &  $\text{-0.119(7)}$  &  $\text{-0.100(7)}$  \
			&  $\text{-0.039(3)}$    \\
			\cline{2-7}
			&[$C_{S_1},C_{S_2}$]  &$\text{[-2.27(21), 0.90(22)]}$&  $\textbf{-0.107(7)}$  &  $\text{-0.163(13)}$  &  $\text{-0.113(9)}$  &  $\
			\text{-0.041(3)}$   \\
			\cline{2-7} 
			&[$C_{V_1},C_T$] &$\text{[0.08(3), 0.04(4)]}$ &  $\text{-0.073(5)}$  &  $\text{-0.153(15)}$  &  $\text{-0.140(19)}$  &  $\text{-0.062(11)}$    \\
			\cline{2-7}
			&[$C_{V_2},C_T$] &$\text{[0.08(5), -0.06(3)]}$ &  $\text{-0.067(3)}$  &  $\text{-0.134(8)}$  &  $\text{-0.117(8)}$  \
			&  $\text{-0.049(4)}$    \\
			\cline{2-7}
			&[$C_{V_1},C_{V_2}$] &$\text{[-0.98(3), 1.05(2)]}$ &  $\text{-0.070(3)}$  &  $\text{-0.143(9)}$  &  $\text{-0.128(9)}$  \
			&  $\text{-0.055(5)}$   \\
			\cline{2-7}
			&[$C_{S_1},C_{V_1}$] & $\text{[0.05(7), 0.03(3)]}$  &  $\text{-0.068(4)}$  &  $\text{-0.138(10)}$  &  $\text{-0.122(10)}$  \
			&  $\text{-0.052(5)}$    \\
			\cline{2-7}
			&[$C_{S_1},C_{V_2}$] & $\text{[0.12(5), -0.03(3)]}$ &  $\text{-0.065(4)}$  &  $\text{-0.130(9)}$  &  $\text{-0.114(10)}$  \
			&  $\text{-0.047(5)}$   \\
			\cline{2-7}
			&[$C_{S_2},C_{V_1}$] & $\text{[0.04(7), 0.04(3)]}$ &  $\text{-0.068(4)}$  &  $\text{-0.140(9)}$  &  $\text{-0.124(9)}$  \
			&  $\text{-0.052(4)}$    \\
			\cline{2-7}
			&[$C_{S_2},C_{V_2}$] &$\text{[0.14(5), -0.05(4)]}$ &  $\text{-0.066(4)}$  &  $\text{-0.132(9)}$  &  $\text{-0.114(10)}$  \
			&  $\text{-0.047(5)}$    \\
			\cline{2-7}
			&[$C_{S_1},C_T$] &$\text{[0.10(5), -0.01(2)]}$ &  $\text{-0.065(4)}$  &  $\text{-0.131(10)}$  &  $\text{-0.114(10)}$  \
			&  $\text{-0.048(5)}$    \\
			\cline{2-7}
			&[$C_{S_2},C_T$] &$\text{[-1.25(6), 0.23(3)]}$ &  $\text{-0.092(5)}$  &  $\textbf{-0.209(14)}$  &  $\text{-0.212(21)}$  &  $\text{-0.113(18)}$    \\
			\hline\hline
		\end{tabular}
	}
	\caption{Bin prediction for $C_F^{\tau}$ observable.}
	\label{tab:obsbinprdall10}
\end{table}
%%%%%%%%%%%%%%%%%%%%%%%%%%%%%%%%%%%%%%%%%%%%%%%%%%%%%%%%%%%%
\begin{table}
	\centering
	\resizebox{0.65\textwidth}{!}{
		\begin{tabular}{|*{8}{c|}}
			\hline\hline
			\multirow{2}{*}{\bf Observables}  & \multicolumn{2}{|c|}{ \bf Scenario } & \multicolumn{4}{|c|}{\bf $\bf q^2$ bin(in \text{$\bf GeV^2$})}\\
			\cline{4-7}
			& \multicolumn{2}{|c|}{\bf (Real $C_i$'s)} &\textbf{$\bf q^2_{min}$-5}  &  \textbf{5-7}  &  $\textbf{7-9}$  &  \textbf{9- $ \bf q^2_{max}$ }\\
			\hline\hline
			&\multicolumn{2}{c|}{\bf SM ($Re(C_i) = 0$)}  &  $\text{-0.96(1)}$  &  $\text{-0.91(1)}$  &  $\text{-0.80(1)}$  &  $\text{-0.51(1)}$    \\
			\cline{2-7}
			$P_{\Lambda_c}$ & \multirow{2}{*}{$C_{S_1}$}  &$\text{[0.1]}$& $\text{-0.96(1)}$  &  $\text{-0.90(1)}$  &  $\text{-0.80(1)}$  &  $\text{-0.51(1)}$    \\
			\cline{3-7}
			&&  $\text{[-0.1]}$  &  $\text{-0.96(1)}$  &  $\text{-0.91(1)}$  &  $\text{-0.81(1)}$  &  $\text{-0.51(1)}$    \\
			\cline{2-7}
			&  \multirow{2}{*}{$C_{S_2}$} &  $\text{[0.1]}$  &  $\text{-0.95(1)}$  &  $\text{-0.89(1)}$  &  $\text{-0.78(1)}$  &  $\text{-0.49(1)}$    \\
			\cline{3-7}
			&&  $\text{[-0.1]}$  &  $\text{-0.97(1)}$  &  $\text{-0.92(1)}$  &  $\text{-0.82(1)}$  &  $\text{-0.52(1)}$    \\
			\cline{2-7}
			& \multirow{2}{*}{$C_{V_1}$}  &  $\text{[0.1]}$  &  $\text{-0.96(1)}$  &  $\text{-0.91(1)}$  &  $\text{-0.80(1)}$  &  $\text{-0.51(1)}$    \\
			\cline{3-7}
			&&  $\text{[-0.1]}$  &  $\text{-0.96(1)}$  &  $\text{-0.91(1)}$  &  $\text{-0.80(1)}$  &  $\text{-0.51(1)}$    \\
			\cline{2-7}
			&  \multirow{2}{*}{$C_{V_2}$}  &  $\text{[0.1]}$  &  $\text{-0.95(1)}$  &  $\text{-0.92(1)}$  &  $\text{-0.85(1)}$  \
			&  $\text{-0.57(1)}$    \\
			&&$\text{[-0.1]}$  &  $\text{-0.93(1)}$  &  $\text{-0.86(2)}$  &  $\text{-0.74(2)}$  &  $\text{-0.45(1)}$   \\
			\cline{2-7}
			&  \multirow{2}{*}{$C_T$} & $\text{[0.1]}$ & $\text{-0.88(3)}$  &  $\text{-0.92(2)}$  &  $\textbf{-0.92(1)}$  &  $\textbf{-0.69(1)}$  \\
			&&  $\text{[-0.1]}$  &  $\textbf{-0.64(3)}$  &  $\textbf{-0.59(3)}$  &  $\textbf{-0.50(2)}$  &  $\textbf{-0.30(1)}$    \\
			\cline{2-7}
			&[$C_{S_1},C_{S_2}$]  &$\text{[-2.27(21), 0.90(22)]}$& $\text{-0.94(2)}$  &  $\text{-0.91(2)}$  &  $\text{-0.84(2)}$  &  $\text{-0.57(4)}$    \\
			\cline{2-7}
			&[$C_{V_1},C_T$] &$\text{[0.08(3), 0.04(4)]}$ &  $\text{-0.99(1)}$  &  $\text{-0.96(3)}$  &  $\text{-0.88(6)}$  &  $\text{-0.58(7)}$    \\
			\cline{2-7}
			&[$C_{V_2},C_T$] &$\text{[0.08(5), -0.06(3)]}$ &  $\text{-0.83(8)}$  &  $\text{-0.78(7)}$  &  $\text{-0.68(6)}$  &  $\text{-0.43(4)}$    \\
			\cline{2-7}
			&[$C_{V_1},C_{V_2}$] &$\text{[-0.98(3), 1.05(2)]}$ &  $\textbf{0.96(1)}$  &  $\textbf{0.91(1)}$  &  $\textbf{0.82(2)}$  &  $\textbf{0.52(2)}$    \\
			\cline{2-7}
			&[$C_{S_1},C_{V_1}$] &$\text{[0.05(7), 0.03(3)]}$  &  $\text{-0.96(1)}$  &  $\text{-0.91(1)}$  &  $\text{-0.80(1)}$  &  $\text{-0.51(1)}$   \\
			\cline{2-7}
			&[$C_{S_1},C_{V_2}$] & $\text{[0.12(5), -0.03(3)]}$ &  $\text{-0.95(1)}$  &  $\text{-0.89(2)}$  &  $\text{-0.78(2)}$  &  $\text{-0.49(2)}$    \\
			\cline{2-7}
			&[$C_{S_2},C_{V_1}$] & $\text{[0.04(7), 0.04(3)]}$ &  $\text{-0.95(1)}$  &  $\text{-0.90(2)}$  &  $\text{-0.80(2)}$  &  $\text{-0.50(2)}$   \\
			\cline{2-7}
			&[$C_{S_2},C_{V_2}$] & $\text{[0.14(5), -0.05(4)]}$ &  $\text{-0.94(1)}$  &  $\text{-0.87(2)}$  &  $\text{-0.75(3)}$  &  $\text{-0.46(3)}$    \\
			\cline{2-7}
			&[$C_{S_1},C_T$] &$\text{[0.10(5), -0.01(2)]}$ &  $\text{-0.93(4)}$  &  $\text{-0.87(5)}$  &  $\text{-0.76(5)}$  &  $\text{-0.48(4)}$    \\
			\cline{2-7}
			&[$C_{S_2},C_T$] &$\text{[-1.25(6), 0.23(3)]}$ &  $\textbf{-0.03(14)}$  &  $\textbf{0.05(12)}$  &  $\textbf{0.11(11)}$  &  $\textbf{0.13(8)}$    \\
			\hline\hline
		\end{tabular}
	}
	\caption{Bin prediction for $P_{\Lambda_c}$ observable.}
	\label{tab:obsbinprdall11}
\end{table}
%%%%%%%%%%%%%%%%%%%%%%%%%%%%%%%%%%%%%%%%%
%%%%%%%%%%%%%%%%%%%%%%%%%%%%%%%%%%%%%%%%%%%%%%%%%%%%%%%%%%%%
\begin{table}
	\centering
	\resizebox{0.65\textwidth}{!}{
		\begin{tabular}{|*{7}{c|}}
			\hline\hline
			\multirow{2}{*}{\bf Observables}  & \multicolumn{2}{|c|}{ \bf Scenario } & \multicolumn{4}{|c|}{\bf $\bf q^2$ bin(in \text{$\bf GeV^2$})}\\
			\cline{4-7}
			& \multicolumn{2}{|c|}{\bf (Real $C_i$'s)} &\textbf{$\bf q^2_{min}$-5}  &  \textbf{5-7}  &  $\textbf{7-9}$  &  \textbf{9- $ \bf q^2_{max}$ } \\
			\hline\hline
			&\multicolumn{2}{c|}{\bf SM ($Re(C_i) = 0$)}  &  $\text{0.09(2)}$  &  $\text{-0.16(2)}$  &  $\text{-0.34(1)}$  &  $\text{-0.47(1)}$   \\
			\cline{2-7}
			$P^{(\Lambda_c)}_{\tau}$ & \multirow{2}{*}{$C_{S_1}$}  & $\text{[0.1]}$  & $\text{0.13(2)}$  &  $\text{-0.10(2)}$  &  $\text{-0.26(2)}$  &  $\text{-0.39(2)}$  \\
			\cline{3-7}
			&&  $\text{[-0.1]}$  &  $\text{0.04(2)}$  &  $\text{-0.22(1)}$  &  $\textbf{-0.41(1)}$  &  $\textbf{-0.55(1)}$    \\
			\cline{2-7}
			&  \multirow{2}{*}{$C_{S_2}$} &  $\text{[0.1]}$  &  $\text{0.11(2)}$  &  $\text{-0.12(2)}$  &  $\text{-0.28(2)}$  &  $\text{-0.40(2)}$   \\
			\cline{3-7}
			&&  $\text{[-0.1]}$  &  $\text{0.07(2)}$  &  $\text{-0.19(1)}$  &  $\text{-0.38(1)}$  &  $\text{-0.54(1)}$   \\
			\cline{2-7}
			& \multirow{2}{*}{$C_{V_1}$}  &  $\text{[0.1]}$  &  $\text{0.09(2)}$  &  $\text{-0.16(2)}$  &  $\text{-0.34(1)}$  &  $\text{-0.47(1)}$   \\
			\cline{3-7}
			&&  $\text{[-0.1]}$  &  $\text{0.09(2)}$  &  $\text{-0.16(2)}$  &  $\text{-0.34(1)}$  &  $\text{-0.47(1)}$   \\
			\cline{2-7}
			&  \multirow{2}{*}{$C_{V_2}$}  &  $\text{[0.1]}$  &  $\text{0.12(2)}$  &  $\text{-0.11(2)}$  &  $\textbf{-0.27(1)}$  &  $\text{-0.39(2)}$   \\
			&&$\text{[-0.1]}$  &  $\text{0.06(2)}$  &  $\text{-0.20(1)}$  &  $\text{-0.39(1)}$  &  $\textbf{-0.54(1)}$    \\
			\cline{2-7}
			&  \multirow{2}{*}{$C_T$} & $\text{[0.1]}$ & $\text{0.11(2)}$  &  $\text{-0.09(2)}$  &  $\text{-0.28(2)}$  &  $\text{-0.46(2)}$    \\
			&&  $\text{[-0.1]}$  &  $\text{0.01(2)}$  &  $\text{-0.17(1)}$  &      $\text{-0.30(1)}$ &  $\textbf{-0.39(1)}$   \\
			\cline{2-7}
			&[$C_{S_1},C_{S_2}$]  &$\text{[-2.27(21), 0.90(22)]}$& $\textbf{-0.39(3)}$  &  $\text{-0.32(6)}$  &  $\text{-0.20(6)}$  &  $\textbf{-0.14(4)}$    \\
			\cline{2-7}
			&[$C_{V_1},C_T$] & $\text{[0.08(3), 0.04(4)]}$ &  $\text{0.11(2)}$  &  $\text{-0.14(2)}$  &  $\text{-0.33(2)}$  &  $\text{-0.49(2)}$   \\
			\cline{2-7}
			&[$C_{V_2},C_T$] &$\text{[0.08(5), -0.06(3)]}$ &  $\text{0.07(2)}$  &  $\text{-0.14(2)}$  &  $\text{-0.28(4)}$  &  $\text{-0.38(6)}$   \\
			\cline{2-7}
			&[$C_{V_1},C_{V_2}$] &$\text{[-0.98(3), 1.05(2)]}$ &  $\text{0.09(2)}$  &  $\text{-0.15(2)}$  &  $\text{-0.32(2)}$  &  $\text{-0.46(3)}$   \\
			\cline{2-7}
			&[$C_{S_1},C_{V_1}$] &$\text{[0.05(7), 0.03(3)]}$  &  $\text{0.11(3)}$  &  $\text{-0.13(5)}$  &  $\text{-0.30(5)}$  &  $\text{-0.43(6)}$   \\
			\cline{2-7}
			&[$C_{S_1},C_{V_2}$] & $\text{[0.12(5), -0.03(3)]}$ &  $\text{0.13(3)}$  &  $\text{-0.10(3)}$  &  $\text{-0.27(4)}$  &  $\text{-0.40(4)}$   \\
			\cline{2-7}
			&[$C_{S_2},C_{V_1}$] & $\text{[0.04(7), 0.04(3)]}$ &  $\text{0.10(2)}$  &  $\text{-0.14(3)}$  &  $\text{-0.31(4)}$  &  $\text{-0.44(5)}$    \\
			\cline{2-7}
			&[$C_{S_2},C_{V_2}$] &$\text{[0.14(5), -0.05(4)]}$ &  $\text{0.10(2)}$  &  $\text{-0.13(3)}$  &  $\text{-0.29(3)}$  &  $\text{-0.41(4)}$    \\
			\cline{2-7}
			&[$C_{S_1},C_T$] &$\text{[0.10(5), -0.01(2)]}$ &  $\text{0.12(3)}$  &  $\text{-0.10(3)}$  &  $\text{-0.27(4)}$  &  $\text{-0.39(4)}$    \\
			\cline{2-7}
			&[$C_{S_2},C_T$] &$\text{[-1.25(6), 0.23(3)]}$ &  $\text{0.02(4)}$  &  $\textbf{0.04(3)}$  &  $\textbf{0.10(5)}$  &  $\textbf{0.22(9)}$   \\
			\hline\hline
		\end{tabular}
	}
	\caption{Bin prediction for $P^{(\Lambda_c)}_{\tau}$ observable.}
	\label{tab:obsbinprdall12}
\end{table}
%%%%%%%%%%%%%%%%%%%%%%%%%%%%%%%%%%%%%%%%%
\newpage
\bibliographystyle{JHEP}
\bibliography{lmblmc_NPv1} 
\end{document}